\documentclass[PhD,english]{sapthesis_modified}
\pdfoutput=1
\usepackage{amsmath}
\usepackage{amssymb}
\usepackage{bbm}
\usepackage{bbold}
\usepackage{enumerate}
\usepackage[style=chem-angew,maxnames=4,articletitle=true,backend=bibtex]{
biblatex}
\usepackage{graphicx}
\usepackage{color}
\usepackage[colorlinks,linkcolor=blue,urlcolor=blue]{hyperref}

\allowdisplaybreaks

\title{Following the evolution of metastable\\ glassy states under external 
perturbations:\\ compression and shear-strain}
\author{Corrado Rainone}
\IDnumber{1207913}
\course[physics]{fisica}
\courseorganizer{Scuola di Dottorato Vito Volterra - \'Ecole Doctorale 564 
\emph{Physique en \^Ile-de-France}}
\cycle{XXVIII}
\submitdate{October 2015}
\copyyear{2015}
\advisor{Prof. Giorgio Parisi} 
\advisor{Dr. Francesco Zamponi}
\authoremail{corrado.rainone@weizmann.ac.il}
\examiner{Prof. Francesco Sciortino}
\examiner{Prof. Federico Ricci-Tersenghi}
\examiner{Prof. Giulio Biroli}
\examiner{Dr. Marco Tarzia}
\examiner{Prof. Giorgio Parisi} 
\examiner{Dr. Francesco Zamponi}
\examdate{21 December 2015}

\renewcommand{\t}{\tau}
\newcommand{\s}{\sigma}
\newcommand{\g}{\gamma}
\newcommand{\D}{\Delta}

\newcommand{\bq}{\mathbf{q}}
\newcommand{\dav}[1]{\overline{#1}}
\newcommand{\dpart}[2]{\frac{\partial #1}{\partial{#2}}}
\newcommand{\dfunc}[2]{\frac{\delta #1}{\delta #2}}
\newcommand{\dtot}[2]{\frac{d#1}{d#2}}
\newcommand{\G}{\Gamma}

\newcommand{\al}{\alpha}
\newcommand{\be}{\beta}
\newcommand{\lam}{\lambda}
\newcommand{\bx}{\boldsymbol{x}}
\newcommand{\by}{\boldsymbol{y}}

\newcommand{\bk}{\mathbf{k}}
\newcommand{\Sc}{\mathcal{S}}

\newcommand{\e}{\epsilon}
\newcommand{\de}{\delta}
\newcommand{\ox}{\overline{\bx}}
\newcommand{\oy}{\overline{\by}}

\newcommand{\ou}{\overline{u}}
\newcommand{\ov}{\overline{v}}

\newcommand{\bA}{\boldsymbol{A}}
\newcommand{\br}{\boldsymbol{r}}

\newcommand{\erf}{\mathrm{erf}}
\newcommand{\Os}{\mathcal{O}}
\newcommand{\C}{\mathcal{C}}

\newcommand{\beq}{\begin{equation}} 
\newcommand{\eeq}{\end{equation}}
\newcommand{\ba}{\begin{eqnarray}}
\newcommand{\ea}{\end{eqnarray}}
\newcommand{\bann}{\begin{eqnarray*}}
\newcommand{\eann}{\end{eqnarray*}}
\newcommand{\thav}[1]{\left<#1\right>}
\newcommand{\la}{\left<}
\newcommand{\ra}{\right>}
\newcommand{\scl}{\D^{\frac{1}{\kappa}}}

\newcommand{\etal}{\emph{et al.}\ }
\let\a=\alpha \let\b=\beta \let\g=\gamma \let\d=\delta
\let\e=\varepsilon \let\z=\zeta \let\h=\eta \let\k=\kappa
\let\l=\lambda \let\m=\mu  \let\x=\xi 
\let\s=\sigma \let\t=\tau \let\f=\varphi \let\c=\chi
 \let\y=\upsilon  \let\G=\Gamma
\let\D=\Delta \let\Th=\Theta  
\let\Si=\Sigma   
\let\ee=\epsilon \let\r=\rho \let\th=\theta \let\io=\infty

\def\FF{{\cal F}} 
 
\def\LL{{\cal L}}  
\def\DD{{\cal D}}\def\AA{{\cal A}}
 \def\SS{{\cal S}}

\def\eee{\mathrm{e}}

\def\to{\rightarrow} \def\la{\left\langle} \def\ra{\right\rangle}
  \def\erf{\text{erf}}
\def\ol{\overline}
\def\wh{\widehat}

\def\de{\mathrm d}

\def\Dg{\Delta^g}
\def\De{\Delta}

\bibliography{bibliografia.bib,HS.bib}

\begin{document}

\frontmatter

\maketitle

\dedication{Alla mia famiglia,\\e ai miei amici a Roma, a Parigi,\\e ovunque 
altro li porter\`a la vita.}

\begin{acknowledgments}
The research work reported in this thesis has been performed over the course of 
a jointly-supervised PhD which has taken place both in the \'Ecole Normale 
Sup\'erieure of Paris and the University of Rome La Sapienza. First of all, I 
want to thank my supervisor in Paris, Francesco Zamponi, for suggesting this PhD 
arrangement which has allowed me to work with some of the best people in the 
field, and to live an experience of both great scientific and personal growth. I 
thank my supervisor in Rome, Giorgio Parisi, for always providing an 
interesting, instructive and original point of view, even on the apparently 
simplest of problems. I thank both my supervisors for directing and following my 
work during these years and for many useful teachings which have greatly helped 
my scientific growth.

I wish to thank all the collaborators that I've worked with during these years 
(among them Patrick Charbonneau, Yuliang Jin, Manuel Sebastian Mariani, Beatriz 
Seoane Bartolom\`e, Hajime Yoshino) for their hard work and invaluable 
contributions. I thank in particular Pierfrancesco Urbani for many stimulating 
discussions, and for giving me precious advice on many occasions, not only on 
scientific matters. I thank as well all the people that I had inspiring 
discussions with over the course of my PhD (among them Marco Baity-Jesi, Giulio 
Biroli, Chiara Cammarota, Ulisse Ferrari, Hugo Jacquin, Luca Leuzzi, Andrea 
Ninarello, Itamar Procaccia, Federico Ricci-Tersenghi, Tommaso Rizzo, Gilles 
Tarjus, Matthieu Wyart), in particular I wish to thank Francesco Turci for a 
very stimulating discussion during a workshop in Mainz. I take the occasion to 
thank the organizers of all the conferences, workshops and schools that I've 
taken part to over the course of my PhD.

I wish to thank all the friends that I've met in Paris, who have pretty much 
been my family during my stay there, for the many beautiful moments spent 
together: Tommaso Brotto, Tommaso Comparin, Matteo Figliuzzi, Silvia Grigolon, 
Alberto Guggiola, Eleonora De Leonardis, Martina Machet, Elena Tarquini, Gaia 
Tavoni, and all the others.

And finally, a special thank to my family. I would have never been able to do 
all this without their love and support.
\end{acknowledgments}

\tableofcontents

\chapter{Introduction}
This thesis is about glass. As stupid as this assertion looks, it is indeed 
important to state this fact loud and clear, at the very beginning.\\
Why is such an assertion necessary or even appropriate? The glass transition is 
one of the great unresolved problems in condensed matter physics (as the 
introduction of pretty much every work on the subject loves to remind) and it 
has been so for decades. And for decades, research has been produced, and still 
is, to investigate its nature. A thesis in the field of the physics of the glass 
transition which says about itself ``this thesis is about glass'' is therefore 
stating an obvious tautology, at the very best. The aim of this introduction is 
to have the reader understand that it is not so, and that indeed the theoretical 
research on the properties of glasses (as opposed to the huge amount of 
experimental and numerical work that has been done, and is still being done) is 
a relatively new subject that we are beginning to explore now.\\
But a pressing question then arises: what were those ``decades of research'' 
referred to above, about? The answer is: not glasses. Or rather, there has been, 
yes, a ponderous amount of \emph{experimental} research about glasses over the 
last decades (Tool's works about fictive temperature are an example), which we 
will reap and use in this thesis. But the \emph{theoretical} research, the 
research aiming to describe glass-related phenomena at first principle level, 
has not been very concerned with glass itself. Rather, most of the theoretical 
efforts carried out up to now are about supercooled liquids, that is, about 
\emph{equilibrium} properties of glass formers.

This distinction is very important, and yet oftentimes forgotten. Despite this, 
it is indeed pretty obvious from an intuitive point of view. Every research 
article about the glass transition will at some point or another contain a 
sentence of the sort ``...it is impossible to obtain data in this regime due to 
the extremely large time needed to equilibrate the sample...'', and indeed, the 
reason why the glass problem is still open lies mainly in the fact that data in 
the deeply supercooled regime are, to state it in an unambiguous way, impossible 
to obtain. And yet, in everyday life, glasses are just everywhere and are indeed 
quite easy to manufacture; they are not a rare and exotic commodity. But despite 
this, the impression that one gets from the literature is that getting new data 
to better understand the glass problem is always sort of a struggle.\\
The distinction above makes it clear why: what researchers have been, and still 
are, mostly concerned about is the supercooled liquid. And supercooled liquids, 
unlike glasses, are indeed very rare and very valuable objects. It is indeed a 
fact that the various theories about the glass transition that are on the table 
today (Random First Order, Dynamic Facilitation, Frustration Limited Domains 
etc.) were conceived first and foremost as theories about supercooled liquids 
rather than glasses, and their most defining predictions concern the supercooled 
regime; this is the reason why it is into that hard-to-reach regime that those 
much needed data are to be searched for. In such a scenario, the glass is at 
best seen as an enemy (interestingly, much like the crystal) who sneaks in 
during your simulation/experiment and ruins your day by pushing out of 
equilibrium your precious supercooled liquid sample.\\
In this thesis, we are concerned with glasses.

The problem with formulating a theory about glass lies in the fact that a glass 
is an intrinsically \emph{out of equilibrium} object, as opposed to the 
supercooled liquid. This simple fact is at the origin of all problems that are 
commonly encountered when trying to conceive a theory of glasses. If the 
theorist is aiming for a first-principle theory, then the obvious starting point 
is of course statistical mechanics, as in all other branches of theoretical 
condensed matter physics. But statistical mechanics is a framework mainly 
concerned with the properties of \emph{equilibrium} systems, whose thermodynamic 
state is stable, and whose lifetime is infinite. Glass has no such property, as 
we enunciated before: its properties depend of the time $t$ and a glass does not 
live forever, but only until the glass former is able to relax and flow again 
like a liquid. There are theoretical tools conceived for the treatment of 
out-of-equilibrium scenarios, but they are all meant to deal with situations 
wherein the system is subject to a drive of some sort (say, an AC current), and 
they are meant for systems with long-range order. Glass is amorphous, and is out 
of equilibrium because it did not have enough time to relax, not because we are 
perturbing it in some way. So those tools are not suitable for our problem.\\
At this point, it looks like a meaningful theory of glass cannot make do without 
a time-based description, a view which the Dynamical Facilitation Theory (DFT), 
for example, embraces heartily; however, the dynamics of generic many-body 
systems, and in particular glass formers, does not enjoy a unified and commonly 
accepted first-principles framework such as the one that statistical mechanics 
is able to provide for systems in thermodynamic equilibrium. We will see over 
the course of this thesis that this weakness is manifest within DFT, whose 
models are necessarily phenomenological in nature and never start from a 
microscopic, first-principles description of the glass-forming liquid. So, it 
looks like one is between the proverbial rock and hard place: to have a 
first-principles theory, one must try to rely on dynamics; but to rely on 
dynamics, the theoretician necessarily has to sacrifice something in terms of 
microscopic description (like in DFT) or simplifying assumptions (like in Mode 
Coupling Theory).

However, if one actually looks at how the properties of a glass change over time 
(for example, its internal energy $U$ as a function of time, or any other 
convenient observable), one can see that the dependence on $t$ is actually 
pretty simple, i.e. the dynamics looks like a \emph{quasi-equilibrium} process 
wherein the observables of the system remain stable over very long time periods, 
of the order of the impossibly long equilibration time needed to observe the 
supercooled liquid. This picture of glass as a system in quasi-equilibrium (or 
\emph{restricted equilibrium}, as we will say more often) is at the root of the 
Random First Order Theory (RFOT) of the glass transition that this thesis is 
based on.\\
The RFOT posits that the glass transition is, yes, a dynamic phenomenon, but 
that it has a static origin. This origin comes in the form of a \emph{Free 
Energy Landscape} (FEL), which is essentially a very rough landscape (think of a 
golf course, for example) of valleys (minima) separated by ridges (saddles), 
wherein single point, representing the glass former, has to navigate towards the 
bottom of the lowest valley in order to attain equilibration. The dynamics of 
the system then unfolds as a series of downhill jumps over the ridges (an 
\emph{activation event}) separated by long persistence times within the valleys 
(referred to as \emph{metastable states}). The large times needed for activated 
jumps to take place delay the onset of equilibration and cause the system to 
behave in a ``glassy'' manner, and as a result of this, the persistence times 
are so large that the system is effectively trapped (or equivalently, 
equilibrated) inside a metastable state for all times which are relevant for 
experimental and practical purposes. 

What is most important about the FEL is that it is a \emph{static} object, in 
the sense that it is uniquely determined by the equilibrium properties of the 
system, with no dynamics or time in play. Despite the fact that it prominently 
affects the dynamics of the glass former, it can be in principle studied with 
suitable static tools. This scenario opens the possibility that the whole 
phenomenology of glass could be in principle described by focusing on the study 
of the valleys (minima) that the system is trapped into during the time regime 
before equilibration, when the glass exists. In particular, since the system is 
equilibrated within a metastable state, one could in principle construct a 
\emph{restricted thermodynamics} by defining a Gibbs measure which only accounts 
for the micro-configurations which are visited by the system as it vibrates 
inside this single minimum. From such a measure one could then compute a 
partition function, a thermodynamic potential, and finally, physical 
observables.\\
Such a construction is referred to as \emph{State Following} construction within 
the theory of generic systems (not only structural glasses) with a rough FEL and 
a consequent RFOT-like behavior. In this thesis we present and apply this 
construction to a realistic model of glass former, namely Hard Spheres (HS). We 
show how it allows to construct a fully analytic theory of glass, entirely from 
first principles, without the need to resort to dynamical tools; we show how it 
allows to obtain results for physical observables which are in agreement with 
the established phenomenology of glasses, and we show how it is also able to 
provide new insights into, and predictions about, the nature of the glass phase.

This thesis is organized as follows: in chapter \ref{chap:supcool}, we give the 
fundamentals of the glass problem (with emphasis on the central phenomenon of 
the glassy slowdown) as a way to properly introduce RFOT and its physical 
picture; in chapter \ref{chap:metglass}, we review the phenomenology of glasses 
as measured in experiments and simulations, in particular Differential Scanning 
Calorimetry and quasi-static shear strain, corresponding to adiabatic changes of 
the temperature $T$ and of the strain parameter $\g$, respectively; in chapter 
\ref{chap:SFconstruction} we present and review in detail the state following 
construction, along with some of the other tools which can be used within RFOT 
to approach the problem of metastability in general; in chapter \ref{chap:SFRS} 
we perform the state following computation for the HS model in the mean-field 
limit, assuming the simplest possible structure for a glassy minimum (i.e. a 
simple paraboloid), and present the results so obtained; in chapter 
\ref{chap:SFfRSB} we dispense with this last assumption and perform a more 
general computation for a arbitrarily complicated structure of the glassy 
minima, and present the results so obtained; in chapter \ref{chap:isocomplexity} 
we provide some comparison with numerics in a simple, modified HS model which 
allows for a simple analytical treatment and is also very easy to simulate; 
finally, in chapter \ref{chap:conclusions}, we summarize our conclusions and 
provide some suggestion for further research in the field of glass physics.

The results presented in this thesis have been for the great part already 
published, but here we present them in a coherent and hopefully self-contained 
manner. We refer the interested reader to
\begin{itemize}
\item Chapter \ref{chap:SFRS}: \cite{corrado2} \fullcite{corrado2}.
\item Chapter \ref{chap:SFfRSB}: \cite{RainoneUrbani15} 
\fullcite{RainoneUrbani15}.
\item Chapter \ref{chap:isocomplexity}: \cite{corrado} \fullcite{corrado}; and 
\cite{CharbonneauJin15} \fullcite{CharbonneauJin15}.
\end{itemize} 
\begin{figure}[h!]
\begin{center}
\includegraphics[width = 0.75\textwidth]{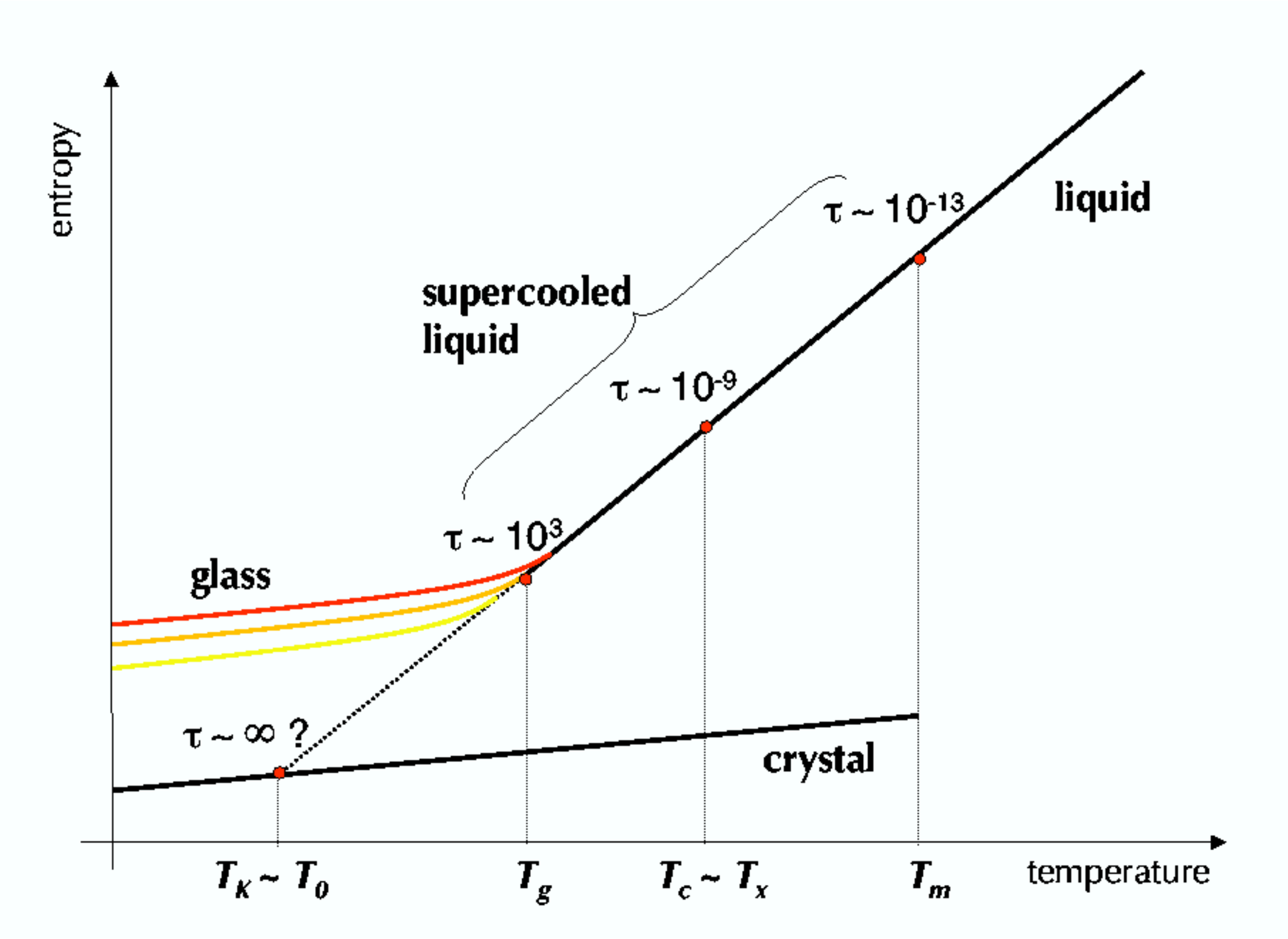}
\caption{The equations of state of a prototypical glass former. We can 
distinguish the crystalline branch, the supercooled branch and, in color, the 
various metastable glass branches. As a guide, in this thesis we will mainly 
focus on cooling protocols wherein one moves down the supercooled branch and 
then into the glassy branches, down to $T=0$. We will not be concerned with the 
deeply supercooled branch (dashed line). Reprinted from 
\cite{CavagnaLiq}.\label{fig:summary}}
\end{center}
\end{figure}

\mainmatter

\chapter{Supercooled liquids and RFOT \label{chap:supcool}}

In this first chapter we focus on the supercooled and deeply supercooled 
branches in figure \ref{fig:summary}, as a way to properly introduce the 
phenomenon of the glassy slowdown, the fundamental notion of \emph{metastable 
glassy state}, and discuss the Random First Order Theory of the glassy slowdown 
upon which this thesis is based. We start by reviewing the basic phenomenology 
of the glass transition, with emphasis on the increase of the relaxation time, 
along with the manifestation of two-step relaxation. From there, we introduce 
the basics of RFOT, which posits that the slowdown of the dynamics in the glassy 
regime can be explained in terms of the insurgence of a great number of 
\emph{metastable glassy states}, which trap the dynamics and hamper structural 
relaxation, thereby forcing the glass former in a metastable, out-of-equilibrium 
glass. We proceed by giving some arguments in support of the RFOT picture, in 
light of the phenomenology of the glassy slowdown discussed earlier. We then 
review summarily the RFOT picture over the course of a conceptual cooling 
experiment on a generic glass former, also discussing the possibility of an 
ideal glass transition. We conclude the chapter with a brief review of some 
other approaches to the glass problem.

\section{The glassy slowdown}

Most liquids (although not all of them \cite{CavagnaLiq}) crystallize upon 
cooling at a certain \emph{melting temperature} $T_m$. However, it is always 
possible, employing some caution, to \emph{supercool} a liquid below its melting 
point, avoiding crystallization and producing a \emph{supercooled liquid}.\\
There are multiple ways to accomplish this. In experiments and industrial 
applications, one usually cools the liquid fast enough that the nucleation and 
growth of the crystal takes place on times much longer that the experimental 
time $t_{exp}$ at which measurements are performed. In simulations, the crystal 
is usually ``killed'' by introducing polydispersity, i.e. by considering a 
liquid whose constituents can have different physical shapes (for example 
spheres with different diameters), so that an ordered, crystalline arrangement 
of the particles is inhibited. We do not delve into this issue and refer the 
reader to the detailed discussion of \cite{CavagnaLiq}.

Once one has managed to obtain a supercooled liquid, it is possible to lower the 
temperature further, always minding the possibility of crystallization. On doing 
so, one can then observe a dramatic increase of the relaxation time (we denote 
it as $\t_R$) (see figure \ref{fig:angell}) over a fairly short range of 
temperature. Besides this sharpness, this sudden growth is also impressive for 
its generality: it manifests in systems that range from atomic liquids, to 
molecular ones, to colloidal compounds and even metallic alloys 
\cite{AngellPlot}: any liquid can form a glass if supercooled fast enough 
\cite{Dy06}. This is already a hint to the fact that the glassy slowdown is a 
fairly general phenomenon, independent of the actual nature of the glass former 
under consideration.\\ 
We remind the reader that the relaxation time can be defined in terms of the viscosity by 
Maxwell's relation \cite{CavagnaLiq,Cates03}
\beq
\eta = G_\infty \t_R,
\eeq
(where $G_\infty$ is the infinite frequency shear modulus) so that the glass 
former becomes more and more sluggish as the temperature is lowered. This 
relation is useful since it allows us to pass from a subtle observable like 
$\t_R$ to a much more tangible physical property like the viscosity.

\subsection{The calorimetric glass transition \label{subsec:glasstrans}}

When the viscosity of a liquid is so high, its ability to flow is severely 
hampered: it takes a time of order $\t_R$ to relax any excitation (for example 
shear) the glass former is submitted to. This means that on experimental 
timescales $t_{exp}<\t_R$ the glass former will effectually respond to an 
external perturbation as if it were an elastic solid, i.e. it will present a 
\emph{shear stress} proportional to the strain \cite{Cates03}
\beq
\s = G_\infty\g.
\eeq
Indeed, if we simply define a solid as any substance that has an elastic 
response, the glass former is effectively a solid on timescales such that 
$t_{exp}<\t_R$.
We stress the fact that this has absolutely nothing to do with the glass 
transition \emph{per se}. The fact that a liquid can respond to shear like a 
solid on short enough timescales is completely general: solidity is indeed a 
timescale-dependent notion \cite{Sausset10}. However, if we put this together 
with the glassy slowdown, we see that the time we would have to wait to see a 
liquid-like response to shear becomes rapidly so large that it becomes 
effectively impossible to do so. When this happens, we get the 
\emph{calorimetric glass transition}, defined as the point where the 
equilibration time of the glass former becomes longer than the experimental 
time, thereby making it a solid from the point of view of the experimentalist. 
We have then the following implicit definition for the calorimetric glass 
transition temperature $T_g$
\beq
\t_R(T_g) \equiv t_{exp}.
\label{eq:Tg}
\eeq
This definition of $T_g$ is the one we are going to follow in the rest of this 
thesis. However, it can be immediately seen that this definition has a problem, 
namely the fact the $t_{exp}$ depends on how our particular experiment (i.e. our 
\emph{protocol}) is designed. It is actually more correct to talk about glass 
transition temperatures, with a plural; but in order to establish a standard, 
the convention is to set $t_{exp}$ to $10^2$ (sometimes $10^3$) 
seconds\footnote{Indeed, the increase of $\t_R$ is so sharp that the actual 
choice of $t_{exp}$ does not make much of a difference.}. This corresponds to 
having for the viscosity
\beq
\eta(T_g) \simeq 10^{13} {\rm Poise},
\eeq
To put this number into perspective, water has a viscosity of about 0.01 Poise, 
and honey's is about between 20 and 100 Poise. A 10cm tall cup containing a 
liquid with a viscosity of $10^{13}$ Poise would take about 30 years to empty 
itself \cite{Dy06}, so this value corresponds by all reasonable standards to a 
solid-like response.
\begin{figure}[htb!]
\begin{center}
\includegraphics[width = 0.5\textwidth]{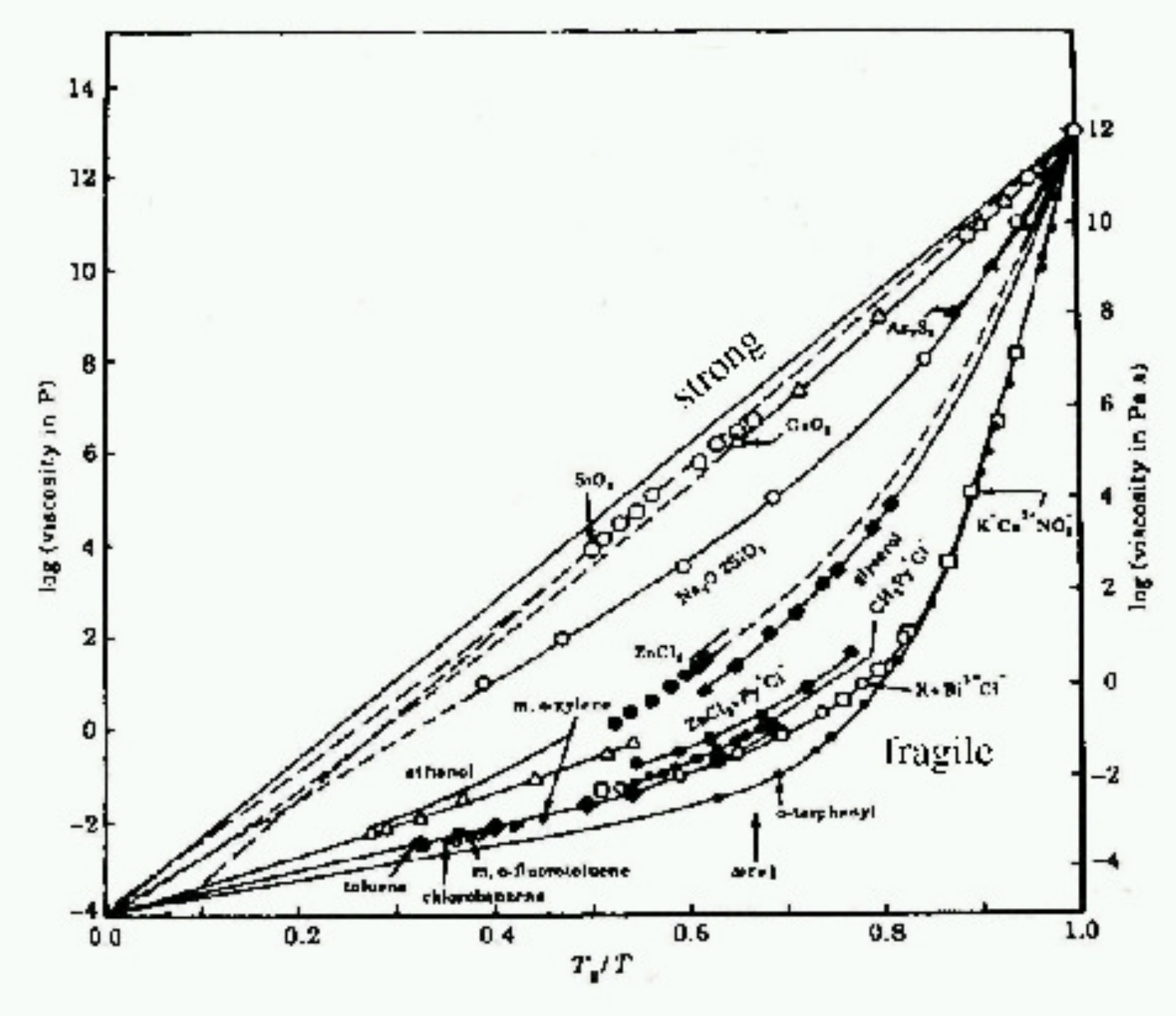}
\caption{Angell's plot. When the ($T/T_g$) ratio is reduced by just one half, the most 
fragile glass formers show an increase of the viscosity (and thus the relaxation 
time) of almost 16 decades, and the viscosity of the strongest ones increases 
anyway of about 10 decades. This stunningly sharp growth is one of the most 
impressive phenomenons in all of low energy physics. Reprinted from 
\cite{AngellPlot}. \label{fig:angell}}
\end{center}
\end{figure}

The definition of $T_g$ allows us to better appreciate the growth of $\t_R$ at 
the onset of the glassy slowdown. We can plot on a logarithmic scale the 
viscosity versus the ratio $T/T_g$ for various glass formers. What we get is the 
plot in figure \ref{fig:angell}, called \emph{Angell's plot}. From Angell's plot 
we can clearly see that the growth of the viscosity (and so of $\t_R$) is at 
least exponential in $T$, and for some glass formers is even sharper. This is 
remarkable especially if one considers that the increase of the viscosity at the 
melting point $T_m$ is much milder \cite{CavagnaLiq}.

The definition of the calorimetric glass transition also allows us to introduce 
a problem which underlies the physics of glasses in general: namely the fact 
that everything has to be defined in a very anthropocentric way. It is 
true, as discussed in \cite{CavagnaLiq}, that the increase of $\t_R$ is so sharp 
that the actual value of $t_{exp}$ doesn't effectively change matters. But this 
does not deny the fact that the only reason why we talk about a calorimetric 
glass transition lies in the fact that we are not patient (or long-lived for 
that matter) enough to observe the equilibration of the glass former below 
$T_g$. One reason for the success of the RFOT theory we are going to discuss 
lies in the fact that it brought all this dynamical, time-based phenomenology 
back to a critical phenomenon with a well defined transition temperature, which 
of course is very appealing to physicists. Nowadays, the actual presence of an 
underlying critical phenomenon is not perceived anymore as a necessity within 
RFOT (an avoided transition would be just as good, as we are going to discuss), 
but it certainly contributed to shaping up the debate in the early days.\\
Nevertheless, we must stress a point: the calorimetric glass transition is not a 
transition, and the only relevant phenomenon is the increase of the relaxation 
time, i.e. the glassy slowdown. This is why this section bears its title.

\subsection{Fragility and the Vogel-Fulcher-Tammann law}
Let us go back to Angell's plot, figure \ref{fig:angell}. As we already said, 
some glass formers have an exponential increase of $\t_R$, while some others 
have an even sharper behavior. When we say ``exponential'', we can't help but 
immediately think about Arrhenius' law
\beq
\t = \t_0 \exp\left(\frac{\D}{k_B T}\right),
\label{eq:arrh}
\eeq
which gives the time needed, for a system at temperature $T$, to overcome an 
energy barrier of height $\D$. The fact that the $\t_R$ vs. $T$ dependence is 
well described by Arrhenius' law already points toward the fact that relaxation 
in supercooled liquids must have something to do with barrier crossing. This is 
the first brick we need to introduce the concept of metastable state.\\
Glass formers which have an Arrhenius-like behavior are referred to as 
\emph{strong} glass formers in glass physics. The champion of strong glass 
formers is undoubtedly Silica (SiO$_2$), namely the ordinary window glass. 
Conversely, those that have a \emph{super-Arrhenius} behavior are dubbed 
\emph{fragile} glass formers. Examples of this class are toluene and 
orto-terphenyl. We remark that the distinction between the two types is not very 
clear-cut (in figure \ref{fig:angell} we can see a variety of behaviors rather 
than two sharply distinct classes), but it's anyway useful.

Because Arrhenius' law is not ok for fragile glass formers, it is automatic to 
ask how we could fit the $\t_R(T)$ dependence for fragile glasses. One possible 
answer has been known, indeed, for quite a long time and is the 
Vogel-Fulcher-Tamman (VFT) law \cite{VFT1,VFT2,VFT3}
\beq
\eta(T) = \eta_\infty \exp\left(\frac{A}{T-T_0}\right).
\eeq
The VFT law is purely phenomenological in nature, nothing more than a fit law 
with three parameters ($\t_0$, $A$ and $T_0$) for viscosity data. However, it 
does a very good job for a great variety of glass formers (see for example 
\cite{VFTtest,Elmatad09} for systematic tests of its validity). We can also see 
that it gives back a pure Arrhenius' law when $T_0=0$, so it also allows to 
interpolate nicely between fragile and strong behavior.

However, we immediately notice that the VFT law says something big: namely, that 
the relaxation time diverges at a temperature $T=T_0$. This is a very strong 
statement, as a divergence of the relaxation time would imply that at $T_0$ 
there is a phase transition. Not a ``calorimetric''  transition with a 
conventional, blurry definition. But a real critical phenomenon with a real 
critical temperature. Such a statement cries for an experimental, unambiguous 
validation.\\
Unfortunately, no such unambiguous validation exist. Although the good job done 
by the VFT law makes it at least reasonable that a divergence exists (and in the 
following we will provide some more arguments in support of its existence), we 
must not forget that the VFT law is just a fit, meant to interpolate data. And 
below $T_g$, by definition, there are no such data. If we follow the convention 
$t_{exp} = 100s$, take for good the VFT law, and choose reasonably $t_{exp} 
\approx 10^{14}\t_0$, $A \approx 10T_0$, we can immediately see that one cannot 
approach the putative critical point more than $\D T\approx \frac{1}{3}T_0$ 
without falling out of equilibrium first. So, since the VFT law is just a fit, 
using it to predict a divergence located so far from the region where there are 
any data to fit looks like an audacious over-stretching. As a matter of fact, 
choosing the ``best'' fit is always a very messy affair. There are indeed 
alternative laws, like B\"assler's law \cite{Bassler87}
\beq
\t_R(T) = \t_0 \exp\left[K\left(\frac{T^*}{T}\right)^2\right],
\label{eq:bassler}
\eeq
which anyway does a comparably good job and contains no divergence whatsoever. 
One could even argue that Arrhenius' law is all we need, since even the $\eta$ vs. $T$ of the most fragile glass formers is approximately a straight line if 
$T$ is close enough to $T_g$, which means that it can be fitted to an Arrhenius' 
law if the range of temperatures is small enough. And there is at least one 
experiment in the literature \cite{Pogna15} wherein, after producing very low-temperature supercooled liquid samples using new techniques\footnote{Namely, 
vapor deposition. We shall discuss it later when we introduce ultrastable 
glasses.}, no super-Arrhenius behavior is actually observed at all.\\
In summary: extrapolations, however appealing they may appear, are insidious. 
The presence or not of a divergence at $T_0$ (see \cite{Hecksher08} for a very 
critical point of view) remains a point of contention to this day.

\subsection{Two-step relaxation \label{subsec:twostep}}

\begin{figure}[htb!]
\begin{center}
\includegraphics[width = 0.75\textwidth]{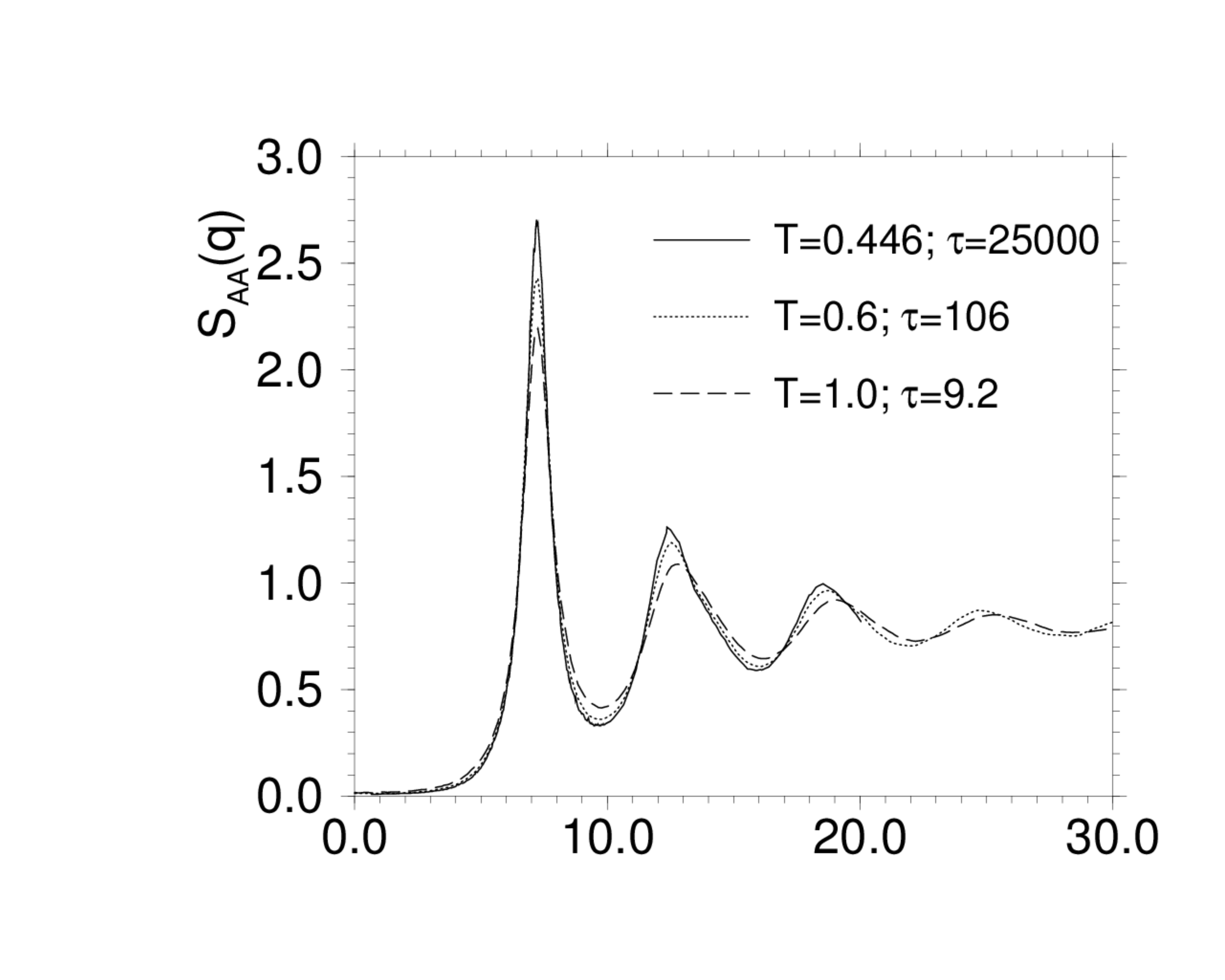}
\caption{The static structure factor of a Lennard-Jones liquid for three different temperatures in the slowdown region. No qualitative changes are observed on lowering $T$. Reprinted from 
\cite{Kob03}.\label{fig:gofr}}
\end{center}
\end{figure}

The growth of $\t_R$ (or equivalently $\eta$) around and below $T_g$ has up to 
now been described with such adjectives as ``dramatic'', ``stunning'', and 
``impressive'', and one can easily verify that the whole literature on the 
subject tends to use similarly grand words when it comes to Angell's plot.\\
When we consider something to be remarkable, it happens because it exceeds our 
expectations. And when we talk about physical quantities, having an expectation 
means having a \emph{scale}, which in our case is $t_{exp}$. However, as we said 
before, $t_{exp}$ is not a fundamental scale in any way, but rather a totally 
anthropocentric and conventional choice. If we were beings with a lifetime such 
that choosing a timescale of, say, $10^{13}$ seconds were reasonable, the growth 
of $\t_R$ would certainly not have appeared as impressive (unless we take a leap 
of faith and believe that a singularity is present at $T_0$), and from a 
qualitative point of view, a glass former would just appear to us as a perfectly 
normal, flowing liquid. So it looks like the phenomenon of the glassy slowdown 
looks exciting merely because we look at it with a built-in timescale that the 
fundamental laws of nature do not share\footnote{My Bachelor's thesis advisor 
was fond of saying ``There only two numbers that matter in physics: zero and 
one''. The reason for this is that we can always choose a scale for the 
phenomena in study and use it to measure the quantities involved. This statement 
applies very well to Angell's plot: nothing 
forbids us to choose a scale such that $\eta(T)\approx O(1)\ \forall T$. And why 
should a constant curve be of any interest for a physicist? From his point of 
view, literally nothing is going on, unless it goes to zero or infinity 
somewhere.}.\\ 
This view seems to be corroborated by the fact that the sudden sluggishness does 
not seem to be accompanied by any structural change whatsoever. If we look at 
the structural properties of a glass former around $T_g$, usually through its 
\emph{static structure factor} $S(\bq)$ \cite{simpleliquids}, no relevant change 
with $T$ is observed \cite{CavagnaLiq,Kob03} (as one can see in figure 
\ref{fig:gofr}), while its relaxation time grows of some 10 orders of magnitude 
in the same temperature range: if we take a snapshot of a glass former near 
$T_g$, which is what the $S(\bq)$ does, it looks exactly like a liquid, with no 
long-range order or Bragg peaks.
This fact is dismaying from a physicist's point of view, considering that the 
wisdom from the theory of critical phenomena suggests that a long relaxation 
time always comes together with a long correlation length (i.e. \emph{critical 
slowing down}) \cite{zinnjustin}. It looks more and more like the glass 
transition is a problem for chemists and material scientists only, certainly not 
for physicists. In this paragraph we explain why it is not so, by showing a 
qualitative fingerprint of glassiness: the two-step relaxation.

Let us consider a model glass former made of $N$ particles. Let us also consider 
a generic time-dependent observable $\Os_i(t)$ relative to particle $i$. We can 
define a \emph{dynamical correlation function}
\beq
C(t,t') = \frac{1}{N}\sum_{i=1}^N \thav{\Os_i(t)\Os_i(t')}.
\eeq
Where $<\bullet>$ denotes an average over the initial condition at $t=0$. Let us 
focus on liquids which are approaching the glass transition, but are still 
equilibrated. In that case the average is carried out using the canonical 
distribution and the dynamics depends only on the difference $t-t'$. This is 
Time Translational Invariance (TTI) \cite{CugliandoloDyn}.\\
For particle glass formers the observable $\Os_i(t)$ is usually the density 
fluctuation (in Fourier space) relative to particle $i$
\beq
\d\rho_i(\bq,t) \equiv \int d\bx\ e^{-i\bq\cdot\bx}\d(\bx-\bx_i(t)) = 
e^{-i\bq\cdot\bx_i(t)}.
\eeq
With this choice, $C(t,0)$ coincides with the intermediate scattering function 
$F_s(\bq,t)$ usually measured in inelastic neutron scattering experiments 
\cite{simpleliquids}.\\
Since to our knowledge the system has only one timescale $\t_R$, we would expect 
for the correlation function a form such as
\beq
F_s(\bq,t) \simeq e^{-\frac{t}{\t_R}}.
\eeq
in principle different $\bq$s may correspond to different $\t_R$s, but the 
variations should only amount to a trivial rescaling that leaves intact the VFT 
dependence on $T$. So we expect a slower and slower, but nonetheless exponential decay on 
approaching $T_g$.
\begin{figure}[htb!]
\begin{center}
\includegraphics[width = 0.6\textwidth]{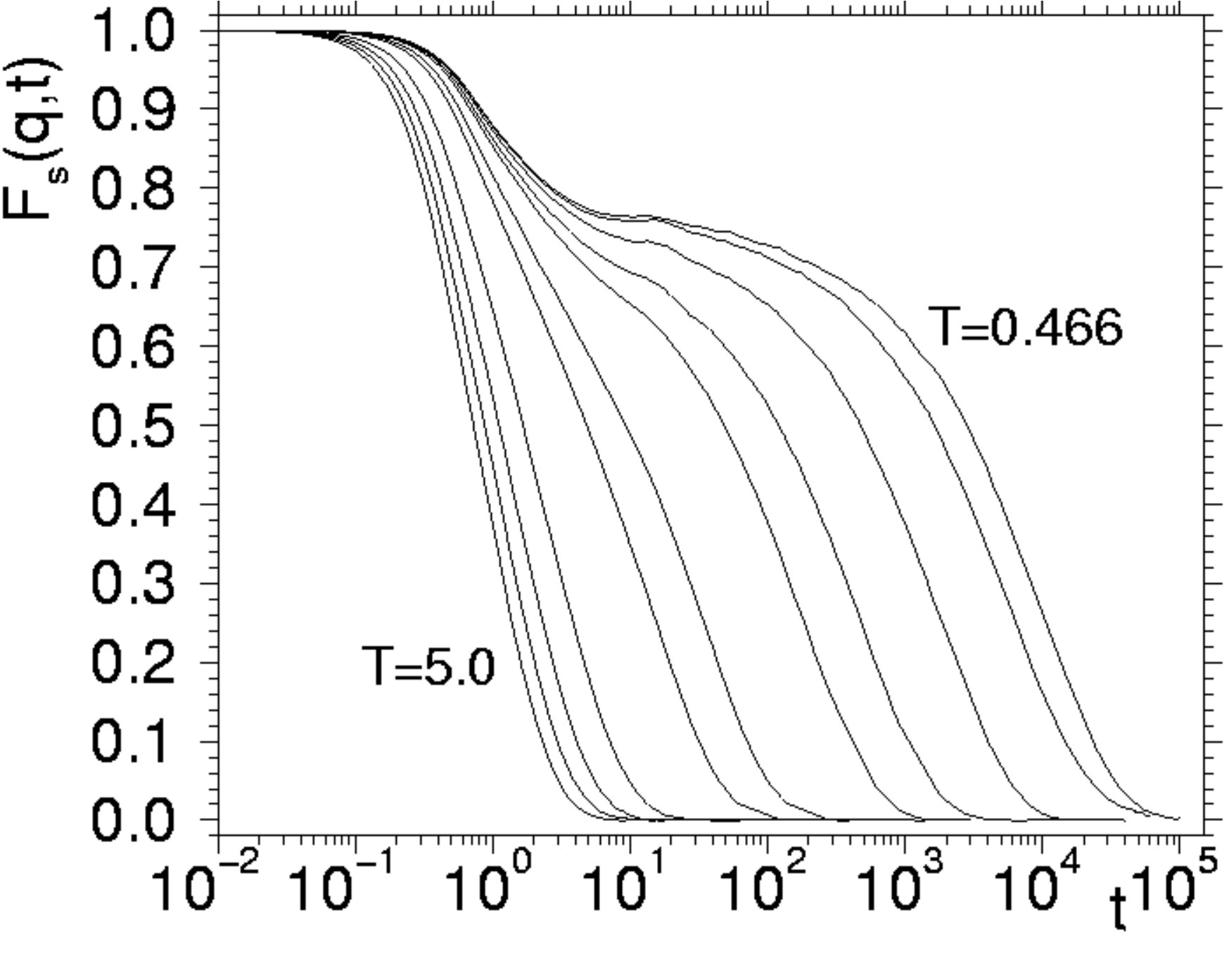}
\caption{The intermediate scattering function obtained from simulation data of a 
Lennard-Jones liquid in the vicinity of $T_g$. As the temperature is lowered, 
two distinct relaxations appear, separated by a plateau of rapidly growing 
length.  Reprinted from \cite{KobAndersen95}.\label{fig:Coft}}
\end{center}
\end{figure}

But this is not what is observed, see figure \ref{fig:Coft}. For high $T$, we 
get the expected exponential relaxation. However, as we approach $T_g$ we can 
see that the correlator changes shape and the relaxation proceeds in two steps: 
firstly a fast relaxation (remember that we are watching things in log time) to 
a plateau with a height $C^*$ different from zero takes place. This first part 
of the relaxation is not very sensible to the onset of the glassy slowdown. 
Then, after another time which grows sharply as the temperature is lowered, we 
get the final structural relaxation. Fittingly, this pattern of decay is called 
\emph{two-step relaxation} and unlike the simple (however impressive) growth of 
$\t_R$, it is indeed a \emph{qualitative} landmark of glassiness, one that will 
stay there even if we change our choice of $t_{exp}$.

So, after all, we were wrong in assuming that the system has only one timescale. 
Indeed, there are two of them, and only one of the two ultimately causes the 
slowdown of the dynamics and the growth of $\t_R$. This means that inside our 
system there is a well defined separation between \emph{fast} processes, which 
yield the initial decay on the plateau and are weakly dependent on $T$, and 
\emph{slow} processes which are on the contrary deeply affected by the onset of 
glassiness. The two different steps are called $\be$ relaxation (for fast 
processes) and $\al$ relaxation (for slow ones), and each of the two has its 
associated timescale, $\t_\be$ and $\t_\al$ respectively. Since structural 
relaxation is of course dominated by the \emph{slowest} processes, we have
\beq
\t_R \simeq \t_\alpha,
\eeq
so that the $\al$-relaxation timescale is the one relevant for equilibration.

The presence of two-step relaxation is our second (an perhaps most important) 
building block towards the concept of metastable state: the presence of the two 
well-separated relaxations, with a long, flat plateau in the middle, seems to 
suggest that a glass can be thought of as a glass former which is 
\emph{partially} equilibrated ($\be$-relaxation has taken place) but still has 
to undergo complete equilibration ($\al$-relaxation), whereupon it becomes a 
supercooled liquid again. This idea of \emph{restricted equilibrium} is the 
fundamental concept behind the State Following construction.\\
The presence of partial relaxation is also important for another reason: if the 
relaxation was a simple (however slow) exponential with a single timescale, then 
every measurement (even of just one-time observables like, say, the pressure) 
made on a timescale $t_{exp}\ll\t_\al$ would have shown a dependence on $t$, 
thereby dooming to fail any idea that glasses can be described by an equilibrium 
(i.e. with no dependence on $t$) approach. The fact that the relaxation of the 
system is effectively frozen on timescales as long as $\t_\al$, however, saves 
us from this problem: there is, of course, a dependence of even one-time 
observables on $t$ (i.e. \emph{aging} \cite{CavagnaLiq,CugliandoloDyn}), but we 
have to wait a very long time to observe it, and before that, one-time 
quantities are effectively constant on timescales which are long, but anyway 
much shorter than $\t_\al \simeq \t_R$ \cite{CavagnaLiq,corrado2}.\\
Now, we can finally be more specific in our glass-supercooled liquid 
distinction. From now on, when we talk about glass, we will mean that we are 
looking at properties of a glass former, below $T_g$, on a timescale such that
\beq
\t_\beta \ll t_{exp} \ll \t_\al,
\eeq
which means that we are looking at the plateau regime. Supercooled liquid 
instead means that we are looking at properties of a glass former when
\beq
t_{exp} \gtrsim \t_\al.
\eeq
Needless to say, and as we anticipated in the introduction, glasses are much 
easier to look at.

\subsection{Real space: the cage\label{subsec:caging}}
How does the relaxation of a glass former \emph{actually} look like? If we take 
a look at the actual movement of the particles, in real space, during 
equilibration, what do we see?\\
In the next two paragraphs we answer this question. Let us define a new 
observable, the \emph{mean square displacement} (MSD) of a tagged particle:
\beq
\langle r^2(t)\rangle \equiv \frac{1}{N}\sum_{i=1}^N 
\thav{|\bx_i(t)-\bx_i(0)|^2},
\eeq
which measures how much a particle is able to move from its initial position as 
time passes. We would expect, for short times, a ballistic regime where $r^2(t) 
\simeq t^2$, followed then by a diffusive regime 
\cite{CavagnaLiq,berthierbiroli11} with $r^2(t) \simeq D\ t$, where $D$ is the 
\emph{diffusion coefficient} \cite{CavagnaLiq}. However, since we already know 
that structural relaxation takes place in a two-step manner, we actually expect 
to see something more interesting.\\
\begin{figure}[htb!]
\begin{center}
\includegraphics[width = 0.6\textwidth]{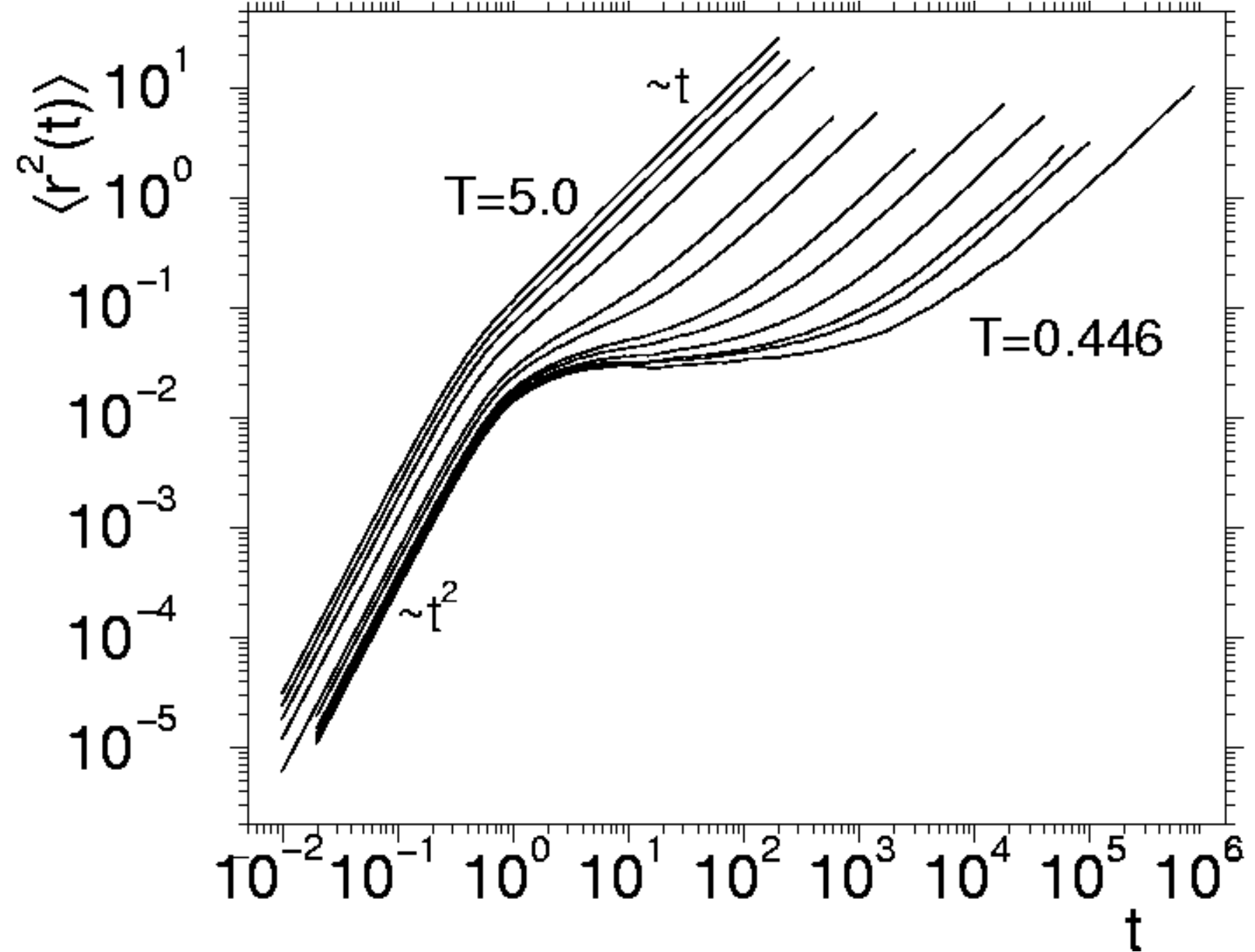}
\caption{The MSD obtained from simulation data of a Lennard-Jones liquid in the 
vicinity of $T_g$. At high temperature, a crossover from ballistic to diffusive 
regime is observed. At the onset of glassiness, a plateau regime in which 
particles are caged appears. Reprinted from 
\cite{KobAndersen95-2}.\label{fig:msd}}
\end{center}
\end{figure}

\begin{figure}[tb!]
\begin{center}
\includegraphics[width = 0.6\textwidth]{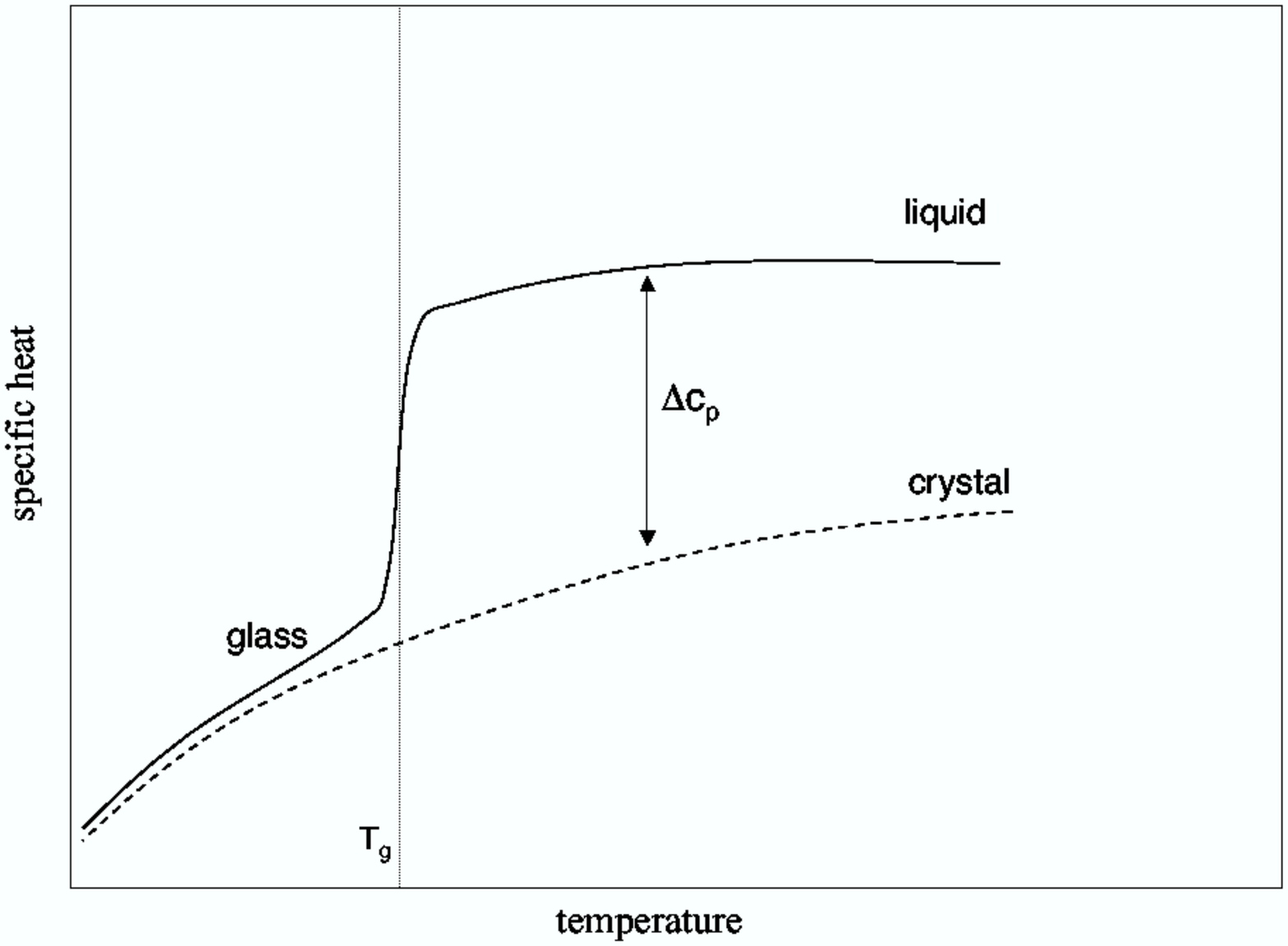}
\caption{Specific heat at constant pressure of a prototypical glass former 
around the calorimetric glass transition. The liquid has a much higher $c_p$, 
but
a sharp drop to the crystalline value is observed as the system becomes glassy. 
Reprinted from \cite{CavagnaLiq}.\label{fig:specheat}}
\end{center}
\end{figure}

And we are not disappointed, see figure \ref{fig:msd}. At high temperature, the 
expected crossover from ballistic to diffusive behavior is observed. As the 
glassy slowdown sets in, a plateau regime, in which particles cannot move, 
manifests between the ballistic and diffusive regimes, similarly to what happens 
for the intermediate scattering function, figure \ref{fig:Coft}. Indeed, the 
timescale necessary to observe diffusion coincides with the $\al$-relaxation 
time $\t_\al$ \cite{KobAndersen95-2}.\\
This provides a picture of two-step relaxation in real space: on a fast 
timescale $\t_\be$, the system undergoes an initial relaxation as particles move 
ballistically. After this, the particles remain stuck for a long time and they 
ability to move is suppressed: this is the \emph{cage 
effect}\cite{CavagnaLiq,berthierbiroli11}. Particles cannot move away because 
they are confined by their neighbors, and thus only vibrate (or rattle) inside 
their respective cages. These vibrations are very small: the MSD in the plateau 
regime is in the range of $10^{-2} -10^{-1}$ particle diameters, and the 
Lindermann ratio, than compares the amplitude of vibrations with the 
intermolecular distance, is only about 10\% in molecular glass formers 
\cite{BB09}.\\ 
This is how a glass looks like when viewed in real space: a system made of 
particles that only vibrate around equilibrium positions which have a disordered 
arrangement in space \cite{Dy06,CavagnaLiq}. A glass is, indeed, an 
\emph{amorphous solid}. It is \emph{solid} because particles are not free to 
move, but rather they can only vibrate around equilibrium positions, like they 
would do in a crystal. But it is also \emph{amorphous} because the equilibrium 
positions have a disordered arrangement in space.\\
This picture is supported by specific heat measurements performed on glass 
formers, see figure \ref{fig:specheat}. At high temperature (that is 
$T_g<T<T_m$), the specific heat of the supercooled liquid is a lot higher than 
the one of the corresponding crystalline solid; this is no surprise, since the 
constituents of the liquid are free to move around and thus they can store much 
more energy than the constituents of the crystal. But when $T_g$ is crossed and 
the glass is formed (remember that we are working at $\t_\beta \ll t_{exp} \ll 
\t_\al$), the specific heat drops and becomes almost equal to that of the 
crystal \cite{CavagnaLiq}: this clearly indicates that the relevant excitations 
in a glass are not too different from the excitations found in 
crystals\footnote{Although they are not the same. We will come back to this when 
we discuss soft modes.}. On timescales $\t_\al$, particles will then be able to 
leave their cages and flow will be 
restored, bringing back the supercooled liquid.

To sum it up, the picture of a glass as a crystal with an amorphous lattice is 
certainly appealing: it is supported by a real-space description of the early 
relaxation in glass formers, and it is also elegant and easy to grasp. 
Nevertheless, we must again take some precaution: first of all, talking about 
``getting out of the cage'' makes it look as if it were a single-particle 
process, while is really a cooperative process (as we will see in the next 
paragraph): all particles are caged and the only way to get out is through 
cooperative motion. Second, a crystal is a stable state of matter, while a glass 
is not: it only lives on timescales much shorter than $\t_\al$, after which 
diffusion sets in and the supercooled liquid comes back.

\subsection{Real space: cooperativity\label{subsec:coop}}

\begin{figure}[htb!]
\begin{center}
\includegraphics[width=0.5\textwidth]{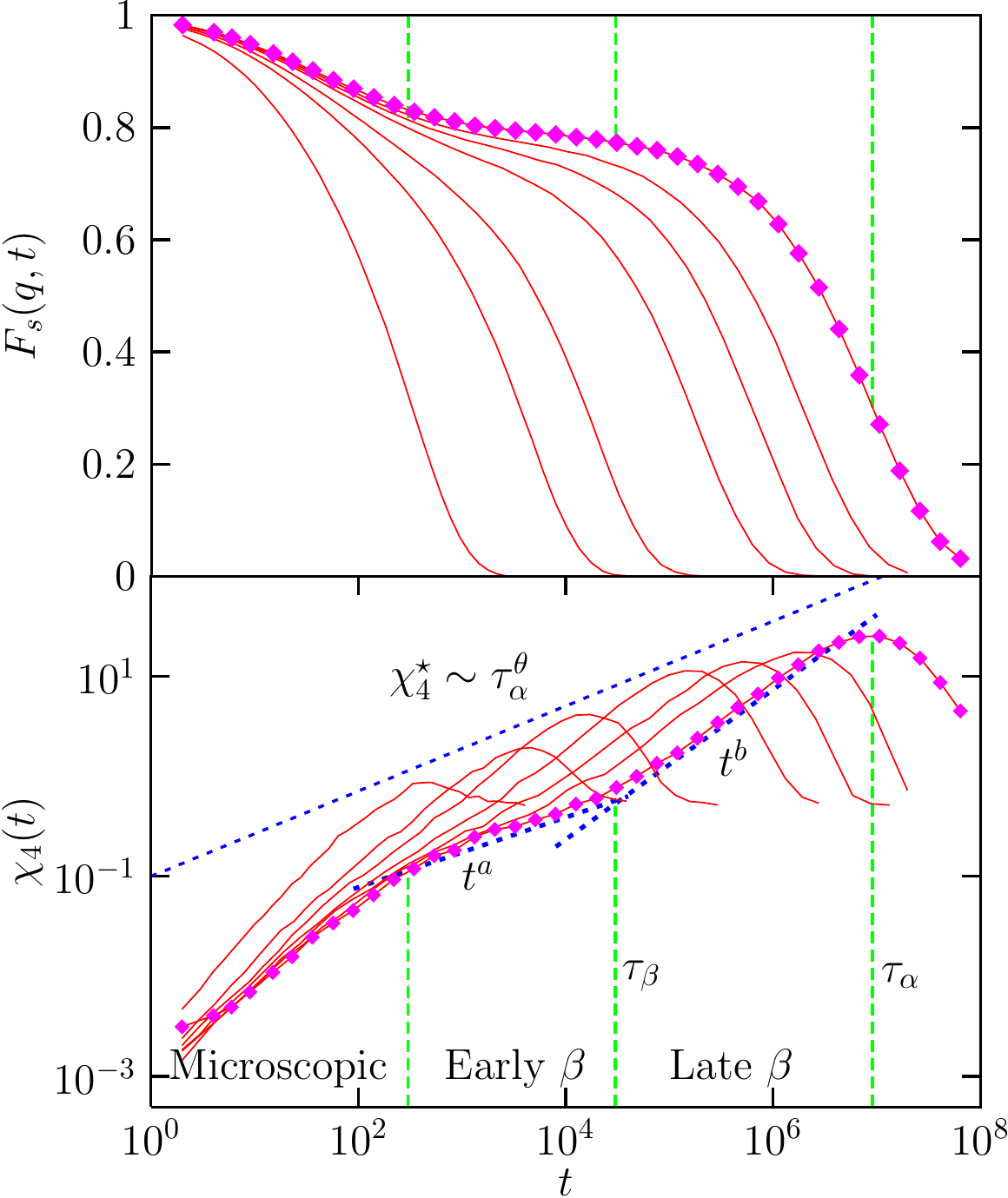}
\caption{The $C(\bx,t)$ (top panel) and the respective $\chi_4(t)$ (lower panel) 
at various temperatures for a supercooled Lennard-Jones mixture in the various 
relaxation regimes. The lowest temperature is highlighted with symbols. The 
maximum of the $\chi_4(t)$ occurs when the $G_4(\bx,t)$ is long-ranged the most, 
which indicates high cooperativity. Unsurprisingly, relaxation is the most 
cooperative at $t^* \simeq \t_\al$. Moreover, the peak of the $\chi_4$ shifts up 
as $T$ is lowered. Reprinted from \cite{berthierbiroli11}.\label{fig:chi4dyn}}
\end{center}
\end{figure}

We focus now on timescales of the order of $\t_\al$, when caging breaks down and 
structural relaxation is reached. Again, how does this process look like in real 
space?\\
We focus again on density fluctuations, but this time in real space: 
\beq
\d\rho(\bx,t) \equiv \sum_{i=1}^{N} \d(\bx-\bx_i(t)) - \rho,
\label{eq:deltarho}
\eeq
where $\rho$ is the number density $N/V$ of the liquid. We want to understand 
how correlated is the motion of particles in the system as relaxation sets in, 
so we have to study how much the correlation in time of the density 
fluctuations, in a certain region of space, is in its turn correlated with the 
same observable, in another point in space. If the correlation is high, it will 
mean that structural relaxation in the first point has to come together with 
relaxation in the other point, which is the definition of cooperativity. To sum 
it up, we have to study the correlations in space of correlations in time.\\
We already know that the correlation in time at a certain point $\bx$ is given 
by the dynamical correlation function
\beq
C(\bx,t) \equiv \thav{\d\rho(\bx,0)\d\rho(\bx,t)},
\eeq
so we just take this definition one step further, but in space, and we define a 
new correlation function, the \emph{four-point correlation function} 
$G_4(\bx,t)$
\beq
G_4(\bx,t) \equiv \thav{\d\rho(0,0)\d\rho(0,t)\d\rho(\bx,0)\d\rho(\bx,t)} - 
\thav{\d\rho(0,0)\d\rho(0,t)}\thav{\d\rho(\bx,0)\d\rho(\bx,t)},
\eeq
which essentially encodes the fluctuations of the dynamical correlation 
function. It is a four-point function because it looks at the correlation 
between two different points in space at two different points in time, rather 
than just two points in space as two-point, ordinary correlation functions do.\\
The necessity of using multi-point correlation functions to detect cooperativity 
in disordered systems was indeed first appreciated in the context of spin 
glasses (we will return to this issue later), rather than supercooled liquids. 
An early discussion about this point can be found in 
\cite{KirkpatrickThirumalai88}. The first study of the $G_4$ in that context is 
reported it \cite{Dasgupta91}, although no interesting results were found at the 
time.
From the $G_4$ we can define a \emph{dynamical susceptibility} in the following 
way
\beq
\chi_4(t) = \int d\bx\ G_4(\bx,t),
\eeq
so that the $\chi_4$ corresponds to the average volume of the regions wherein 
dynamics is cooperative. As those regions grow in size, and the $G_4$ has thus a 
slower decay in space, the $\chi_4(t)$ is supposed to grow. So, if a maximum of 
the $\chi_4$ shows up at a certain time, say $t^*$, then we will know that $t^*$ 
is the time when the relaxation is cooperative the most. Studies of the $\chi_4$ 
in numerical simulations (see for example \cite{FranzDonati99}, and 
\cite{ToninelliWyart05} for a review) and even experiments 
\cite{BerthierBiroli05} have indeed confirmed these expectations, as shown in 
figure \ref{fig:chi4dyn}.
If one superimposes the dynamic correlation function and its corresponding 
$\chi_4$, it can be seen clearly that the maximum of the $\chi_4$, which is the 
fingerprint of cooperativity, manifests during the $\al$-relaxation regime. As 
it was reasonable to expect, the decaging process is highly cooperative and 
requires all particles (or a least an extensive fraction of them) to move, 
destroying the amorphous lattice which had caused the slowdown in the earlier 
phases of relaxation. In addition to this, the value $\chi_4(t^*)$ at the 
maximum shifts up with decreasing temperature (one can in fact see that 
$\chi_4(t^*) \simeq (\t_\al)^\theta$ \cite{berthierbiroli11,ToninelliWyart05}), 
indicating that as temperature is lowered, more and more cooperativity is 
required for the system to attain relaxation and flow. This again is no 
surprise, since more cooperativity requires more time, producing the glassy 
slowdown. This qualitative behavior is remarkably general 
\cite{ToninelliWyart05}.\\
Once the time $t^*$ whereupon cooperativity manifests the most is known, one can 
take the corresponding $G_4(\bx,t=t^*)$ and define a \emph{lengthscale} $\xi_d$, 
called the \emph{dynamical} lengthscale, which gives the average size of clusters 
of cooperative motion in the system. This is a more difficult task than the 
study of the $\chi_4$ since finite-size effects can spoil the result unless 
sufficiently large systems are considered 
\cite{berthierbiroli11,FlennerSzamel10,KarmakarDasgupta09}, but it can be 
carried out nonetheless, see for example 
\cite{FranzDonati99,FlennerSzamel10,KarmakarDasgupta09}.

The study of the phenomenology of these clusters, called \emph{dynamical 
heterogeneities}, is a very rich and active field that reaches far beyond the 
glass transition problem. But it being a feature of the $\al$-relaxation regime 
(and thus, of the supercooled liquid), it is pretty tangential to our subject 
and we will not cover it in this thesis. For the interested reader, we can refer 
to a review on the subject \cite{Ediger00} and a book \cite{Berthier2011}.

At the end of this section, we hope that the reader has been convinced of the 
fact that relaxation in glasses happens on two well-defined timescales, and that 
he has a clear visual representation of how these two phases unfold in real 
space. First, a fast relaxation ($\be$-relaxation) whereupon particles are caged 
and only vibrate around equilibrium positions arranged in an amorphous fashion. 
Then, on timescales $t^* \simeq \t_\al \simeq \t_R$, a second relaxation 
($\al$-relaxation) whereupon particles decage and the whole structure rearranges 
cooperatively.

\section{From the slowdown to RFOT \label{sec:RFOT}}
In the preceding section we have detailed the main signatures of the glassy 
slowdown. Of course a lot more could be said (stretched exponential relaxation, 
Stokes-Einstein violation, etc.), but all these things tend to happen in the 
supercooled liquid, that is on timescales such that $t \simeq \t_\al$ and so 
they are out of our scope. For the interested reader we reference the 
pedagogical review of \cite{CavagnaLiq}, the more technical one of 
\cite{berthierbiroli11}, along with some more reviews and textbooks 
\cite{EAN96,Kob03,Dy06,Gotze91,DeBenedetti97,BK05,leuzziglassy,}. We will 
balance this lack of focus on the supercooled liquid with a more detailed 
treatment of the actual glass, when $\t_\be \ll t \ll \t_\al$.

The problem of formulating a theory of the glassy slowdown has been open for at 
least three decades, and by all appearances is still far from being solved. 
There are at least two reasons for this. The first one, of course, is that 
experiments, simulations and the like are very hard to perform because of the 
impossibly large experimental time which would be needed. As we said in the 
introduction, the theories of the glass transition which are in competition 
today were born as theories of supercooled liquids, and so their most relevant 
predictions, where with ``relevant'' we mean \emph{predictions that could 
actually enable us to validate one theory and falsify the others}, always kick 
in inside a deeply supercooled regime which is experimentally and numerically 
unreachable. A second reason, which is more subtle, is that it is not even very 
clear what a theory of glass transition is actually supposed to do. Since the 
main phenomenon is the glassy slowdown, a theory has at least to explain why the 
slowdown happens and propose a coherent theoretical picture for it. So it has, 
at the very least, to allow one to get back the VFT law, or some alternative 
law, like B\"assler's, to fit the Angell's plot with. Already at this point we 
can see how fishy the situation is: there is not even agreement on which 
predictions the theory is supposed to produce; fits are just fits, after all. As 
a result of this, competition between the various theories is mainly based on 
criteria of theoretical consistency and predictive power 
\cite{berthierbiroli11}, rather than quantitative, stringent tests that are 
impossible to perform and, even when they are performed, always leave some room 
for interpretation wherein incorrect theories could settle and thrive (the 
debate on the VFT law is a good, but definitely not the only, example).\\
If the aim of the theorist is to formulate a ``universal'' theory \emph{\`a la 
Landau} (which is the declared goal and philosophy of RFOT, for example), then 
there is only one universal quantity that such a theory can be able to predict: 
namely, a critical exponent. Unfortunately, apart from the critical exponents of 
MCT \cite{modecoupling} (which are anyway relative to a nonexistent dynamical 
arrest transition, as we are going to discuss), no such critical exponent has 
ever been measured, and indeed, since the glass transition is no transition at 
all, one even wonders where to look for such an exponent. As a matter of fact, 
the greatest, recent success of RFOT consisted in the prediction of the critical 
exponents of the jamming transition in hard spheres \cite{nature}, the jamming 
transition \cite{LNSW10,liunagel98,Biroli07,} being a problem which initially 
was not related to glass forming liquids, if not in a tangential manner. The 
RFOT, as all other theories of the glass transition, was initially conceived as 
something that lives on the equilibrium, supercooled branch in figure 
\ref{fig:summary}. And yet it had to go all the way to $T=0$ on the glassy 
branches (from a very pedestrian standpoint, the jamming transition is basically 
what happens to a large class of glasses when they are quenched down to zero 
temperature) to produce a quantitative, falsifiable prediction of an exponent.\\
This strange fact can however teach us a lesson: theories on the glass 
transition \emph{can} make falsifiable predictions, if only one bothers to look 
at the actual glass, which is the \emph{only} thing we are actually able to look 
at, and that we can experiment on. One of the aims of this thesis is to convince 
the reader that, even tough RFOT struggles (as all other theories do), to affirm 
itself when it must describe supercooled liquids, it is definitely superior to 
(and has a lot more potential than) other theories when it comes to the 
treatment of the metastable glass. In the following we will explain RFOT with 
added focus on the central concept of metastable state, as a way to get this 
point across.

\subsection{The foundations of RFOT \label{subsec:RFOTfoundations}}

The Random First Order Theory of the glass transition is based on three 
conceptual pillars:
\begin{enumerate}
\item The glassy slowdown is caused by the emergence, at low temperature, of a 
large collection of metastable states. The dynamics has to proceed as a series 
of activated barrier jumps between those states, causing the slowdown.
\item These states have a thermodynamic origin, in the sense that they can be 
identified with the minima of a static free-energy functional.
\item These states are exponentially many in the system size $N$, with their 
number given by $\mathcal{N} = e^{\Sigma N}$, where $\Sigma$ is a static 
quantity called \emph{configurational entropy} or \emph{complexity}.
\end{enumerate}
Summarizing, RFOT says that the slowdown of the relaxation dynamics of a liquid 
close to glassiness is due to the fact that it takes place in a very rough 
\emph{free energy landscape} (FEL), characterized by the presence an exponential 
number of minima. RFOT started essentially as a mean-field approach to the study 
of the free-energy landscape in generic disordered systems 
\cite{berthierbiroli11}.

To fix ideas, let us consider the topical example, namely the Curie-Weiss theory 
of ferromagnetism. In that context, one is able to compute the Gibbs Free energy 
$f(m,T)$ of the system, as a function of the global magnetization $m$ 
\cite{replicanotes}. The Helmoltz free energy at zero external magnetic field  
is the Legendre transform of the $f$, so it will be given by the Gibbs free 
energy evaluated in its stationary points $m^*$, $f(m^*(T),T)$.
\begin{figure}[tb!]
\begin{center}
\includegraphics[width=0.5\textwidth]{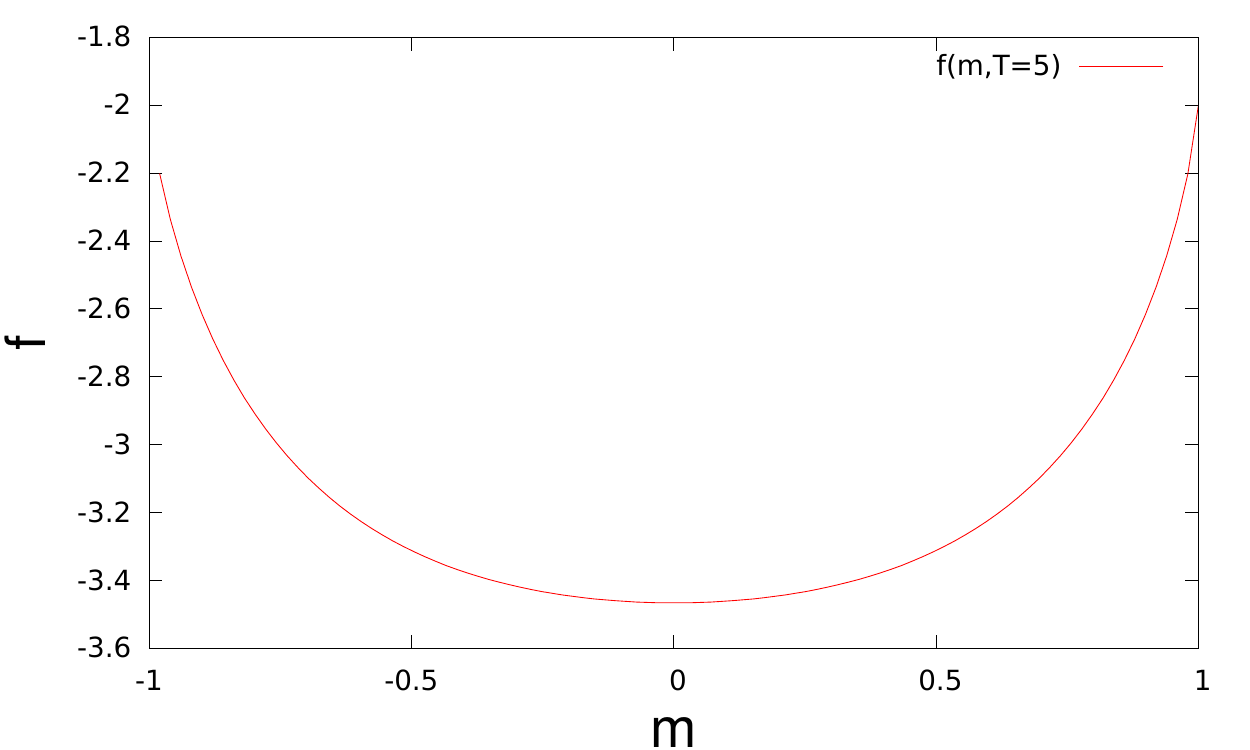}\nolinebreak[4]
\includegraphics[width=0.5\textwidth]{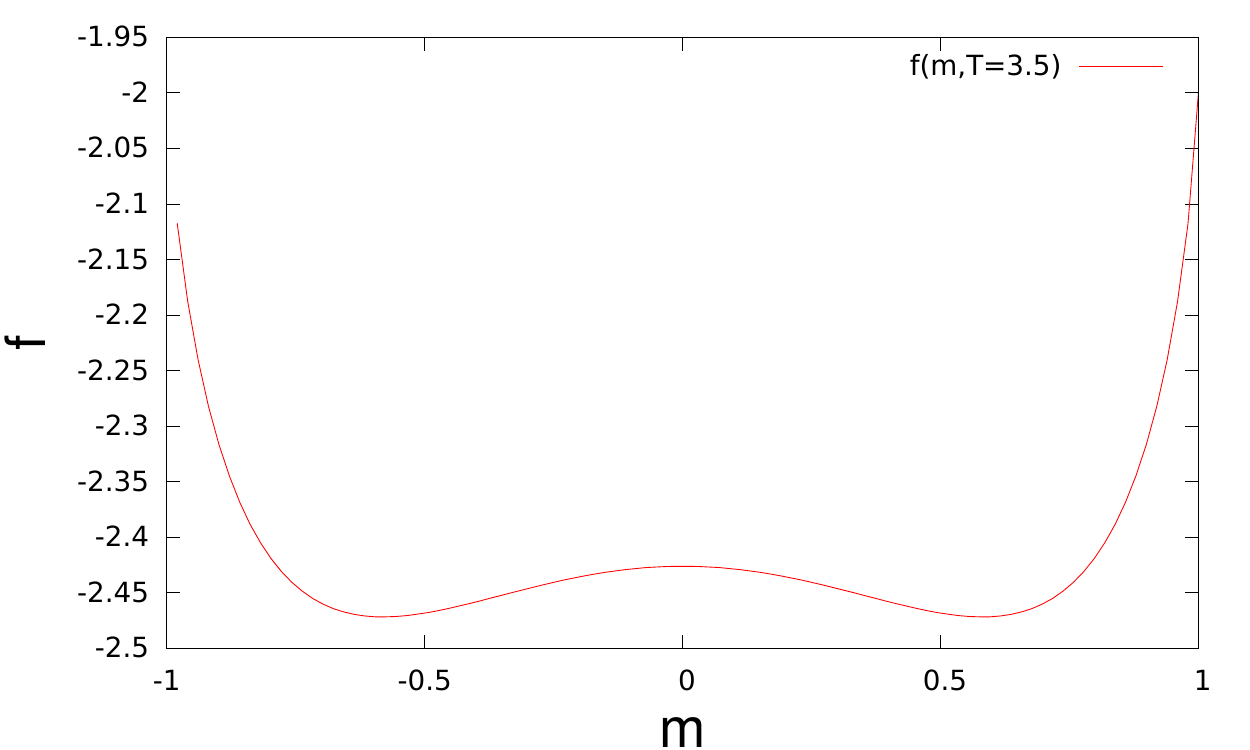}
\caption{Gibbs free energy as a function of the magnetization $m$ for the 
ferromagnetic Curie-Weiss model above (left) and below (right) $T_c$.}
\label{fig:curieweiss}
\end{center}
\end{figure}
At high temperature, only one minimum with zero magnetization is present and the 
system is paramagnetic and ergodic, i.e. it can visit all of the microscopic 
configurations that are allowed by conservation laws. But below a certain 
temperature $T_c$ (see figure \ref{fig:curieweiss}), the paramagnetic minimum 
splits in two distinct minima with $m^* \neq 0$, which correspond to two 
different \emph{states} with opposite magnetizations. A phase transition takes 
place: in the thermodynamic limit, the system cannot go from one state to the 
other because it would have to surmount extensively ($\sim N$) large barriers to 
do so, and ergodicity is broken. From now on, the expressions ``minimum of the 
energy landscape'' and ``metastable state'' are to be considered 
interchangeable.

RFOT follows this basic nucleus, with only one (but very game-changing) 
modification. Since we are considering a disordered system, the Gibbs free 
energy gets replaced with a more complicated free energy functional which is a 
function of a \emph{local} order parameter, rather than a global one. In the 
case of spin systems, it will be a function of all the single-site 
magnetizations $m_i$ and is referred to as Thouless-Anderson-Palmer (TAP) free 
energy \cite{TAP77}. In the case of liquids, it is usually a functional of the 
local density profile \cite{simpleliquids}. For a lattice gas we can for example 
define the Landau potential \cite{berthierbiroli11}
\beq
\Omega(\{\mu_i\},T) \equiv -\frac{1}{\beta}\log\sum_{n_i}\exp \left[-\beta 
H(\{n_i\}) -\beta\sum_{i=1}^N n_i\mu_i \right],
\eeq
where $n_i$ is the site occupation number, $H$ is the Hamiltonian and $\mu_i$ a 
local chemical potential. The free-energy functional $F(\{\rho_i\})$ will then 
be the Legendre transform of the $\Omega$ with respect to all the $\mu_i$s:
\beq
F(\{\rho_i\},T) = \Omega(\{\mu^*_i\}) + \sum_{i=1}^N \mu^*_i\rho_i,
\label{eq:free_en_functional}
\eeq
with the $\mu^*_i$s determined by the condition
\beq
\dpart{\Omega}{\mu_i} + \rho_i = 0.
\eeq
This definition can be generalized to the case of a density profile $\rho(\bx)$ 
in the continuum \cite{simpleliquids,replicanotes}, as we are going to see in 
the following. The FEL is the hyper-surface obtained by scanning the 
$F(\{\rho_i\})$ over all possible values of the local order parameters 
$\rho_i$.\\
Its stationary points, in particular, have cardinal importance. As the $F$ is 
the Legendre transform of the $\Omega$, this means that
\beq
\Omega(\{\mu_i=0\},T) = \underset{\{\rho_i\}}{\rm min}\ F(\{\rho_i\},T),
\eeq
which means that the thermodynamics of the system in absence of external 
chemical potentials is given by the free energy functional computed on its 
stationary points (as it happens if the Curie-Weiss model where the Helmholtz 
free energy is given by the Gibbs free energy computed in its stationary points 
in $m$).\\
With these definitions, the analogy with magnetic systems is clear: if we 
consider local density fluctuations
\beq
\d\rho_i \equiv \rho_i-\rho,
\eeq
where $\rho$ is again the number density, we can see that the homogeneous, high 
temperature liquid ($\d\rho_i = 0$) corresponds to the paramagnet, while the 
glass, with its amorphous nature, would correspond to a disordered ferromagnet 
with a rough free energy landscape. A crystal would not be homogeneous, but it 
would anyway have a periodic $\d\rho_i$ profile, so it would be analogous to an 
anti-ferromagnet \cite{replicanotes}. We can appreciate how the idea of a glass 
as a system with a rough free energy landscape is indeed very reasonable (see 
for example \cite{SW84,SSW85,CKDKS05,ChaudhuriKarmakar08}).

We now proceed to explain why the three tenets of RFOT are coherent with the 
phenomenology presented in the preceding section.

\subsection{Dynamics: MCT and Goldstein's picture \label{subsec:dynamics}} 
Let us start with points 1 and 2. Since those points make assertions about 
dynamics, we consider the theory that has been, up to very recently, the only 
first-principles theory for the dynamics of glass formers: the Mode Coupling 
Theory (MCT) \cite{BengtzeliusGotze84,Leutheusser84,modecoupling}.

\subsubsection{Mode Coupling Theory and the $p$-spin}
The aim of Mode Coupling Theory is to write a closed equation for the 
intermediate scattering function (or equivalently, the dynamical structure 
factor) $F_s(\bq,t)$ for an equilibrated liquid close to glassiness.  Let us 
consider the Newtonian (deterministic, without noise) dynamics of a generic 
liquid made of N particles with positions $\bx_i$ and momenta $\mathbf{p}_i$. 
Every macroscopic, time-dependent observable for such a system will be a 
function of the positions and momenta, $A(t) \equiv 
A(\{\bx_i(t)\},\{\mathbf{p}_i(t)\})$, like the density fluctuations in equation 
\eqref{eq:deltarho}. From Hamilton's equations, one can derive the equation of 
motion for a generic observable $A(t)$
\begin{equation}
\frac{dA}{dt} = \{A(t),H\} \equiv i\mathcal{L}A(t),
\label{eq:liouville}
\end{equation}
where $\{A,B\}$ is the Poisson bracket
\begin{equation}
\{A,B\} = 
\sum_{i=1}^{N}\left(\dpart{A}{\boldsymbol{x}_{i}}\cdot\dpart{B}{\mathbf{p}_{i}} 
- \dpart{B}{\boldsymbol{x}_{i}}\cdot\dpart{A}{\mathbf{p}_{i}}\right),
\end{equation}
and we have defined the Liouville operator
\beq
\mathcal{L}(\bullet) \equiv -i\{\bullet,H\}.
\eeq
We want to write an equation of motion for a correlator $C(t) \equiv 
\thav{A(t)A(0)}$, where $A(t) = \d\rho_i(\bq,t)$, and $\thav{\bullet}$ denotes 
an average over the initial conditions $\bx_i(0)$ and $\mathbf{p}_i(0)$ carried 
out with the canonical distribution. The original derivation of MCT carries out 
this program using Zwanzig's projection operator formalism \cite{Zwanzig01} 
(although field-theoretic derivations are available, see 
\cite{bouchaudcugliandolo}). We skip directly to the final result
\begin{equation}
\frac{d^{2}F(\bq,t)}{dt^{2}} + \frac{\bq^{2}k_{B}T}{mS(q)}F(\bq,t) + 
\frac{m}{Nk_{B}t}\int_{0}^{t}du\ 
\thav{R_{-\mathbf{q}}R_{\mathbf{q}}(t)}\frac{d}{dt}F(\bq,t-u) = 0, 
\label{eqmctesatta}
\end{equation}
where
$$
R_{\bq}(t) = \dtot{J_{\mathbf{q}}^{L}}{t} - 
i\frac{i|\bq|k_{B}T}{mS(\bq)}\delta\rho_{\mathbf{q}}.
$$
and $J^L_\bq$ is the longitudinal current \cite{reichmancharbonneau}. This 
result is exact and does not require any simplifications. The nucleus of MCT 
consist in two uncontrolled approximations that are made on the memory kernel 
$\thav{R_{-\bq}R_\bq(t)}$ in order to get a closed, soluble equation. For the 
sake of brevity, we do not discuss them here and refer the interested reader to 
\cite{reichmancharbonneau}.
space of bilinear density products, and the second in expressing the resulting 
four-point dynamical correlation function as a product of two two-point 
functions $F_s(\bq,t)$ \cite{berthierbiroli11,reichmancharbonneau}.  
At the end of the day, one gets for the memory kernel  
\beq
\frac{m}{Nk_{B}t}\thav{R_{-\mathbf{q}}R_{\mathbf{q}}(t)} = \frac{\rho 
k_{B}T}{16\pi^{3}m}\int d\bk\ 
|\tilde{V}_{\mathbf{q}-\mathbf{k},\mathbf{k}}|^{2}F(k,t)F(|\mathbf{k}-\mathbf{q}
|,t),
\eeq
where the vertex $\tilde{V}_{\mathbf{q}-\mathbf{k},\mathbf{k}}$ has the 
definition
\beq
\tilde{V}_{\mathbf{q}-\mathbf{k},\mathbf{k}} \equiv 
\left\{(\hat\bq\cdot\bk)c(|\bk|)+\hat\bq\cdot(\bq-\bk)c(|\bq-\bk|)\right\},
\label{eq:MCTvertex}
\eeq
and $c(|\bq|)$ is the direct correlation function \cite{simpleliquids}. With 
this expression, one can get a closed integro-differential equation for the 
intermediate scattering function, which can be solved easily once the static 
structure factor $S(\bq)$ is known. MCT is thus capable of predicting the 
relaxation patterns of glass formers from exclusive knowledge of static 
information. Despite the fact that the approximations involved are uncontrolled 
(which means that still today there is no idea as to what we are actually 
discarding in imposing them), this is anyway a remarkable result and MCT has 
enjoyed a lot of success since its inception.

\begin{figure}[htb!]
\begin{center}
\includegraphics[width = \textwidth]{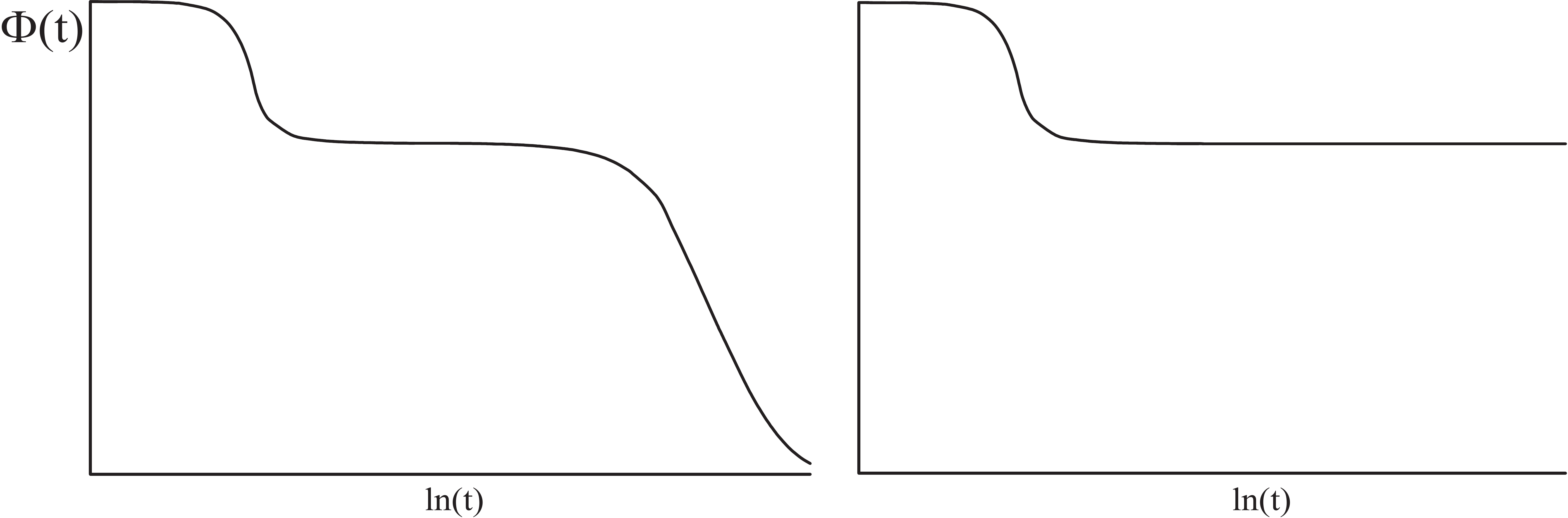}
\caption{The various relaxation patterns predicted by MCT, near the MCT 
transition (left panel) and below it (right panel). Notice the logarithmic scale 
in $t$. Reprinted from \cite{reichmancharbonneau}.}
\label{fig:reichchar}
\end{center}
\end{figure}

The main prediction of MCT is undoubtedly the one of \emph{dynamical arrest}: at 
high temperature the dynamical correlator decays exponentially to zero, as one 
would expect from the discussion in subsection \ref{subsec:twostep}. However, at 
a certain temperature $T_{MCT}$ (sometimes denoted simply as $T_c$), the 
correlator, after an initial fast relaxation, will remain stuck on a plateau and 
the system will never attain equilibrium: an ergodicity breaking takes place. At 
temperatures $T \gtrsim T_{MCT}$, one can observe a two-step decay reminiscent 
of the one discussed in section \ref{subsec:twostep} (see figure 
\ref{fig:reichchar}). In fact, the length of the plateau, which as we already 
know corresponds to the $\al$-relaxation time, goes to infinity on approaching 
$T_{MCT}$ as
\beq
\t_\al \propto \frac{1}{(T-T_{MCT})^\gamma},
\eeq
where the $\gamma$ exponent can be computed from the memory kernel. This sharp 
transition can be interpreted as one from liquid to solid, and since no 
information about a crystalline state was used in the derivation, the solid the 
system freezes into must be non-crystalline, i.e., a \emph{glass} 
\cite{reichmancharbonneau}.

This looks amazing: we have a first principles theory for the dynamics of glass 
formers which is able to predict the two-step relaxation patterns observed in 
simulations and experiments, and also predicts a sharp transition, with 
ergodicity breaking and divergence of the relaxation time, at a certain 
temperature. There is only one small problem, namely that, in pretty much all 
cases
$$
T_{MCT} > T_g,
$$ 
so one can easily go and see if the transition is actually there, and this is 
not the case: the  $\al$-relaxation time does grow sharply, but it stays finite 
and the system remains ergodic at temperatures below $T_{MCT}$. The MCT 
transition does not exist in real glass formers and must therefore be an 
artifact of the theory and its approximations 
\cite{CavagnaLiq,berthierbiroli11}. 

The reason for such a spectacular discrepancy would have been apparent some 
years after. Already in the original papers of Bentgzelius \etal 
\cite{BengtzeliusGotze84} and Leutheusser \cite{ Leutheusser84}, the authors 
proposed a ``schematic'' approximation of the MCT equation, which consisted in 
simplifying the wave vector dependence of the memory kernel, replacing the 
integral over $\bk$ with its value at a certain wave vector $\bk_0$ where the 
static structure factor has a strong peak. With this simplification one gets the 
``schematic'' MCT equation
\beq
\frac{d^{2}\phi(t)}{dt^{2}} + \Omega_{0}^{2}\phi(t) + \lambda\int_{0}^{t}du\ 
\phi^{2}(t-u)\dot{\phi}(u) = 0.
\label{eq:mctschematic}
\eeq 
This equation happened to have the same form as the equation that would have 
been derived in \cite{kirkpatrickthirumalai} for the equilibrium dynamics for a 
certain class of schematic models of disordered ferromagnets, i.e. spin glasses 
(SG)\cite{parisiSGandbeyond}. The generic Hamiltonian for these models was first 
introduced in \cite{GM84} and has the form
\beq
H = \sum_{i_1<i_2<\dots<i_p}J_{i_1,\dots,i_p}\s_{i_1}\s_{i_2}\dots\s_{i_p},
\label{eq:Hpspin}
\eeq
where the couplings $J_{i_1,\dots,i_p}$ are identically, independently 
distributed random variables (usually with a Gaussian probability distribution). 
They are called $p$-spin spherical models (PSMs) because of the $p$-body 
interaction involved (with $p > 2$), and because the spins are soft spins which 
must obey the spherical constraint
\beq
\sum_{i=1}^N \s_i^2 = N
\eeq
Indeed, it was this analogy between MCT and the dynamics of $p$-spin SG that 
gave the original impulse for the formulation of RFOT as a theory for describing 
the glass transition 
\cite{KW87,KW87b,kirkpatrickthirumalai,KT88,KTW89}.

The PSM, although idealized and decidedly far from being a realistic model of a 
glass former, has numerous advantages: both its statics \cite{crisantistatics} 
and dynamics \cite{crisantisommers} can be exactly solved and the properties of 
its free-energy landscape can be studied in great detail. In particular, the 
presence, in a certain range of temperatures, of a great number of metastable 
minima with nonzero magnetization and free energy higher than the paramagnetic 
one (as postulated by RFOT) can be proven analytically \cite{cavagnagiardina98} 
(see \cite{pedestrians} for a review).\\ 
If one looks at the Hamiltonian \eqref{eq:Hpspin}, in can be seen immediately 
that all spins interact with one another: the model is \emph{fully connected} 
and has no space structure, so it is a mean field (MF) model in the traditional 
sense. Because of this, the barriers between minima in the free-energy landscape 
scale as the system size $N$; in the thermodynamic limit, $N$ goes to infinity 
and barriers become in turn infinite: the system is unable to nucleate from one 
state to the other and remains forever stuck in the one it started from, 
producing an hard ergodicity breaking like the one observed in MCT 
\cite{pedestrians}.\\
This invites us to rationalize the ergodicity breaking predicted by MCT as a 
MF-born artifact: in the real world, barriers are always finite and the system 
can always escape from the state is in, even though an extremely large time, of 
the order of $\t_\al$, is needed to do so. Because of this, the glass former 
spends as extremely long time partially equilibrated inside a metastable state 
(and a plateau regime is consequently observed), but it eventually escapes and 
relaxes, restoring ergodicity and bringing back the supercooled liquid. But if 
we accept MCT to have a mean-field nature, because of the analogy with the 
dynamical equations for a MF model (where activated barrier crossing is 
forbidden by construction), then this activated scenario cannot take place: as 
soon as the system finds itself in a state, it cannot escape and ergodicity is 
broken. Today, the status of MCT as a mean-field theory of glassy dynamics, 
although not apparent from direct inspection of the MCT equations, is pretty 
much an accepted and established fact \cite{CavagnaLiq,berthierbiroli11}. 
Furthermore, the recent derivation of the dynamics of hard spheres in the limit 
of infinite spatial dimensions (which as we are going to see corresponds to the 
MF limit) \cite{infdynamics} has shown that the \emph{exact} dynamical equations 
do have an MCT-like form. We refer to \cite{andreanovbiroli,FPRR11} for further 
reading on the MF nature of MCT.

\subsubsection{Goldstein's picture}
As a matter of fact, the idea of the equilibrium dynamics of glass formers as a 
process dominated by activation was not very new even back then (even the 
exponential form of the $\t_R(T)$ growth points towards this), as it had already 
been formulated by Goldstein \cite{Go69} in 1969. Goldstein pictured the 
dynamics as taking place in the \emph{potential energy landscape} (PEL) of the 
liquid, i.e. the hyper-surface obtained by scanning the interaction potential of 
the system over all values of particle coordinates. We can visualize is as a 
very rugged landscape of hills separated by narrow valleys 
\cite{CugliandoloDyn}, at the bottom of which lie the minima of the potential 
energy of the system, called \emph{inherent structures} \cite{CavagnaLiq}. 
Adding thermal energy (i.e. raising the temperature) can be seen as a flooding 
of this landscape, with the level of the water higher the higher the temperature 
and the system can be seen as a boat that has to navigate the landscape 
\cite{CugliandoloDyn}. \\
When the temperature is low, only a few, disconnected lakes of water are 
present, and to sail them all ergodically, the boat must be transported by land 
over the ridges that separate the lakes: this is an activation event, ruled by 
Arrhenius' law. Goldstein postulated that energy minima differed only by a 
change of a subextensive number $n$ of degrees of freedom, and thus could be 
surmounted by a system equipped with a thermal energy of order $k_B T$. This 
way, the system as a whole (as described by a single point in configuration 
space) would always have been in the process of transition, but on the local 
level the jumps would have been separated by a timescale that would grow in 
temperature in an (at least) Arrhenius fashion as required by an 
activation-dominated mechanism. Thus Goldstein's picture provided a good 
explanation for the two-step relaxation observed in glass formers and for the 
exponential growth of the relaxation time: the short $\be$-relaxation would 
correspond to our boat sailing inside a single lake on a short timescale, while 
the $\al$-relaxation would correspond to a much longer transport by land of the 
boat, over a ridge and down into the next lake.\\
What happens if we keep flooding? Pictorially speaking, at a certain point the 
flood should become so severe that the water arrives at the level of the highest 
ridges which separate the valleys i.e. at the level of \emph{saddles}, that is 
stationary points which have at least one unstable direction. When this happens, 
the boat only sees a large body of water wherein it is able to sail ergodically 
without the need for land transport: activation ceases to be the main mechanism 
of relaxation and Goldstein's scenario breaks down. This should happen for an 
high enough temperature $T_x$,  and analogy with MCT suggest the identification
$$
T_x \sim T_{MCT}
$$
which has indeed been verified both in simulations \cite{SchroderSastry00} and 
experiments \cite{Sokolov98}. This identification bolsters the picture of glassy 
dynamics below $T_{MCT}$ as an activation-dominated process.\\ 
The idea of the glass transition as a phenomenon ruled by a topological change 
in the energy landscape has indeed been very fruitful. In the PSM it can be 
proven analytically \cite{CavagnaGiardina01} that the stationary points of the 
energy landscape are minima (i.e. they have no unstable directions) only up to a 
certain threshold energy $E_{th}$, above which saddles take over. It can be also 
seen that $T_{MCT}$ corresponds to the temperature such that the typical 
stationary points visited by the system are exactly those with $E = E_{th}$, 
thereby providing an exact realization of Goldstein's scenario.\\
In real glass formers, analytic calculations of the sort are not possible and 
one must rely on numerics. Nevertheless, multiple studies (see for example 
\cite{Angelani00,Broderix00,GrigeraCavagna02} and the discussion in 
\cite{CavagnaLiq}) seem to confirm this picture. These results are very welcome, 
since they prove that even though in real liquids the MCT transition is wiped 
out by activation mechanisms, the topological transition is still present and 
fuels the fundamental analogy with the PSM even for out of MF glass formers.\\
Since the MCT temperature is the one where a crossover to activated dynamics 
takes place, and metastable states responsible for the slowdown appear, it is 
commonly taken as the reference temperature where the onset of ``glassiness'' is 
located, also because it has a fundamental and unambiguous definition, contrary 
to $T_g$. From now on, when we say ``low temperature'', we mean that we are 
below $T_{MCT}$.

There is however an important warning to give: potential energy landscapes and 
free energy landscapes are not the same thing and one must not confuse the two: 
the energy landscape is defined in the configuration space of the liquid and is 
independent of the temperature, while the free energy landscape is defined in 
the space of local order parameters and changes when the temperature is varied. 
Of course the two are the same when $T=0$, and it could make sense to keep the 
identification as long as the temperature is very low, but attention must always 
be paid and the idea of identifying states with minima in the potential energy 
landscape is just plain wrong 
\cite{BiroliMonasson00,CavagnaLiq,berthierbiroli11}.

At the end of this paragraph, we hope that the reader is convinced that the 
two-step relaxation observed in glass formers is reasonably interpreted as 
originated by the appearance, in the free energy landscape of the system, of a 
collection of metastable minima (states) which exert a trapping effect on the 
dynamics for a stretch of time $t \simeq \t_\al$ and keep the liquid from 
attaining relaxation and flowing. And that the identification of these states 
with minima of the free energy landscape appears reasonable in light of the 
analogy between MCT and the dynamics of the PSM, and by the presence of a 
topological transition in the PEL in both cases. Summarizing, we hope that he 
now believes that the two first tenets of RFOT appear at least a reasonable 
starting point for a theory of the glass transition.

\subsection{Complexity: Kauzmann's paradox}
\begin{figure}[htb!]
\begin{center}
\includegraphics[width = 0.4\textwidth]{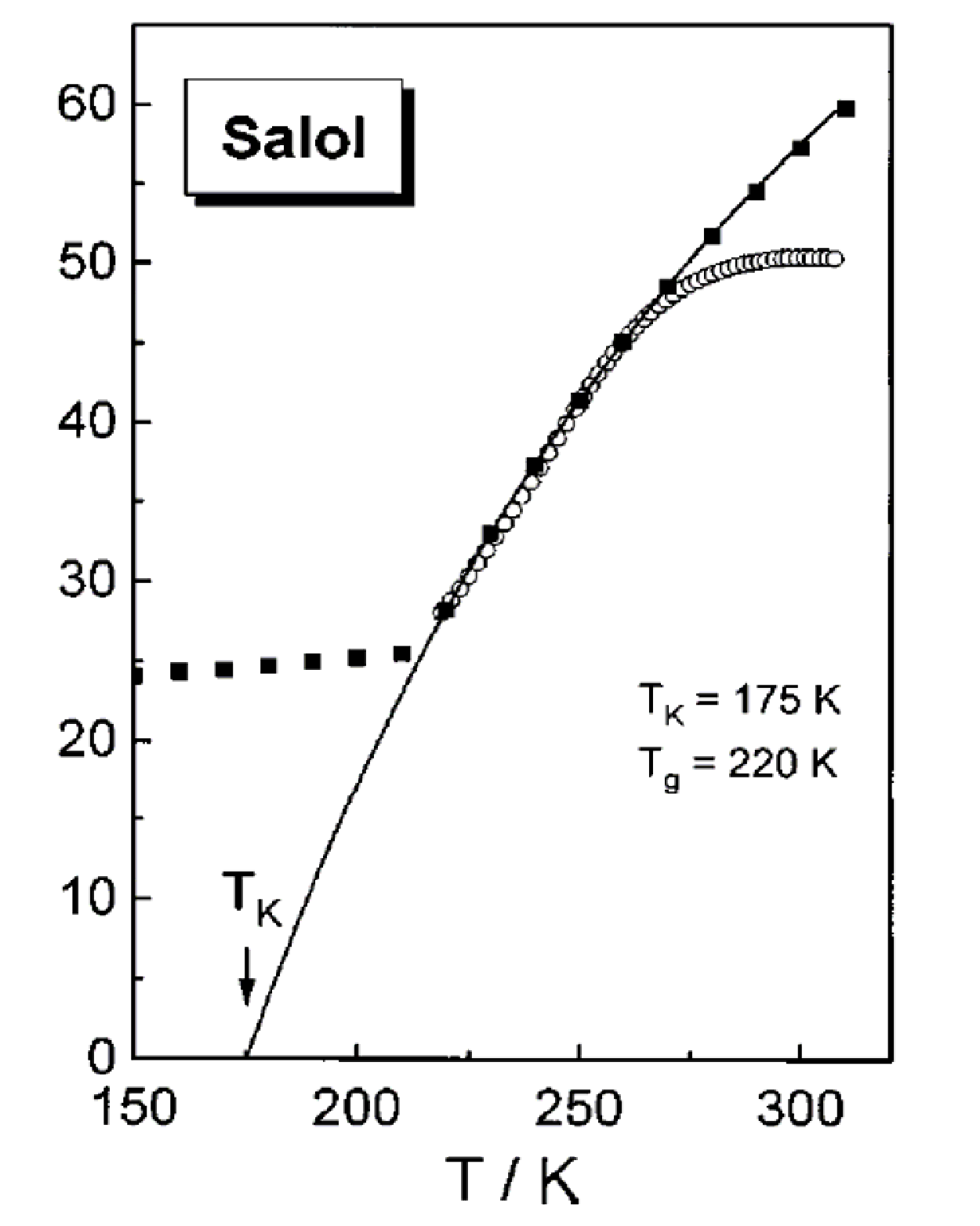}
\caption{Excess entropy as a function of temperature for salol, in $kJ/(K\times 
\mathrm{mol})$. Dots are experimental data, while the line is a fit of the form 
$S_{exc}(T) = A(1-T_K/T)$. Reprinted from \cite{BB09}.}
\label{fig:salol}
\end{center}
\end{figure}

Let us now turn to point 3. We have to ask ourselves the question ``how many 
states can one have?''\\
How can we label a state? As we said, a state is a minimum of the FEL identified 
by a set $\{\rho_i^*\}$ of local densities (in liquids), or a set $\{m_i^*\}$ of 
local magnetizations (in spin models). Thus a glassy state in a realistic glass 
former corresponds to an amorphous density profile the liquid is frozen into. We 
had already encountered this picture when discussing caging: particles can only 
vibrate around equilibrium positions arranged in an amorphous structure, thus 
every state corresponds to such a structure.\\
Since these structures are amorphous, we can already surmise that a lot of them 
should exist. There are not many ways of arranging particles in an ordered 
structure, but there sure are a lot of possible disordered arrangements. 
Intuitively, the number of such arrangements should be equal to the number of 
configurations the liquid has at its disposal, divided by the number of 
configurations visited by the glass during the vibration around the amorphous 
structure.\\
The thermodynamic potential that logarithmically counts the configurations 
available to the system is the entropy, and we already know from the discussion 
in section \ref{subsec:caging} that the vibrational excitations of particles in 
a glass are not too different from the ones found in crystals. We could thus 
hope to count the number of amorphous structures by taking the supercooled 
liquid entropy and subtracting from it the entropy of the corresponding crystal, 
as a reasonable proxy for the vibrational entropy of the glass\footnote{A 
justification is that vibrational contributions are given by an harmonic 
expansion around a potential energy minimum, and the fact that the minimum 
corresponds to an ordered (crystal) or disordered (glass) arrangement of the 
particles should not change matters much. This is reasonable, but it does not 
make sense for systems where a harmonic expansion does not exist, like hard 
spheres \cite{CavagnaLiq}.}. This leads to the definition of the \emph{excess 
entropy}
\beq
S_{exc}(t) \equiv S_{liq}(T) - S_{cr}(T),
\eeq
which we can measure by exploiting the relation between entropy and specific 
heat \cite{CavagnaLiq}
\beq
\dtot{S}{T} = \frac{c_p(T)}{T}.
\eeq
In figure \ref{fig:salol} we show the excess entropy as a function of the 
temperature for salol, a fragile glass former.
We can see that the excess entropy freezes at $T_g$ to the value it had in the 
supercooled liquid: this is due to the fact that there is no latent heat at the 
glass transition (differently from what happens at the melting point $T_m$, see 
figure \ref{fig:summary}) , so the entropy is continuous at $T_g$. On further 
cooling, the excess entropy stays pretty much constant. This is of no surprise, 
as its derivative is proportional to the difference between liquid and 
crystalline specific heat, and we already mentioned in section 
\ref{subsec:caging} that $c_p^{liq} \simeq c_p^{cr}$. Nevertheless, the excess 
entropy at $T_g$ is of the order of $3k_B$ per molecule, which is large 
\cite{BB09}: the number $\mathcal{N}$ of possible amorphous configurations 
scales exponentially with the size of the system
\beq
\mathcal{N} \simeq e^{\Sigma N}
\eeq
where $\Sigma$ is the \emph{complexity} (some prefer to call it 
\emph{configurational entropy} and denote it $S_c$ or $s_c$, but it is a matter 
of taste), and is the central static quantity of RFOT.\\
In glass formers, the presence of an exponentially large number of metastable 
glassy states (and thus the possibility to define a configurational entropy) is 
a reasonable hypothesis (at least we hope that this discussion made it easier to 
accept), but it is an incontrovertible fact in the PSM, where the complexity can 
be analytically computed \cite{cavagnagiardina98,pedestrians} starting directly 
from the stationary points of the TAP free energy \cite{TAP77}. Again the PSM 
furnishes us with a setting wherein the basic ideas of RFOT are exactly 
realized.

\subsubsection{Kauzmann's entropy crisis}
\begin{figure}[htb!]
\begin{center}
\includegraphics[width = 0.5\textwidth]{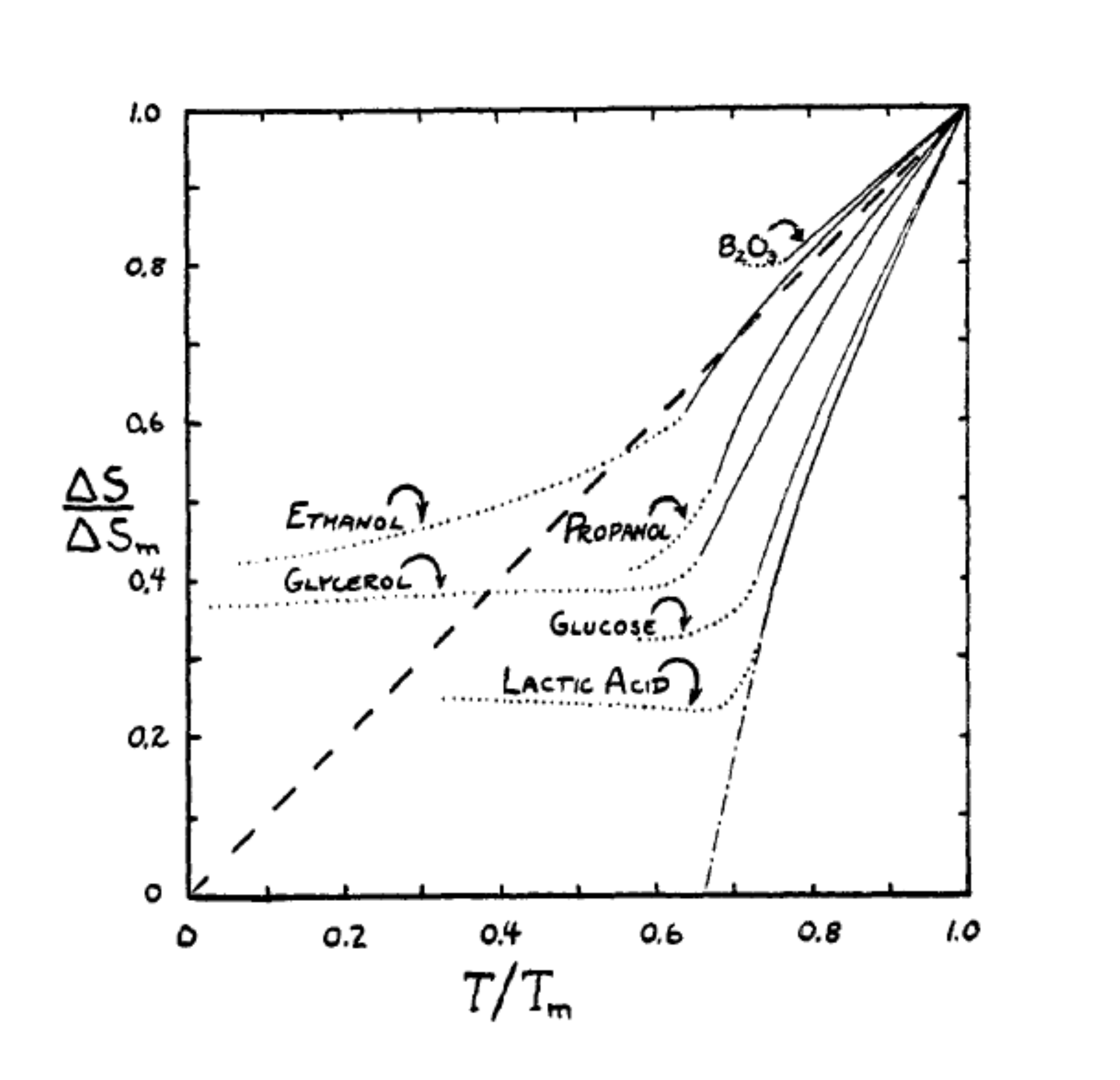}
\caption{The original figure from Kauzmann's paper. For various glass formers, 
the temperature seems to vanish at $T=T_K$, $T_K \neq 0$. Reprinted from 
\cite{Ka48}.}
\label{fig:kauzmann}
\end{center}
\end{figure}

If we look closely at figure \ref{fig:salol}, we can see a curious thing. The 
extrapolation of the excess entropy to temperatures below $T_g$ goes to zero at 
about 175 K, a value far above absolute zero. So there is a finite temperature 
where the entropy of the supercooled liquid would become equal to that of the 
crystal, a very counter-intuitive phenomenon. We would expect the liquid entropy 
to be always above the crystalline one (a liquid is disordered, a crystal is 
not) for any finite temperature.\\
This vanishing of the excess entropy for finite temperature had indeed been 
known for quite some time, as it was first described by Kauzmann in 1948 
\cite{Ka48}. What Kauzmann did was to extrapolate below $T_g$ the data for 
various observables (enthalpy, free volume, energy, etc.), including the excess 
entropy. From figure \ref{fig:kauzmann}, we can observe that the excess entropy 
seems to vanish for temperature different from zero in various glass formers. 
This temperature has been christened $T_K$ in honor of Kauzmann, and the 
vanishing of the excess entropy is referred to as \emph{Kauzmann's paradox}, or 
\emph{Kauzmann's entropy crisis}.

We must immediately state a fact: Kauzmann's paradox was a paradox only back 
then. As strange as the vanishing of the excess entropy can appear, there is no 
law of nature that actually forbids it. The crystallization transition in hard 
spheres for example is precisely due to the fact that the crystalline entropy 
becomes larger than the liquid one at high enough density \cite{AW57}, so 
nowadays we know that the entropy crisis does not violate any fundamental laws 
and its presence is not a serious problem. Nevertheless, it appeared paradoxical 
back then, and Kauzmann himself, in his paper, was eager to find a way out of 
it.\\
There are two possible interpretations. If we believe the fact that the excess 
entropy is a proxy for the configurational entropy $\Sigma$, and that it is 
possible, at least in principle, to equilibrate the supercooled liquid down to 
$T_K$, then a phase transition must be located there. If the configurational 
entropy vanishes linearly at $T_K$ as figure \ref{fig:kauzmann} suggests, then 
the corresponding specific heat has a step at $T_K$, signature of a second-order 
transition \cite{berthierbiroli11}. This transition has been dubbed the 
\emph{ideal glass transition} (sometimes \emph{Kauzmann transition}).\\
Kauzmann himself did not believe this, and he proposed an alternative argument: 
it is not possible to equilibrate the supercooled liquid down to $T_K$, because 
the relaxation time grows so much that, in the end, it becomes larger than the 
crystal nucleation time. So either the glass former goes out of equilibrium and 
forms a glass, or nucleation will kick in and crystallize our sample. In any 
case, the supercooled liquid ceases to exist and the paradox at $T_K$ is 
avoided. We refer to \cite{CavagnaLiq} for a discussion on this point.

\subsubsection{Is the ideal glass transition necessary?}
Unsurprisingly, the interpretation of $T_K$ as the locus of a phase transition 
had a lot more fortune. The idea that the glassy slowdown is a manifestation of 
an underlying critical point at $T_K$ is indeed very appealing and conceptually 
elegant. It also brings back the glass problem to a context, the one of critical 
phenomena, that physicists are very familiar with, and for which a lot of 
theoretical tools are at their disposal.\\
An argument in favor of this idea is that the temperature $T_K$ is always very 
near to the temperature $T_0$ where the VFT fit has a divergence, and this 
applies to glass formers which have $T_g$s that vary from 50K to 1000K (see 
\cite{MartinezAngell01} for a compilation of data and 
\cite{DebenedettiStillinger01} for a discussion). The coincidence is indeed 
remarkable, so much that for many proponents of RFOT, it cannot be a coincidence 
and they see it as incontrovertible proof that the ideal glass transition 
exists. However, there are glass formers where $T_K$ and $T_0$ can differ as 
much as 20 \% \cite{Tanaka03}, so such unshakable certainty is ill-advised, at 
least for now.

The idea of an ideal glass transition at $T_K$ is indeed so powerful and 
fascinating, that over the years it has come to be identified as the main 
prediction of RFOT. This is so much true, that most research articles that go 
and try to disprove RFOT focus on disproving the existence of the ideal glass 
transition (see for example \cite{Stillinger88} and \cite{Garrahan14}). This 
misunderstanding is also probably due to the fact that most models used in RFOT 
theory do have an ideal glass transition, starting from the paradigmatic PSM 
\cite{pedestrians}.\\
We argue here that identifying the ideal glass transition (and also the VFT law) 
with RFOT means missing the point: none of the three tenets that we formulated 
at the beginning of this section has anything to do with the ideal glass 
transition. For the RFOT picture to hold, we only need that the insurgence of 
metastable states cause the glassy slowdown, and that those metastable states 
have a \emph{static} origin in the sense that they can be identified with the 
minima of a suitable free energy functional. This scenario can unfold 
independently from the presence or not of an ideal glass transition at finite 
temperature.\\
In summary, we argue that even an avoided transition ($T_K=0$) is good enough 
for RFOT \cite{berthierbiroli11}.

\subsection{Summary of RFOT: for $T_{MCT}$ to $T_K$.\label{subsec:RFOTsum}}
In the preceding sections we have provided enough (at least we hope) arguments 
to convince the reader that the RFOT theory of the glass transition is a good 
starting point for a description of the physics of glass formers. Let us now 
give a more unified perspective, and summarize what happens during a cooling 
experiment of a glass former according to RFOT.

To be completely general, we focus on a generic system whose 
micro-configurations are denoted as $\C$ and has a Hamiltonian $H(\C)$. The 
partition function is
\beq
Z = \int d\C\ e^{-\be H(\C)}.
\eeq
From the discussion of \ref{subsec:caging} we know that states can be visualized 
as ``patches'' of configurations, namely those configurations which are visited 
by the system as particles vibrate around the amorphous structure that 
identifies the state. Assuming that each configuration can be unambiguously 
assigned to a single state, and that the ``tiling'' so generated covers the 
whole space of configurations\footnote{This is a very strong assumption, but it 
can be rigorously proven to be true in the PSM, see \cite{talagrand}.}, we can 
write the partition function as
\beq
Z = \sum_\al \int_{\C\in\al}d\C\ e^{-\beta H(\C)} = \sum_{\al} e^{-\be N f_\al}.
\eeq
Where $\al$ is an index that identifies a state, and we have defined the 
intensive free energy $f_\al$ of a state. We can transform the sum over $\al$ in 
an integral using Dirac delta functions
\beq
\sum_{\al} e^{-\be N f_\al} = \int df\ \sum_\al\d(f-f_\al)e^{-\be Nf}.
\eeq
We now notice that
\beq
\sum_{\al} \d(f-f_\al) = \mathcal{N}(f) = e^{N\Sigma(f,\beta)},
\eeq
so that we are able to easily introduce the $f$-dependent complexity, which 
logarithmically counts the number of states with have the same in-state free 
energy $f$, or equivalently, the number of minima in the FEL who have the same 
height $f$. We get
\beq
Z = \int df\ e^{-\beta N[f-T\Sigma(f,\beta)]}.
\eeq
In the thermodynamic limit, we can evaluate this integral with the saddle-point 
(or steepest-descent) method \cite{AdvancedMethods}, getting
\beq
Z = e^{-\beta N[f^*-T\Sigma(f^*,\beta)]},
\eeq
where $f^*$ is determined by the condition
\beq
\dtot{\Sigma}{f} = \frac{1}{T},
\label{eq:feq}
\eeq
which means that the partition function is dominated by the states with $f=f^*$ 
only, while the others do not have any impact on the thermodynamics of the 
system. The states with $f = f^*$ are referred to as \emph{equilibrium states} 
for this reason, and the complexity of those states
\beq
\Sigma(f^*(\beta),\beta) \equiv \Sigma(\beta),
\eeq
is accordingly called the \emph{equilibrium complexity}, the one that is 
measured in experiments and simulations.\\
\begin{figure}[tb!]
\begin{center}
\includegraphics[width = 0.6\textwidth]{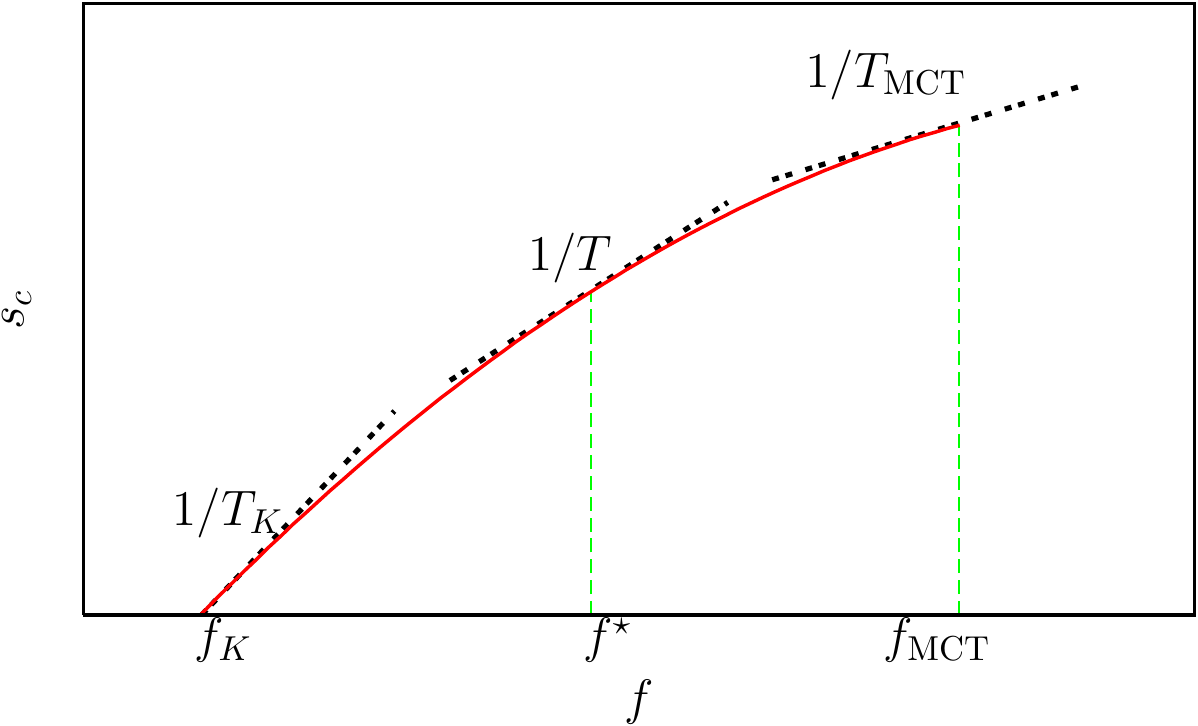}
\caption{The typical form of the complexity as a function of the in-state free 
energy $f$. It is a monotonically increasing function in an interval 
$[f_{min},f_{MCT}]$, and zero otherwise. For $f>f_{MCT}$, the FEL is dominated 
by unstable stationary points and the complexity is accordingly zero, as it 
happens in the PEL for $E>E_{th}$. At $f_{MCT}$ stable minima, which can be 
found everywhere in the interval $[f_{min},f_{MCT}]$, start to appear. At 
$f_{min}$, the complexity vanishes continuously as the number of minima becomes 
sub-exponential. The states with $f = f_{min}$ and those with $f=f_{MCT}$ 
correspond to the typical equilibrium states visited by the system at $T=T_K$ and 
$T=T_{MCT}$ respectively. Reprinted from \cite{berthierbiroli11}.}
\label{fig:RFOT}
\end{center}
\end{figure}
The typical form of the complexity for a system with an RFOT-like FEL is shown 
in figure \ref{fig:RFOT}. Let us now perform an infinitely slow cooling (such 
that the system is always equilibrated) and discuss the various regimes that 
take place as the system scans different regions of the FEL while $f^*(T)$ 
changes with temperature.
\begin{itemize}
\item $\mathbf{T>T_{MCT}}$: At high temperature, the minimization of the the 
free energy functional yields only the homogeneous solution $\rho^*_i = \rho\, 
\forall i$, with the corresponding free-energy $F(\{\rho\}) = F_{liq}$. The 
system is ergodic and liquid.
\item $\mathbf{T_K<T<T_{MCT}}$: At $T  = T_{MCT}$, $f^*=f_{MCT}$, states start 
to have an impact on the system and the relaxation time starts to increase. 
Those states are metastable since $f^*>F_{liq}$, but one can see that $F = 
f^*-T\Sigma(f^*,T) = F_{liq}(T)$ for every $T$ in this interval, and that the 
free energy is analytic at $T_{MCT}$.\\
This can be interpreted as follows: for $T_K<T<T_{MCT}$, the equilibrium liquid 
splits up in a collection of states, each identified by an amorphous structure 
and a set of vibration modes around it. On timescales $\t_\be<t<\t_\al$, the 
system remains trapped in one of the equilibrium states with $f=f^*$, producing 
the plateau regime observed in the dynamics. When at $t\simeq\t_\al$ relaxation 
approaches, the system starts to visit the other equilibrium glassy states 
gaining an entropic advantage in the form of $-T\Sigma(f^*,T)$, bringing back 
the supercooled liquid with its free energy $F_{liq}(T) =  f^*-T\Sigma(f^*,T)$. 
This way, although states appear at $T_{MCT}$, they only impact the dynamics of 
the system and its equilibrium state remains always the supercooled liquid (as 
experimentally observed). In fact, if we compute the probability, at 
equilibrium, to find the system in one particular equilibrium state $\al$, we 
get
\beq
P_{\al} = \frac{\int_{\C\in\al}d\C\ e^{-\beta H(\C)}}{\int d\C\ e^{-\beta 
H(\C)}} = \frac{e^{-\beta N f^*}}{e^{-\beta N (f^*-T\Sigma(f^*))}} = 
e^{-N\Sigma(f^*)} \underset{N \to \infty}{\longrightarrow} 0,
\eeq
so the system ``hops'' seamlessly between all possible equilibrium states.\\
On lowering $T$, $f^*$ will decrease and the system will sample states in lower 
and lower regions of the energy landscape. We stress the fact that such a 
protocol is very difficult to realize in practice as the relaxation time starts 
to grow sharply below $T_{MCT}$. If one performs an infinitely rapid quenching 
dynamics (like the MCT one), the system will not have time to descend in the FEL 
and will remain stuck in the highest states with maximal complexity.\\
Of course, an hopping process like the one we described can take place only in 
real systems. In MF models, once the system is blocked in a state, it can never 
get out, as MCT predicts.
\item $\mathbf{T \leq T_{k}}$ At $T = T_K$, the equilibrium states become the 
ones with $f = f_{min}$ and the complexity vanishes. We have still 
$f_{min}-T\Sigma(f_{min}) = F_{liq}$, and the total entropy $\Sigma + s_{vib}$ 
is continuous, but the specific heat ($c_p = \dtot{S}{\log T}$) has a jump 
induced by the vanishing of $\Sigma$: the entropy vanishing scenario of Kauzmann 
is realized and the ideal glass transition takes place.\\
Below $T_K$, the number of states becomes sub-exponential and the system can now 
be found, at equilibrium, inside a single glassy state, thus yielding a stable, 
thermodynamic glass. 
\end{itemize}

This is how the glassy slowdown happens according to RFOT. One must admit that 
the picture is quite elegant and brings together nicely many different inputs 
and observations, from MCT to excess entropy to Kauzmann's paradox. Needless to 
say, the PSM realizes this scenario exactly \cite{pedestrians}.

\subsection{Beyond mean field: scaling and the mosaic}
We mentioned at the start of this section that a theory of the glass transition 
has at least to describe well the glassy slowdown, which means that it has to 
reproduce the VFT fit (or some other fit) of the $\t_R$ vs. $T$ dependence. 
Within RFOT, this means that we need to compute $\t_\al$ as a function of $T$, 
and to do so, we must focus on the dynamics, in particular on long timescales 
comparable with $\t_\al$, when relaxation occurs.\\
But this brings to the surface the great weakness of RFOT 
\cite{berthierbiroli11}, namely its reliance over the concept of metastable 
state, and consequently on a mean-field description. The MF nature of RFOT 
models (starting from the PSM, but there are many others) makes it, on one side, 
an ideal playground to study metastability: states are sharply defined, they 
have an infinite lifetime, and their properties can be studied analytically even 
for realistic models of glass formers as we are going to see in the following of 
this thesis. But on the other side, the failure to take into account activation 
mechanisms and non-MF effects in general, means that the theory is found wanting 
when attempting to describe the regime wherein non-MF effects come into play, 
which is also the regime wherein relaxation occurs (remember the wrong MCT 
prediction of a transition at $T_{MCT}$). This difficulty in going beyond MF 
(see for example \cite{Rizzo14} for a dynamics-based attempt) is the great 
weakness of RFOT and a solution to this problem does not seem to be forthcoming 
\cite{berthierbiroli11}. As a result of this, the connection between RFOT and 
the dynamics of the system on long ($t\simeq \t_{\al}$) timescales comes from a 
bundle of phenomenological scaling arguments, which goes under the name of 
\emph{mosaic theory} \cite{KW87b,KTW89,LW07,BB09}, and was essentially conceived 
as a reworking of the old Adam-Gibbs theory \cite{AGDM} to include the notion of 
complexity.

Mosaic theory is an attempt to bring out of MF the MF-based concept of 
metastable state: in the real world, the $\t_\al$ timescale is always finite, so 
the concept of metastable state must become local in time. Indeed, since 
metastable states are, well, metastable, they are intrinsically 
out-of-equilibrium objects and so any attempt at a rigorous definition must 
start from the dynamics (see for example \cite{Gaveau98,BiroliKurchan01}). 
Another problem comes from the fact that states, being MF objects, do not take 
into account at all the notion of \emph{real space}. In MF models, there is no 
space structure, so it makes sense to talk about the system globally being in a 
``state'' and hopping to another global state at the onset of relaxation as we 
said in the preceding paragraph. Once real space comes into play, this picture 
clearly makes no sense: hopping from on state to another takes place through a 
nucleation mechanism which can only be \emph{local} in space: as soon as a 
sufficient time passes for activation to happen, droplets will start to form at 
certain points in the sample, and each of them will be in a certain ``state''. 
There is absolutely no reason for them to be all in the same state, since all 
states with $f=f^*$ (including the one the system is about to leave) have the 
same free energy and thus they are completely degenerate. 
a free energy gain in passing from one state to another, it would mean that the 
system is not at equilibrium, in contrast to the RFOT picture which asserts that 
the equilibrated, liquid system is restored by the seamless hopping process 
between states.
Thus, we can expect that the original state will break up in a collection of 
tiles (a mosaic), each of them in a different state. The fact that 
rearrangements must be local should not come as a surprise, considering that the 
local nature of rearrangements in glass formers had been already pointed out by 
Goldstein \cite{Go69}. In summary, states must be defined locally both in space 
and time, i.e. they are characterized both by a \emph{timescale} $\t_\al$ and a 
\emph{lengthscale} $\xi$.

We are interested in computing the timescale as a proxy for $\t_R$. What we can 
do is work out the lengthscale, get from it an estimate of the barrier size to 
rearrangement, and get from it the timescale using Arrhenius' formula. Since we 
have said that rearrangement should take place through nucleation, let us assume 
that our glass former is in a state $\gamma$. When a time $t \simeq \t_\al$ has 
passed, thermal fluctuations will form a droplet of linear size $R$, typically 
in another state $\d$. On the boundary, the mismatch between the two states will 
induce a free energy \emph{cost} in the form of a surface tension
\beq
\Upsilon R^\theta,
\eeq 
where theta is a generic exponent, $\theta \leq d-1$. Usually it is equal to 
$d-1$ (where $d$ is the dimension of space) since it represents a surface term, 
but this not need be the case in general \cite{CavagnaLiq}. This free energy 
price has to be balanced by a free energy \emph{gain} of some sort. In 
nucleation theory it is usually given by the free energy difference between the 
two coexisting phases, but in this case the two states have the same free energy 
$f=f^*$. Thus the complexity comes into play: the droplet can be found in 
$\mathcal{N} = e^{R^d\Sigma(T)}$ different states, so it is of course invited to 
explore them all and gain an entropic advantage in the form of $-T\Sigma(T)R^d$ 
rather than staying in state $\gamma$. The total energy barrier for forming the 
droplet is
\beq
\D F =  \Upsilon R^\theta -T\Sigma(T)R^d,
\eeq
which means that large droplets will tend to survive, while small ones will tend 
to go back to the background state. The crossover between the two happens at a 
lengthscale $\xi$ which can be obtained by setting $\D F$ to zero. We get:
\beq
\xi = \left(\frac{\Upsilon}{T\Sigma(T)}\right)^{\frac{1}{d-\theta}}.
\eeq
Now, following \cite{KTW89}, we fix the barrier height $\D$ as
\beq
\D \equiv \underset{R}{\rm max}\ \D F(R) =  
\frac{\Upsilon^{\frac{d}{d-\theta}}}{[T\Sigma(T)]^{\frac{\theta}{d-\theta}}},
\eeq
and using Arrhenius' formula we get
\beq
\t_R = \t_0\exp\left(\frac{\Upsilon^{\frac{d}{d-\theta}}}{k_B 
T[T\Sigma(T)]^{\frac{\theta}{d-\theta}}}\right).
\eeq
Now we know that near to $T_K$
\beq
\Sigma(T) \simeq A(T-T_K),
\eeq
so we can plug this into the formula for $\t_R$, getting
\beq
\t_R = \t_0\exp\left(\frac{\Upsilon^{\frac{d}{d-\theta}}}{k_B T 
[AT(T-T_K)]^{\frac{\theta}{d-\theta}}}\right),
\label{eq:taurmosaic}
\eeq
so we get a law which looks like the VFT one, although not the same. However, in 
\cite{KTW89} it was claimed that $\theta = d/2$, which would give for $\t_R$
\beq
\t_R = \t_0\exp\left(\frac{\Upsilon^2}{k_B T [AT(T-T_K)]}\right) \underset{T \to 
T_K}{\simeq} \t_0\exp\left(\frac{B}{T-T_K}\right),
\eeq
exactly the VFT law. We must however state again that the VFT law is just a fit, 
not a fundamental law or the result of a first-principles computation, so 
perhaps it is not worth it to fiddle with $\theta$ and make assumptions about 
its value just for the sake of getting it back. Using the \eqref{eq:taurmosaic} 
with $\theta$ as a fitting parameter would  probably do an even better job than 
the VFT and there are alternative laws which provide anyway a good fit of the 
data.

Summarizing, even if we had to use some arguments (and some common sense) to get 
the results, RFOT indeed passes the test, in the sense that it does provide a 
good explanation for the super-Arrhenius increase of $\t_R$, in the form of the 
\emph{complexity}: the barrier size scales with the inverse of a power of 
$\Sigma$ and thus increases when $T_K$ is approached, causing a sharper increase 
that the simple Arrhenius' one (where the barrier would be constant).\\
There is an alternative formulation of the mosaic theory which does not use 
nucleation and is conceptually more robust than the one we just presented, we 
refer to \cite{BB04} and \cite{CavagnaLiq} for details. In \cite{BB04}, the 
authors also propose a method to measure the lengthscale $\xi$, through the 
definition of a special correlation function, the \emph{point-to-set} 
correlation function. The measurement of $\xi$ using this tool has indeed given 
encouraging results and a growth of $\xi$ on supercooling is observed 
\cite{BBCGV08}, so the mosaic picture, and RFOT with it, does seem to be on the 
right track when it comes to the description of supercooled liquids. We refer to 
\cite{CavagnaLiq} for an in-depth discussion on the point-to-set lengthscale.

We conclude here our exposition of RFOT, and we hope that we managed to make it 
looks at least as a reasonable starting point for a theory of glasses. For 
further reading, we refer to \cite{CavagnaLiq} for a pedagogical approach, to 
\cite{berthierbiroli11} for a more technical point of view, and to \cite{BB09} 
for a critical assessment. We refer also the reviews \cite{MP09}, \cite{LW07}, 
and a book \cite{RFOTbook}.

\section{Other approaches}
A much beloved quote by prof. D. Weitz says ``There are more theories of the 
glass transition than there are theorists who propose them''. While it is 
certain that prof. Weitz was a little exaggerating, it is true that there are 
indeed many different theoretical pictures for the glassy slowdown. This is not 
necessarily a bad thing, since it shows that this field of research has still 
many open problems, there is still a lot of work to do, and the debate is fluid 
and lively (tellingly, prof. Weitz's quote is much beloved by glass theorists 
themselves), so there it no shortage of ``other approaches'' that we could talk 
about.\\
However, since they are so many, we will focus here only on the ones that are 
most popular at the moment, and that are more suitable of an analytic, 
statistical-mechanical treatment. Reviews on the approaches we will leave out 
can be found for example at \cite{De96,Sciortino05,Dy06}, and we refer to 
section IV.A of \cite{berthierbiroli11} for more references.

\subsection{Dynamic facilitation theory}
The Dynamic Facilitation Theory (DFT) \cite{ChandlerGarrahan10,RS03,GST10} 
picture is in many ways completely specular to the RFOT approach. Whereas RFOT 
posits a static explanation (in the form of metastable states) for the glassy 
slowdown, DFT favors a completely dynamical approach and postulates that 
thermodynamics plays absolutely no role. While RFOT relies mainly on MF-born 
concepts (like global metastable states and the FEL) and on MF models, DFT is 
firmly rooted in real space and its paradigmatic models all are 
finite-dimensional. While RFOT is at pain when it comes to link its 
thermodynamic foundations to dynamics, in DFT dynamics is the very cornerstone 
of the approach.\\ 
These stark differences between the two approaches come from their differing 
views about what is the distinguishing phenomenological fingerprint of 
glassiness. For RFOT, the fingerprint of glassiness is activation, \emph{\`a la 
Goldstein}, so RFOT naturally stages itself into the FEL, and as a result of this 
it naturally relies on a static MF description of the FEL. According to DFT, the 
fingerprint is \emph{cooperativity}, which takes place during the dynamics of 
the system when viewed in real space. So DFT, accordingly, stages itself in real 
space and naturally relies on dynamical tools \cite{CavagnaLiq}.

The philosophy of DFT is that diffusion and relaxation can only be achieved 
through cooperativity: for a particle to escape its cage, all particles around it must 
also decage and move away, and this in turn will stimulate other particles to 
move \cite{Glarum60}. So, we can see decaging process as the creation of a new 
\emph{defect} (a cluster of mobile particles), which is in turn susceptible of 
inducing (\emph{facilitating}) mobility in nearby regions, creating other 
defects. This picture is undoubtedly reasonable, as we have seen in paragraph 
\ref{subsec:coop} that facilitation does play a role in the dynamics of glass 
formers close to $T_g$. It looks even more reasonable in light of the fact that 
it is possible to define models based exclusively on the idea of dynamic 
facilitation, called Kinetically Constrained Models (KCMs) \cite{RS03}.

\subsubsection{Kinetically Constrained Models}
KCMs can come in different flavors, but they all have two things in common: 
their thermodynamics is trivial, and their dynamics is constrained by rules 
which mimic the facilitation picture. A first example is the Kob-Andersen (KA) 
lattice gas  \cite{KAModel}, wherein a particle can hop from one site to the 
other only if i) it is empty, and ii) if there are less than $m$ neighbors 
around it ($m=6$ on a cubic lattice would correspond to the unconstrained gas). 
It is basically a model that enforces the notion of caging in a strict sense, 
and can be studied for various values of $m$ and different lattice topologies, 
from cubic to Bethe \cite{RS03}.\\
Another champion of KCMs is the Fredricksen-Andersen (FA) model \cite{FAModel}, 
which opts for a specular philosophy to the one of the KA model, focusing on 
holes rather than particles. Each site on the lattice can be mobile $n_i=1$ or 
not $n_i=0$. The Hamiltonian, again, contains no interaction
\beq
H_{FA} = J\sum_{i} n_i
\label{eq:FAH}
\eeq 
and $\thav{n_i} \propto \exp(-\beta J)$, so that mobility is suppressed at low 
temperatures (as one would expect). The dynamics, for its part, takes place with 
the usual Glauber rules, but a site can make a transition from mobile to 
non-mobile only if there are at least other $k$ neighboring mobile sites. A 
variation on the FA is the East model \cite{EastModel}, where only sites on the 
left in each space dimension can facilitate the dynamics.

\subsubsection{Strengths and weaknesses}
The main advantage of DFT lies in the fact that dynamics is the very core of the 
approach, so, for example, their predictions on the relaxation time can be 
easily obtained. Some models show a strong, Arrhenius-like behavior, like the FA 
with $k=1$, while others have a more fragile character. The East model for 
example has $\log \t_\al \simeq \frac{1}{T^2}$, which quite reminisces 
B\"assler's law, eq. \eqref{eq:bassler}, and the FA on a square lattice with 
$k=2$ (the original version of the model from \cite{FAModel}) is even more 
fragile, with $\t_\al \simeq \exp[\exp(c/T)]$. The great majority of KCMs do not 
predict a divergence of the relaxation time at finite temperature, but it is 
possible to define a KCM in such a way that a divergence is present 
\cite{ToninelliBiroli06}. Despite this, their predictions for $\t_\al$ can be 
shown to fit experimental viscosity data quite well, see for example 
\cite{Hecksher08,Elmatad09}. So DFT does pass the test, perhaps in an even more 
convincing manner than RFOT.

Another great advantage of KCMs lies in the fact that they naturally reproduce 
the phenomenon of dynamical heterogeneities 
\cite{BerthierGarrahan03,Harrowell93}, so much that some studies on dynamical 
heterogeneities were actually motivated by their observation in the context of 
KCMs. They also give the possibility to study in great detail multi-point 
correlation functions (like the $G_4$), even enabling researchers to get 
rigorous scaling relations between susceptibilities, lengthscales and timescales 
\cite{FranzMulet02,ToninelliWyart05}. The DFT picture provides also a natural 
explanation of the violation of the Stokes-Einstein relation for viscosity and 
diffusion \cite{JungGarrahan04}.
We stress that all these predictions come easily from DFT, while a 
dynamics-based formulation of RFOT is still lacking (see \cite{infdynamics} for 
a possible starting point), so there is no denying that the DFT picture is 
clearly superior to RFOT when it comes to the study of dynamics in the 
supercooled liquid regime.

However, there are some weaknesses. The biggest, conceptual weakness lies in the 
exclusive role attributed to facilitation, which is seen as the \emph{only} 
possible mechanism for relaxation. To make this more clear, this means that 
mobility and defects cannot be created in any way: a region of the system which 
is not mobile cannot relax unless a drifting defect visits it, and no 
spontaneous motion is possible (mobility is conserved). This is a very strong 
assumption, which may not be true, see for example \cite{Candelier10}.\\
On the other hand KCMs cannot make do without the assumption of mobility 
conservation, or at least without assuming that violations to this assumption 
become more and more rare when $T$ is lowered. If mobility can be created and 
destroyed, KCMs immediately become trivial models and their glassy phenomenology 
is wiped out \cite{berthierbiroli11}, which is a huge and undefended weak point.

In addition to this, as much as RFOT suffers from an over-reliance on MF-based 
statics, DFT seems to suffer from an over-reliance on dynamics. In particular, 
the fact that KCMs all have a trivial thermodynamics does not even seem a 
necessity, and it constitutes no proof that glass formers share this feature. 
Moreover, there is a subtle point, namely the fact that having a trivial 
thermodynamics does not mean that all the RFOT based phenomenology of metastable 
states cannot take place in KCMs. The thermodynamics of RFOT models is indeed 
trivial, in a way, since from $T_{MCT}$ down to $T_K$ equilibrium is always 
given by the supercooled liquid; and indeed, the presence of metastable states 
cannot be detected using standard statistical mechanical tools, requiring to use 
of the TAP approach \cite{TAP77} or the replica method \cite{replicanotes}. In 
summary, we argue that the triviality of the standard thermodynamics of KCMs 
does not imply the triviality of their replica-based (or state-following based) 
thermodynamics.

A less severe and more taste-related weakness is the one of predictive power. 
Intuitively, for a theory to be very predictive we should have to put a few 
things into it, and get a lot in return. DFT, as we just detailed, does give you 
a lot, but it also requires you to put a lot inside. As interesting as KCMs are, 
there is at present no way of linking them to microscopic models of glass 
formers \cite{berthierbiroli11}. Their dynamical rules in particular are just 
imposed from the outside without any microscopic of first-principles 
justification, and they cannot be generally derived (see \cite{Garrahan02} for 
an exception) from an interaction Hamiltonian.\\
This means that in order to get quantitative predictions about real glass formers, 
KCMs must usually be suitably ``tuned'' using experimental or numerical data, 
which is unpleasant. We will return to this point in the following when 
discussing the DFT approach to metastable glasses. We will see that, while DFT 
needs extensive tuning to work well, the RFOT-based state following construction 
enables us to get predictions from first principles, at the only price of an 
Hamiltonian.

\subsection{Frustration limited domains}

According to the Frustration Limited Domains (FLD) \cite{TarjusKivelson05} 
picture, the fingerprint of glassiness is disorder and the amorphous nature of 
the glass phase, which is rationalized as a consequence of geometric 
frustration. Frustration can be broadly defined as the incompatibility between a 
\emph{locally} preferred arrangement, and the symmetry of the space it finds 
itself in, thereby rendering the local structure incapable of tiling the whole 
space forming a global periodic structure.\\
A pedagogical example is a triangle of spins with anti-ferromagnetic 
interactions: the local preferred structure would be a +1 spin and a -1 spin at 
the end of each bond, but the triangular topology renders this arrangement 
impossible: at least one bond is not satisfied (which means that the relative 
energy cannot be minimized), and as a result there are three possible (and 
equivalent) optimum frustrated arrangements. This example illustrates how the 
frustration is caused by geometry: on a square lattice, there would have been no 
problem. Moreover, it also illustrates how frustration is also a source of 
degeneracy and multiplicity, an important ingredient in the context of amorphous 
materials. This ideas can also apply to particle systems. It is for example 
known that the local preferred order for packings of spheres in $d=3$ is the 
icosahedral one, which is however incompatible with periodic ordering 
\cite{Frank52}.

According to FLD, the glassy slowdown is a manifestation of a second-order 
critical point at a certain temperature $T^*>T_m$, which is destroyed by 
frustration \cite{Kivelson95}. When the liquid is cooled down, it starts to form 
locally preferred structures (LPS) in preparation for the transition at $T^*$, 
but those structures are incapable of tiling the whole space due to frustration, 
so they end up forming domains of size $\xi$ separated by topological defects, 
where a surface tension will be located due to the mismatch. The rearrangement 
of these domains, not much differently from what happens in the mosaic picture, 
will then have to proceed by activation \cite{Kivelson95}, producing a slowdown 
of the dynamics like the one observed in glass formers \cite{ViotTarjus00}.

\subsubsection{Models}
The scenario suggested by FLD can be implemented in statistical-mechanical 
models. A particularly elegant realization in the case of spheres is the one 
proposed by Nelson \cite{Nelson02}: the idea is to embed the spheres in a 
spherical manifold with $d=3$, in such a way that the local icosahedral order is 
now compatible with extension in space. The energy of the system can then be 
minimized and a ``reference'' configuration obtained. The curvature of space is 
then reduced, up to the point when the euclidean flat space is recovered. This 
way one can observe how the ordered configuration on the sphere changes to a 
disordered configuration rife with FLDs, separated by a complex network of 
defects. However, this approach is also technically very hard to implement and 
it is almost impossible to get quantitative predictions \cite{berthierbiroli11}.

There are two other possible ways: either a phenomenological scaling treatment, 
like the one implemented in \cite{Kivelson95}, or coarse grained lattice models. 
The Hamiltonian of such models is always made up of two terms: one that 
reproduces the unfrustrated system and yields the second-order transition at 
$T^*$, and another which acts as a source of frustration. A paradigmatic 
Hamiltonian is 
\beq
H = -J\sum_{\thav{ij}} S_iS_j + Q\sum_{i\neq j} \frac{S_iS_j}{|x_i-x_j|},
\label{eq:Hfld}
\eeq
where we can clearly see the competition between ferromagnetic, local 
interaction (which will tend to favor local ordering) and the long-ranged 
Coulombic interaction which acts as a frustrating term. Models of this sort can 
be defined with soft spins, Ising spins, $O(N)$ spins and Potts variables. In 
the case of $O(N\to \infty)$ and soft spins one can see that the transition is 
killed as soon as $Q \neq 0$ \cite{Chayes96}, while in the case of Ising spins 
the transition becomes first order \cite{Brazovskii75}. Anyway, in both cases, a 
disordered phase is found at low $T$ wherein dynamics is slowed down in a glassy 
manner \cite{TarjusKivelson05}. The size $\xi$ of FLDs in particular is found to 
have the scaling
\beq
\xi \simeq \sqrt{\frac{J}{Q}} \xi_0^{-1},
\label{eq:xifld}
\eeq
where $\xi_0$ is the correlation length of the avoided transition at $T^*$. 
Since it decreases as the system is cooled below $T^*$, the size of FLDs is 
found to increase on lowering $T$, which in turn means that the barrier to 
rearrangement of the FLD must increase upon supercooling. So FLD theory passes 
the test as RFOT and DFT, since it does provide a good explanation for the 
super-Arrhenius increase of the relaxation time.\\
We can also observe from the \eqref{eq:xifld} that more frustration means bigger 
FLDs. This is reasonable, since frustration keeps the FLDs from tiling the whole 
space, so having more of it should correspond to smaller domains. This in turn 
implies a direct relation between fragility and frustration, which is probably 
the most original prediction of FLD theory. As a result, FLD can account for a 
wide range of different behaviors, from strong to fragile, just by tuning 
suitably the strength of frustration.

\subsubsection{Strengths and weaknesses}
FLD theory has the great merit of actually posing deep questions about the 
nature of cooperative regions seen in glassy systems. RFOT just postulates them 
as originated by an underlying disordered FEL, and DFT only focuses on how they 
move. FLD instead describes them explicitly, in a very practical and very 
grounded way, and proposes a coherent and elegant picture for their origin. 
FLD has also the merit of bringing back the attention of researchers on a fact 
so obvious it is often forgotten: glasses behave like solids, and the rigidity 
of solids is indeed due to structure, not dynamics, so perhaps it is too soon to 
rule out the existence of any structure in glassy materials: they may actually 
hide more order than we think (see for example \cite{Royall15}).

This grounded and real-space bound description of FLD theory is, however, also 
its weakness: as of now there have not been any direct observations of FLDs 
\cite{berthierbiroli11,CavagnaLiq}. We have focused before on how the static 
structure factor $S(\bq)$ does not seem to capture anything unusual on crossing 
$T_g$, although the fragmentation in FLDs could be so severe that the residual 
order eludes a bulk tool like the $S(\bq)$. In principle it could be possible to 
observe FLDs and their related LPSs through numerical simulations, but this is 
more complicated than it looks \cite{MossaTarjus06}, and there is not even 
agreement on which is the locally preferred order one should look for. For 
example in \cite{CoslovichPastore07}, the growth of icosahedral order is found 
to be more pronounced in fragile liquids as expected from the FLD approach, but 
in \cite{Tanaka12} it is argued that everything can be understood in terms of 
bond-orientational order, rather than icosahedral, so that the situation looks 
very convoluted.

Moreover, one would also appreciate to go beyond scaling arguments and 
coarse-grained models like the one defined in equation \eqref{eq:Hfld}, and 
perform calculations on microscopic models of glass formers, but this does not 
look easy. It is indeed possible to implement numerically a Nelson-like 
treatment for a Lennard-Jones mixture \cite{SaussetTarjus08} with very 
encouraging results, but at present there is no apparent way of translating this 
into a statistical-mechanical calculation. As a matter of fact, models like the 
one in equation \eqref{eq:Hfld}, when treated with the replica method, show a 
Kauzmann transition like the one predicted in RFOT (see for example 
\cite{SchmalianWolynes01}), so it could very well be that a first-principles 
treatment of FLD will end up giving back RFOT results, which could be a very 
interesting turn of events.

\chapter{Metastable glasses \label{chap:metglass}}
In this chapter we focus on the glassy branches in figure \ref{fig:summary}, 
when our glass former is frozen, for a time $\t_\al$, in an amorphous solid 
called a glass. We will mainly focus on two different experimental designs: in 
the first one, the preparation of the glass is followed by a waiting time during 
which the glass is left at rest to age, followed by experimental measurements of 
its thermodynamic properties; in the second one, the glass is prepared and then 
subjected to an external mechanical drive, during which its response to the 
drive is characterized. In both cases we will present the usual phenomenology as 
observed in simulations and experiments, and also the theoretical tools up to 
now used to approach the problem. The aim of this chapter is to detail how 
glasses, despite being out of equilibrium systems, are anyway long-lived states 
of matter endowed with well defined physical properties, that can be measured 
and hopefully computed from a first-principles theory.

\section{Thermodynamics and aging \label{sec:aging}}
In the preceding chapter we have discussed the properties of glass formers, as 
observed in experiments and simulations designed in such a way that 
$$
t_{exp} \gtrsim \t_\al,
$$
on a range of temperatures which goes roughly from $T_K$ (we do not care if 
$T_K$ is zero or not) to $T_{MCT}$. This range of temperature corresponds to the 
one wherein the glassy slowdown takes place, and phenomenology beyond the one 
observed in simple liquids can manifest.\\
In this chapter, on the other hand, we talk about glasses. This means that the 
typical experiment/simulation will be designed so that
$$
\t_\be \ll t_{exp} \ll \t_\al.
$$ 
Let us be more specific, and perform an idealized experiment of length $t_{exp}$ 
on a glass former. The simplest possible experiment consist of at least two 
phases: first a preparation of the sample, which takes a time $t_{prep}$, and 
then a measurement of some sort, carried out at $t_{exp}$. The time that elapses 
between the end of preparation and the experimental time will be called the 
\emph{waiting time}, and denoted as $t_w$. 
The simplest preparation protocol one can consider is a very rapid quench of the 
glass former down to a target temperature $T$, $T_K<T<T_{MCT}$ \cite{Struik78}. 
This is actually the process by which glasses are canonically made, and we will 
refer to such procedures as \emph{quenching protocols}; however, one could also 
consider more creative protocols. For example, one could equilibrate the glass 
former down to a temperature $T_f$, $T_K<T_f<T_{MCT}$ (the notation will be 
clear later), \emph{then} quench it rapidly down to a target temperature $T$ 
\cite{Kovacs63} and perform measurements. Such a protocol will be referred to as 
an \emph{annealing protocol}.\\
To equilibrate the system at $T_f$, one can choose any convenient protocol; the 
most straightforward one is a step-like quench down to $T_f$, after which the 
sample is left at rest until equilibrated. The annealing time needed for such a 
protocol is then
\beq
t_{ann} = \t_{R}(T_f) = \t_0\exp\left(\frac{A}{T_f - T_0}\right) 
\label{eq:anntime}
\eeq
so it obviously grows very rapidly with $T_f$. Of course, in a real laboratory 
it is impossible to achieve infinite quench rates (the thermal conductivity of 
any substance is finite) and more complicated protocols could be needed in order 
to avoid crystallization \cite{CavagnaLiq}, but since the sample is equilibrated 
at $T_f$, the actual protocol used does not matter: the system loses memory of 
its history once equilibrated and its properties do not depend on time, a 
situation referred to as \emph{Time Translational Invariance} (TTI), as we 
already discussed \cite{CugliandoloDyn}. Once the annealing phase has been 
completed, we quench the glass down to $T$, a process which will take a time 
$t_{qu}$ such that $t_{ann}+t_{qu}=t_{prep}$, and then wait a time $t_w$ before 
performing our measurements. After $t=t_{ann}$ the system is out of equilibrium, 
so from $T_f$ downward its equation of state will deviate from the supercooled 
liquid one as shown in figure \ref{fig:summary}: $T_f$ is the temperature 
whereupon the system forms a glass. This temperature has been defined by Tool in 
\cite{Tool46} and called the \emph{fictive temperature}. In any reasonable 
experimental setting it is practically equal to the glass transition temperature 
$T_g$, so we will consider them as being interchangeable from now on.

One can easily understand the difference between the two protocols we described 
in light of the summary of RFOT we made in section \ref{subsec:RFOTsum}: in the 
first protocol we do not let the system equilibrate in any way, so it remains 
arrested in the threshold states in the highest regions of the FEL 
(corresponding to $f=f_{th}$ such that $\Sigma(f,T)$ is 
maximized\footnote{Technically, the definition of threshold state it that of a 
minimum which is infinitesimally near to being a saddle. However, it is 
reasonable that such stationary points will be in the highest part of the FEL 
(ridges are always located higher than valleys) and thus have maximal 
complexity, as long as the complexity is a monotonously increasing function of 
$f$. If these conditions are met, the identification between threshold and 
maximal complexity is safe (and is rigorously true in the PSM).}), and will 
still be there when we perform our measurement at $t_{exp}$, unless the waiting 
time is very long. In the second protocol, the system remains anyway blocked in 
the threshold states, but then we give it enough time $t_{ann}\simeq \t_R(T_f)$ 
to descend in the FEL (thanks to activation), down to the equilibrium states 
with $f=f^*$ determined by the condition \eqref{eq:feq} with $T=T_f$. This way, 
when we will quench it again down to $T$, the system will remain trapped in one 
of the equilibrium states selected at $T_f$, and again, it will still be there 
when we perform our measurements at $t_{exp}$.\\
In summary, having a lower temperature $T_f$ (and thus a longer protocol) means 
that the phenomenology observed at $t_{exp}$ will be ruled by deeper and deeper 
minima in the FEL.
 

\subsection{Protocol dependence}
From the description we just gave of the simplest protocols for glass 
preparation, the reader can immediately understand that glass is an 
intrinsically \emph{out of equilibrium} object. In practical terms, this means 
that a typical glass has not had enough time to forget its history and so the 
properties measured at $t_{exp}$ can depend, in principle, both from the 
preparation protocol, and on the time that has passed between the end of 
preparation and the measurement. In summary, any measurable observable $\Os$ can 
depend on $t_{ann}$, $t_{qu}$ and $t_w$, and on the details of whatever happened 
during these time periods.

We already said that the dependence on $t_{ann}$ is no big deal: the system is 
left at rest until equilibration sets in at $T_f$ (or $T_g$), so it has forgot 
all its past history, including the annealing protocol employed to get at $T_g$, 
which is effectively the only trace left of the past history of the glass. The 
dependence on $t_w$ is very weak as well, at least for one-time observables: in 
the case of quenching protocols, power laws with small exponents or even 
logarithmic laws, such as
\ba
\Os(t_w) &=& \Os_\infty + \left(\frac{\t_0}{t_w}\right)^\al,\\
\Os(t_w) &=& \Os_\infty +\log(1+\t_0/t_w),
\ea
are usually reported \cite{Struik78}. The dependence should be even weaker for 
the annealing protocols that reach deeper and more stable states in the FEL. 
This also implies that the dependence on $t_{qu}$ is as well weak: a rapid 
quenching protocol followed by a long waiting time is the exact same thing as a 
long quench followed by a short $t_w$.\\ 
In summary, the properties measured in a glass annealed at a temperature $T_g$ 
and then quenched at a temperature $T$ do not depend much on the protocol needed 
to get from $T_g$ to $T$\footnote{This is true as long as the glassy metastable 
state the system is in does not undergo any further in-state phase transition 
(like the minimum splitting in two or more sub-minima). We will see that this is 
exactly the case with the Gardner transition.}, and only depend on the 
temperature $T_g$ (or equivalently $T_f$), and on $t_{ann}$ through it:
$$
\Os(t_{w},t_{ann},t_{qu}) \longrightarrow \Os(T_g,T).
$$ 
which was indeed Tool's original idea for $T_f$ \cite{Tool46}. Despite the fact 
that a glass is not at equilibrium, its one-time properties are effectively 
independent of time  (at least until it relaxes again on an impossibly long 
timescale $\t_\al(T)$): it actually makes sense to talk about a thermodynamics 
of glasses.

\begin{figure}[t!]
\begin{center}
\includegraphics[width = 0.6\textwidth]{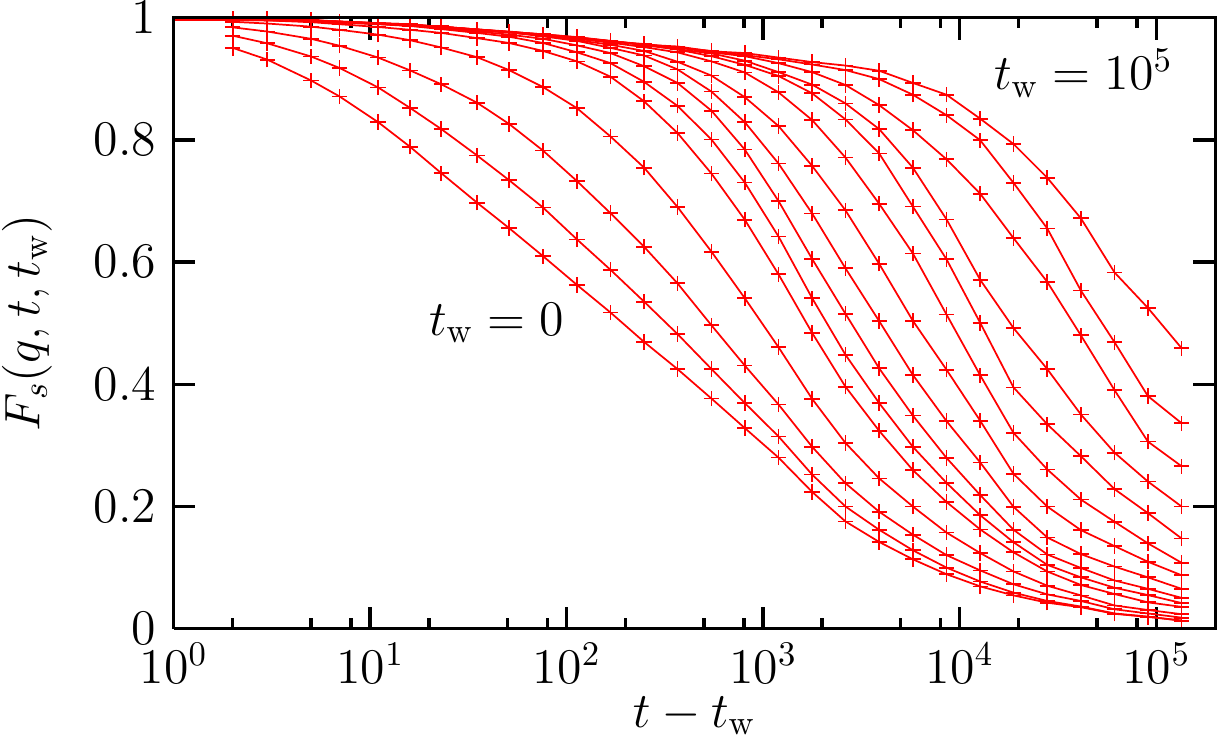}
\caption{The intermediate scattering function of a prototypical glass former 
after a quenching protocol, in an aging situation. We can see how the relaxation 
time $\t_0$ increases when the waiting time increases, a typical aging effect. A 
quasi-equilibrium regime is observed at small times. Reprinted from 
\cite{berthierbiroli11}.\label{fig:aging}}
\end{center}
\end{figure}

To detect actual signatures of aging, one must focus on two-time quantities, 
like the already mentioned dynamical correlation functions. In an out of 
equilibrium, aging scenario, they will depend on both times $t$ and $t'$
\beq
C(t',t) \longrightarrow C(t_w,t_w+\t).
\eeq 
In figure \ref{fig:aging} we show a typical aging-like relaxation pattern for 
the dynamical correlator, plotted for various waiting times after a quenching 
protocol. At small times the curves are almost superimposed and the dependence 
on $t_w$ is weak, so that TTI is restored. At longer times, a stronger 
dependence on $t_w$ is observed, and the correlator decays to zero on a 
timescale $\t_0$ which depends strongly on $t_w$. This dependence on $t_w$, 
which can also be seen as the \emph{age} of our glass, is the definition of 
aging \cite{berthierbiroli11,CugliandoloDyn}. 

The fact that $\t_0$ increases with $t_w$ may seem counter-intuitive: the system 
has to eventually decorrelate on a fixed timescale $\t_\al(T)$, so one would 
naively expect $\t_0(t_w) = \t_\al - t_w$. This however is wrong: the system has 
first to relax inside a steady state, and then decorrelate again. When $t_w = 0$ 
is zero, the preparation of the sample has just ended and the system has not yet 
had time to settle it its new state. So when we observe it after an interval 
$\t$, is has decorrelated significantly as part of this partial equilibration 
process. But if we wait a time $t_w$, take a snapshot of the system, and then 
take another at $t_w+\t$, we will observe less decorrelation, since part of the 
process has taken place before the first snapshot at $t_w$: the more we wait, 
the less decorrelation we observe, until at $t_w \simeq \t_\al$ the dynamics 
crosses over to the equilibrium one where TTI is valid again, an \emph 
{interrupted aging} scenario \cite{CugliandoloDyn}.

All this phenomenology seems well tailored to RFOT: at $T_g$, the system is at 
equilibrium, visiting a great number of degenerate states whose free energy 
$f^*$ only depends on $T_g$ through equation \eqref{eq:feq}. When the system is 
quenched down to $T$, is has not enough time to equilibrate again, but it can 
equilibrate inside the state with a relaxation time of order $\t_\be$ weakly 
dependent of the temperature, generating a TTI dynamics on short timescales. 
After this, the system will stay equilibrated inside the state and its 
observables will reach a restricted equilibrium value, independent of time and 
dependent only on the thermodynamics  (and so on $T$) of the metastable state. 
Of course, if the experimental time is comparable to $\t_\al$, one can actually 
observe aging effects, as the glass relaxes and its observables go back to their 
equilibrium, supercooled liquid values. But this is possible only when $T\simeq 
T_g$, very near to the equilibrium line, so those effects are effectively 
negligible.

\subsection{Ultrastable glasses \label{subsec:ultrastable}}

The protocol dependence properties of glasses open exciting scenarios, wherein 
one could manufacture materials whose physical properties could be ``tuned'' 
just by changing their preparation protocol. However, such a program is not easy 
to realize, at least until recently: the time $t_{ann}$ needed to get a low 
$T_g$ grows very steeply with $T_g$ (equation \eqref{eq:anntime}), so only a 
very limited range of $T_g$s is within practical reach. This is the reason why 
the brutal quenching protocols like the one described at the beginning of this 
chapter have been for much time the standard for studying glasses 
\cite{Struik78}: doing better is very time-consuming and difficult.\\
However, in recent years, a new experimental and numerical protocol has allowed 
researchers to get, in a few hours, glasses with $T_g$s that would correspond to 
canonical preparation protocols up to decades long, allowing us to go much 
beyond $T_{MCT}$. The glasses so produced are fittingly called 
\emph{ultra-stable} glasses 
\cite{ParisiSciortino13,Sw07,SEP13,SEP13corr,LEDP13,SepulvedaTylinski14,Pogna15}
.

Ultrastable glasses are prepared via a special protocol that goes under the name 
of \emph{vapor deposition}: the glass former is slowly deposited onto a 
substrate whose temperature is controlled by a thermostat, and is of course 
lower that the calorimetric glass transition temperature defined in section 
\ref{subsec:glasstrans}. In such a setup, the particles near the free surface 
enjoy enhanced mobility and are able to look for equilibrated configurations 
much faster than they would in the bulk of the sample, as it would happen in a 
canonical annealing protocol. This creates, layer by layer, glasses which are 
exceptionally stable and correspond to ordinary glasses with impossibly long 
annealing times (around 40 years for some samples described in \cite{Sw07}). 
This procedure can be implemented both in experiments 
\cite{Sw07,SepulvedaTylinski14} and simulations \cite{SEP13,LEDP13}, and can be 
shown to work well both for fragile and strong glass formers 
\cite{SepulvedaTylinski14}.

\begin{figure}[t!]
\begin{center}
\includegraphics[width = \textwidth]{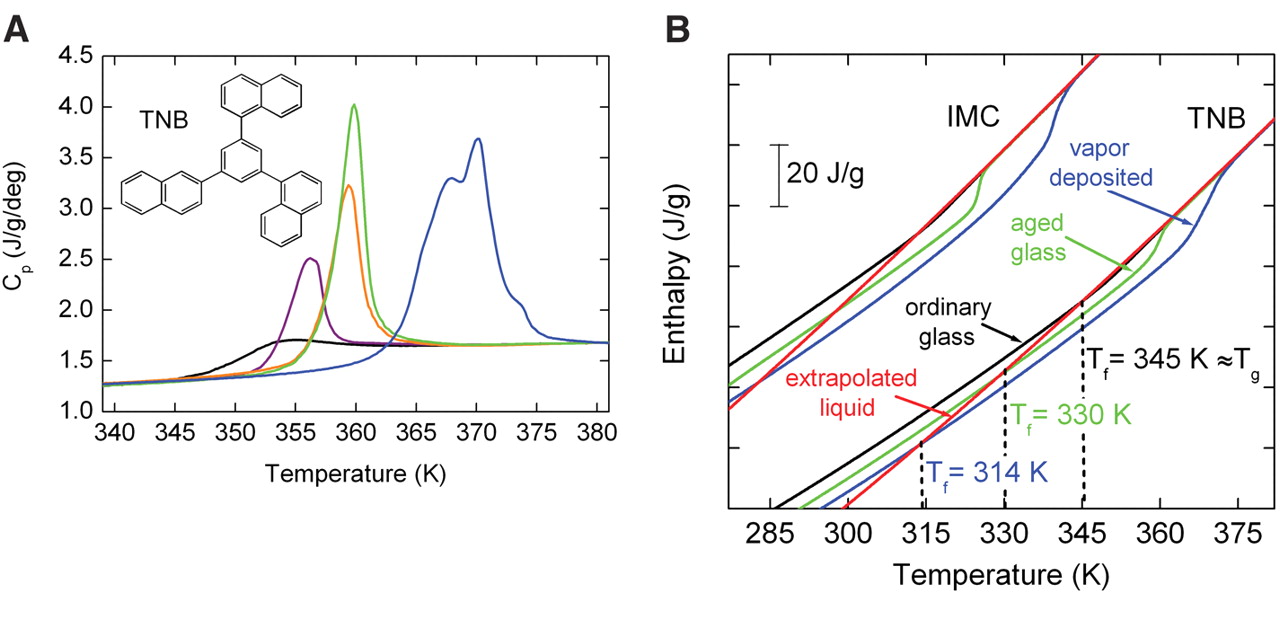}
\caption{\emph{(Right panel)}: enthalpy for an ultrastable glass (blue), a glass 
with an annealing time of 15 days (green) and an ordinary glass produced by 
cooling the liquid at about 40K/min (black). In a cooling scan, the enthalpies 
branch away from the liquid value at the fictive temperature $T_f$, and remain 
always higher than the liquid enthalpy (since $C_p^{gl} < C^{liq}_p$ as we 
already mentioned). In a heating scan, the enthalpy of the glass becomes lower 
than the liquid enthalpy, up to a temperature $T_{on}$ when the glass melts back 
into the liquid and the enthalpy regains its liquid value. Both $T_f$ and 
$T_{on}$ quantify the stability of the corresponding glass. \emph{(Left panel):} 
heat capacity at the onset transition. A maximum is observed, which separates 
the glass at low $T$ from the liquid at high $T$. Obviously $C_p^{gl} < 
C^{liq}_p$ as expected. Reprinted from \cite{Sw07}.\label{fig:ultraexp}}
\end{center}
\end{figure}

Once the sample has been prepared, it can be studied and characterized in a 
variety of ways. In experiments, Differential Scanning Calorimetry (DSC) is 
usually employed: heating and cooling scans are performed on the sample and its 
properties (heat capacity at constant pressure and enthalpy in the case of 
\cite{Sw07}) are measured as a function of $T$ (see figure 
\ref{fig:ultraexp}).\\
The cooling scans in the right panel of figure \ref{fig:ultraexp} correspond 
essentially to the equations of state in figure \ref{fig:summary}: the enthalpy 
branches away from the liquid EOS at a fictive temperature $T_f$ which depends 
on the preparation protocol. In addition to this, one can also perform an 
heating scan, wherein the sample is prepared and then (rapidly) heated up. In 
this case, hysteresis is observed: the enthalpy becomes lower than the liquid 
one, up to an onset temperature $T_{on}$ whereupon the glass melts back into the 
liquid.\\
The stability of the glass is characterized by the two temperatures $T_f$ and 
$T_{on}$: lower $T_f$ and larger $T_{on}$ correspond to higher stability (and a 
larger hysteresis cicle). A lower $T_f$ means that the glass is arrested in 
lower minima in the FEL, and correspondingly, more thermal energy must be 
supplied to the system to dislodge it from these minima and melt it back into 
the liquid, yielding a higher onset temperature \cite{Sw07}. 

In numerical simulations \cite{SEP13}, one has access to positions and momenta 
of all particles, allowing a more refined analysis. In particular, the 
properties of the PEL that underlies the liquid can be studied. Such a study can 
be carried out in the following way: one takes an equilibrium configuration 
(essentially a set of $3N$ positions) of the supercooled liquid at a certain 
temperature $T$ below $T_{MCT}$. One can then minimize the potential energy of 
the liquid, using for example a gradient descent method \cite{CavagnaLiq}, using 
the equilibrium configuration as a starting point for the algorithm. This 
procedure (which essentially corresponds to a sudden quench to zero temperature) 
will produce a configuration corresponding to a minimum of the potential energy 
(an inherent structure) with a certain energy $E_{is}$, which will be one of the 
minima that the system typically vibrates around at temperature $T$ 
\cite{Sciortino05}. As we mentioned in section \ref{subsec:dynamics}, lower 
minima of the PEL can only be reached through activation, so they correspond to 
larger annealing times and thus to higher stability.\\
\begin{figure}[t!]
\begin{center}
\includegraphics[width = 0.5\textwidth]{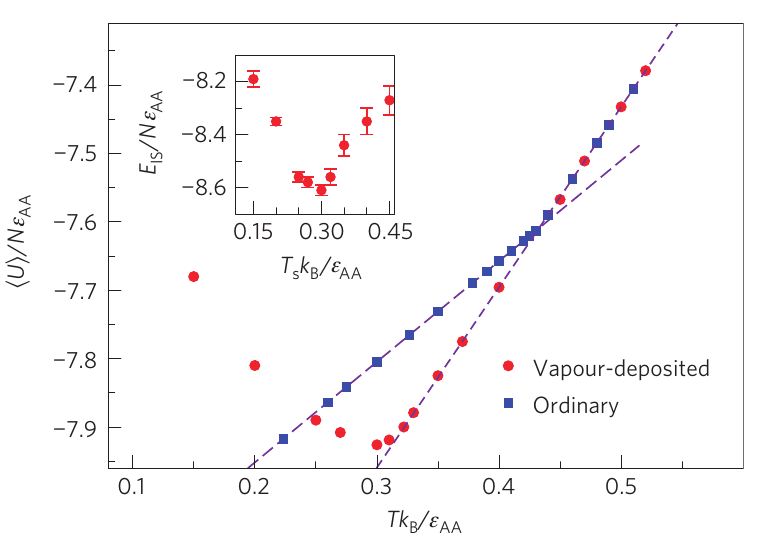} \nolinebreak[4]
\includegraphics[width = 0.5\textwidth]{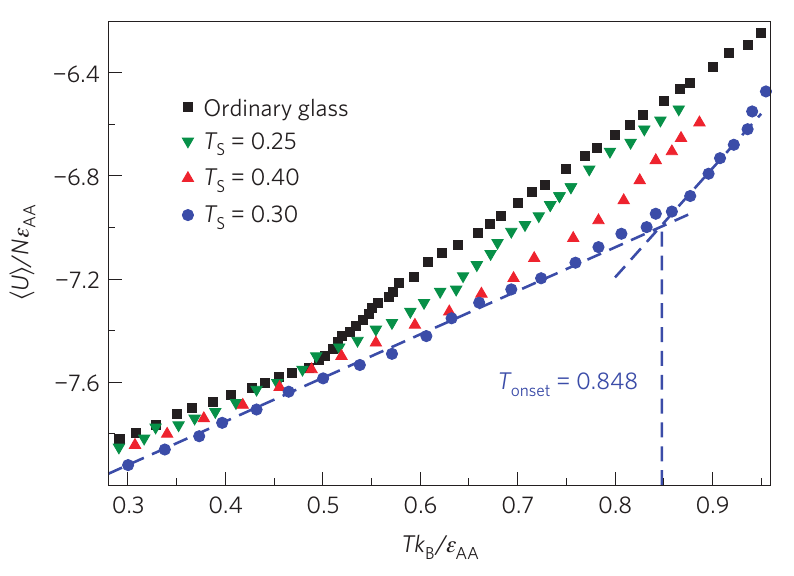}
\caption{\emph{(Left panel)}: Average potential energy $\thav{U}$ of the 
vapor-deposited glasses (red symbols) as a function of the substrate temperature 
$T_s$, compared with the same observable for an ordinary glass (blue symbols). 
The vapor-deposited glasses show supercooled liquid behavior (which means that 
$T_f = T_s$) for temperatures much below the glass transition temperature for 
the ordinary glass, with a lower limit at $T_s\simeq 0.3$, about the value of 
$T_K$ for this system. In the inset, the inherent structure energy is plotted as 
a function of $T_s$: a minimum is observed, again at $T=0.3$, corresponding to 
optimal substrate temperature. \emph{(Right panel)}: heating and cooling scans 
on glasses with three different substrate temperatures, with $T_f = T_s$ for all 
the glasses, except for the one with $T_s = 0.25$ (as one can see in the left 
panel). Again an onset transition is observed, with $T_{on}$ shifting up for 
more stable glasses. All data are plotted in Lennard-Jones units. Reprinted from 
\cite{SEP13}.\label{fig:ultranum}}
\end{center}
\end{figure}
In figure \ref{fig:ultranum} we show some results from \cite{SEP13}, for a glass 
produced by vapor deposition \emph{in silico} of Lennard-Jones particles: one 
can see that a lower substrate temperature allows the system to reach lower 
minima in the PEL, as expected. This provides one more way to quantify the 
stability of the glasses produced by vapor deposition. In particular, the glass 
with the lowest $E_{is}$ corresponds to an optimal substrate temperature 
approximately equal to the Kauzmann temperature for the glass former in study 
\cite{SEP13}, a very intriguing coincidence.\\
In addition to this, one can also study the average potential energy $\thav{U}$ 
as a function of $T$, and perform cooling and heating scans similarly to 
\cite{Sw07}. Again, an onset transition is observed, with higher $T_{on}$ 
corresponding to higher stability as expected.\\
In numerical simulations one can also study very precisely the structural 
properties of the glass, for example employing Voronoi tesselation. We do not 
delve into the details, but we summarize that the more stable glasses do show 
structural peculiarities, with a homogeneous (albeit always amorphous) structure 
almost totally devoid of defects (such as polycrystallites), in contrast with 
ordinary glasses which present bigger defects, and in larger number 
\cite{SEP13}.  This is in agreement with the fact that ultrastable glasses also 
show suppressed diffusion properties with respect to ordinary glasses 
\cite{Sw07}, another index of enhanced stability.

In summary, ultrastable glasses are certainly going to be studied more and more 
in the next years: they are relatively easy to produce and allow researchers to 
bypass annealing protocols which would be decades long, one of the big 
difficulties in the study of the glass transition. Moreover, we mentioned in the 
introduction and in section \ref{sec:RFOT} that most of the relevant predictions 
of theories of the glass transition are found at temperatures much lower than 
$T_{MCT}$, so that only experiments performed at those temperatures, on an 
\emph{equilibrated} glass former, can hope to unveil ``smoking gun'' evidence in 
favor of one particular approach. Ultrastable glasses seem to give a big help in 
this direction, so a new influx of experimental and numerical data from their 
study is to be expected in the future.

\subsection{The jamming transition \label{subsec:jamming}}
Let us now make a detour to talk about a phenomenon which at first glance is 
completely unrelated to glasses: the jamming transition 
\cite{LNSW10,liunagel98,Biroli07}. The aim of this paragraph is to convince to 
reader that it is not at all a detour, at least within RFOT.

The jamming transition is probably one of the most ubiquitous phenomena one can 
conceive. The canonical example of jamming system is a fistful of sand: when it 
is not compressed, it responds to stresses by flowing, more or less like a 
liquid would. However, if we clench our fist, at a certain point we will not be 
able to squeeze the sand anymore and the response will be solid-like: the grains 
of sand are mechanically in contact, forming an amorphous, tight packing. The 
jamming transition is a transition between a loose, liquid-like system to a 
jammed, solid-like one.\\
This may look like a calorimetric glass transition, but it is not the same 
thing. In the glass transition the solid that originates from the glass former 
is due to caging and vibrations inside the cage, which render it capable of 
bearing loads and respond like a solid; however, the system is still 
compressible and pressure is finite. In the jamming case, there is no 
temperature and the solid-like behavior is due to the forming of a network of 
mechanical contacts between the grains; and if those grains can be reasonably 
modeled as mechanically undeformable, hard particles, the resulting solid is 
incompressible: its pressure is infinite.
\begin{figure}[t!]
\begin{center}
\includegraphics[width = 0.4\textwidth]{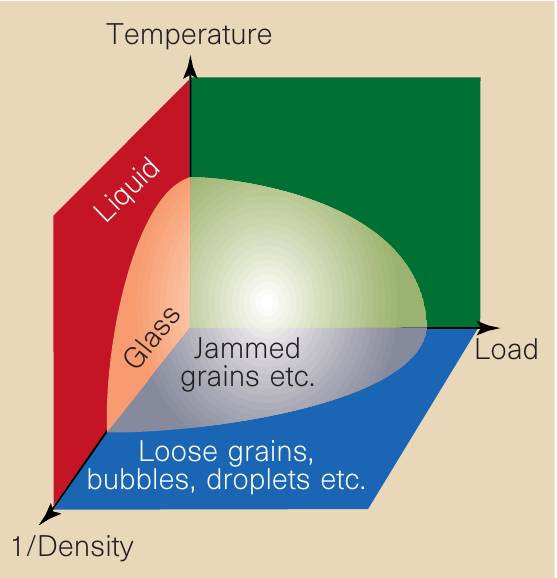} 
\caption{The proposed unified phase diagram for the glass and jamming 
transitions. Reprinted from \cite{liunagel98}. \label{fig:jamming}}
\end{center}
\end{figure}

To study jammed packings, one usually constructs them using a certain protocol 
(the choice of words is not casual). In experiments, one can for example throw 
the grains in a shaking box one at a time until the packing jams \cite{SK69}. In 
simulations, a very popular algorithm is the one by Lubachesky and Stillinger 
\cite{LS90} (LS), wherein the packing is created by inflating the spheres at a 
fixed rate $\gamma$ during a molecular dynamics run. Another possibility is to 
consider soft particles with a potential that vanishes outside the particles 
(tennis balls, essentially): one starts from a random configuration, compresses 
it, then minimizes the potential energy, and then compresses it again, until a 
zero energy configuration cannot be found anymore \cite{OLLN02,OSLN03}. The 
jamming problem can then be formulated as: \emph{``Given a procedure to 
construct an amorphous packing, what are the properties of the packing so 
obtained? First of all, what is its jamming density 
$\varphi_j$}\footnote{$\varphi$ is defined as the fraction of volume occupied by 
the particles, $\varphi \equiv \frac{N}{V_{par}}{V} = \rho V_{par}$.}? \emph{How 
does the contact network behave? Which properties depend on the actual 
procedure, and which ones do not?''}. A ponderous research effort has ensued, in 
the last years, to answer these questions.

Luckily for us, this effort has been successful (see for example the reviews 
\cite{TS10,He10}), at least for frictionless spherical particles,  so we can 
reap and summarize the most relevant results:
\begin{enumerate}
\item The jamming density $\varphi_j$ does depend on the protocol used. In the 
paradigmatic case of the LS algorithm for hard spheres, it can be seen that a 
lower rate $\gamma$ corresponds to a lower $\varphi_j$ \cite{SDST06,CBS09}. In 
3$d$, an fcc crystal is produced for small rates ($\varphi_j = \varphi_{FCC} 
\approx 0.74$), while for a fairly large range of intermediate rates an 
amorphous packing with $\varphi_j = \varphi_{GCP} \approx 0.68$ is 
produced\footnote{We mention that this numerical estimate is subject to an error 
of about 10\%, so caution is advised, as always.}. For even higher rates, 
$\varphi_j$ goes down smoothly with $\g$.
\item \emph{All} amorphous packings of frictionless particles at the jamming 
threshold $\varphi_j$ are \emph{isostatic} \cite{Mo98,Ro00}, which means that 
the average number of contacts $z$ in the packing is just the one needed to 
ensure mechanical stability, $z_{iso}=2d$\footnote{$d$ is the dimension of 
space.}, in agreement with Maxwell's criterion \cite{Maxwell64}. 
\item The probability distribution of the absolute value of forces in a packing 
$P(f) \equiv \sum_{i=1}^{N_c} \delta(f-f_i)$ (where $N_c$ is the number of 
contacts) has a power-law behavior (a pseudogap) for small forces, $P(f) \simeq 
f^\theta$, where the exponent $\theta$ is apparently the same for every $d\geq 
2$ \cite{LDW13,CCPZ12,CharbonneauCorwin15}.  
\item The pair distribution function $g(h)$\cite{simpleliquids}, $h \equiv 
(r-D)/D$ (where $D$ is the diameter of a sphere) is singular at small $h$: $g(h) 
\simeq h^{-\gamma}$, with the exponent $\g$ again independent of dimension 
\cite{SDST06,CCPZ12,LDW13}. Indeed a scaling relation can be derived to link 
$\g$ with $\theta$ \cite{Wy12,LDW13}. 
\item The Debye-Waller factor (or equivalently the MSD) of a packing of hard 
spheres subject to agitation (like in the LS algorithm), and near the jamming 
threshold, scales with the reduced pressure\footnote{$p \equiv \frac{\beta 
P}{\rho}$, where $P$ is the pressure.} as $\D \simeq p^{-\kappa}$ 
\cite{IBB12,nature}.
\item The density of states (DOS) $D(\omega)$ of vibrational modes in a jammed 
packing of spheres (soft or hard) has a plateau down to zero frequency 
\cite{OSLN03,SLN05}, a property referred to as \emph{marginality}. It can be 
shown that this property is intimately connected with isostaticity, as the 
frequency $\omega^*$ that delimits the plateau at low frequencies can be shown 
to obey the scaling $\omega^* \simeq z-z_{iso}$  \cite{WNW05}. This means that 
the packing is mechanically stable (is in a stationary point of the PEL where no 
negative modes can be found), but only marginally so: it can be destabilized 
without paying an energy cost.\end{enumerate}
In summary, jammed packings surprisingly show properties with a remarkable 
degree of universality, in the sense that they are both protocol-independent, 
and also apparently independent of the dimension $d$ of space as  long as $d\geq 
2$. The isostaticity and marginality of jammed packings point toward the fact 
that the jamming transition may be a phenomenon governed by a critical point in 
the Landau sense (albeit at $T=0$), with an associated set of critical 
exponents. As a matter of fact, a whole scaling description of the jamming 
transition can be derived just by assuming marginality \cite{Wy12,LDW13,BW09b}.
Moreover, we point out the fact that these critical properties can be measured 
and characterized very accurately, differently from what happens in the case of 
glasses where no transition is present and everything is built around elusive, 
subtle observables, like the complexity $\Sigma$. 

What do jammed packings have in common with glasses? They have in common the 
most important thing, namely the characteristic that makes them both hard to 
approach theoretically: protocol dependence.\\
Glasses are protocol dependent because they are not able to equilibrate on human 
timescales, while jammed packings are protocol dependent because they are 
\emph{athermal} systems made of macroscopic objects, such that the relevant 
energy scale is far above $k_B T$. In both cases, the system is unable to forget 
its history and treating it with the usual tools of statistical mechanics is 
impossible. But at the end of the day, the jamming problem (``\emph{Given a 
procedure to construct an amorphous packing, what are the properties of the so 
obtained packing?''}) and the glass problem (\emph{``Given a protocol to produce 
a glass, what are the properties of the so obtained glass?''}) have the exact 
same nature, which is the reason why we put them in the same chapter.

\begin{figure}[tb!]
\begin{center}
\includegraphics[width = 0.5\textwidth]{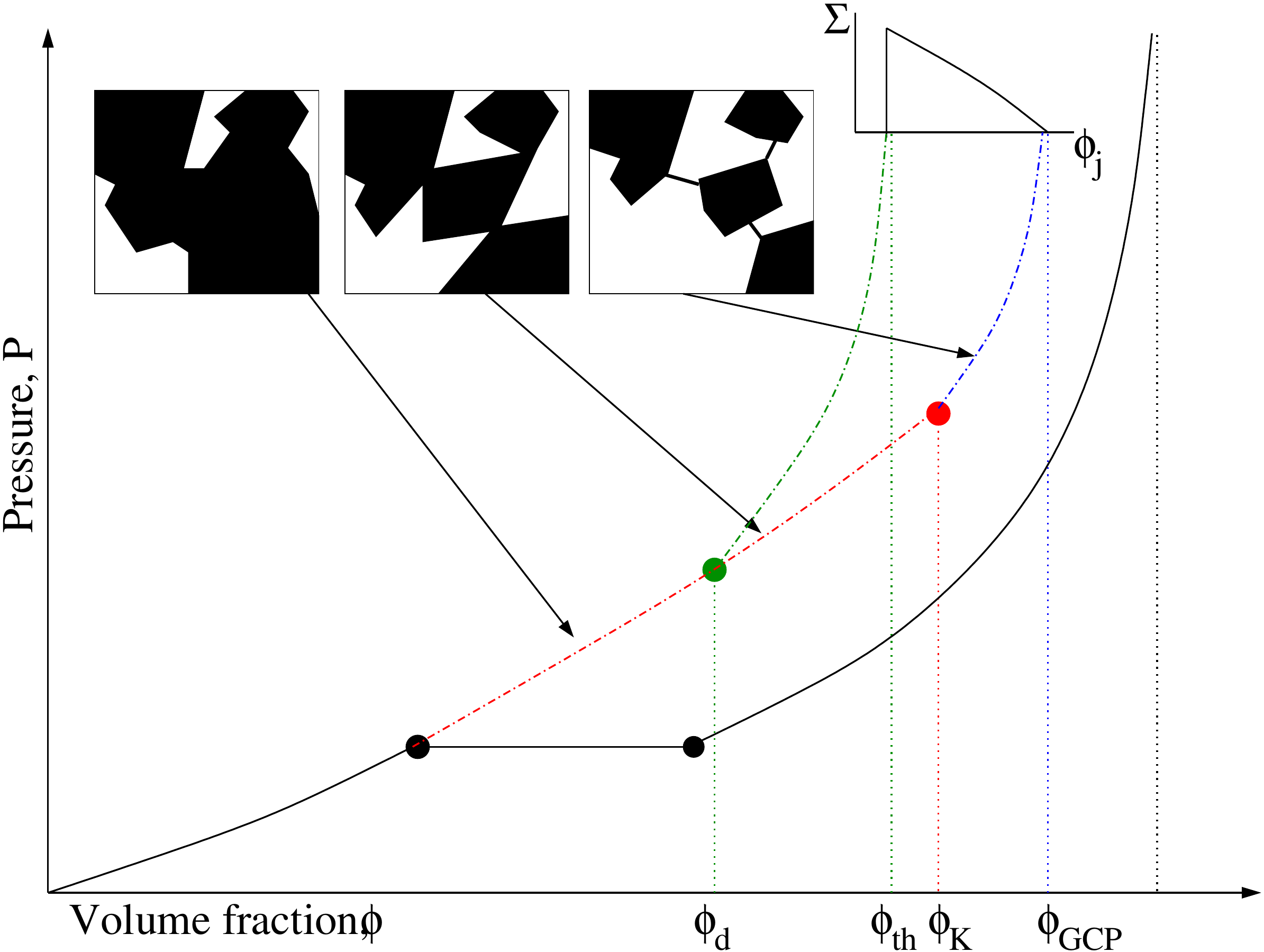} \nolinebreak[4]
\includegraphics[width = 0.5\textwidth]{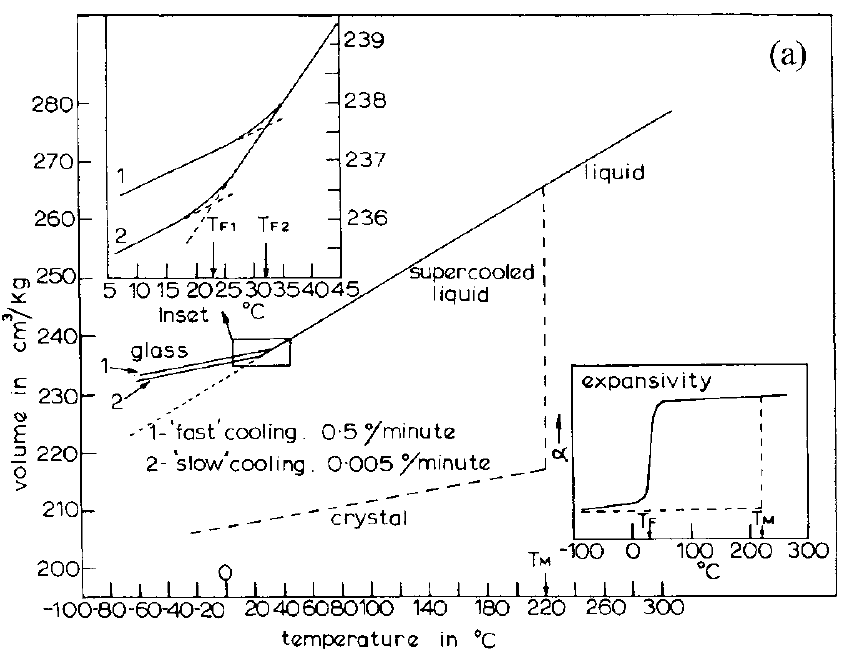}
\caption{\emph{(Left panel)}: pressure vs. packing fraction phase diagram of 
hard spheres as proposed in \cite{parisizamponi}. The slowest compression rates 
lead to a crystalline packing with maximal density $\varphi_{FCC}$. Higher, but 
still fairly low rates allow one to go into the ``supercooled'' regime all the 
way down to the Kauzmann transition, and then to $\varphi_{GCP}$ which is the 
jamming point for the ideal glass and corresponds to the maximal density for an 
amorphous packing. Higher rates select a glassy state between $\varphi_d$ 
(analogous to $T_{MCT}$) and $\varphi_K$ in analogy with an annealing protocol 
for a glass, and these states have jamming densities in a continuous interval 
between $\varphi_{th} = \varphi_j(\varphi_d)$ and $\varphi_{GCP} = 
\varphi_j(\varphi_K)$. \emph{(Right panel)}: specific volume vs. temperature 
phase diagram of a generic glass former. The similarity with the left panel is 
manifest if one identifies $1/\varphi = v$ and $P=1/T$. Reprinted from 
\cite{Dy06}.\label{fig:diagcomparison}}
\end{center}
\end{figure}

The idea that the protocol dependence of glasses and jamming systems could be 
treated on the same footing was first introduced in \cite{liunagel98}, where a 
unified phase diagram was proposed, see figure \ref{fig:jamming}. In this 
picture, jamming systems could be viewed as glasses of repulsive 
particles\footnote{The presence of an at least soft-core repulsion is of course 
necessary to induce a jamming transition.} quenched down to zero temperature and 
then compressed until mechanically rigid. This would allow for a unified 
treatment of the two problems.\\
Within the RFOT approach, the implementation of this program consist in 
identifying jammed packings with the endpoints of metastable glassy states at 
$T=0$, $\varphi=\varphi_j$, and infinite pressure (see figure 
\ref{fig:diagcomparison}). This is more easily done in the celebrated hard 
sphere (HS) \cite{simpleliquids} model where temperature plays no role and the 
only control parameter is the packing fraction $\varphi$, although it is 
possible to apply it also to soft harmonic spheres. This approach to the jamming 
transition was first reviewed in \cite{parisizamponi}, and allows one to treat 
jammed packings in a purely static fashion, thanks to the static nature of 
metastable glassy states which is at the very heart of RFOT.\\
In last few years, this line of research has produced remarkable results. It was 
shown in particular that the HS model could be solved exactly in the infinite 
dimensional, MF limit $d\to \infty$ \cite{KPZ12}, employing the replica method 
which is the natural tool for treating RFOT models \cite{replicanotes,MP09} 
(more on this in the following), and that this exact solution could accurately 
predict the critical exponents $\gamma, \theta$ and $\kappa$ along with the 
isostaticity property of packings \cite{KPZ12,KPUZ13,CKPUZ13,nature}. All these 
results were obtained only from first principles and to this date embody the 
greatest accomplishment of RFOT, showing that the initial conjecture of 
Kikpatrick, Thirumalai and Wolynes 
\cite{KW87,KW87b,kirkpatrickthirumalai,KT88,KTW89} is 
indeed realized in a strong sense in the MF limit.

Since the jamming transition and the RFOT predictions about it concern the 
properties of metastable glassy states at infinite pressure, we are going to 
talk more, and with more detail, about them in the following.

\subsection{Theoretical approaches to aging}
As we mentioned in the previous paragraph, a satisfactory ``theory of aging'' 
must be able, given a certain preparation protocol, to predict the properties of 
the glass so produced. Since a glass is out of equilibrium, using standard 
statistical mechanics would only give back trivial results relative to the 
equilibrated supercooled liquid, which means that one must in principle resort 
to off-equilibrium dynamical tools. In this case, to predict the properties of 
an aged glass, one must 
\begin{enumerate}
\item Write the equations for the dynamical process that reproduces the protocol 
under consideration,
\item solve them and compute the values of observables from the solution.
\end{enumerate}
Specifying to our case, in the case of brutal quenching protocols 
\cite{Struik78} one needs to study a dynamical process starting from a random 
initial configuration, while in the case of the annealing protocols (like those 
employed for ultrastable glasses) \cite{Kovacs63} one must consider a dynamics 
starting from an initial configuration equilibrated at $T_f$ 
\cite{CugliandoloDyn}.

This program does not look easy, especially considering the fact that a theory 
for the dynamics of glass formers is still lacking. MCT is by construction a 
theory for equilibrium dynamics, so it cannot be employed in this setting, and 
the recently proposed MF approach of \cite{infdynamics} is still in its infancy. 
Nevertheless, something has been done, and we review briefly those efforts.

\subsubsection{RFOT \label{subsec:RFOTaging}}
We mentioned in section \ref{subsec:dynamics} that the MCT equations are 
formally identical to the equations for the equilibrium dynamics of the PSM 
\cite{pedestrians}. However, while the MCT equations are derived assuming 
equilibrium \emph{ab initio}, in the case of the PSM the derivation is 
completely general, which means that so obtained dynamical equations have 
general validity and can be employed for aging studies.\\
Those equations read \cite{pedestrians}
\ba
\dpart{C(t,t_w)}{t} &=& \mu(t)C(t,t_w) + 2TR(t_w,t) + \frac{p}{2} 
\int_{-\infty}^{t_w}dt'\ C^{p-1}(t,t')R(t_w,t') \label{eq:agingPSM1}\\
&& + \frac{1}{2}p(p-1)\int_{-\infty}^t dt'\ 
R(t,t')C^{p-2}(t,t')C(t',t_w),\nonumber\\
\dpart{R(t,t_w)}{t} &=& \mu(t)R(t,t_w) + \delta(t-t_w) \label{eq:agingPSM2} \\
&& + \frac{1}{2}p(p-1)\int_{t_w}^t dt'\ 
R(t,t')C^{p-2}(t,t')R(t',t_w),\nonumber\\
\mu(t) &=& T + \frac{p}{2} \int_{-\infty}^t dt'\ C^{p-1}(t,t')R(t,t'), 
\label{eq:agingPSM3}
\ea
where the correlation function $C$ and \emph{response function} $R$ have been 
defined
\beq
C(t,t_w) \equiv \frac{1}{N}\sum_{i=1}^N\dav{\thav{\sigma_i(t)\sigma_i(t_w)}} 
\qquad R(t,t') \equiv \frac{1}{N}\sum_{i=1}^N 
\dfunc{\dav{\thav{\s_i(t)}}}{h_i(t')},
\eeq
($h_i(t)$ is a space and time dependent perturbing magnetic field) and 
$\dav{(\bullet)}$ denotes an average over the quenched disordered couplings 
\cite{pedestrians}. In an equilibrium situation, the correlation and response 
functions are linked by the fluctuation-dissipation theorem (FDT) 
\cite{CugliandoloDyn}
\beq
R(t,t_w) = \frac{1}{T}\dpart{C(t,t_w)}{t_w},
\label{eq:FDT}
\eeq
while in an aging situation they have to be treated separately. The solution for 
these equations in the aging regime was first described in 
\cite{CugliandoloKurchan93} and is reviewed in great detail in 
\cite{CugliandoloDyn}, so here we just summarize the most relevant points.\\
The solution exhibits an aging phenomenology just like the one reported in 
figure \ref{fig:aging}: at short times $\t = t-t_w$, the dynamics is TTI and the 
FDT holds. On intermediate timescales the dynamics remains arrested around a 
plateau whose height can be shown to be equal to the ``size'' of the threshold 
states at temperature $T$ \cite{CugliandoloKurchan93}, after which the system 
then decorrelates on a timescale $\t \simeq \t_0(t_w)$. However, one can also 
observe that the length of the plateau $\t_0(t_w)$ increases indefinitely with 
$t_w$, so that the system is effectively aging forever and the dynamics never 
crosses over to an equilibrium one (which would require that $\t_0(t_w)$ 
saturate to a $t_w$ independent value, restoring TTI). This is no surprise, 
since the system needs activation the leave the threshold level in the FEL, 
which is forbidden by construction in the PSM. In fact, the aging dynamics of 
\cite{CugliandoloKurchan93} corresponds to the slow descent of the system 
towards a threshold minimum in the FEL \cite{KurchanLaloux96}, which however is 
reached only asymptotically in $t_w$. The system descends effectively forever, 
moving along ridges and visiting stationary points with a lower and lower number 
of unstable directions, thereby slowing down more and more as time passes, 
causing the increase of $\t_0$ with $t_w$.\\
The most relevant prediction of the aging dynamics defined above consists in the 
fact that, on timescales much larger that $t_w$, a generalized version of the 
FDT holds \cite{CugliandoloKurchan93}. One can define a 
\emph{fluctuation-dissipation ratio} (FDR) in the following way 
\cite{berthierbiroli11}
\beq
R(t,t_w) = \frac{X(t,t_w)}{T}\dpart{C(t,t_w)}{t_w}.
\label{eq:FDR}
\eeq
This can be done in general, but in MF spin glass models like the PSM, one can 
observe that
\beq
X(t,t_w) \simeq x(C(t,t_w))
\eeq
where $x$ is a generic function. Since in the case of the PSM (and RFOT systems 
in general) the decay of the correlator is two-step, the function $x$ 
effectively reduces to two numbers: $x=1$ when $C$ is large and both TTI and the 
FDT are valid, and $x=X_{\infty}$ when $C$ is small and decorrelation sets in. 
This invites us to define an \emph{effective temperature}
\beq
T_{\rm eff} = \frac{T}{X_{\infty}},
\eeq
so that the aging system can be visualized as a system at equilibrium, only with 
a temperature $T_{\rm eff}$ different from the bath temperature $T$ 
\cite{CugliandoloKurchan97,FranzVirasoro00}.

\begin{figure}[tb!]
\begin{center}
\includegraphics[width = 0.7\textwidth]{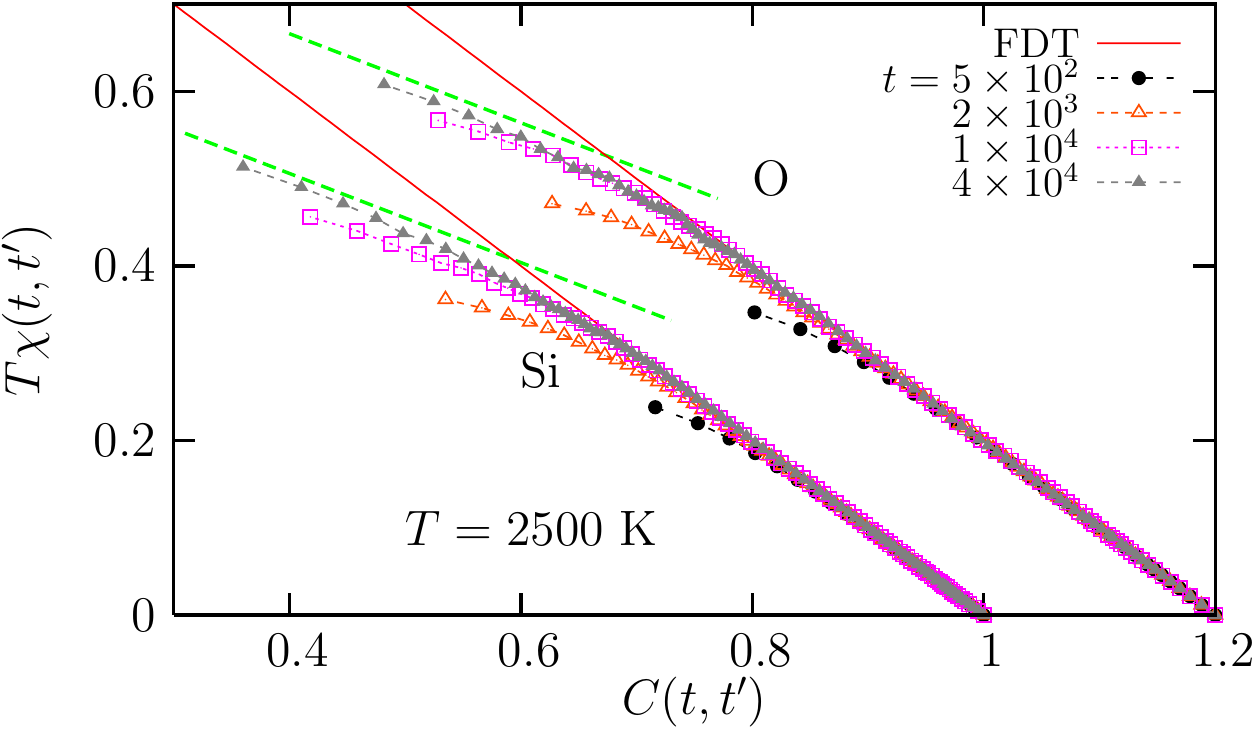}
\caption{Susceptibility vs. correlation plot for a simulated SiO$_2$ glass. For 
large times, the numerics for both atomic species converge smoothly to a two 
straight-line plot with $X_\infty \approx 0.51$, which yields $T_{\rm eff} 
\approx 4900K$. Reprinted from \cite{Berthier07}.\label{fig:teff}}
\end{center}
\end{figure}

To measure the FDR, one can define a susceptibility
\beq
\chi(t,t_w) \equiv \int_{t_w}^t dt'\ R(t,t'),
\eeq
which together with the \eqref{eq:FDT} would imply
\beq
T\chi(t,t_w) = C(t,t)-C(t,t_w),
\eeq
so that a plot of $\chi(t,t_w)$ vs. $C(t,t_w)$ is a straight line with slope 
$-1$. If the FDT is not valid, but an effective temperature can be defined, one 
would expect to see a slope of  $-X_\infty$ when $C$ is small. This prediction 
has been validated in numerical simulations of realistic models of glass 
formers, see for example \cite{BarratKob99,Berthier07} and figure 
\ref{fig:teff}.

The temperature defined through the \eqref{eq:FDR} can be shown to have a lot 
more physical meaning that one could assume at first sight. In particular, it 
can be shown to possess all the properties required from a temperature as a 
state variable: it can be measured with a suitable thermometer and controls the 
direction of heat flows, as a ``real'' temperature is supposed to do 
\cite{CugliandoloKurchan97}. Moreover, a whole thermodynamics for metastable 
glasses can be in principle built around the concept of effective temperature, 
see \cite{leuzziglassy}.\\
The presence of an effective temperature is intimately linked with the presence 
of two relevant timescales for equilibration: the fast degrees of freedom are 
able to equilibrate (on a timescale $\t_\beta$),  with the bath at temperature 
$T$, but the slow degrees of freedom are unable to do so since their 
equilibration time $\t_\al$ is too large. However, they can be conceived as 
``quasi-equilibrated'' at $T_{\rm eff}$ \cite{Kurchan05}. We will see that the 
replica method allows one to give more solidity to the notion of effective 
temperature.

This effective temperature approach has been for much time the standard RFOT 
approach to aging (we refer to \cite{CrisantiRitort03} for a review), but it 
clearly suffers from some problems. First of all, not all experiments reveal the 
presence of a well-behaved effective temperature (see for example 
\cite{Greinert06}), and we will also see that the presence of activation 
mechanisms can lead to negative effective temperatures, a clearly paradoxical 
result.\\
However, the biggest weakness is of conceptual nature: since activation is 
prohibited, the system is always and forever stuck at the threshold level, 
without ever being able to penetrate below it. This means that this kind of 
dynamics is only capable of reproducing brutal quenching protocols, and fails 
completely when one tries to describe annealing protocols such as the ones we 
defined in section \ref{sec:aging}. A satisfactory theory of aging must be able 
to account for those protocols and the appearance, in recent years, of 
ultrastable glasses analogue to glasses prepared with very slow annealing 
protocols \cite{Sw07,SEP13} now makes it even more of a mandatory task, for 
models and theorists, to go ``beyond the threshold'' \cite{CugliandoloDyn}.

Since within RFOT the system cannot go beyond the threshold by itself, we must 
put it there ourselves, namely we must consider a dynamics whose initial 
configuration is not random, but sampled from the canonical Boltzmann-Gibbs 
distribution at $T_f$. This program can actually be implemented in the case of 
the PSM, wherein the equations for such a dynamics have been derived in 
\cite{BarratBurioniMezard}. Their solution shows, unsurprisingly, an equilibrium 
dynamics (with FDT and TTI both valid) much like the MCT one, with the only 
difference that the system relaxes inside a single metastable state selected by 
the \eqref{eq:feq} instead of equilibrating in the ergodic supercooled liquid. 
The system of course never gets out (since activation is forbidden), so one has 
a well-defined plateau regime for long times, from which the interesting 
in-state observables can be computed\footnote{The quenching and annealing 
dynamics can be treated in a unified manner within the dynamical TAP formalism, 
see \cite{Biroli99}.}.\\
This is exactly what we want, but anyway there is a problem, namely that 
implementing the program of \cite{BarratBurioniMezard} in real liquids looks 
like a difficult task. In principle a generalization of the formalism employed 
in \cite{infdynamics} could be developed, but it doesn't look at all easy, since 
the derivation of \cite{infdynamics} for just an equilibrium, MCT-like dynamics 
is already quite computationally heavy. In this thesis we propose a much 
simpler, but equally predictive alternative.

\subsubsection{DFT}
\begin{figure}[tb!]
\begin{center}
\includegraphics[width = 0.6\textwidth]{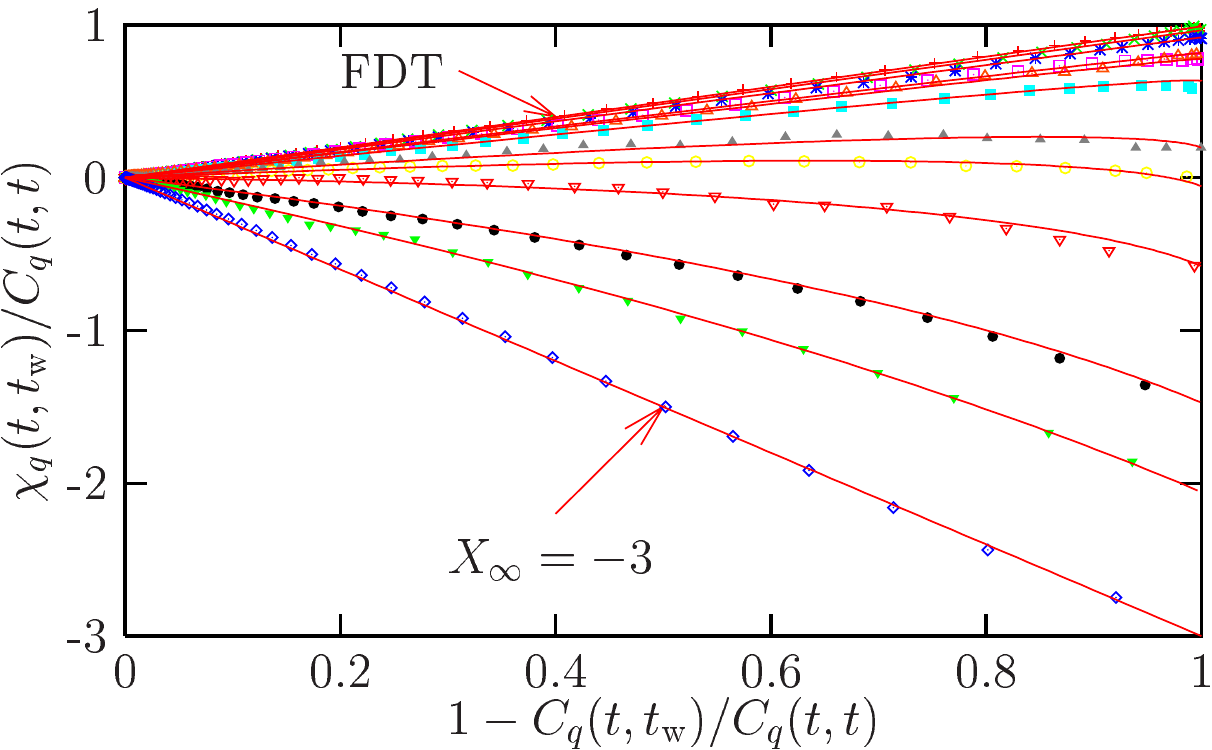}
\caption{Susceptibility vs. correlation plots for the FA model in $d=3$. Symbols 
are numerical data while lines are theoretical calculations. Wave-numbers 
decrease from top to bottom, with the bottonmost corresponding to $q=0$. A 
negative FDR is observed. Reprinted from \cite{Mayer06}.\label{fig:TeffKCM}}
\end{center}
\end{figure}

The aging dynamics of KCMs has been extensively reviewed in \cite{Leonard07}, so 
we will be brief. As we mentioned before, they have the big advantage of being 
well defined models which can be treated analytically and have a real space 
structure, and yet exhibit a remarkably rich phenomenology. In particular, they 
provide an ideal playground to study activation mechanisms.\\
The possibility of activation in the dynamics of KCMs can lead to very 
interesting results in terms of FDT violation. The case of the FA model, 
equation \eqref{eq:FAH}, in particular, has been extensively studied 
\cite{Mayer06}. The dynamical correlations considered in the FA are the Fourier 
transforms of the local mobility correlation functions
\beq
C_q(t,t_w) = \sum_i\sum_j \thav{n_i(t)n_j(t_w)}_c e^{-i q(r_i-r_j)}.
\eeq
That for $q=0$ correspond to the dynamical correlation function of the energy 
\cite{Mayer06}
\beq
C_0(t,t_w) \propto \thav{H(t)H(t_w)}_c.
\eeq
Remarkably, the structure of FDT violations is again found to be very simple. In 
particular, one has, for $d>2$ and $q=0$, a well-defined long-time FDR 
$X_\infty$, see figure \ref{fig:TeffKCM}. Notice how the FDR depends on the 
wave-number (and thus on the lengthscale) considered, an effect which could not 
be observed in MF models where space plays no role. Unsurprisingly, to find a 
well defined FDR one must consider the $q=0$ limit, corresponding to global 
observables.\\
However, one can see from figure \ref{fig:TeffKCM} thet
$$
X_\infty = -3,
$$
for every $d \geq 2$. This yields a negative effective temperature, quite a 
paradoxical result.\\
Despite being apparently paradoxical, a simple argument \cite{Mayer06} shows 
that this behavior is indeed to be expected, due to the possibility of 
activation. As we said before, the $C_0(t,t_w)$ is related to the fluctuations 
of the energy, whose conjugated variable is temperature. If one raises the 
temperature of the system, the timescale for activation goes down following 
Arrhenius' law, equation \ref{eq:arrh}, speeding up relaxation and allowing the 
system to descend in the PEL. So the response of the energy to a temperature 
step is negative, causing the negative FDR observed in \cite{Mayer06}. This 
example shows how the effective temperature is a MF-bound concept, and how it 
may not be viable for the description of aging in real glassy systems 
(especially strong ones) wherein activation is expected to play a role. This 
view is corroborated by the fact that relaxation mechanisms of KCMs seem to be a 
lot more mean-field like for more ``fragile'' KCMs, like the East Model 
\cite{EastModel}. In that case it can be seen that the relaxation proceeds in a 
step-like manner with multiple time sectors, each of whom can be associated to 
an effective temperature \cite{Leonard07}, as argued in \cite{Kurchan05}.

\begin{figure}[t!]
\begin{center}
\includegraphics[width = \textwidth]{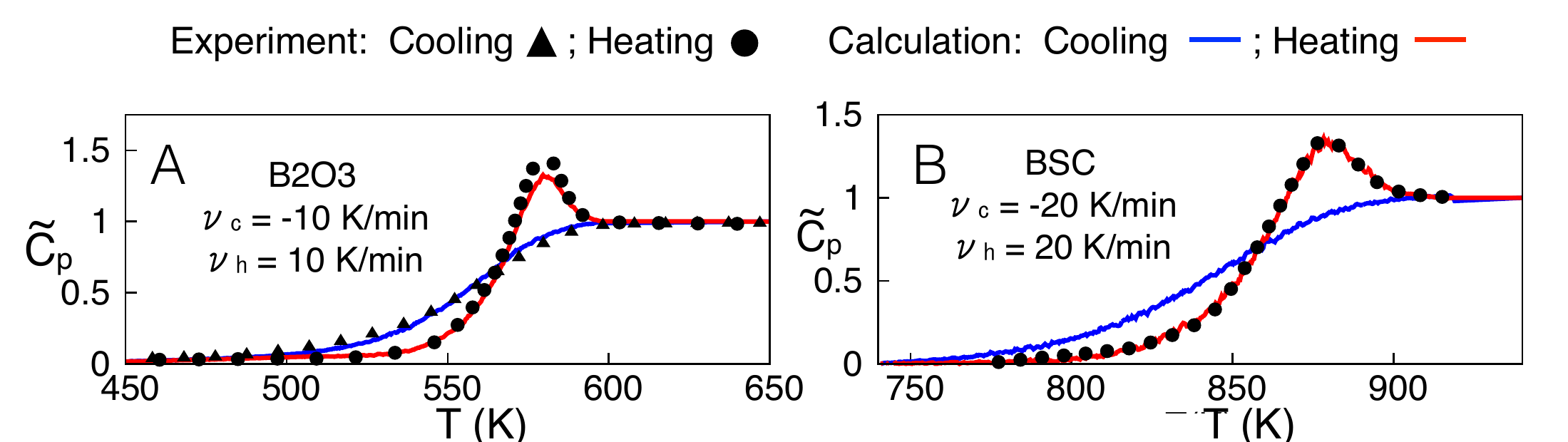}
\caption{Heat capacities of fragile glass formers (boron oxide in the left 
panel, borosilicate in the right panel) measured from DSC experiments, compared 
with predictions from the non-equilibrium dynamics of the East model 
\cite{EastModel}. $\nu_{c/h}$ are the cooling/heating rate respectively. A very 
good agreement between experiment and theory can be observed. More plots can be 
found in \cite{KeysGarrahan13}.\label{fig:calorimetricKCM}}
\end{center}
\end{figure}

Besides the description of activation mechanisms, Dynamical Facilitation Theory 
can also well describe the properties of glasses in the plateau regime. In 
particular, it has not remained silent after the appearance of ultrastable 
glasses: in \cite{KeysGarrahan13}, the non-equilibrium dynamics of the East 
model for a cycling temperature protocol is studied, with the express purpose of 
reproducing the results of \cite{Sw07} (in particular the left panel of figure 
\ref{fig:ultraexp}) and DSC experiments in general. We report a sample of the 
results in figure \ref{fig:calorimetricKCM}.
A very good agreement between theory and experiment is found, but we warn that 
this comes at a price: a rigorous mapping between actual glass formers and KCMs 
is still lacking, so to obtain an agreement like the one in figure 
\ref{fig:calorimetricKCM} the East model must be ``tuned'' suitably. In 
particular, the values the energy scale $J$ for mobility, the onset temperature 
$T_0$ whereupon the glassy slowdown starts to manifest, the microscopic 
relaxation time $\t_0$ and the fractal dimension of heterogeneities $d_f$ must 
all be fixed by hand, either from reversible transport data \cite{Elmatad09} or 
from atomistic simulation results; and even at that point, an additional 
optimization over one remaining free parameter is required 
\cite{KeysGarrahan13}.

In summary, KCMs do have a more true-to-life nonequilibrium dynamics with 
respect to the RFOT approach, but they also suffer from their nature of 
effective models, whose physics is imposed from the outside instead of being 
derived from first principles. Our state following approach, by contrast, can 
produce qualitatively accurate results starting just from the microscopic 
interaction potential of a glass former of choice, without the need for 
phenomenological arguments or scaling treatments.

\section{Driven dynamics and rheology \label{sec:shear}}

After a glass has been prepared, one can of course do a lot more than just let 
it age. Namely, one can also perturb the glass with some external drive and 
measure its response to such a drive. There is no shortage of drives one could 
supply to a glass: electric currents, electromagnetic fields, scattering 
particles etc., but here we are concerned with mechanical drives, the simplest 
being shear strain. The study of the rheology of glasses, especially for low 
temperatures deep in the metastable glass phase ($T \ll T_g$), ties with the 
study of mechanical response and plasticity in amorphous solids in general (see 
for example 
\cite{LN01,RTV11,IBS12,MaloneyLemaitre06,NicolasMartens14,NicolasBocquet14}) and 
also in pastes, foams, colloids etc. \cite{Cates03}, so it goes without saying 
that it is a field relevant for multiple engineering and material science 
applications.

\begin{figure}[tb!]
\begin{center}
\includegraphics[width = 0.6\textwidth]{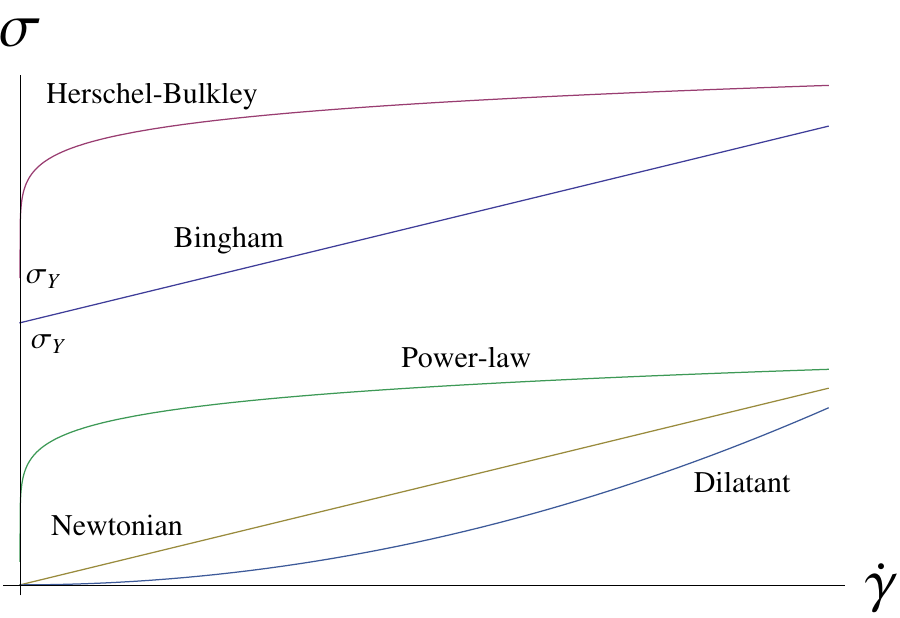}
\caption{The most common models of flow in fluids and plastics. Plastics 
(Herschel-Bulkley and Bingham) are distinguished by the presence of a 
\emph{yield stress} $\s_Y$, namely a finite level of stress that must be 
supplied to the system in order to observe flow, while fluids flow for any value 
of the stress. \label{fig:flowcurve}}
\end{center}
\end{figure}

Since we are focusing on glasses, which are solids, let us consider a cube of 
glass, and apply a shear displacement on its topmost face, along the $x$-axis. 
Every point $\bx$ in the cube is transformed in another point $\bx'$ the 
following way
\ba
x' &=& x+\gamma y,\nonumber \\
y' &=& y, \label{eq:affinetransf}\\
z' &=& z, \nonumber
\ea
Where $\g = \D x/L$, i.e. the displacement of the topmost face divided by the 
side $L$ of the cube. The difference $\bx'-\bx$ is a vector field, called the 
\emph{displacement field} \cite{Cates03} and denoted as $\mathbf{u(\bx)}$. One 
then has
\beq
\dpart{u_x}{y} \equiv \g_{xy} = \gamma
\eeq
where $\gamma_{\al\be}$\footnote{In the following we will denote space 
directions with Greek indices $\al,\be,\g,\d \dots$ and particles with Latin 
indices $i,j,k\dots$.} is the \emph{strain tensor}. Such a shear geometry is 
referred to as \emph{simple shear}, and corresponds to the simplest possible 
structure of the strain tensor
\beq
\gamma_{\al\be} = \begin{pmatrix}
0 & \g & 0\\
0 & 0 & 0\\
0 & 0 & 0
\end{pmatrix}
\eeq
Of course, one could apply deformations in many different ways, each 
corresponding to a different choice of shear geometry (a very popular one is the 
Couette geometry \cite{Larson99}) and more complicated displacement fields, 
which in turn requires a full tensorial description of the strain 
\cite{Cates03}, but in this thesis we focus on simple shear.\\
If our material is a solid, we reasonably expect from it an \emph{elastic 
response} when the shear is small enough: the material responds to the 
displacement with a \emph{shear stress}\footnote{In reality, the independent 
variable is actually the stress (essentially the force we supply to the 
material), and the displacement (essentially the strain) depends on it. Most 
simulations are however performed under strain control, although stress control 
can in principle be employed, see for example \cite{Dailidonis14}.} proportional 
to the strain \cite{Cates03}
\beq
\s_{xy} = \s = \mu\g,
\eeq 
which is essentially Hooke's law cast in a shear-strain setting. The quantity 
$\mu$ is called the \emph{shear modulus} or \emph{elastic modulus}.\\
All these relations are valid at equilibrium. To be general, and remembering the 
discussion of section \ref{subsec:glasstrans} and \cite{Sausset10}, we can 
include time into the description and write
\beq
\s(t) = G(t-t')\gamma,
\eeq
where $t'$ is the time whereupon the strain is applied, and we have defined a 
response function $G(t)$. With this definition we can treat both fluids and 
solids: in a fluid, the $G(t)$ will vanish after some time (intuitively, the 
relaxation time $\tau_R$) and the fluid will absorb the stress (think of honey), 
while for a solid one will have a nonzero limit for large times, 
$\lim_{t\to\infty}G(t) = \mu$.\\
Let us now assume that we perform an experiment with time-dependent strain. We 
approximate this strain as formed by a succession of strain steps, each one at a 
time $t_i$, $\d\g(t_i) \equiv \g(t_{i+1})-\g(t_i)$, so that
\beq
\s(t) = \sum_i G(t-t_i)\d\g(t_i) \simeq \sum_i G(t-t_i)\dot{\g}(t_i)\d t_i
\eeq
which in the continuum limit becomes 
\beq
\s(t) = \int_0^t\ G(t-t')\dot{\g}(t')dt',
\eeq
where $\dot{\g}(t)$ is the \emph{strain rate}. This relation was derived in the 
elastic, low-strain limit, and fittingly has a strong linear response flavor. In 
the general case, the stress response will be a more complicated functional of 
the shear rate
\beq
\sigma(t) = \mathcal{F}[\dot{\g}(t)].
\eeq
This last relation is called a \emph{constitutive equation}. The aim of a 
rheology experiment (theory) is to measure (compute) the constitutive equation 
for a given material, given a shear protocol $\dot{\g}(t)$ 
\cite{Cates03,Larson99}.\\
In the case of steady shear $\dot{\g} = const$ in particular, one gets
\beq
\s(t) = \dot{\g}\int_0^t G(t')dt',
\eeq
and the ratio of shear stress over shear rate is just the viscosity $\eta$, so 
in the long time limit one has \cite{Cates03}
\beq
\eta = \frac{\s}{\dot\g} = \int_0^\infty G(t)dt.
\eeq
This linear relation is valid in the limit of small rates. In general the curve 
$\s(\dot{\g})$, called the \emph{flow curve}, will have a non-linear form. See 
figure \ref{fig:flowcurve} for the most common flow curves found in plastics and 
fluids \cite{Larson99}.\\
The flow curve characterizes the rheology of the material in the long time 
steady state \cite{Cates03}: for every value of the stress (essentially the 
force we supply to the material) the flow curve tells us the shear rate 
(essentially the velocity of flow) obtained with that stress. Most glasses and 
amorphous materials in general exhibit a \emph{Herschel-Bulkley} \cite{Larson99} 
flow curve, namely
\beq
\s(\dot{\g}) = \s_Y + K\dot{\g}^n,
\eeq
with an exponent $n$ close to $1/2$ 
\cite{NicolasMartens14,PrincenKiss89,LemaitreCaroli09}. Since the flow curve 
exists only for $\s>\s_Y$, a glass has a finite \emph{yield stress}: it flows 
only when the stress is high enough to deform it. For lower stresses, it 
responds elastically like a solid \cite{RTV11}.\\
If then one remembers that $\eta(\dot\g) = \s(\dot\g)/\dot{\g}$, we can see that 
glass formers typically exhibit \emph{shear thinning}: viscosity (and thus the 
relaxation time) goes down with the shear rate \cite{Cates03}, a behavior common 
to a vast class of fluids \cite{berthierbiroli11}. Another model with 
yield-stress behavior is the \emph{Bingham plastic} \cite{Larson99}, where $n=1$ 
and viscosity is constant.

\subsection{Athermal startup shear protocols \label{subsec:qsshear}}
Most experiments and simulation on glasses employ a \emph{startup shear} 
protocol, where a simple shear strain is quasi-statically applied to the glass 
and the stress response is measured, until the glass reaches the yielding point 
whereupon it starts to flow steadily. Interestingly, the steady state usually 
depends only on the driving, which means that the material has forgotten its 
history and aging has stopped, a process referred to as \emph{rejuvenation} 
\cite{berthierbiroli11,BerthierBarrat00}.

To be more specific, the glass is typically prepared by quenching an 
equilibrated configuration down to zero temperature, dropping the system in an 
inherent structure. Once preparation is complete, a strain is applied in small 
steps, either by implementing a transformation like the \eqref{eq:affinetransf} 
on particle coordinates, or by keeping the glass former between two walls moving 
in opposite directions \cite{RTV11,Kou12}; in both cases, the strain acts as an 
effective deformation of the PEL. After each step, the potential energy is 
minimized and the system allowed to settle in a new inherent structure (which 
may or may not be different from the initial one), before applying shear again. 
Such a protocol is called an Athermal Quasi Static (AQS) protocol since thermal 
fluctuations play no role and the system is allowed to equilibrate in the PEL 
after each step ($\dot{\g} \to 0$) \cite{MaloneyLemaitre06}.

\begin{figure}[htb!]
\begin{center}
\includegraphics[width = 0.6\textwidth]{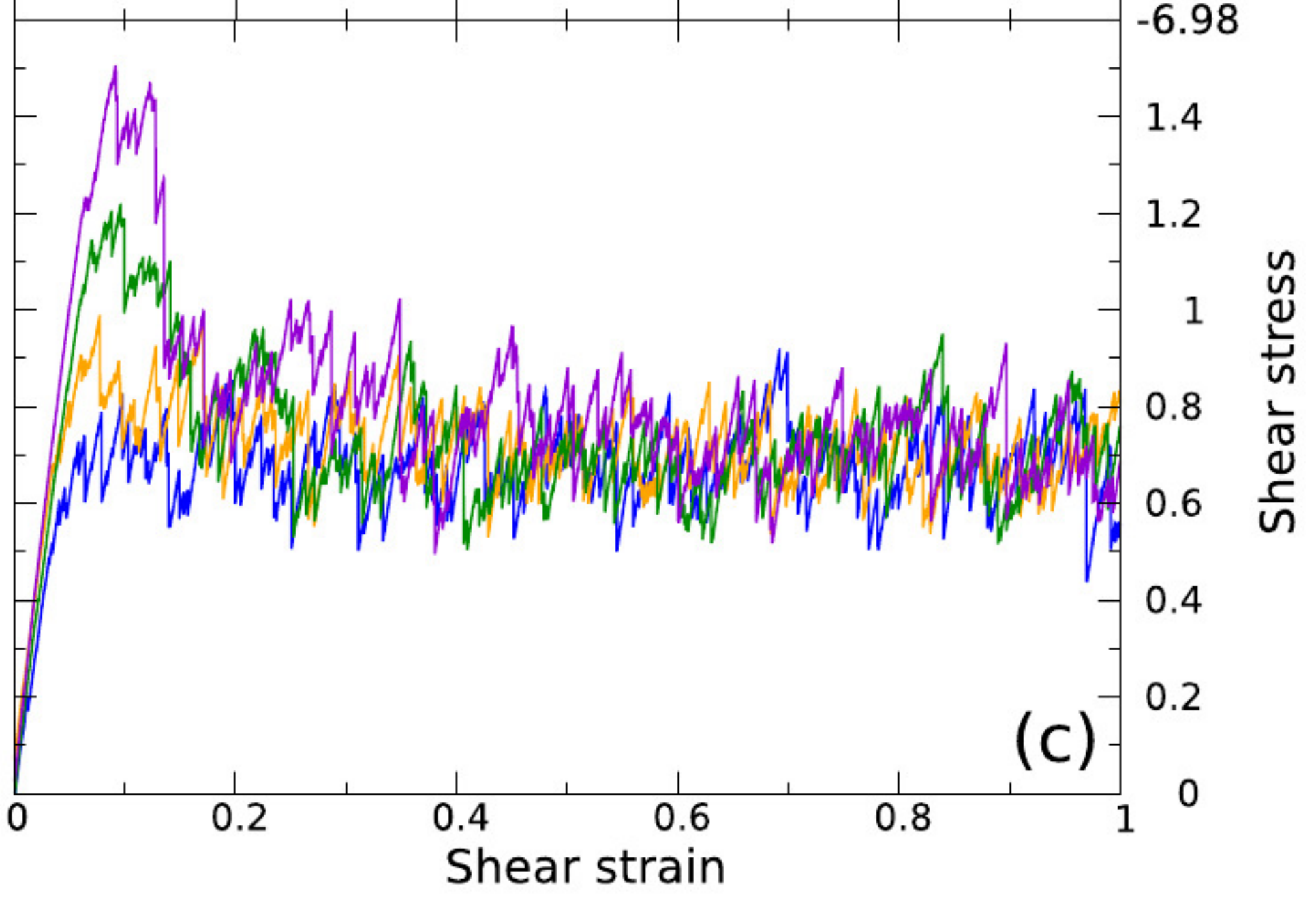}
\caption{Stress-strain curve for a Lennard-Jones system with Wahnstr\"om 
parameters in an AQS protocol. The quench rates, in Lennard-Jones units, are 
$2\times10^{-2}$ (blue), $2\times10^{-3}$ (yellow), $2\times10^{-4}$ (green), 
$2\times10^{-5}$ (violet). A linear regime is observed for small strain, 
followed by a stress overshoot and then a steady flow. A slower quench 
corresponds to higher shear modulus and a more prominent stress overshoot. 
Reprinted from \cite{RTV11}. \label{fig:shearstrain}}
\end{center}
\end{figure}
The observed stress-strain curves are reported in figure \ref{fig:shearstrain}. 
One can observe, for small strain, a linear regime wherein response is elastic 
and a shear modulus can be defined, followed by an overshoot in stress, and then 
a flowing steady state \cite{RTV11,Petekidis03}. The magnitude of the 
overshoot\cite{Kou12,RTV11} and the shear modulus \cite{Laurati11,RTV11} both 
increase when the quench is slower, which is reasonable: slower quenches 
correspond to deeper minima in the PEL and thus more stability and rigidity. 
This qualitative behavior is remarkably general 
\cite{RTV11,Kou12,MaloneyLemaitre06}: it has been observed in systems that range 
from polymer glasses \cite{Wang09} to colloidal gels \cite{Laurati11} to 
metallic glasses \cite{FalkLanger04}. Moreover, in granular materials a 
phenomenon referred to as \emph{dilatancy} is observed: even though shear strain 
transformations such as the \eqref{eq:affinetransf} are supposed to preserve 
volume (and thus pressure), the pressure is found to increase quadratically with 
the strain \cite{Reynolds85,Tighe14}.

The origin of the stress overshoot can be better understood if one looks more 
closely at the stress-strain curves in figure \ref{fig:shearstrain}.
\begin{figure}[htb!]
\begin{center}
\includegraphics[width = 0.6\textwidth]{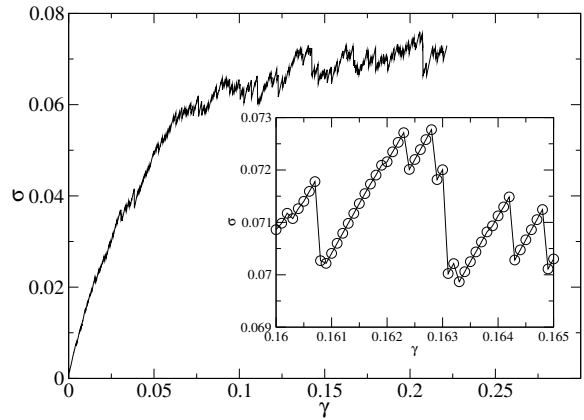}
\caption{Stress-strain curve for a simulated system of harmonic disks. In the 
inset a ``zoom in'' is reported, wherein one can observe that the curve is 
really made up of roughly linear, elastic segments separated by plastic events, 
or avalanches. Reprinted from \cite{MaloneyLemaitre06}. 
\label{fig:shearstrainlemaitre}}
\end{center}
\end{figure}
We report such a ``close up'' in figure \ref{fig:shearstrainlemaitre}. One can 
see that a typical AQS stress-strain curve is actually made up of short 
segments, wherein the response is almost perfectly linear and elastic energy is 
loaded into the material; these segments are separated by catastrophic events 
whereupon the stress drops down sharply and the energy is dissipated; these are 
called \emph{plastic events}, or \emph{avalanches}. In a PEL perspective, one 
can visualize the elastic part as a deformation of the inherent structure the 
system is in, which however maintains its identity. At a certain point, the 
inherent structure opens up along an unstable mode, stability is lost, and the 
system is kicked away, producing an avalanche until it finds a new 
minimum\footnote{Such a process is referred to as \emph{saddle node bifurcation} 
and is somewhat analogous to a spinodal point. Indeed, one can describe it using 
a simple cubic theory wherein the potential energy of the system is projected 
along the (almost) unstable mode, and interesting scaling predictions can be 
obtained, see \cite{KarmakarLerner10}.}. Interestingly, a stress overshoot is 
again observed at the end of elastic segments \cite{MaloneyLemaitre06}.

In the elastic segments, the system ``follows'' the inherent structure in 
strain, so the motion of particles will be generally made up of two 
contributions: the affine transformation due to the strain and the 
\emph{nonaffine} contribution  necessary to track the minimum and maintain 
mechanical stability \cite{LemaitreMaloney06}:
\beq
\bx'_i = \bx_i + \boldsymbol{\gamma}\cdot \bx_i + \by_i,
\label{eq:nonaffinetransf}
\eeq
where $\bx_i$ are the coordinates of particle $i$ in the unstrained inherent 
structure, $\bx'_i$ its coordinates in the strained one, and $\by_i$ is the 
nonaffine contribution. Stability within the original minimum requires that the 
force on each particle $i$, $\boldsymbol{f_i} = -\dpart{V}{\bx_i}$ be zero. The 
nonaffine contributions can be worked out by requiring that the forces stay zero 
when the strain is applied, with corresponds to imposing
\beq
\dtot{\boldsymbol{f}_i}{\g} = \dpart{\boldsymbol{f}_i}{\g} + 
\dpart{\boldsymbol{f}_i}{\bx_j} \cdot \dtot{\by_j}{\g} = 0.
\eeq
Where the repeated indices are understood to be summed over. One can then define 
two quantities
\beq
\boldsymbol{\Xi}_{i} \equiv \dpart{\boldsymbol{f}_i}{\g} \qquad \boldsymbol{v}_i 
\equiv \dtot{\by_j}{\g},
\eeq
called the \emph{mismatch force}\footnote{The mismatch force can be seen as the 
contribution of particle $i$ to the shear stress, derived with respect to its 
position, $\boldsymbol{\Xi}_i = \frac{\partial^2V}{\partial\g\partial\bx_i}$.} 
and the \emph{nonaffine velocity}, respectively \cite{LemaitreMaloney06}. 
Interestingly, the nonaffine forces are nonzero only for amorphous inherent 
structure configurations \cite{LemaitreMaloney06}, so they can be used as a 
measure of disorder. With these definitions one can solve for the nonaffine 
velocities:
\beq
\boldsymbol{v}_i = -\boldsymbol{\mathcal{H}}^{-1}_{ij}\cdot\boldsymbol{\Xi}_j 
\longrightarrow v_{i\al} = -\mathcal{H}^{-1}_{i\al j \be} \Xi_{j\be},
\label{eq:nonaffinev}
\eeq
where $\mathcal{H}_{i\al j\be}$ is the Hessian matrix (or dynamical matrix), 
that we had already encountered in the context of jammed packings.\\
Once the nonaffine velocities have been determined, the coefficients of the 
elastic theory \cite{LandauElasticity} for the glass can be readily obtained. 
Their expression is
\beq
\mu_n \equiv \frac{1}{n!}\frac{d^n}{d\g^n}\sigma(\{\bx(\g)\}) = 
\frac{1}{n!}\left(\dpart{ }{\g} + \boldsymbol{v_i}\cdot\dpart{ }{\bx_i}\right)^n 
\dtot{V(\{\bx(\g)\})}{\g}.
\label{eq:elastic}
\eeq
One can immediately observe that the presence of nonaffine velocities has no 
influence on the stress thanks to the minimum condition $\dpart{V}{\bx_i}=0$. 
However it impacts all elastic coefficients, in particular the shear modulus:
\beq
\mu = \dpart{\s}{\g} - \Xi_{i\al}\mathcal{H}^{-1}_{i\al j \be}\Xi_{j\be} = 
\mu_{a} - \mu_{na},
\label{eq:nonaffinemu}
\eeq
where $\mu_a$ is the \emph{Born term} for a pure affine deformation 
\cite{Born98}, and $\mu_{na}$ is the nonaffine contribution.

The presence of the shear overshoot can now be understood: a loss of stability 
is by definition associated with the appearance of a zero 
mode\footnote{Obviously the Hessian will always contain zero modes associated to 
symmetries, i.e. Goldstone modes. But they can be identified and removed 
easily.} in the Hessian matrix, whose inverse appears in the nonaffine 
contribution in the \eqref{eq:nonaffinemu}. So the nonaffine contribution will 
grow as the endpoint is approached, while the Born contribution stays always 
finite: at a certain $\g$ the shear modulus will be zero (and the stress-strain 
curve flat) and then drop until it becomes infinitely negative at the endpoint, 
causing the stress to go down in a steeper and steeper manner. The stress 
overshoot is thus an inevitable consequence of the loss of stability.\\
This is valid for the small segments associated with a single inherent structure 
as just described. In one wants to ``coarse grain'' this picture to the whole 
strain-stress curve, then one may argue that a basin containing different 
inherent structures (called a \emph{metabasin} \cite{heuer2008exploring}) should 
replace the inherent structure, and then the yielding transition to the steady 
state would correspond to a loss of stability in the whole metabasin. This 
picture is easy to grasp at a pictorial level, but it is obviously difficult to 
translate in a theory: what does exactly mean that a metabasin loses its 
stability and opens up? Is there a zero mode somewhere again, relative to the 
whole metabasin? Can one define and compute the associated Hessian matrix? Does 
the loss of stability of the metabasin imply a loss of stability for all the 
inherent structures therein contained?\\
We will see that these notions can be made more precise within replica theory.

\subsection{The yielding transition \label{subsec:yielding}}
The transition between the elastic regime and the steady state is referred to as 
``yielding transition'' and has lately been the subject of extensive study (see 
for example 
\cite{IlyinProcaccia15,KarmakarLerner10,KarmakarLerner10b,linlerner14,Kou12}). 
We already notice the fact that there is no clear definition for the yielding 
transition (much like the glass transition, in fact). A rheology-bound 
definition would suggest to choose the maximum of the stress $\s_Y$ and its 
associated $\g_Y$ as the yielding point. Others opt for a definition in terms of 
onset of energy dissipation (and consequently, avalanches) \cite{FioccoFoffi13}, 
or for a definition in terms of qualitative changes in the structure of the PEL 
(namely, in the statistics of barriers between inherent structures) after the 
transition \cite{KarmakarLerner10b}, but there is no general agreement. Even 
though it is evident, from figures \ref{fig:shearstrain} and 
\ref{fig:shearstrainlemaitre}, that there is a qualitative change in behavior 
between the ``elastic'' regime and the steady state, it is not easy to pin the 
exact point whereupon it happens.

\begin{figure}[htb!]
\begin{center}
\includegraphics[width = 0.5\textwidth]{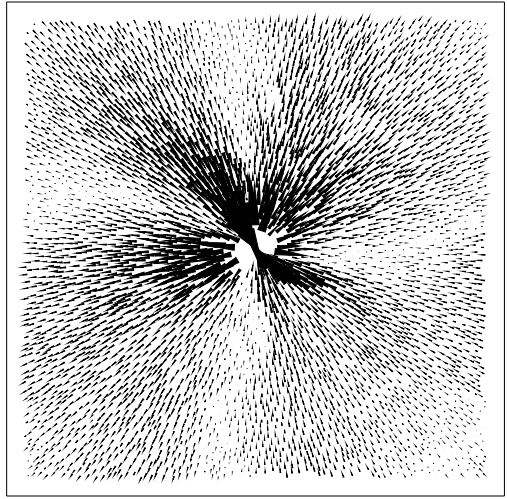}
\caption{The (almost) critical mode associated with one of the instabilities (or 
equivalently, stress drops) in figure \ref{fig:shearstrainlemaitre}. The 
quadrupolar structure is clearly discernible. Reprinted from 
\cite{MaloneyLemaitre06}. \label{fig:criticalmode}}
\end{center}
\end{figure}

In the next paragraph we will see that there are theoretical approaches that 
ensure a good \emph{macroscopic} description of yielding, in the sense that they 
do reproduce flow curves (figure \ref{fig:flowcurve}) with a yield-stress form. 
However, a ``theory of yielding'' has to to do more, in the sense that it has to 
provide a good \emph{microscopic} (or mesoscopic, at least) description of the 
transition and the ensuing flow.\\
Nowadays, it is generally agreed upon that flow can be thought of as a sequence 
of elementary, mesoscopic rearrangements that take place at well-defined points 
in the sample, called Shear Transformation Zones (STZ) \cite{ArgonSTZ}. Such a 
shear transformation will then induce a stress that will propagate elastically in 
the sample, in analogy with the nucleation of an Eshelby inclusion 
\cite{EshelbyInclusion}, inducing other STZs in a cascade and producing an 
avalanche. This picture is confirmed by a normal mode analysis of the Hessian 
matrix: the eigenvector (which contains the particle displacements associated 
with the mode\footnote{If one writes the inverse Hessian in spectral form, 
$\boldsymbol{\mathcal{H}}^{-1} = \sum_i 
\left|\boldsymbol{\psi}_i\right>\frac{1}{\lam_i}\left<\boldsymbol{\psi}
_i\right|$, the solution for the nonaffine velocities \eqref{eq:nonaffinev} 
becomes $\boldsymbol{v} = - \sum_i 
\left|\boldsymbol{\psi}_i\right>\frac{1}{\lam_i}\left<\boldsymbol{\psi}
_i\right|\left.\boldsymbol{\Xi}\right>$, so the nonaffine velocity is dominated 
by the term proportional to the critical mode when the system is near enough to 
the instability.}) of the critical mode associated with the instability shows 
strong localization properties \cite{HentschelKarmakar11} and a quadrupolar 
angular symmetry \cite{MaloneyLemaitre04,MaloneyLemaitre06} (see figure 
\ref{fig:criticalmode}), just like the displacement field induced by an Eshelby 
inclusion. Indeed, it has been argued in \cite{HentschelKarmakar11} that only 
low-lying localized modes contribute to plasticity, in the sense that they are 
the only ones that can produce divergencies in the coefficients of the elastic 
theory \eqref{eq:elastic}. This would exclude delocalized modes such as phonons 
from the excitations relevant for yielding, a very strong assertion.\\
The idea that flow is initiated in definite points is not new. The mechanism for 
failure in crystalline solids is indeed ruled by a population of topological 
defects, known as dislocations \cite{dislocations}, where structural failure 
manifests at the onset of flow. Amorphous solids, however, lack an analogue of 
dislocations because of their disordered nature, and as a result of this it is 
difficult (or perhaps impossible) to predict where failure occurs 
\cite{GendelmanJaiswal15}. Indeed, most research on yielding focuses on finding 
a way to overcome this difficulty.\\
The first approach that comes to mind is a normal mode analysis of the Hessian, 
in particular of the soft modes at lowest frequency \cite{manningliu11}: from an 
intuitive point of view, one would expect the perspective critical mode to be 
the lowest frequency one, unless the instability is too far away. One could then 
identify the mode, extract from it the polarization vectors of particles in the 
mode, and from it deduce where the ST will occur. Sadly, it is not that simple, 
as the dynamics of the modes as strain is applied is highly chaotic 
\cite{manningliu11} and the critical mode coincides with the lowest frequency 
one just for a small interval of strains before the instability 
\cite{MaloneyLemaitre06}.

Another difficult point is the phenomenon of \emph{shear banding}, namely the 
tendency of flow in amorphous materials to concentrate in well defined bands, 
leaving the rest of the material unperturbed \cite{berthierbiroli11}. This 
phenomenon is of great relevance since it is responsible for the brittleness 
typical of glasses: instead of deforming plastically, a glass usually breaks, 
because the flow at the onset of deformation concentrates in bands, producing 
fractures. This behavior seems to depend on the amorphous nature of the glass 
and not on its actual chemical composition: metallic glasses 
\cite{Johnson99,Ashby06}, for example, are usually brittle and have a strong 
tendency to shear band, while their crystalline counterparts do not 
\cite{RTV11}.\\
Shear bands are nothing but a special form of dynamical heterogeneity 
\cite{Berthier2011}, which points towards the fact that a real-space description 
is needed to characterize them, like shown in 
\cite{DasguptaHentschel12,DasguptaGendelman13}. This however looks like a 
challenging task, especially for MF-bound approaches like RFOT.

\subsection{Theoretical approaches}
We conclude the chapter with a brief review of some theoretical approaches to 
glassy rheology. A ``perfect'' theory or rheology, as we said before, would be a 
theory capable of predicting the constitutive equation for a generic shear rate 
protocol $\dot{\g}(t)$, and it goes without saying that such a program can be 
implemented only within a dynamical approach. This is usually very difficult, 
although not impossible.\\
However, one does not necessarily have to consider time-dependent shear 
protocols. For the purpose of determining the flow curve (and the yield stress 
$\s_Y$), just the capacity to treat steady shear protocols $\dot{\g}=const$ 
would anyway be sufficient. A startup shear protocol with quasi-static strain 
would enable us to obtain the shear modulus and the yield stress, and in 
principle it does not require to resort to dynamics, as we are going to see.

\subsubsection{MCT}
We mentioned before the fact that a glass subject to shear can rejuvenate, which 
means that its aging stops and ergodicity is restored \cite{berthierbiroli11}. 
Thanks to this fact, MCT does a lot better in the context of driven dynamics 
than it does in aging situations, and a lot of research efforts have been 
dedicated to the derivation of MCT-based rheological equations. The first of 
such derivations, for the steady shear case, was reported in 
\cite{MiyazakiReichman02}, wherein a field-theoretic formalism  was employed. A 
different derivation \cite{FuchsCates09} uses the projection operator formalism, 
although the physics is the same in both cases. Above $T_{MCT}$ the viscosity is 
found to obey the scaling law
\beq
\eta(\dot{\g},T) = \eta(T)[1+\dot{\g}/\dot{\g}_0]^{-\nu}
\label{eq:shearthinning}
\eeq
with $\nu=1$, a shear thinning behavior. Below $T_{MCT}$, a Bingham plastic-type 
\cite{Larson99} behavior is predicted, with constant viscosity after the 
yielding point.\\
This approach can even be generalized to arbitrary shear protocols, as shown in 
\cite{BraderCates08,BraderCates09}. Again, in the liquid phase a shear-thinning 
behavior is predicted, with a flow curve more or less of the Herschel-Bulkley 
type \cite{BraderCates09}. If a step shear protocol is employed, $\dot{\g}(t) = 
\g\d(t)$, one can again observe the usual stress overshoot, and the yield stress 
and shear modulus can be computed. The shear modulus is found to increase when 
temperature decreases, as reasonably expected. Oscillatory shear protocols have 
also been studied within the MCT approach, see for example \cite{Miyazaki06}.

In summary, MCT provides a good qualitative description of flow in glass 
formers, and is also remarkably flexible, allowing researchers to consider many 
different shear histories. All of this despite being a first principle (for much 
time, the \emph{only} first principle) approach to glassy dynamics, which is 
always a very welcome feature.\\
However, one must not forget that MCT has an intrinsically MF nature, and as a 
consequence of this it predicts a nonexisting glass transition (as discussed in 
section \ref{subsec:dynamics}) which renders it unable of providing a good 
description of glass-formers for temperatures much below $T_{MCT}$. As a result 
of this, the rheological variants of MCT do nor better nor worse than their 
equilibrium counterpart for temperatures deep in the glass phase.

\subsubsection{RFOT}
Within RFOT every equilibrium glass is associated to a glassy metastable state. 
As a result of this, properties relative to the state, before yielding occurs 
(first of all the shear modulus), can be computed from first principles without 
resorting to dynamics, see \cite{yoshinomezard,YoshinoZamponi14}, and 
\cite{Yoshino12} for a review. In \cite{yoshinomezard}, in particular, it is 
shown that the arrest of the system within a metastable state is associated with 
the appearance of a finite shear modulus that can be analytically computed from 
replica theory; to this day, this is the only first-principle prediction of the 
manifestation of a finite shear modulus (which we recall is the hallmark of 
solidity \cite{Sausset10}) at the calorimetric glass transition.\\
However, the description of flow after yielding still requires a dynamical 
approach, which unsurprisingly relies again the analogy between MCT and PSM 
dynamics.

In \cite{BerthierBarrat00} the dynamics of a PSM-like model is considered, which 
the addiction of a \emph{driving force} which mimes the shear driving. The 
Langevin equation for a single spin is
\beq
\dot{\s_i}(t) = -\mu(t)\s_i(t) - \dpart{H}{\s_i} + f^{\rm drive}_i(t) + 
\eta_i(t),
\eeq
with the driving force defined as
\beq
f_i^{\rm drive}(t) = \epsilon(t)\sum^* 
\tilde{J}^{j_1,\dots,j_{k-1}}_i\s_{j_1}(t)\dots\s_{j_{k-1}},
\label{eq:drivingforce}
\eeq
where 
\beq
\sum^* \equiv \sum_{i<j_1<j_2<\dots j_{k-1}} +  \sum_{j_1<i<j_2<\dots j_{k-1}} + 
\dots +  \sum_{j_1<j_2<\dots j_{k-1}<i},
\eeq
and the couplings $\tilde{J}$ are quenched random variables symmetrical with 
respect to a permutation of the $j$ indices, but uncorrelated with respect of 
permutations of the $i$ index with any $j$ index. As a result of this, this 
force cannot be written as the derivative of a potential.

The role of the shear rate is played by $\epsilon(t)$ and a steady flow with 
$\epsilon = const$ is considered in \cite{BerthierBarrat00}. In the fluid phase 
above $T_{MCT}$ a shear thinning behavior much like the one in equation 
\eqref{eq:shearthinning} is found, although the exponent $\nu$ is now a function 
of the temperature, with $\nu(T_{MCT}) = 2/3$ and $\nu(T \to 0)\to 1$. In the 
glass one has
\beq
\eta(\dot\gamma,T) \simeq \dot{\g}^{-\nu(T)},
\eeq
so that the behavior is Herschel-Bulkley for all temperatures below $T_{MCT}$.

It's interesting to note that these results are different from the MCT ones, 
despite the fact that the two approaches are supposed to have the same physical 
content. This may be due to the chosen form of the driving force, that despite 
being reasonable is still far away from being a description of a real, three 
dimensional flow. In any case, this approach suffers from the same shortcomings 
of the MCT approach in terms of MF-bound description.\\
Very little effort has been up to now dedicated to the description of flow 
beyond mean-field \cite{berthierbiroli11}, even at the level of the scaling 
description provided by the mosaic (see \cite{BB09} for a discussion and some 
first predictions). Anyway, this lack of focus is shared also by other 
approaches such as Dynamical Facilitation and Frustration Limited Domains 
\cite{berthierbiroli11}.

\subsubsection{Effective models}
The rheology of glassy materials beyond MF has been up to now studied mainly 
through phenomenological models. The most popular of them is undoubtedly the 
Soft Glassy Rheology (SGR) model \cite{SollichLequeux97,Sollich98}, which 
consists in an adaptation of the trap models for aging originally proposed by 
Bouchaud \cite{Bouchaud92,MonthusBouchaud96}.\\
In this class of models the system is described as a single point that moves in 
a complex PEL through activation, just like in Goldstein's picture \cite{Go69}. 
The PEL itself is idealized as a host of traps with a certain depth, described 
by a distribution $\rho(E)$ of traps depths, which in turn induces through 
Arrhenius' law a distribution $\rho(T,\t)$ of persistence times within the 
traps. The distribution of times is usually chosen in such a way that its first 
moment diverges for low temperatures \cite{berthierbiroli11}, so to reproduce 
ergodicity breaking.\\
The SGR model implements strain as an effective lowering of the barriers between 
traps \cite{Cates03}. We stress the fact that such a model has anyway a strong 
MF spirit, since only the PEL is considered and every detail regarding actual 
real space structure is neglected. However, the SGR model also allows 
activation, a trait which is not shared by the MCT and RFOT approaches, allowing 
the study of the interplay between activation and driving. It is also remarkably 
flexible: arbitrary shear protocols can be considered and the flow curve easily 
obtained: in particular, the model behaves as a Herschel-Bulkley plastic below 
the ergodicity breaking (glass transition) temperature, while in the fluid phase 
the behavior goes from Newtonian to power-law as the temperature is lowered 
\cite{Cates03}.

Other phenomenological models have been defined over the years, mainly with the 
description of the yielding transition in mind. As a result of this, the all use 
STs as their building blocks, and mainly differ in how the interaction between 
STs is modeled. Two classes of such models exist: in the first class, the 
interaction between STs is modeled as thermal noise, in a MF kind of way 
\cite{meanfieldSTZ,HebraudLequeux98}. In the second class 
\cite{elastoplasticSTZ1,elastoplasticSTZ2}, STs are put on a lattice (in a 
spirit similar to cellular automatons \cite{linlerner14}) and their interaction 
is usually mediated by an elastic propagator of the form
\beq
\mathcal{G}(\phi,r) \simeq \frac{\cos(4\phi)}{r},
\eeq
which reproduces the quadrupolar structure shown in figure 
\ref{fig:criticalmode}.

These models however have many shortcomings. The MF models, while exactly 
solvable, all suffer from their neglect of real space structure, a choice which 
seems to be very ill advised in the case of the yielding transition, whose 
phenomenology possesses a strong characterization in real space as we discussed 
in paragraph \ref{subsec:yielding}. In addition to this, the modeling of 
interactions between STs as an effective thermal bath (a trait also shared by 
the SGR model, in fact) requires the definition of an associated ``mechanical 
noise'' effective temperature $T_{\rm eff}$, whose nature is all but clear 
\cite{Cates03,NicolasMartens14,BerthierBarrat00}, and that of course suffers 
from all the shortcomings of effective temperatures in general as discussed in 
paragraph \ref{subsec:RFOTaging}. We refer to \cite{IlyinProcaccia15} for a very 
critical point of view on the subject.\\
Elasto-plastic lattice models \cite{elastoplasticSTZ1,elastoplasticSTZ2}, on the 
other hand, are more true-to-life and do take real space structure into account, 
but they are anyway idealized \cite{Berthier11}: for example, they are limited 
to two dimensions, and usually do not take into account the displacement of STZs 
as the material is deformed (see \cite{NicolasBocquet14}); besides this, they 
are not analytically solvable. As a result of these problems, usually they need 
some tuning to satisfactorily reproduce flow curves \cite{NicolasBocquet14}.

\chapter{The State Following construction \label{chap:SFconstruction}}

In the preceding chapter we have detailed how glasses are endowed with well 
defined and time-independent physical properties: they have a specific heat, a 
compressibility, a Debye-Waller factor, a shear modulus, etc.: our aim is to 
compute analytically those quantities from a first-principles theory. In this 
chapter we present some of the tools that can be used to approach the problem 
within RFOT, all of which rely in some way or another on the replica method 
originally developed in the context of spin glass theory. In particular we 
present the State Following construction that this thesis is based on, along 
with a possible generalization of it that could in principle allow to model more 
complicated protocols than the ones considered in this work. The reader should 
appreciate how all the computation schemes that we present in this chapter are 
centered around the aim of treating metastability in a purely static fashion, 
without having to solve the dynamics, in accordance with the RFOT picture of the 
glass transition as a phenomenon with a static origin.

\section{The real replica method \label{sec:realreplica}}
We are interested in a metastable glass, be it obtained with a quenching 
\cite{Struik78} or an annealing protocol \cite{Kovacs63}. We want to compute its 
physical properties as they would manifest themselves in a DSC experiment (paragraph 
\ref{subsec:ultrastable}) or in a quasi-static simple shear experiment 
(paragraph \ref{subsec:qsshear}). In both those cases, the perturbation 
(temperature change or shear) is applied adiabatically to the material, which 
means that the system is in \emph{restricted equilibrium} inside a metastable 
state during the experiment. Within RFOT, as we mentioned, states have a 
\emph{static} origin: they are not just born from the plateau regime in the 
dynamics, but they are minima of a suitable, static FEL. As a result of this, 
and differently from what happens in other approaches to the glass problem, they 
have a static free energy
\beq
f_\al(T,\g) = -\frac{1}{N\beta}\log\int_{X \in \al}dX\ e^{-\beta V_\gamma(X)} = 
F(\{\rho^\al(\beta,\gamma)\},\beta)
\eeq 
where $X \equiv (\bx)_{i=1}^N$ is a generic configuration of the glass former, 
$F$ is the free-energy functional, $\rho^\al$ the amorphous profile that 
corresponds to the state, and $\g$ a generic perturbation that the glass is 
subjected to. This free energy rules the whole thermodynamics of the metastable 
state, and the physical observables of the system at restricted equilibrium can 
be computed from it using standard equilibrium thermodynamic relations, like 
Maxwell's: within RFOT, to study a metastable glass, we have to compute its free 
energy $f_{\al}$.\\
It is however obvious that computing the $f_\al$  is not sufficient, because we 
also need to know which state the system is in, i.e. we need to know $\al$. 
However, as we detailed in paragraph \ref{subsec:RFOTsum}, this knowledge comes 
to us in the form of the \emph{complexity}. If the glass has been prepared 
through a quenching protocol, the state will be one of the threshold states such 
that 
\beq
f_\al = f_{th}:\ \underset{f}{\rm max}\ \Sigma(f,\beta) = \Sigma(f_{th},\beta),
\label{eq:fthreshold}
\eeq
while if the glass has been made with an annealing protocol down to a glass 
temperature $T_g$, the state will be selected by the condition \eqref{eq:feq}, 
that we recall here
\beq
f_\al = f^*:\ \frac{1}{T_g} = \left.\dtot{\Sigma(f,\beta)}{f}\right|_{f=f^*},
\eeq
and the system will remain in this state during the quench down to the target 
temperature $T$. In both cases, knowledge of both the in-state free energy $f$ 
and the complexity $\Sigma$ is required to study the thermodynamics of state 
$\al$.

In principle, the computation of the free energy and complexity would require 
starting from the free energy functional $F$: one has to compute the functional, 
then find its stationary points as a function of $T$, and then count them to get 
the complexity; quite a challenging program. This ``hands-on'' approach can 
nevertheless be implemented in the case of the $p$-spin spherical model (PSM), 
wherein the free energy is the TAP free energy \cite{TAP77} of all local 
magnetizations $m_i$ as we already discussed (we refer again to 
\cite{pedestrians} for further reading); but it is not viable in the case of 
real glass-formers, also considering the fact that the free energy functional in 
real liquids is usually a functional of the local density profile in the 
continuum \cite{simpleliquids} $\mathcal{F}[\rho(\bx)]$, unless we limit 
ourselves to lattice gases.\\
The solution to this problem was proposed in \cite{monasson} and goes under the 
name of \emph{real replica} method. The basic idea is to introduce $m$ replicas 
of the original system, with the condition that they all live in the same 
metastable state. This can be accomplished, for example, by introducing a weak 
coupling $\epsilon$ between replicas \cite{monasson}. The partition function for 
the replicated system reads then
\beq
Z_m = \sum_\al\ e^{-\beta Nmf_\al}.
\eeq
now we can again introduce the complexity
$$
\frac{1}{N}\log\left[\sum_\al \d(f-f_\al)\right] \equiv \Sigma(f,\beta),
$$
and again we get 
\beq
Z_m = \int df\ e^{-\beta N[mf-T\Sigma(f,\beta)]} = e^{-\beta 
N[mf^*-T\Sigma(f^*,\beta)]} \equiv e^{-\beta N\Phi(m,\beta)},
\eeq
where we have defined the free energy of the replicated system $\Phi(m,\beta)$ 
and the condition \eqref{eq:feq} on $f^*$ has now been ``upgraded'' to
\beq
\frac{m}{T} = \left.\dtot{\Sigma(f,\beta)}{f}\right|_{f=f^*}.
\label{eq:feqreplica}
\eeq
Then one than easily prove that
\ba
f^*(m,\beta) &=& \dpart{ }{m}[\Phi(m,\beta)],\label{eq:mdepf}\\
\Sigma(f^*(m,\beta),\beta) &=& m^2 \dpart{ }{m}[m^{-1}\beta\Phi(m,\beta)]. 
\label{eq:mdepcompl}
\ea
So, once we are able to compute the free-energy $\Phi$ of the replicated system 
and perform its analytic continuation to real values of $m$, the real replica 
method enables us to compute the free energy $f^*(m,\beta)$ of the equilibrium 
states fixed by the \eqref{eq:feqreplica} and their complexity 
$\Sigma(f^*(m,\beta),\beta)$. The full complexity function can be then computed 
by inverting the \eqref{eq:feqreplica} to get $m(f^*,\beta)$ and plugging it 
into the $m$-dependent complexity, equation \eqref{eq:mdepcompl}) to get $\Sigma(f,\beta)$ 
\cite{monasson,zamponi,replicanotes}.\\
Summarizing, the basic idea of the real replica method is to introduce a 
parameter $m$ conjugated to the in-state free energy $f$, allowing us to compute 
it just by taking a derivative of the replicated free energy, in a standard 
thermodynamic fashion \cite{monasson}. From the \eqref{eq:feqreplica}, we can 
see that choosing a different $m$ selects a different group of metastable 
states, which are different from the equilibrium states of the system unless 
$m=1$: the presence of the parameter $m$ allows us to choose $f^*$ at our 
leisure, and study different groups of states according to our needs: within 
this formalism, choosing a dynamical protocol corresponds to choosing a function 
$m(T)$.\\
Indeed, observing more closely the \eqref{eq:feqreplica} and comparing it with 
the \eqref{eq:feq}, we are immediately prompted to define an effective 
temperature in the following way
\beq
\frac{1}{T_{\rm eff}} = \frac{m}{T} = 
\left.\dtot{\Sigma(f,\beta)}{f}\right|_{f=f^*},
\label{eq:teffreplica}
\eeq
which means that the states we select by choosing $m$ at temperature $T$ are 
effectively those that would be equilibrated at $T=T_{\rm eff}$ defined in the 
\eqref{eq:teffreplica}: we can appreciate how the real replica method is nothing 
but a static (and more flexible) incarnation of the effective temperature 
picture \cite{CugliandoloKurchan97,FranzVirasoro00,Kurchan05}.

\subsection{Quenching: the threshold}
In the case of a quenching protocol down to a target temperature $T$, the system 
will remain stuck in the threshold states fixed by the \eqref{eq:fthreshold}, so 
the function $m(T)$ will simply be $m(T) = m_{th}(T)$ such that
\beq
\frac{m_{th}}{T} = \left.\dtot{\Sigma(f,\beta)}{f}\right|_{f=f_{th}},
\eeq
and unsurprisingly, one has 
\beq
\frac{m_{th}}{T} = \frac{1}{T_{\rm eff}}
\eeq
where $T_{\rm eff}$ corresponds \cite{FranzVirasoro00} to the effective 
temperature computed from the dynamical solution \cite{CugliandoloKurchan93}. 
The real replica method allows one to compute all interesting long-time 
observables relative to a generic quenching dynamics.

\subsection{Annealing: isocomplexity \label{subsec:isocomplexity}}
In the case of an annealing dynamics, things are more complicated. We know that 
the state of the system is selected at equilibrium through the \eqref{eq:feq}, 
so we know that $m(T_g) = 1$ and $f_\al = f(1,T_g)$. However, we still need to 
determine the rest of the function $m(T)$ as the system is quenched below $T_g$ 
and the original equilibrium state is ``followed'' in temperature. Summarizing, 
we need a criterion to choose a function $m(T)$ consistent with the requirement 
that the system remain in that same state as $T$ changes 
\cite{LopatinIoffe02}.\\
Following \cite{montanariricciisocomplexity}, one can assume that, as $T$ is 
changes, states do not coalesce, or merge, or cross. This means that the number 
(and thus the complexity) of states at each free energy level $f$ is a conserved 
quantity, and can be used as a label for the states. This method is usually 
referred to as isocomplexity \cite{LopatinIoffe02,montanariricciisocomplexity}. 
The function $m(T)$ will be then determined by the condition
\beq
\Sigma(1,T_g) = \Sigma(m(T),T), \label{eq:isocompl}
\eeq
where $\Sigma$ is the $m$-dependent complexity computed in the 
\eqref{eq:mdepcompl}. In chapter \ref{chap:isocomplexity} we will present some 
results on glassy state following obtained with the isocomplexity assumption.

\subsection{Summary}
The real replica method provides us with a set of tools to treat metastability 
without having to resort to the TAP approach. 
This method is thus well suited to the study of a quenching dynamics like the one considered in 
\cite{CugliandoloKurchan93}. It constitutes a computational tool conceived for 
the treatment of all systems with an RFOT transition (from structural glasses, 
to spin glasses \cite{parisiSGandbeyond} and even optimization problems in 
computer science \cite{mezardmontanari}). 
Because of this, it has been for much time the standard method for the treatment 
of structural glasses within RFOT, from the very first papers 
\cite{parisimezardjchem99,parisimezardprl99} to the more recent efforts of the 
series \cite{KPZ12,KPUZ13,CKPUZ13,nature}.

Nonetheless, as we detailed in the previous chapter, in recent years the 
experimental and numerical focus has moved away from quenching protocols;
and within the real replica method, the only way to treat annealing protocols is 
the isocomplexity assumption, which despite being reasonable fails in all models 
except the PSM: the states do actually cross, merge and coalesce, so assuming 
that their number is conserved is just plain wrong. The isocomplexity assumption 
is thus, at best, an approximation of the real dynamics of the system, even at 
mean-field level.\\
This weakness should not come as a surprise considering how the real replica 
method is basically a static recasting of the effective temperature concept, 
which postulates that the typical configurations visited during a 
non-equilibrium, aging dynamics are just the configurations the system would 
visit if it were equilibrated at $T=T_{\rm eff}$. This is a very strong 
assumption that is false in most cases: the configurations visited by the system 
during aging may very well have nothing to do with equilibrium ones, even at MF 
level, and as a result of this they may have a vanishing weight in the 
equilibrium probability distribution, and be missed completely by the real 
replica computation.\\
In summary, we need a more refined formalism to treat annealing dynamical 
protocols like the ones considered in \cite{BarratBurioniMezard}. In the 
following section we introduce that formalism, and we refer to 
\cite{zamponi,MP09,replicanotes} for further reading on the real replica method.

\section{The two-replica Franz-Parisi potential \label{sec:FP}}
\begin{figure}[t!]
\begin{center}
\includegraphics[width = 0.5\textwidth]{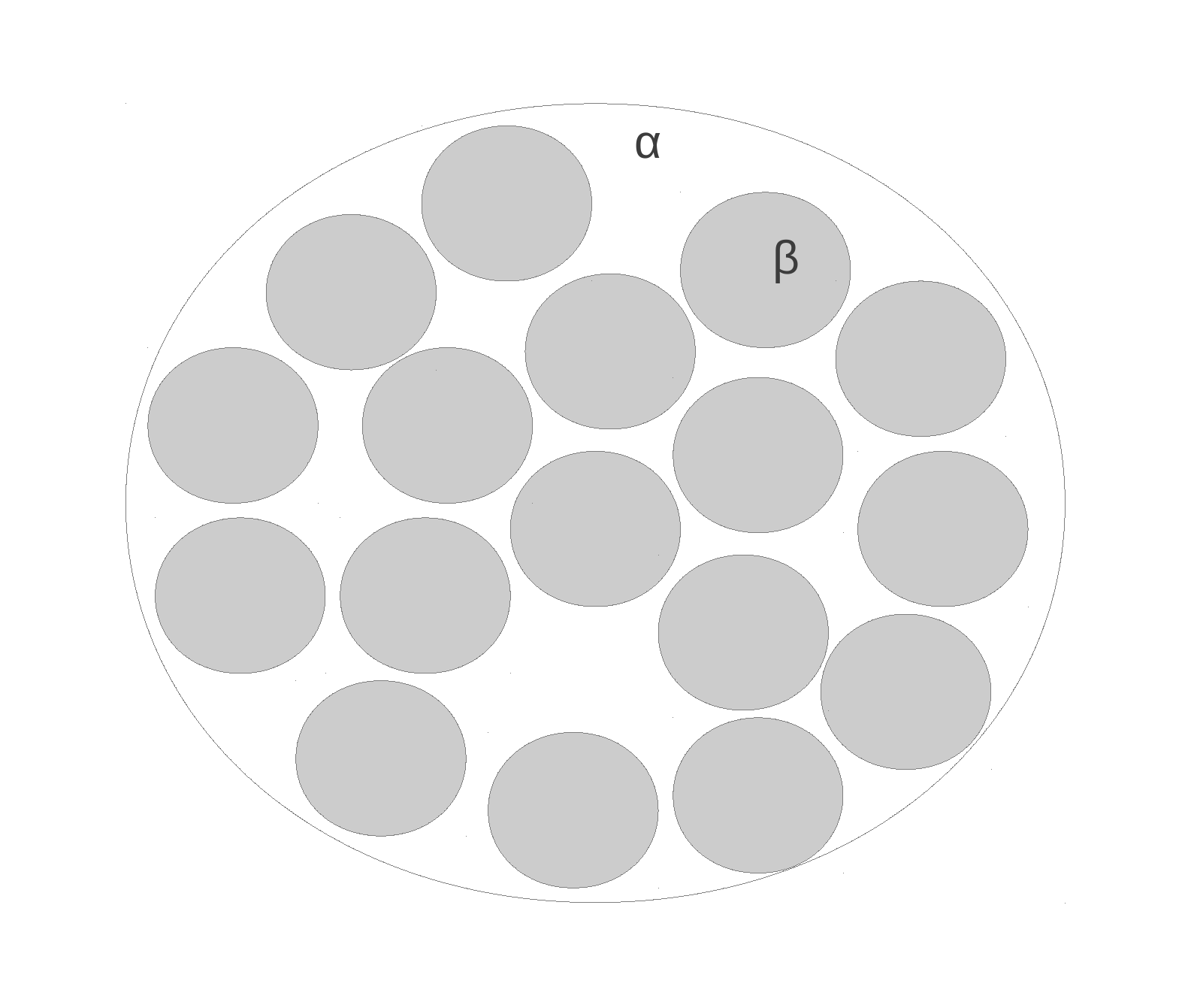}
\caption{States as patches in the space of configurations of the 
glass-former.\label{fig:states}}
\end{center}
\end{figure}

Let us go back to the definition of the in-state free energy $f_\al$
\beq
f_\al(T,\g) = -\frac{1}{\beta N}\log\int_{X \in \al}dX\ e^{-\beta V_\gamma(X)}.
\eeq 
We have detailed in paragraph \ref{subsec:RFOTsum}  how states are essentially 
``patches'' in the configuration space of the system (see figure 
\ref{fig:states}), which in this case is the $Nd$-dimensional space of vectors 
$X$. We have to somehow find a way to make sure that only configurations 
belonging to state $\al$ are included in the partition function above, which 
means that we have to define a Gibbs measure somehow \emph{constrained} inside 
the state.\\
According to the amorphous solid picture \ref{subsec:caging}, the configurations 
belonging to state $\al$ will consist of the fixed amorphous configuration the 
particles vibrate around, which is given by a set of positions $R \equiv 
(\boldsymbol{r})_{i=1}^N$, and all configurations visited during the vibration. 
We can thus hope to implement the constraint in state $\al$ by accepting in the 
partition function only configurations which are not too far away from $R$. We 
can thus write
\beq
f(T,\g;R) = -\frac{1}{\beta N}\log \int dX\ e^{-\beta V_\g(X)}\theta[\D_r - 
\D(X,R)],
\label{eq:FPR}
\eeq
where $\D(X,R)$ is the rescaled MSD between $X$ and $R$
\beq
\D(X,R) \equiv \frac{d}{N}\sum_{i=1}^N (\bx_i - \boldsymbol{r}_i)^2
\eeq
and $\theta$ is the Heaviside theta function \cite{Abramowitz}. We have 
essentially chosen $\theta[\D_r - \D(X,R)]$ as the characteristic function of 
state $\al$.\\
The $f(T,\g;R)$ is the free energy of the glass selected by the amorphous 
configuration $R$. However, it is clear that such a thing cannot be computed: on 
technical level, the presence of the constraint breaks translational invariance 
and prevents us from using standard statistical-mechanical methods; on a 
conceptual level, we do not know $R$.\\
To circumvent this difficulty we can assume that the properties of the glass do 
not depend much on the actual realization of the amorphous configuration $R$, 
i.e., that the observables of the glass possess a \emph{self-averaging} 
\cite{pedestrians} property. As a rule of thumb, this is usually true for 
extensive observables like the free energy. Thus we can average the $f$ over all 
possible amorphous configurations $R$, which will be distributed with a certain 
probability distribution $P(R)$.\\
We then have to find out which is $P(R)$, but this is simple: the configuration 
$R$ is by definition the last configuration visited by the system before falling 
out of equilibrium, so it will be distributed with the usual canonical 
distribution at $T=T_g$,
\beq
P(R) = \frac{e^{-\beta_gV(R)}}{Z(T_g)}.
\eeq
So in the end we can define the free energy of the glass
\beq
f_g(T,\g;T_g) = \overline{f(T,\g;R)} = \int dR\ \frac{e^{-\beta_gV(R)}}{Z(T_g)} 
f(T,\g;R), 
\label{eq:FP}
\eeq
which is the free-energy of a glass at temperature $T$, subjected to a generic 
perturbation $\gamma$, and prepared through an annealing protocol such that the 
system falls out of equilibrium at $T_g$\footnote{We stress again that the 
self-averaging property is not true for all observables. In particular, it is 
not true for observables strongly related to the structure $R$ of the glass, 
such as the refractive index: the structure $R$ is sampled from the distribution 
$P(R)$, so it not uniquely determined by $T_g$, and as a result of this glasses 
with the same fictive temperature can have different refractive indexes as 
discovered in \cite{Ritland56}.}: it allows us to follow a state in temperature 
from when it is selected by the \eqref{eq:feq} at $T_g$ to a temperature $T$ 
whereupon a measurement is performed. This free energy, dubbed the 
\emph{Franz-Parisi potential} (FP), has been formalized in \cite{franzparisipot} 
in the context of spin glass models, with the express purpose of studying the 
long time limit of annealing protocols like the ones considered in 
\cite{BarratBurioniMezard}, and is the centerpiece of State Following (SF) 
formalism. Up to now it has been only employed in the context of schematic spin 
glass models \cite{BFP97,KZ10,KZ10b,KZ13}; in this thesis we apply it, for the 
first time, to a realistic model of glass former.

This is exactly what we need, but there is clearly a missing ingredient: which 
is the value of $\D_r$? We have to somehow fix its value in such a way that the 
whole glass state, and nothing more than that, is sampled.\\
To understand how to choose $\D_r$, let us rewrite the \eqref{eq:FPR} in the 
following way:
 \beq
\frac{1}{Z} \int dX\ e^{-\beta V_\g(X)}\d[\D_r - \D(X,R)] = \thav{\d[\D_r - 
\D(X,R)]} \equiv e^{-\beta N (V(\D_r)-V(\infty))},
 \label{eq:FPRlargedev}
 \eeq
wherein we have replaced the Heaviside theta with a Dirac delta\footnote{In the 
thermodynamic limit, the two choices are completely equivalent: if the theta is 
the characteristic function of the state, then the delta is the characteristic 
function of its boundary. But in the thermodynamic limit the dimensionality of 
the configuration space goes to infinity, and the volume of any compact set 
inside it concentrates on its own boundary \cite{KurchanLaloux96}. It is a 
purely geometrical fact.} and used the fact that choosing $\D_r = \infty$ means 
considering the whole configuration space. The \eqref{eq:FPRlargedev} is nothing 
but the probability\footnote{We recall that the prescription to compute the 
probability distribution of a generic observable $\mathcal{O}(\mathcal{C})$ is 
$P(\Os) = \thav{\d(\Os-\Os(\mathcal{C}))}$.} that, if we throw a random 
configuration $X$ with the canonical distribution, we find it to be at a 
distance $\D_r$ from the configuration $R$, written in a large deviation form 
through the function $V(\D_r) \equiv f(T,\g;R)$.

How does $V(\D_r)$ behave? The probability to find the configuration $X$ just on 
top of the configuration $R$ is obviously zero, so one must have $V(\D_r = 0) = 
\infty$. Conversely, when $\D_r = \infty$, we are accepting all configurations, 
so the probability is one. Between these two values, we expect that at high 
temperature the probability will just monotonically increase (and $V(\D_r)$ 
decrease), since we are considering a larger and larger region of the space of 
configurations and the Boltzmann-Gibbs distribution is effectively a uniform 
distribution at high $T$.\\
However, this will not be true below $T_{MCT}$, whereupon glassy states appear 
(see figure \ref{fig:states}). The configuration $R$ will belong to one of the 
states and the probability will go up monotonically only as long as we consider 
through $\D_r$ a region contained within the state: when the region becomes 
bigger, one starts to sample configurations on the boundary, which are unlikely 
configurations that the system only visits when barrier crossings take place: 
the probability will start to go down and $V(\D_r)$ to increase. When $\D_r$ is 
increased further, other states are taken into consideration and $P(\D_r)$ will 
increase anew.\\
Summarizing, below $T_{MCT}$, the function $V(\D_r)$ will have a minimum for 
$\D_r^* \neq 0$, and that value of $\D_r$ will correspond to the optimal 
sampling of configurations inside the state. This picture is not changed by the 
introduction of the average over $R$ \eqref{eq:FP}: the $f_g(T,\g;T_g)$ must be 
minimized over $\D_r$ to obtain the free energy of the glass.

\begin{figure}[ht!]
\begin{center}
\includegraphics[width = 0.7\textwidth]{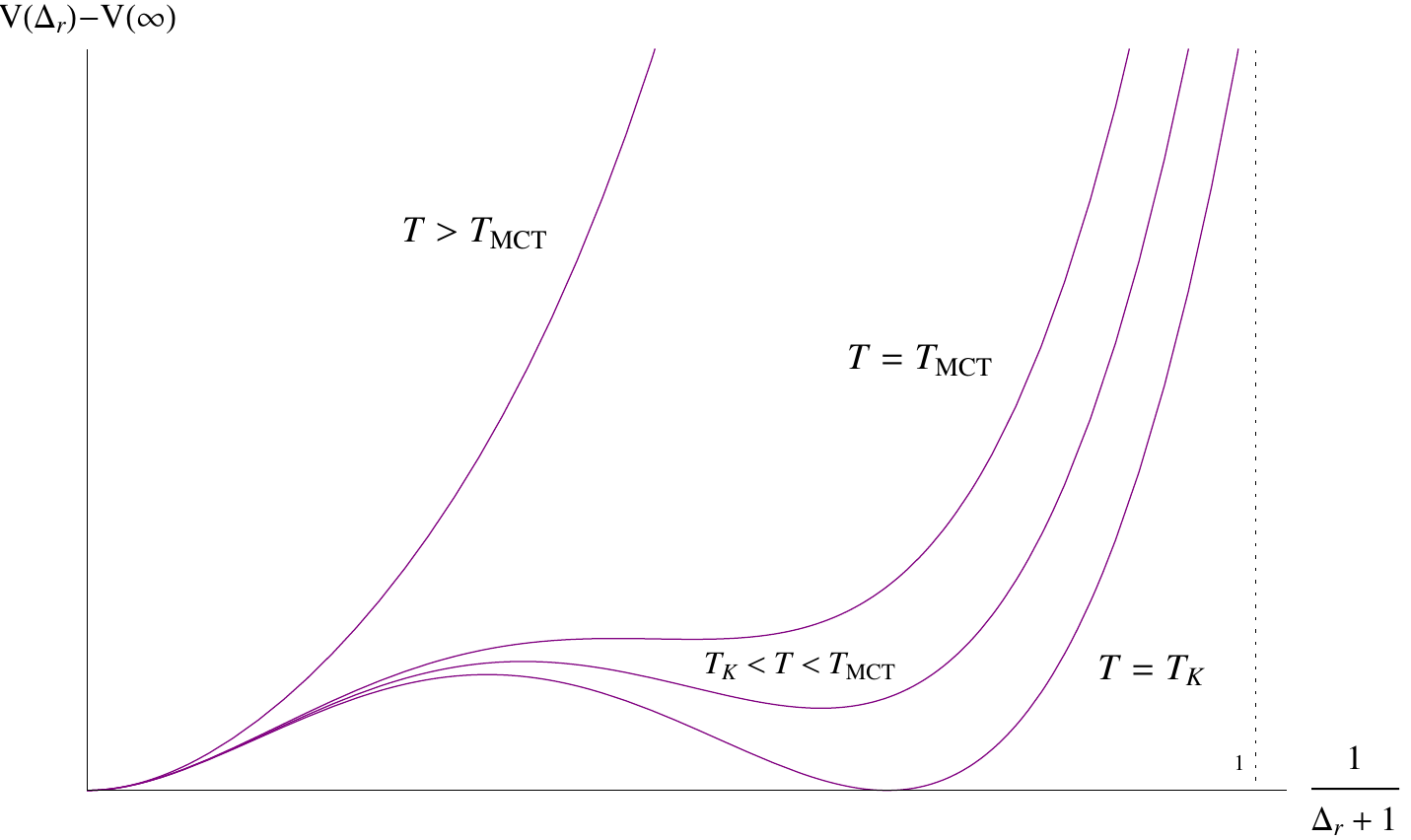}
\caption{The FP potential as a function of $\D_r$ for $T=T_g$, as the 
temperature is lowered as detailed in paragraph 
\ref{subsec:RFOTsum}.\label{fig:franzparisi}}
\end{center}
\end{figure}

If one chooses $T=T_g$, then the FP potential corresponds to the free energy of 
an equilibrium state at $T_g$, the same one can compute from the real replica 
method. In that case, since $V(\D_r = \infty) = F_{liq}$, one has by definition
\beq
V(\D_r^*) - V(\infty) = T\Sigma(T) 
\eeq
as sketched in figure \ref{fig:franzparisi}.
From the \ref{fig:franzparisi} the reader can appreciate the first order 
character of the glass transition according to RFOT: $T_{MCT}$ corresponds to 
the spinodal point whereupon metastable states lose stability; below $T_{MCT}$, 
a glassy minimum is present with $\D_r = \D_r^* \neq 0$, but always metastable 
with respect to the liquid minimum with $\D_r = \infty$, with a free energy gap 
between the two equal to $T\Sigma(T)$; at $T=T_K$, the glassy minimum becomes 
stable with respect to the liquid one and the ideal glass transition takes 
place.

In chapters \ref{chap:SFRS} and \ref{chap:SFfRSB} we present the results 
obtained applying the SF construction to hard spheres in the MF limit. We refer 
to \cite{KZ13} for a comparison of the isocomplexity and SF approaches in the 
context of $p$-spin glasses.

\section{Beyond two replicas: the replica chain and pseudodynamics 
\label{sec:chain}}

The FP construction can be generalized to larger numbers of replicas. Its 
generalization to three replicas was for example used in 
\cite{CavagnaGiardina13} to study barriers between metastable states, and its 
generalization to an arbitrary number of replicas was first sketched in 
\cite{KrzakalaKurchan07}.

Let us suppose that we have a generic system with configurations $\mathcal{C},\ 
\mathcal{C'}\dots$, a Hamiltonian $H(\mathcal{C})$ and a notion of 
``similarity'' between configurations $q(\mathcal{C},\mathcal{C'})$, which has 
to be conveniently chosen depending on the system. Above, we have used the MSD 
between configurations, but if one starts from a density-functional theory of a 
liquid, the choice could be \cite{FranzParisi15}
\beq
q(\mathcal{C},\mathcal{C'}) = \int d\bx d\by\ 
w(\bx-\by)\rho_{\mathcal{C}}(\bx)\rho_{\mathcal{C'}}(\by),
\eeq
where $\rho_\mathcal{C}(\bx)$ is the density profile relative to the 
configuration $\mathcal{C}$ \cite{simpleliquids} and $w$ a coarse-graining 
function whose details have no relevance. For spin models one usually uses 
\cite{FranzParisi13}
\beq
q_{\mathcal{C}\mathcal{C'}} = \frac{1}{N}\sum_{i=1}^{N} 
\sigma^{\mathcal{C}}_i\sigma^{\mathcal{C'}}_i.
\eeq
One starts from replica 1, which is chosen to be equilibrated at a temperature 
$T_1$
\beq
P(\C_1) = \frac{1}{Z}e^{-\beta_1H(\C_1)}.
\eeq
Replica 2 is chosen to be equilibrated at a temperature $T_2$, and constrained to 
be near to replica 1 by using the $q(\C_1,\C_2)$
\beq
P(\C_1,\C_2) = 
\frac{1}{Z(\C_1)}e^{-\be_2H(\C_2)}\d(q(\C_1,\C_2)-C(1,2))\frac{1}{Z}e^{
-\beta_1H(\C_1)}, 
\eeq
where 
\beq
Z(\C_1) \equiv \int d\C_2 e^{-\beta_2 H(\C_2)} \d(q(\C_1,\C_2)-C(1,2)).
\eeq
This corresponds to the two-replica case; in general, one can define a 
transition probability
\beq
M(\C_{s+1}|\C_s) \equiv \frac{1}{Z(\C_{s})}e^{\beta_{s+1}H(\C_{s+1})} 
\d(q(\C_s,\C_{s+1})-C(s,s+1)),
\label{eq:replicachain}
\eeq
in such a way that the probability of a ``trajectory'' can be written as
\beq
P(\C_L,\C_{L-1},\dots,\C_1) = \prod_{i=1}^{L-1}M(\C_{s+1}|\C_s)P(\C_1). 
\eeq
for a chain of length $L$. The transition rate defined in the 
\eqref{eq:replicachain} defines a Markov stochastic process, and with it, a 
dynamics. Within this dynamics, the system samples the phase space with an 
equilibrium Boltzmann-Gibbs distribution at each step, so it is allowed to 
equilibrate, but not too far away from the configuration at the preceding step, 
because of the $\d$ constraints: the replica chain defined above implements 
formally the RFOT idea of the dynamics of glass formers as a process made of 
activated jumps between states. Because the system equilibrates at each step, 
the dynamics defined in \eqref{eq:replicachain} is fittingly referred to as a 
\emph{Boltzmann pseudodynamics} \cite{FranzParisi13}.

The replica chain can be interesting to study for a finite number $L$ of bonds 
(for example, in \cite{KZ10,SunCrisanti12} it was argued that the two-replica 
potential may be not sufficient for following states adiabatically in the whole 
range of temperatures), but its most interesting application is in the limit $L 
\to \infty$ of infinite bonds and constant chain length. In that case, the 
values of the index $i$ are promoted to continuous variables and the parameters 
$C(s,s+1)$ to functions $C(t,t')$. It can then be shown \cite{FranzParisi13} 
that in this limit, and with $\beta_s = const = \beta$ one gets back, for the 
PSM, the slow part of the dynamical equations \eqref{eq:agingPSM1}, 
\eqref{eq:agingPSM2} and \eqref{eq:agingPSM3} for a quenching dynamics. This is 
no surprise: in the pseudodynamics defined above, the system equilibrates within 
a state at each step, and as a result of this only the details of the slow part 
of relaxation are reproduced, while the fast part is neglected by construction. 
An annealing dynamics \cite{BarratBurioniMezard} can be modeled by choosing 
$\beta_1 = \beta_g$ and $\beta_s = \beta\ \forall s \neq 1 $.\\
The main advantage of the pseudodynamics approach is the usual advantage of RFOT 
tools in general (real replica method, FP potential, etc.): since the system is 
equilibrated at each step, the actual computation of the properties of the chain 
is completely \emph{static} in nature, and corresponds, roughly, to the problem 
of computing the static properties of a replicated system, where each replica 
corresponds to a node in the chain. Again, details of the long-time dynamical 
properties of the system can be computed without ever touching the dynamics 
itself.

In \cite{FranzParisi15} the replica chain is applied to a generic model of glass 
former in a liquid theory setting. Within liquid theory, the computation of the 
statics of a liquid typically reduces to the computation of its pair 
distribution function $g(r)$ \cite{simpleliquids}. To implement the chain 
formalism, one must consider a mixture of different particle species, each of 
them corresponding to one replica in the chain, with the pair distribution 
function generalized to $g_{ab}(r)$ where $a$ and $b$ are species-labeling 
indexes. Within liquid theory for particle mixtures, the $g_{ab}(r)$ is related 
to the direct correlation function $c_{ab}(r)$ by the Ornstein-Zernike relation 
\cite{simpleliquids}
\beq
c_{ab}(\bx,\by) = h_{ab}(\bx,\by) - \rho\sum_{c} \int d\boldsymbol{z}\ 
h_{ac}(\bx,\boldsymbol{z})c_{cb}(\boldsymbol{z},\by), 
\eeq
where $h_{ab}(\bx) \equiv g_{ab}(\bx) - 1$. To solve the statics of the mixture, 
one has to find another relation to link the $c_{ab}(\bx)$ with the 
$g_{ab}(\bx)$, so to get a closed system of integral equations, as per usual 
practice in liquid theory \cite{simpleliquids}. In \cite{FranzParisi15}, two 
closure schemes are studied: the first is the well known Hypernetted Chain (HNC) 
closure \cite{simpleliquids}:
\beq
\log[h_{ab}(\bx)+1]+\beta v_{ab}(\bx) = h_{ab}(\bx) - c_{ab}(\bx),
\eeq
where $v_{ab}(\bx)$ is the interaction potential between species $a$ and $b$;  
the other is a closure scheme proposed by Szamel in \cite{Szamel10}:
\beq
c_{ab}(k) = \int dq\ V(k,q)h_{ab}(q)h_{ab}(k-q),
\eeq
where $V(k,q)$ is the MCT vertex defined in the \eqref{eq:MCTvertex}. Very 
interestingly, once one imposes TTI and FDT, in both cases the slow part of the 
MCT equation is recovered \cite{FranzParisi15}. This result bolsters the RFOT 
picture of dynamics as a hopping process between states.

The replica chain is not the main focus of this thesis so our treatment of it 
stops here. Up to now, it has only been used in the case $\beta_s = const$ which 
reproduces a quenching dynamics, but its potential goes far beyond that. On one 
hand, it could be interesting to consider general protocols wherein $\beta(t)$ 
is a full function of the time, but the most interesting application could come 
from its generalization to a shear strain situation, instead of simple aging. 
For example, it could be in principle used to get a complete RFOT-born 
rheological theory of glass formers for a generic shear protocol $\dot{\g}(t)$.

\chapter{The replica symmetric ansatz \label{chap:SFRS}}

In this chapter we perform the computation of the Franz-Parisi potential for the 
hard sphere (HS) model in the MF limit. As we anticipated, the perturbations we 
focus on will be adiabatic compression/decompression and quasi-static shear 
strain, and we compute the response of glassy states to these perturbations. As 
we detail in the following, the computation requires the formulation of an 
\emph{Ansatz} about the structure of the metastable state the system is trapped 
in. In this chapter we focus on the so-called Replica Symmetric (RS) ansatz, 
which means that we assume the state to be a simple minimum of the FEL without 
any further internal structure. Since all calculations are performed with the 
saddle-point method \cite{AdvancedMethods}, the results obtained from this 
ansatz and its relative saddle point must be checked for stability, i.e. the 
Gaussian fluctuations around the saddle point value of the integral must be 
negative. We verify that beyond a certain value of both the compression and the 
strain parameter, it is not so: the RS Ansatz is unstable and a more complicated 
structure manifests inside the glassy minimum in study, requiring a more 
complicated Ansatz.

\section{Computation of the FP potential}
We want to compute:
\beq
F_g(T,\g;T_g,\D_r) = \frac{1}{Z_m} \int dR^1 \cdots dR^m \, e^{-\b_g 
\sum_{a=1}^m V(R^a)} F(T,\g;R^1,\D^r),
\eeq
with
\beq
F(T,\g;R,\D^r) \equiv -T \log \int dX  e^{-\b V_\g(X)}\delta(\D^r-\D(X,R)),
\label{eq:defF}
\eeq
where to be general we average the $F$ over $m$ replicas instead of just one. 
This way, one can in principle take a state outside of the equilibrium line by 
choosing $m$ appropriately as explained in section \ref{sec:realreplica}, a 
protocol which could for example correspond to a quench followed by an adiabatic 
perturbation. However, in this thesis we only focus on the equilibrium, $m=1$ 
case, corresponding to an annealing protocol as previously discussed.\\
The average above defined can be computed using the \emph{replica trick} 
\cite{pedestrians,replicanotes}. If one defines
\beq
Z_g \equiv \int dX  e^{-\b V_\g(X)}\delta(\D^r-\D(X,R)),
\eeq
and
\beq\begin{split}
-\b N F_{\rm FP}
 &=  \log \int dR^1 \cdots dR^m dX^{1} \cdots dX^{s} e^{-\b_g \sum_{a=1}^m 
V(R^a) - \b \sum_{b=1}^{s} V_\g(X^b)} \\
&= \log \int dX^1\cdots dX^m e^{-\b_g \sum_{a=1}^m V(X^a)} (Z_g)^s = \log[ Z_m 
\overline{(Z_g)^s} ] \ ,
\end{split}\eeq
then we have, at leading order for small $s$
\beq\begin{split}
-\b N F_{\rm FP} &= \log \left[ Z_m  \overline{ (Z_g)^s } \right] \sim  \log 
\left[ Z_m  (1 + s \overline{ \log(Z_g)} + O(s^2) ) \right]\\
&= \log Z_m + s \, \overline{ \log Z_g } + O(s^2)  \\ 
&= -\b F_m - s \b F_g(T,\g;T_g) + O(s^2)
\ .
\end{split}\eeq
Therefore we have to
compute the free energy of $m+s$ replicas; $m$ ``reference'' ones and $s$ 
``constrained'' ones, that are at different temperature or density,
and then we have to expand it around $s=0$; the leading order gives the 
replicated free energy $F_m$ \cite{monasson}, 
while the linear order in $s$ will yield the FP free energy $F_g(T,\g;T_g)$ we 
want to compute.

\subsection{Perturbations}
The construction above can be performed for any model $V(X)$ of glass former. In 
the following, we focus only on the hard sphere (HS) \cite{simpleliquids} 
interaction potential, whose definition we recall here
\beq
v_{HS}(\bx) \equiv \begin{cases}
                    0 &|\bx|>D\\
                    \infty &|\bx| \leq D
                   \end{cases}
\eeq
hence temperature plays no role and the packing fraction is the only relevant 
control parameter \cite{simpleliquids}. Furthermore, the energy is zero, 
therefore the free energy contains only the entropic term and $-\b F = S$.\\
For technical reasons, it is convenient to fix the packing fraction through the 
sphere diameters, while assuming that the number density $\rho$ be constant,
as in the LS algorithm~\cite{LS90}.
We consider the $m$ reference replicas to have diameter $D_g$ and packing 
fraction $\f_g$,
while the $s$ constrained replicas have the same number density but $D = D_g ( 1 
+ \h/d)$. 
Following~\cite{parisizamponi,KPZ12} we also define a rescaled packing fraction 
$\wh \f = 2^d \f/d$ that has a finite limit when $d\to\io$.
Note that the packing fraction of the constrained replicas is therefore
$\f = \f_g (D/D_g)^d \sim \f_g e^\h$ and similarly
$\wh\f = \wh\f_g e^\h$.

Following~\cite{yoshinomezard,Yoshino12,YoshinoZamponi14}, we also apply a shear 
strain $\g$ to the constrained replicas, which is obtained by deforming linearly
the volume wherein the system is contained. 
Following the discussion in section \ref{sec:shear}, we call $x'_\m$, with 
$\mu=1,\cdots,d$, the coordinates in the original reference frame, in which the 
shear strain is applied.
In this frame, the cubic volume is deformed because of shear strain.
To remove this undesirable feature, we introduce new coordinates $x_\m$ of a 
``strained'' frame wherein the volume
is brought back to a cubic shape.
If the strain is applied along direction $\mu=2$, then
all the coordinates are unchanged, $x_{\mu} = x_{\mu}'$, except the first
one which is changed according to
\beq
x_{1}' = x_{1} + \g x_{2} \ , \hskip30pt  x_{1} = x_{1}' - \g x_{2}' \ .
\eeq
Let us call $S(\g)$ the matrix such that $\bx' = S(\g) \bx$.
In the original frame (where the volume is deformed by strain), two particles of 
the slave replica
interact with the potential $v(|\bx'-\by'|)$.
If we change variable to the strained frame (where the volume is not deformed), 
the interaction is 
\beq
v_{\g}(\bx-\by) = v(|S(\g)(\bx-\by)|) \ . 
\eeq
An important remark is that $\det S(\gamma) =1$ meaning 
that the simple strain defined above does not
change the volume and thus the number density $\rho=N/V$ of the system.

In summary, to follow a glass state under a compression and a strain, we have to 
compute the Franz-Parisi potential
where the constrained replicas have a diameter $D = D_g (1 + \h/d)$ 
and interact with a potential $V_\g(X) = \sum_{i<j} v_\g(\bx_i - \bx_j)$. The 
control parameter of the reference replica is their density $\f_g$,
while the control parameters of the constrained replicas are the compression 
parameter $\h = \log(\f/\f_g)$ and shear strain $\g$.
The replicated entropy of this system can be computed through a generalization 
of the methods of refs.~\cite{KPZ12,KPUZ13},
which we sketch below.

\subsection{The replicated entropy and the RS ansatz \label{subsec:replicateds}}

The \emph{exact} expression of the replicated entropy of HS in the MF limit, and for a completely generic 
replica structure has been derived in~\cite{KPUZ13}:
\beq\label{eq:gauss_r}
\begin{split}
s[\hat \a] =\ &1 - \log\r + d \log (m+s) + \frac{(m+s-1)d}{2} \log(2 \pi e 
D_g^2/d^2)\\
&+ \frac{d}2 \log \det(\hat \a^{m+s,m+s}) -  \frac{d}2 \wh \f_g \,
 \FF\left( 2 \hat \a \right) \ ,
 \end{split}
\eeq
where $\hat\a$ is a $(m+s)\times(m+s)$ symmetric matrix defined in appendix 
\ref{app:infinited} and $\hat \a^{a,a}$ is the matrix obtained from $\hat\a$
by deleting the $a$-th row and column. We refer to appendix \ref{app:infinited} 
for a sketch of the derivation of the $\infty$-dimensional solution.\\
As explained in appendix \ref{app:infinited}, the matrix $\hat\a$ encodes the 
fluctuations of the replica displacements $u_a \equiv x_a - X$ around the center 
of mass of all replicas. Because $\sum_a u_a=0$,
the sum of each row and column of $\hat\a$ is equal to zero, i.e. $\hat\a$ is a 
{\it Laplacian} matrix. 
Here we used $D_g$ as the unit of length, and for this reason $D_g$ and
$\wh\f_g$ appear in eq.~\eqref{eq:gauss_r}. We call the last term in 
Eq.~\eqref{eq:gauss_r} the ``interaction term'', while the one containing the 
determinant  of $\a^{m,m}$ will be called the ``entropic term''.\\
It is usually more convenient to use a different matrix, denoted as $\hat \D$
\beq
\D_{ab} \equiv \frac{d}{D_g^2}\la(u_a-u_b)^2\ra = \al_{aa} + \al_{bb} - 
2\a_{ab},
\eeq
which encodes the MSDs between replicas; the matrix $\hat\D$ has a more 
straightforward physical interpretation and is more suited to the definition of 
the parameter $\D^r$ in section \ref{sec:FP}, but it is completely equivalent to 
the $\hat\al$.

\begin{figure}
\begin{center}
\includegraphics[width = 0.5\textwidth]{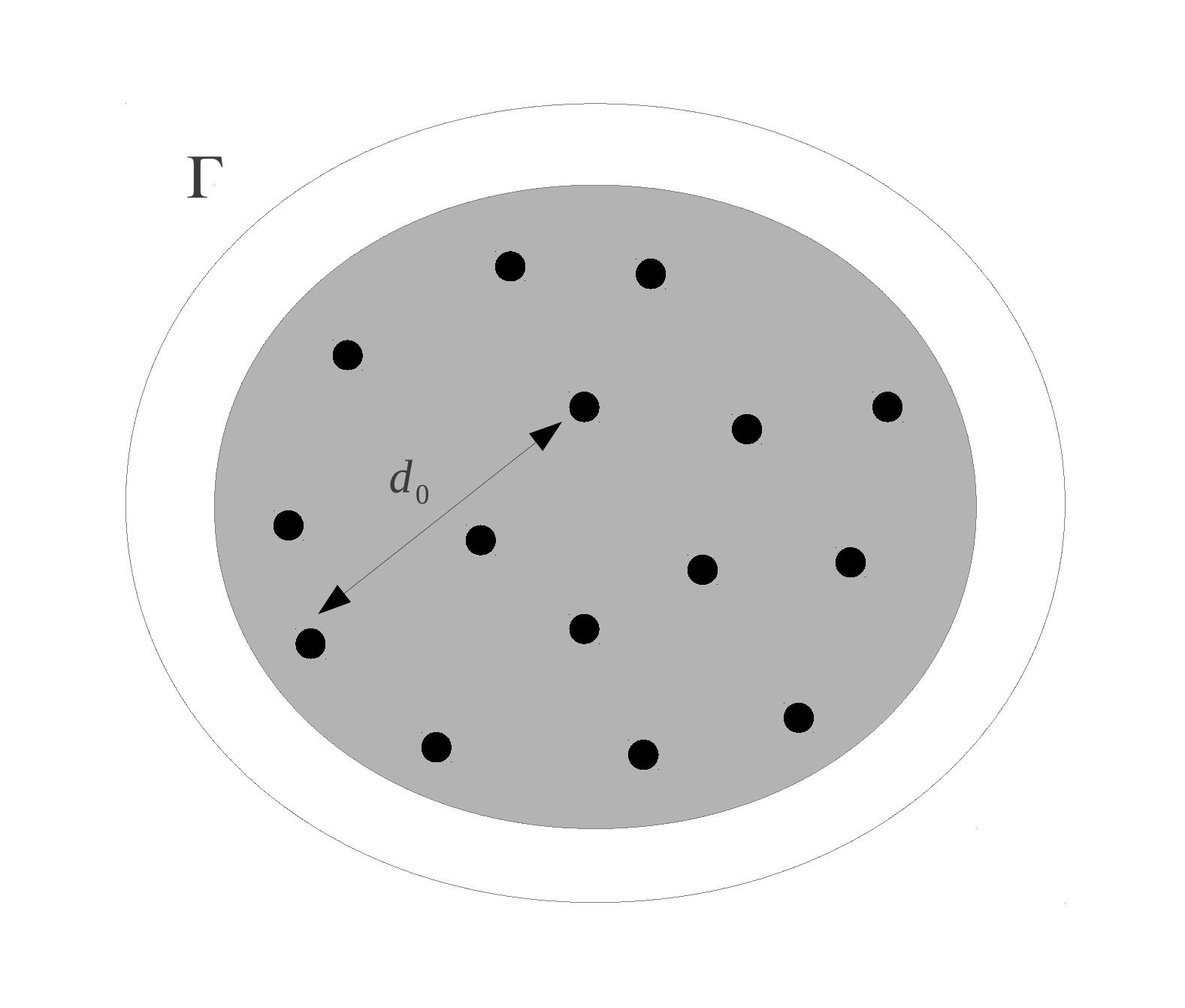}\nolinebreak[4]
\includegraphics[width = 0.5\textwidth]{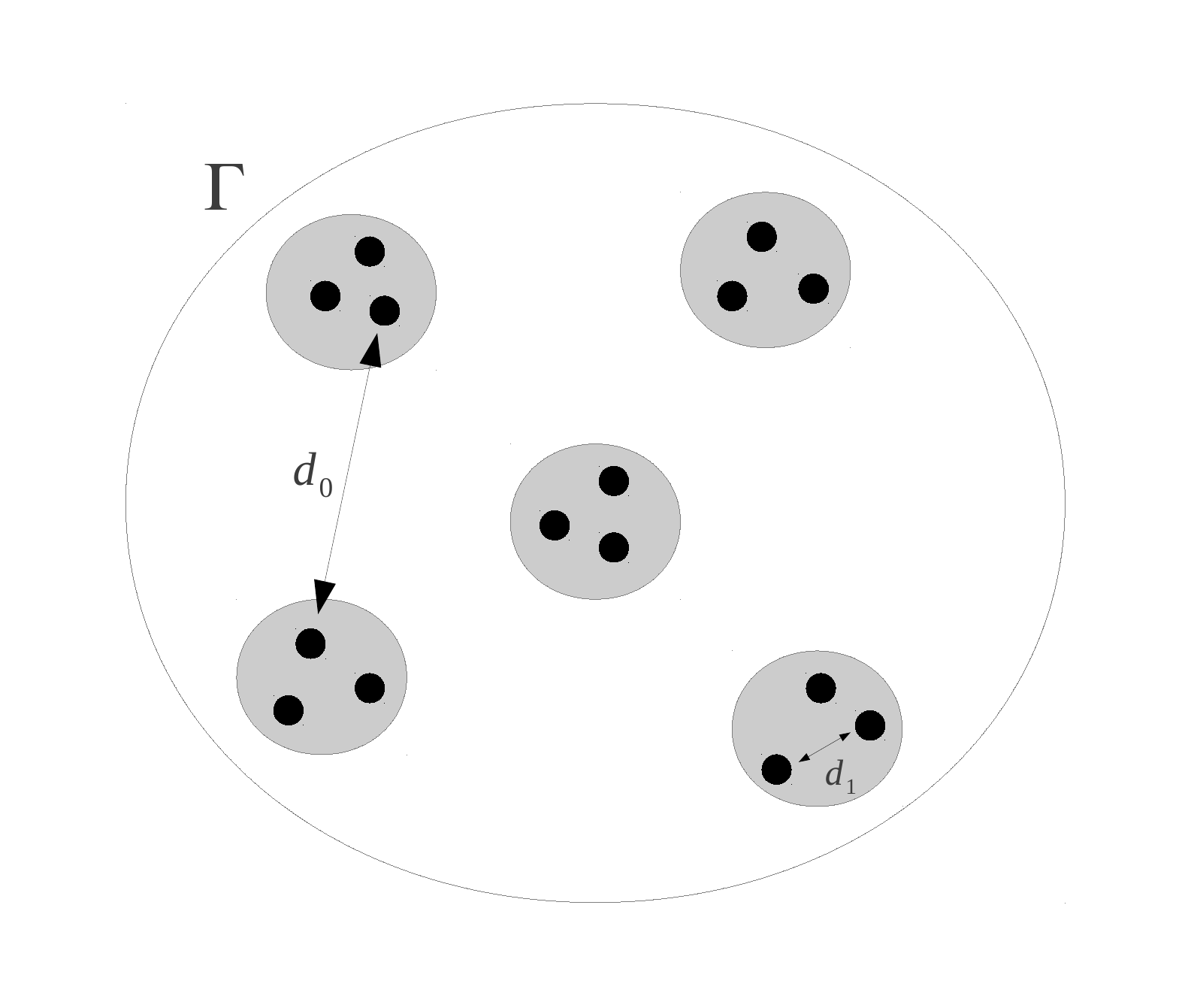}
\caption{RS and RSB structures for the glassy state. Grey blobs are sub-states 
and black dots are replicas. \label{fig:replicasymmetry}}
\end{center}
\end{figure}

Following \cite{KPUZ13}, the \eqref{eq:gauss_r}, once optimized over $\hat \a$ 
(or equivalently $\hat \D$), yields the entropy of the replicated system of hard 
spheres. We must then perform its analytic continuation to real $s$ and then 
take the linear order in $s$ to get the FP potential. In order to perform this 
computation, we must make a choice, an \emph{ansatz}, for the matrix $\hat \D$, 
which encodes the replica structure of the problem, and therefore its physical 
content in terms of structure of the FEL. The simplest choice is the replica 
symmetric (RS) ansatz, which reads
\beq\label{eq:DeltaRS}
\hat \D = 
\begin{bmatrix}
0  & \Dg   & \cdots            & \Dg   & \D^r   & \multicolumn{4}{c}{  \cdots }  
        & \D^r \\
\Dg   & 0   & \cdots            & \Dg   &        & \multicolumn{4}{c}{         } 
         &      \\
\vdots & \ddots & \ddots            & \vdots & \vdots & \multicolumn{4}{c}{      
   }          & \vdots \\
\Dg   & \cdots & \Dg              & 0   & \D^r   & \multicolumn{4}{c}{  \cdots } 
         & \D^r \\
\D^r   & \multicolumn{2}{c}{\cdots} & \D^r   & 0  & \De   & \multicolumn{3}{c} { 
\cdots } & \De \\
       &      &                     &        & \De   & 0   & \multicolumn{3}{c} 
{ \cdots } & \De \\
\vdots &      &                     & \vdots & \vdots & \vdots &      & \ddots & 
              & \vdots \\
       &      &                     &        & \De   & \multicolumn{3}{c} { 
\cdots } & 0   & \De \\
\D^r   & \multicolumn{2}{c}{\cdots} & \D^r   & \De   & \multicolumn{3}{c} { 
\cdots } & \De   & 0\\
\end{bmatrix} 
\eeq
where $\D^g$ is internal to the block of $m$ replicas, $\De$ to the $s$ 
replicas, and $\D^r$ is the relative
displacement between the $m$-type and $s$-type replicas. We also define the 
parameter
\beq\label{eq:Deltaer}
\begin{split}
\D^f \equiv\ &2 \D^r - \D^g -\D.
\end{split}\eeq
Let us examine the \eqref{eq:DeltaRS}. The $m$-block encodes the MSDs between 
the master replicas, the $s-$block of the slave replicas, and the off-diagonal 
blocks the MSDs between the master and slave replicas: the MSDs in the 
off-diagonal blocks are all equal to $\D^r$ as prescribed by the $\d$-constraint 
in the \eqref{eq:defF}; as discussed before, we assume the replicas in the 
$m$-block to be at equilibrium in the liquid phase, so there is no reason for 
the MSDs between them to have any special structure.\\
The physical content of the RS ansatz is contained in the $s-$block. Suppose 
that we have a glassy minimum in the FEL, and we throw $s$ replicas inside it at 
random using the Boltzmann-Gibbs distribution; those replicas probe the 
structure of the bottom of the glassy minimum as we follow it under compression 
or shear. If the minimum is just a plain paraboloid, there is no reason for the 
MSDs between any couple of the slave replicas to depend on the actual couple we 
choose: once they equilibrate inside the state, all replicas are equivalent and 
we can permutate them as we please without changing the physics of the problem: 
we are indeed in a replica symmetric scenario.\\
Let us now assume that bottom of the state actually contains three different 
sub-minima: the situation changes completely. For simplicity let us assume that 
$s=9$, and replicas 1-3, 3-6 and 6-9 end up in minimum 1, 2 and 3 respectively: 
it is clear that the permutation symmetry between replicas has been broken; 
replicas inside the same sub-minimum (like replicas 1 and 3) will be close 
together and will have a low mutual MSD $\D_{13}$, but replicas in two different 
sub-minima (for example replicas 3 and 5) will be farther apart and will have a 
MSD $\D_{35}$ higher than $\D_{13}$: we are in a one-step replica symmetry 
broken (1RSB) scenario \cite{replicanotes}, and the $s-$block of matrix $\hat 
\a$ will have to contain two MSD parameters, $\D_2$ for replicas inside the same 
sub-minimum and $\D_1$ for replicas in different ones \cite{parisiSK}. In figure 
\ref{fig:replicasymmetry} we represent pictorially the difference between the RS 
and 1RSB scenarios.\\ 
The breaking of replica symmetry described above can be iterated: the sub-minima 
could contain sub-sub minima (2RSB) and so on. The process can in principle go 
on indefinitely, and we will see that in fact it does. However, in this chapter 
we limit ourselves to the simplest RS structure, encoded by matrix 
\eqref{eq:DeltaRS}.

The entropy~\eqref{eq:gauss_r} must now be computed for the choice 
\eqref{eq:DeltaRS} of the matrix $\hat \a$ and then optimized with respect to 
$\Dg,\ \De$, and $\D^r$. The computation in quite long and not particularly 
instructive, so we give the details in appendix \ref{app:replicatedRS}, and we 
skip directly to the final result. The entropic term is
\beq
\begin{split}
s_{entr} =&
\log\det\alpha^{m,m} = (1-m-s)\log2-2\log (m+s) + (m-1)\log\Delta^g\\
&+ (s-1)\log\Delta + \log[ms\Delta^f + s\Delta^g + m\Delta],
\end{split}
\label{eq:entrRS}
\eeq
and for the interaction term one has
\beq\label{eq:FF_final}
\FF(\Dg,\De,\D^f)  =
\int \frac{ d\z }{\sqrt{2\pi}}
e^{ - \frac{\z^2}2 }
\FF_0\left( \Dg,\De,\D^f + \z^2 \g^2 \right) \ .
\eeq
with $\FF_0$ equal to
\beq\label{FF0_final}
\begin{split}
\FF_0(\Dg,\De,\D^f)  
= &\int dy \, e^{y} \, \Bigg\{  1 
- \Th\left(  \frac{y + \Dg/2}{\sqrt{2\Dg}} \right)^m\\
&\times \int dx \,  \Th\left(  \frac{ x+ y - \h +\De/2}{\sqrt{2 \De}} \right)^s 
\frac{ e^{- \frac1{2 \D^f} \left( x - \D^f/2  \right)^2  } }{\sqrt{2\pi \D^f}} 
\Bigg\} \ ,
\end{split}\eeq
where we have defined
\beq
\Th(x) \equiv \frac{1}{2}(1+\erf(x)),
\eeq
and $\erf(x)$ is the error function \cite{Abramowitz}. We notice that the 
compression parameter $\eta$ and the shear $\g$ both enter only in the 
interaction term. This completes the computation of the entropy of the $m+s$ 
replicas.

\subsection{Final result for the entropy of the glassy state}

Now that we have obtained the replicated entropy, we have to expand it for small 
$s$ and take the leading order in $s$, in order to get the Franz-Parisi entropy 
of the glassy state.\\
For $s\to 0$ we obtain the real replica entropy for $m$ 
replicas~\cite{parisizamponi,KPZ12}:
\beq\begin{split}
\lim_{s\to 0} s[\hat \a] = s_{m}(\Dg) =\ & 1 -\log\r + \frac{d}2 (m-1) + 
\frac{d}2 \log m + \frac{d}2 (m-1) \log( \pi \Dg/d^2 ) \\& -\frac{d}2 \wh\f_g 
 \int dy \, e^y \, \left[1 - \Th\left(  \frac{y + \Dg/2}{\sqrt{2\Dg}} \right)^m 
\right] \ .
\end{split}\label{eq:seref}\eeq

The linear order in $s$ gives, finally, the in-state entropy of the glassy state 
(Franz-Parisi entropy):
\beq\label{eq:seint}
\begin{split}
\lim_{s\to 0} \partial_s \{ s[\hat \a] \} = s_{g}  
 =\ & \frac{d}2 + \frac{d}2 \frac{\Dg  + m \D^f }{m \De} + \frac{d}2 \log(\pi 
\De/d^2) \\
&+\frac{d \wh\f_g}2
\int \DD\z
  \int  dy \,e^{y} \, \Th\left(  \frac{y + \Dg/2}{\sqrt{2\Dg}} \right)^m \\
  &\times\int dx \, \log\left[ \Th\left(  \frac{ x+ y - \h +\De/2}{\sqrt{2 \De}} 
\right) \right] 
\frac{ e^{- \frac1{2 \D_\g(\z)} \left( x - \D_\g(\z)/2  \right)^2  } 
}{\sqrt{2\pi \D_\g(\z)}}  \ ,
\end{split}\eeq
where $\D_\g(\z) = \D^f + \z^2 \g^2$ and we recall that $\DD \z = \frac{ d\z 
}{\sqrt{2\pi}}
e^{ - \frac{\z^2}2 }$.
It will be often convenient to make a change of variable $x' = (x - \D_\g(\z)/2 
) / \sqrt{ \D_\g(\z) }$ in the integral,
which leads to (dropping the prime for convenience):
\beq\label{eq:seint2}
\begin{split}
 s_{g} =\ & \frac{d}2 + \frac{d}2 \frac{\Dg  + m \D^f }{m \De} + \frac{d}2 
\log(\pi \De/d^2)+\frac{d \wh\f_g}2
  \int  dy \,e^{y}\ \Th\left(  \frac{y + \Dg/2}{\sqrt{2\Dg}} \right)^m\\
  &\int \DD\z \, \DD x \, \log\left[ \Th\left(  \frac{ \sqrt{ \D_\g(\z) } x 
+\D_\g(\z)/2   + y - \h +\De/2}{\sqrt{2 \De}} \right) \right] 
 \ .
\end{split}\eeq
From this expression of the internal entropy, we can obtain the saddle point 
equations for $\De$ and $\D^f$ and study the behavior
of glassy states. We notice that the parameter $\D_g$ is contained only in the 
\eqref{eq:seref}, so its saddle point equation is independent of both $\eta$ and 
$\g$ and only depends on the glass transition density $\wh{\varphi}_g$.

\subsection{Saddle point equations}
As already detailed, the FP entropy \eqref{eq:seint2} must be optimized over 
$\D$, $\D^f$ and $\D^g$ in order to get the entropy of the metastable glassy 
state. The equation for $\Dg$ is obtained by maximizing eq.~\eqref{eq:seref}. We 
have
\beq\label{eq:Dg}
0 = \frac{m-1}{m\Dg} + \frac{\wh\f_g}2 \int dy \ e^y 
\Th\left(\frac{y+\Dg/2}{\sqrt{2\Dg}}\right)^{m-1}
 \frac{e^{-\frac{(y + \Dg/2)^2}{2\Dg}}}{\sqrt{2\pi \Dg}} \left(   \frac12 - 
\frac{y}\Dg  \right) \ .
\eeq
For a fixed reference density $\wh\f_g$ (and fixed $m$, here we are mostly 
interested in $m\to 1$), one can solve this equation to obtain $\Dg$.
Then,
the entropy in eq.~\eqref{eq:seint2} must be maximized with respect to $\De$ and 
$\Delta^f$ to give the internal entropy of a glass state prepared
at $\wh\f_g$ (the value of $\Dg$ is the equilibrium one corresponding to 
$\wh\f_g$) and followed at a different state point parametrized by $\h$ and 
$\g$.
As usual in replica computations \cite{parisiSGandbeyond}, the analytical 
continuation to $s\to 0$ induces a change in the properties of the entropy, and 
as a consequence
the solution of the equations for $\De$ and $\Delta^f$ is not a maximum, but 
rather a saddle-point. However, the correct prescription is not
to look at the concavity of the entropy, but to check
that all the eigenvectors of the Hessian matrix of the $s[\hat \a]$ remain 
negative, as we discuss in the next section.

The equations for $\De$ and $\Delta^f$ are obtained from the conditions 
$\frac{\partial s_{g}}{\partial \De} = 0$ and
$\frac{\partial s_{g}}{\partial \D^f} = 0$. Starting from eq.~\eqref{eq:seint2} 
and taking the derivatives, we get
\begin{eqnarray}
\label{eq:eqD}
 0 &=& \frac{m\De-\Dg-m\Delta^f}{m \De^2}\\
 & &+ \frac{\wh\f_g}2 \int dy \DD x  \DD\z \ 
e^y\frac{\Th\left(\frac{y+\Dg/2}{\sqrt{2\Dg}}\right)^m}{\Th\left(\frac{\x}{\sqrt
{2\De}}\right)}
 \frac{e^{-\frac{\x^2}{2\De}}}{\sqrt{2\pi \De}}  \left( 1 - \frac{\x 
}{\D}\right),\nonumber \\
\label{eq:eqDf}
 0&=& \frac{1}{\De} +\frac{\wh\f_g}2 \int dy \DD x \DD\z \ 
e^y\frac{\Th\left(\frac{y+\Dg/2}{\sqrt{2\Dg}}\right)^m}{\Th\left(\frac{ 
\x}{\sqrt{2\De}}\right)}
 \frac{e^{-\frac{\x^2}{2\De}}}{\sqrt{2\pi \De}} \left( 1 + \frac{ x 
}{\sqrt{\Delta_\gamma(\zeta)}}\right),
 \end{eqnarray}
with
\begin{eqnarray}
 \label{eq:xidef}
 \x &=&  \sqrt{ \D_\g(\z) } x +\D_\g(\z)/2 +y-\h+\De/2 \ ,
 \end{eqnarray}
 wherein again $\D_\g(\z) = \D^f + \g^2 \z^2$.
 In some cases, it might be useful to perform an additional change of variables 
from $y$ to $\xi$. Those saddle point equations must be solved for varying 
$\eta$ and $\gamma$, and the entropy \eqref{eq:seint} must be computed along the 
solution in order to get the physical observables of the glass that we compute 
in the next paragraph.

\subsection{Physical observables}
We now compute the pressure and the shear stress, that are the responses of the 
glassy state to compression and shear-strain, respectively. We recall that 
$\De$ and $\Delta^f$ are obtained by setting the derivatives of $s_{g}$ with 
respect to them equal to zero, which means that when we take for example the 
derivative
of $s_{g}$ with respect to $\g$, it is enough to take the partial derivative 
instead of the total one.

For a system of hard spheres, the reduced pressure $p = \b P/\r$ is the response 
of the system to compression and is given by~\cite{parisizamponi,simpleliquids}
\beq
p_g = - \wh\f \frac{\partial s_g}{\partial \wh\f} = - \frac{\partial 
s_{g}}{\partial \h} \ ,
\eeq
and we get from eq.~\eqref{eq:seint2}:
\beq
\frac{p_g}{d} = \frac{\wh\f_g}{2}\int dy \DD x \DD \z \ 
e^y\frac{\Th\left(\frac{y+\Dg/2}{\sqrt{2\Dg}}\right)^m}{\Th\left(\frac{\xi}{
\sqrt{2\De}}\right)}\frac{e^{-\frac{\xi^2}{2\De}}}{\sqrt{2\pi \De}}
\label{eq:eqpressure}
\eeq
recalling Eq.~\eqref{eq:xidef}.
The $p_g/d$ vs. $\wh\f$ (or $\h = \log(\wh\f/\wh\f_g)$) curve is the equation of 
state of the corresponding metastable glass.

The response to a shear strain is given by the shear stress, which is defined 
as~\cite{YoshinoZamponi14}
\beq
\b \s = - \frac{\partial s_g}{\partial \g},
\eeq
and we get from eq.~\eqref{eq:seint2}:
\beq
\frac{\b \s}{d} = - \g \frac{ \wh\f_g  }2 \int dy \DD x \DD\z \  
e^y\frac{\Th\left(\frac{y+\Dg/2}{\sqrt{2\Dg}}\right)^m}{\Th\left(\frac{\x}{\sqrt
{2\De}}\right)}
\frac{e^{-\frac{\x^2}{2\De}}}{\sqrt{2\pi \De}} 
\left(1 + \frac{x}{\sqrt{\Delta_\gamma(\zeta)}}\right) \z^2 \ .
\label{eq:eqshear}
\eeq

\subsubsection{Shear modulus and dilatancy}

It is interesting to consider as a particular case the response of the glass to 
an infinitesimal strain, $\g\to 0$.
In that case, we have that both $\D_\g(\z) \to \D^f$ and $\x \to \sqrt{ \D^f } x 
+\D^f/2 +y-\h+\De/2$ become independent of $\z$.
We have thus
\beq\label{eq:mu}
\frac{\b \mu}{d} = \lim_{\g\to 0} \frac{\b \s}{d \g} = - 
\frac{ \wh\f_g  }2 \int dy \DD x  \  
e^y\frac{\Th\left(\frac{y+\Dg/2}{\sqrt{2\Dg}}\right)^m}{\Th\left(\frac{\x}{\sqrt
{2\De}}\right)}
\frac{e^{-\frac{\x^2}{2\De}}}{\sqrt{2\pi \De}} 
\left(1 + \frac{x}{\sqrt{\Delta^f}}\right) \int \DD\z \ \z^2 = \frac{1}{\De} \ ,
\eeq
where the last equality is obtained by noticing that $\int \DD\z \ \z^2 = 1$ and 
using eq.~\eqref{eq:eqDf} in the limit $\g\to 0$, where
again the integral over $\z$ disappears because $\x$ and $\D_\g$ become 
independent of $\z$. In this way we see that 
$\s/\g \to \mu$, where $\mu$ is the shear modulus of the glass and it is 
inversely proportional to the cage radius. This provides
an alternative derivation of the results 
of~\cite{yoshinomezard,YoshinoZamponi14}.

From eq.~\eqref{eq:mu} we deduce that for small $\g$ the physical entropy is
\beq
s_g(\h,\g) = s_g(\h,\g=0) - \frac{d}2 \g^2 \frac{1}{\De(\h,\g=0)} + \cdots \ , 
\eeq
where $\De(\h,\g)$ is the solution of eqs.~\eqref{eq:eqD}-\eqref{eq:eqDf}. 
Therefore we have
\beq
\begin{split}
p_g(\h,\g) =&\ -\frac{\de s_g(\h,\g)}{\de \h} = p_g(\h,\g=0) + 
\frac{d}2 \g^2 \frac{\de}{\de\h} \frac{1}{\De(\h,\g=0)} + \cdots\\
= &\ p_g(\h,\g=0) + \g^2 (\b R(\h)/\r) + \cdots \ ,
\end{split}
\eeq
from which we deduce the expression of the dilatancy $R$ as
\beq\label{eq:R}
\frac{\b R(\h)}{\r} = \frac{d}2 \frac{\de}{\de\h} \frac{1}{\De(\h,\g=0)} \ .
\eeq

\section{Stability of the RS ansatz}
In the preceding section we have detailed the computation of the replicated 
entropy and the FP potential, employing the simplest possible RS ansatz. The 
computation has been made with the saddle point, or steepest descent method 
\cite{AdvancedMethods}, wherein an integral in the form $\int dx\ e^{Nf(x)}$ is 
approximated as $e^{Nf(x^*)}$, with $x^*$ a point of maximum of the $f(x)$. It 
can be shown that the error committed with this approximation vanishes in the 
limit $N\to 0$ as long as the point $x^*$ is a maximum, i.e. if $\left.\frac{d^2 
f}{dx^2}\right|_{x=x^*}<0$.\\
We stress the fact that this is not only a mathematical problem, but also a 
physical one: in statistical mechanics, the function $f(x)$ is usually the Gibbs 
free energy as a function of the order parameter, like the Gibbs free energy 
$f(m)$ of the Curie-Weiss model as discussed in paragraph 
\ref{subsec:RFOTfoundations}, and its stationary points $x^*$ correspond to the 
possible phases the system can be found in. If the second derivative of the 
$f(x)$ becomes zero, this means that a phase transition is taking place: the 
stable phase (or equivalently, the saddle point) the system is in becomes 
critical (flat), an infinite susceptibility manifests, and below the transition 
the stable saddle point shifts to a different value $x^*$, while the stationary 
point that was stable above the transition ($m^* = 0$ in the case of the 
Curie-Weiss model) becomes unstable, as shown in figure \ref{fig:curieweiss} for 
the ferromagnetic transition.

In the case of the replicated entropy, we have a much more complicated ``Gibbs 
free energy'' $s[\hat\a]$ and ``order parameter'' $\hat\a$, but the spirit is 
exactly the same: checking the concavity of the replicated entropy $s[\hat\a]$ 
is one and the same with studying the behavior of the glassy state in terms of 
in-state phase transitions wherein the glassy minimum would ``split'' in 
sub-minima as in figure \ref{fig:curieweiss}. In summary, we have to check that 
the saddle point given by equations \eqref{eq:Dg}, \eqref{eq:eqD} and 
\eqref{eq:eqDf} is a maximum of the replicated entropy.

It is however clear that the question requires some caution. First of all, we 
have restricted ourselves to a fixed form of the matrix $\hat \a$, i.e. the RS 
one. As a result of this, checking the concavity of the replicated entropy with 
respect to $\D_g$, $\D$ and $\D_f$ would only tell us if there is an instability 
within the RS ansatz, but would miss instabilities towards saddle points with 
more RSBs, which as we discussed before is the interesting case corresponding to 
a phase transition inside the minimum. As a result of this, the general Hessian 
matrix of the $s[\hat\a]$,
\beq
M_{a < b;c < d} \equiv \frac 2d \frac{\d^2 s[\hat \a]}{\d \a_{a<b}\d \a_{c<d}}
\label{eq:defhessian}
\eeq
must be calculated, and only \emph{then} we can compute it on the RS solution at 
the saddle point.\\
Another issue is the fact that the $s[\hat \a]$ is also a function of the 
parameter $s$, which we send to zero in the end to compute the FP potential. So 
we are considering the Hessian of a function of a matrix which has one block of 
size $s\times s$ with $s$ going to zero, a clearly pathological situation. To 
illustrate the problems that one could have, a pedagogical example 
\cite{parisiSGandbeyond} is a $s\times s$ matrix with zeros on the diagonal and 
an off-diagonal parameter $q$
\beq
Q = \begin{pmatrix} 0 & q & q\\
q & 0 & q\\
q & q & 0
\end{pmatrix} \qquad s=3
\eeq
And a function $s[Q]$ defined as
\beq
s[Q] \equiv -{\rm Tr}(Q^2) = -s(s-1)q^2.
\eeq
As long as $s\geq1$, it is clear that $q=0$ is a maximum of $s$. But if we send 
$s\to 0$, it evidently becomes a minimum: changing the value of $s$ can change 
the nature of stationary points of the $s[\hat \a]$, as we precedingly 
discussed. As a result of this, the usual prescription in replica theory is that 
the Hessian \eqref{eq:defhessian} must be computed for general $s$, and then one 
must verify that the analytic continuation of its eigenvalues for $s\to 0$ be 
negative, in order to check stability \cite{parisiSGandbeyond,zamponi}.

\subsection{The unstable mode \label{subsec:unstablemode}}
Now that we have settled these questions we can proceed with the calculation of 
the Hessian \eqref{eq:defhessian}. If we observe it closely, we see that we are 
checking the fluctuations of the $s[\hat \a]$ with respect to all the elements 
in the $\hat \a$ matrix, i.e. we are considering fluctuations in all the blocks. 
But already from an intuitive point of view, it is clear that only fluctuations 
in the sector of slave replicas should matter: the $s$ replicas are the ones 
that probe the bottom of the glass state as we follow it, while the $m$ ones 
only select the state and remain at equilibrium in the liquid phase. Only the 
$s$ replicas are able to detect a transition within the state like the one shown 
in figure \ref{fig:curieweiss} (this intuitive argument is made more formal in 
appendix \ref{app:stabilityRS}).\\
This means that we can focus just on the ``reduced Hessian''
\beq
M^s_{a \leq b;c\leq d} \equiv \frac 2d \frac{\d^2 s[\hat \a]}{\d \a_{a<b}\d 
\a_{c<d}}, \qquad a,b,c,d\  \in [m+1,m+s]. 
\eeq
and compute it on the RS solution. This reduced Hessian enjoys the same exact 
replica permutation symmetries of the Hessians typically considered within the 
real replica method \cite{crisantisommers,BM79,KPUZ13,ferrarileuzzi2}, wherein 
the matrix $\hat\a$ is just an RS matrix without any block structure. In 
particular, one can show that because of replica symmetry, the Hessian $M^s$, 
when computed on the RS solution, must necessarily have the general form
\beq
M^s_{a < b;c < d} 
=M_1\left(\frac{\d_{ac}\d_{bd}+\d_{ad}\d_{bc}}{2}\right)+M_2\left(\frac{\d_{ac}
+\d_{ad}+\d_{bc}+\d_{bd}}{4}\right)+M_3,
\label{stab_matr}
\eeq
so it effectively depends only on three parameters $M_1$, $M_2$ and $M_3$. As 
shown in \cite{crisantisommers,BM79}, this operator has only three independent 
eigenvalues,
\ba
\l_R &=& M_1 \\
\l_L &=& M_1 + (s-1)(M_2 + sM_3)\\
\l_A &=& M_1 + \frac{s-2}{2}M_2
\ea
called the \emph{replicon}, \emph{longitudinal} and \emph{anomalous} modes, 
respectively. Each of these modes is relative to a subspace of the vector space 
the $s-$block of the matrix $s$ lives in. Of these three modes, the replicon 
mode is the one which is linked to instabilities towards $s$-blocks with mode 
RSBs (the state splits up as in figure \ref{fig:curieweiss}), while the 
longitudinal one, for example, gives information relative to spinodal points 
(the state opens up along an unstable direction and becomes a saddle), so it is 
linked to the MCT transition and threshold states \cite{zamponi}, and also with 
yielding as we are going to see. However, here we are only interested in the 
replicon mode.

In summary, we must compute the replicon mode $\l_R = M_1$, perform its analytic 
continuation for $s\to 0$, and check its sign along the solution given by 
equations \eqref{eq:eqD}, \eqref{eq:eqDf} and \eqref{eq:Dg}. The computation of 
the replicon follows the same lines as the one performed in \cite{KPUZ13} with 
only a few modifications, so we skip, again, directly to the final result and 
refer to appendix \ref{app:stabilityRS} for the details. The replicon mode, for 
$s=0$, is:
\beq\label{eq:lR}
\l_R = \frac{1}{\D^2}(-16 -8 \wh\f_g -8\wh\f_g  \langle \Th_0(\l)^{-1}  
\LL_0(\l)
\rangle),
\eeq
With
\beq
\begin{split}
\LL_s(\l) = \left[\left(\frac{\Th_1(\l)}{\Th_0(\l)}\right)^2 - \l 
\frac{\Th_1(\l)}{\Th_0(\l)}\right]
\left[(2 - 2 \l^2) + (s-4) \left(\frac{\Th_1(\l)}{\Th_0(\l)}\right)^2 + 
(6 - s) \l \frac{\Th_1(\l)}{\Th_0(\l)}\right],
\end{split}
\eeq
and 
\ba
\Th_0(x) &\equiv& \Th(-x/\sqrt{2}),\\
\Th_1(x) &\equiv& e^{-\frac{1}{2}x^2}/\sqrt{2\pi},
\ea
and the average $\thav{\bullet}$ is defined as
\beq
\langle O(\l) \rangle=\int_{-\infty}^\infty \mathcal D \l \, G(\l) O(\l) = 
\int_{-\infty}^\infty \mathcal D \l \bar{\mathcal D}_m \l' \,K(\l',\l) \, O(\l). 
\label{eq:Glambdadef}
\eeq
with the kernel $K(\l',\l)$ defined in the \eqref{eq:Kdef} and the measure 
$\bar{\mathcal D}_m \l'$ in the \eqref{eq:measuredef}.

Now, we must solve the equations \eqref{eq:eqD}, \eqref{eq:eqDf} and 
\eqref{eq:Dg} for varying $\eta$ and $\g$, and then compute along the solution 
the pressure \eqref{eq:eqpressure}, the shear strain \eqref{eq:eqshear}, and the 
replicon mode \eqref{eq:lR}, which will allow us to draw phase diagrams for the 
glass like the ones measured experimentally in figures \ref{fig:ultraexp} and 
\ref{fig:shearstrain}. In the following section we present the results so 
obtained.

\section{Results}

In this section we solve the saddle point equations derived above, and compute 
the observables of the glass. We proceed as follows: first we choose a planting 
density $\wh \f_g$, and solve the \eqref{eq:Dg} to get the cage radius $\D$. We 
then solve the \eqref{eq:eqD} and \eqref{eq:eqDf} for varying $\eta$ and $\g$, 
and compute the value of observables along the solution. We treat separately the 
compression-decompression and shear strain case, and we repeat the procedure for 
different $\wh\f_g$s, corresponding to different annealing protocols as 
discussed in paragraph \ref{subsec:ultrastable}.\\
The equations cannot be solved analytically in the general case, so a numerical 
algorithm is required. Our choice is to use the iteration method, and we compute 
numerically the integrals appearing in the expressions using the simplest 
rectangle method; the integrands (which are essentially error functions or 
suitable combinations of them) are implemented numerically using the Faddeeva 
Library for 
C++\footnote{\url{http://ab-initio.mit.edu/wiki/index.php/Faddeeva_Package}}.\\
Before reporting the full solution, we first focus on two special limits wherein 
an analytical solution can be obtained, providing a check of the numerics.

\subsection{Special limits \label{subsec:speclimits}}

\subsubsection{Equilibrium limit}

When $\h = \g = 0$, the constrained replicas sample the glass basins in the same 
state point as the reference replicas.
Therefore, it is reasonable to expect that eqs. \eqref{eq:eqD} and 
\eqref{eq:eqDf} admit
$\De = \D^r = \Dg$, hence $\D^f=0$, as a solution. 

To check this, we first analyze eq.~\eqref{eq:eqD}. For $\h=\g=0$, $\De=\Dg$ and 
$\D^f=0$, we have $\D_\g(\z)=0$ and $\xi = y + \De/2$.
Therefore the integrand does not depend on $x$ and $\z$ and $\int \DD x \DD 
\z=1$. Then it is clear that eq.~\eqref{eq:eqD} becomes
equivalent to eq.~\eqref{eq:Dg} and is satisfied by our conjectured solution.

The analysis of Eq.~\eqref{eq:eqDf} is slightly more tricky. Setting $\h=0$, 
$\gamma=0$, $\De = \Dg$ we get, with a change of variable $x' = x \sqrt{\D^f} + 
\D^f/2$
(and then dropping the prime for simplicity):
\beq
\begin{split}
-\frac{2}{\wh\f_g\De} =\ &\int dy\ 
e^y\Th\left(\frac{y+\De/2}{\sqrt{2\De}}\right)^m\\
&\times\int dx\, \frac{e^{-\frac{(x-\Delta^f/2)^2}{2\Delta^f}}}{\sqrt{2\pi 
\Delta^f}}\frac{1}{\Th\left(\frac{x+y+\De/2}{\sqrt{2\De}}\right)}\frac{e^{-\frac
{(x+y +\De/2)^2}{2\De}}}{\sqrt{2\pi \De}} 
\left(\frac{x+\Delta^f/2}{\Delta^f}\right) \ .
\label{eq:eqaf1}
\end{split}
\eeq
We now observe that
\beq
 \left(\frac{x+\Delta^f/2}{\Delta^f}\right) 
\frac{e^{-\frac{(x-\Delta^f/2)^2}{2\Delta^f}}}{\sqrt{2\pi \Delta^f}}
 =  \left( -\frac{d}{dx} + 1 \right) 
\frac{e^{-\frac{(x-\Delta^f/2)^2}{2\Delta^f}}}{\sqrt{2\pi \Delta^f}} 
\xrightarrow[\D^f\to 0]{ } -\delta'(x) + \delta(x)
\eeq
where $\d(x)$ is the Dirac delta distribution.
Therefore eq.~\eqref{eq:eqaf1} becomes, with some manipulations
\beq
\begin{split}
 -\frac{2}{\wh\f_g\De} =\ &\int dy\ 
e^y\Th\left(\frac{y+\De/2}{\sqrt{2\De}}\right)^m\int dx \, \left[ -\delta'(x) + 
\delta(x) \right]
\frac{1}{\Th\left(\frac{x+y+\De/2}{\sqrt{2\De}}\right)}\frac{e^{-\frac{(x+y 
+\De/2)^2}{2\De}}}{\sqrt{2\pi \De}} \\
=\ &\int dy\ e^y\Th\left(\frac{y+\De/2}{\sqrt{2\De}}\right)^m  \left( 
\frac{d}{dy} + 1 \right) \left[
\frac{1}{\Th\left(\frac{y+\De/2}{\sqrt{2\De}}\right)}\frac{e^{-\frac{(y 
+\De/2)^2}{2\De}}}{\sqrt{2\pi \De}}  \right] \\
=\ &\frac{m}{m-1} \int dy\, e^y\Th\left(\frac{y+\De/2}{\sqrt{2\De}}\right)^{m-1} 
 \left(\frac12 - \frac{y}{\De} \right) \frac{e^{-\frac{(y 
+\De/2)^2}{2\De}}}{\sqrt{2\pi \De}} \ ,
\end{split}
\eeq
So, also the equation for $\D_f$ becomes equivalent equivalent to 
eq.~\eqref{eq:Dg}. We conclude that for $\h=\gamma=0$, the solution of 
Eqs.~\eqref{eq:eqD}-\eqref{eq:eqDf} is $\Delta^f=0$ and $\Dg=\De$ for all $m$, 
which furnishes us with a starting point for the iteration algorithm.

A particularly interesting case is when $m=1$. In this case the reference 
replicas are in equilibrium in the liquid phase, and the
constrained replicas therefore sample the glass basins that compose the liquid 
phase {\it at equilibrium}, like in the real replica method 
\ref{sec:realreplica}. From eq.~\eqref{eq:eqpressure} it is quite easy to see 
that for $m=1$, $\g=\h=0$, $\D^f=0$, $\Dg=\De$, one has 
$\D_\g(\z)=0$ and $\xi = y + \De/2$ and
\beq
p_g = \frac{d \, \wh\f_g}{2}\int dy \, 
e^y \frac{e^{-\frac{(y+\De/2)^2}{2\De}}}{\sqrt{2\pi \De}} = \frac{d\, 
\wh\f_g}{2} = p_{\rm liq} \ .
\eeq
Where we have used the fact that the equation if state of HSs in the MF limit is 
just the Van Der Waals one \cite{simpleliquids}. This shows in particular that 
the pressure of the glass {\it merges continuously with the liquid pressure} at 
$\wh\f_g$, as it should be.
\ref{chap:SFconstruction}. The difference $\Si = s_{\rm liq} - s_g$ then gives 
the equilibrium complexity $\Sigma(\hat\f)$ of the supercooled liquid.

\subsubsection{The jamming limit}
The other interesting limit is of course the jamming limit, whereupon the 
internal pressure of the glass state diverges as detailed in paragraph 
\ref{subsec:jamming} and
correspondingly its MSD $\De\to 0$~\cite{parisizamponi}.\\
To investigate this limit, we specialize to the case $\g=0$ and we consider the 
limit $\De\to 0$ of equations \eqref{eq:eqD} and \eqref{eq:eqDf}. 
Using the relation
\beq
\lim_{\mu\to 0} \Th(x/\sqrt{\mu})^\mu = e^{-x^2 \th(-x)} \ ,
\eeq
the leading order of eq.~\eqref{eq:seint2} is
\beq\label{eq:sgj}
\begin{split}
 s_g  
 \simeq\ &\frac{d}2 \frac{\Dg  + m \D^f }{m \De} -\frac{d \wh\f_g}{4 \De}  \int  
dy \,e^{y} \, \Th\left(  \frac{y + \Dg/2}{\sqrt{2\Dg}} \right)^m\\
&\times\int_{-\io}^{\h-y} dx \, (  x+ y - \h )^2  \frac{ e^{- \frac1{2 \D^f} 
\left( x - \D^f/2  \right)^2  } }{\sqrt{2\pi \D^f}}\\
 =\ &\frac{d}{2 \De} \left\{ \frac{\Dg}m  +  \D^f  
-\frac{ \wh\f_g}{2 }  \int  dy \,e^{y} \, \Th\left(  \frac{y + 
\Dg/2}{\sqrt{2\Dg}} \right)^m \int_{-\io}^{0} dx \, x^2  
\frac{ e^{- \frac1{2 \D^f} \left( x -y + \h - \D^f/2  \right)^2  } }{\sqrt{2\pi 
\D^f}} \right\}.
\end{split}\eeq
Hence, we obtain that $s_g \sim C/\D + \log\D + \cdots$ when $\D \to 0$, where 
the term $\log\D$ is explicitly present in eq.~\eqref{eq:seint2}.
Next, we observe that: 
\begin{itemize}
\item The coefficient $C$ should vanish at jamming. This is because $s_g = 
\De^{-1} C + \log \De + \cdots$, hence the equation for $\De$
is $-\De^{-2} C + \De^{-1} + \cdots = 0$, or equivalently $-C + \De + \cdots 
=0$,
which shows that when $C  \to 0$, also $\De = C \to 0$. The jamming point is 
therefore defined by $C\to 0$.
Note by the way that this condition guarantees that $s_g \sim \log \De$ when 
$\De \to 0$, which is the physically correct behavior of the glass entropy 
because
particles are localized on a scale $\D$~\cite{parisizamponi}.
\item The derivative of $C$ with respect to $\D^f$ should also vanish, because 
it determines the equation for $\D^f$ at leading order in $\De$.
\end{itemize}
The two conditions $C=0$ and $dC/d\D^f=0$ give two equations that determine the 
values of $\h$ and $\D^f$ at the jamming point, for a fixed glass 
(i.e. at fixed $\wh\f_g, \Dg, m$).
These two equations read
\beq\label{eq:appj}
\begin{split}
0 &= \frac{\Dg}m  +  \D^f  
-\frac{ \wh\f_g}{2 }  \int  dy \,e^{y} \, \Th\left(  \frac{y + 
\Dg/2}{\sqrt{2\Dg}} \right)^m \int_{-\io}^{0} dx \, x^2  
\frac{ e^{- \frac1{2 \D^f} \left( x -y + \h - \D^f/2  \right)^2  } }{\sqrt{2\pi 
\D^f}}  \ , \\
0 &= 1 - \frac{ \wh\f_g}{2 } \frac{d}{d\D^f}  \int  dy \,e^{y} \, \Th\left(  
\frac{y + \Dg/2}{\sqrt{2\Dg}} \right)^m \int_{-\io}^{0} dx \, x^2  
\frac{ e^{- \frac1{2 \D^f} \left( x -y + \h - \D^f/2  \right)^2  } }{\sqrt{2\pi 
\D^f}} \  .
\end{split}\eeq
Note that one could get
the same equations by taking directly the $\De \to 0$ limit of 
Eqs.~\eqref{eq:eqD} and \eqref{eq:eqDf}.
This system of two equations determines the values of the jamming density $\wh 
\f_j = \wh \f_g e^{\h_j}$ 
and the corresponding $\Delta^f_j$. Note that in general $C \sim |\h - \h_j|$ 
and therefore $\De = C \sim |\h - \h_j|$ vanishes linearly at jamming, as we 
will see.\\
Moreover, from the \eqref{eq:eqpressure} one can see that in the jamming limit
\beq
p \simeq \D^{-1}.
\eeq
This scaling of the pressure is predicted also by the real replica method 
\cite{KPUZ13}, always with the RS ansatz. However, it does not coincide with the 
scaling $p\simeq \D^{-\kappa}$ reported in paragraph \ref{subsec:jamming} and in 
\cite{IBB12}. This already points towards the fact that the simplest RS ansatz 
is not sufficient in the jamming limit and a phase transition to a more 
complicated internal structure of the state is present, as we will show in the 
following.

\subsection{Compression-decompression}
We report here the result for compression and decompression protocols, with 
varying $\eta$ and $\g=0$. This protocols mimic, for HSs, the DSC experiments 
discussed in paragraph \ref{subsec:ultrastable}.

\subsubsection{Mean Square displacements}
\begin{figure}[t!]
\includegraphics[width=.5\textwidth]{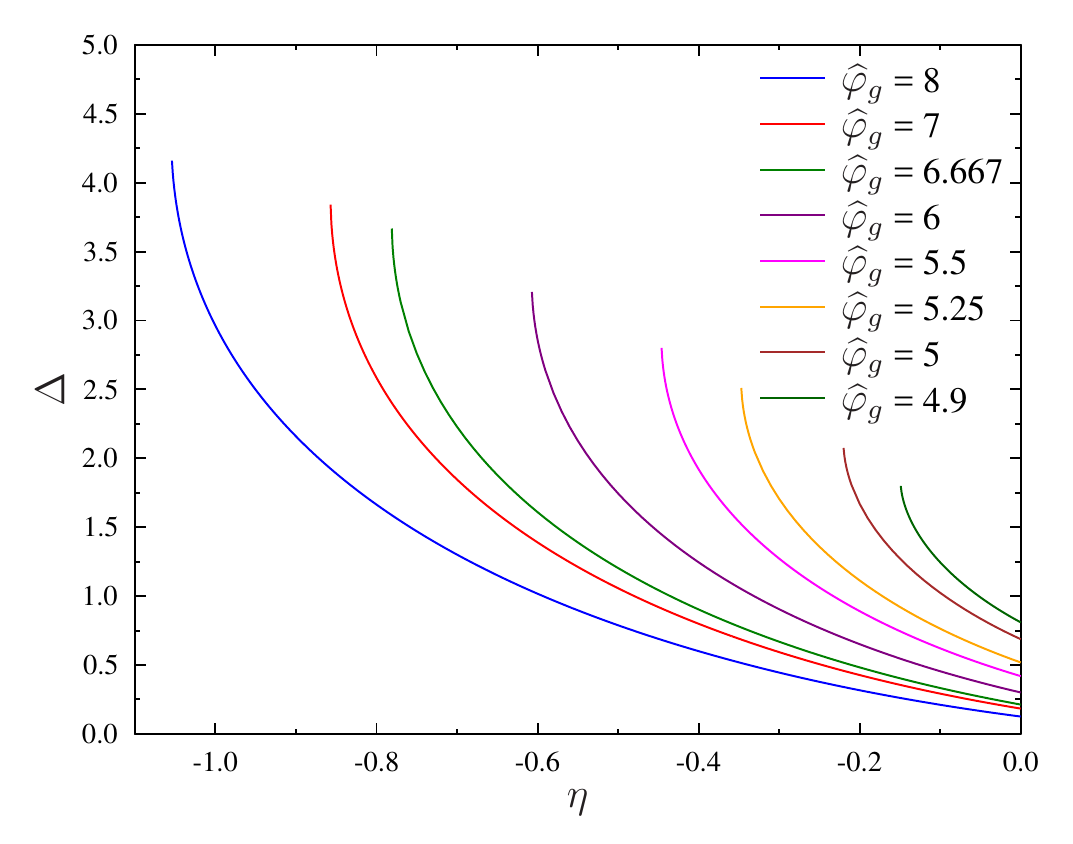}\nolinebreak
\includegraphics[width=.5\textwidth]{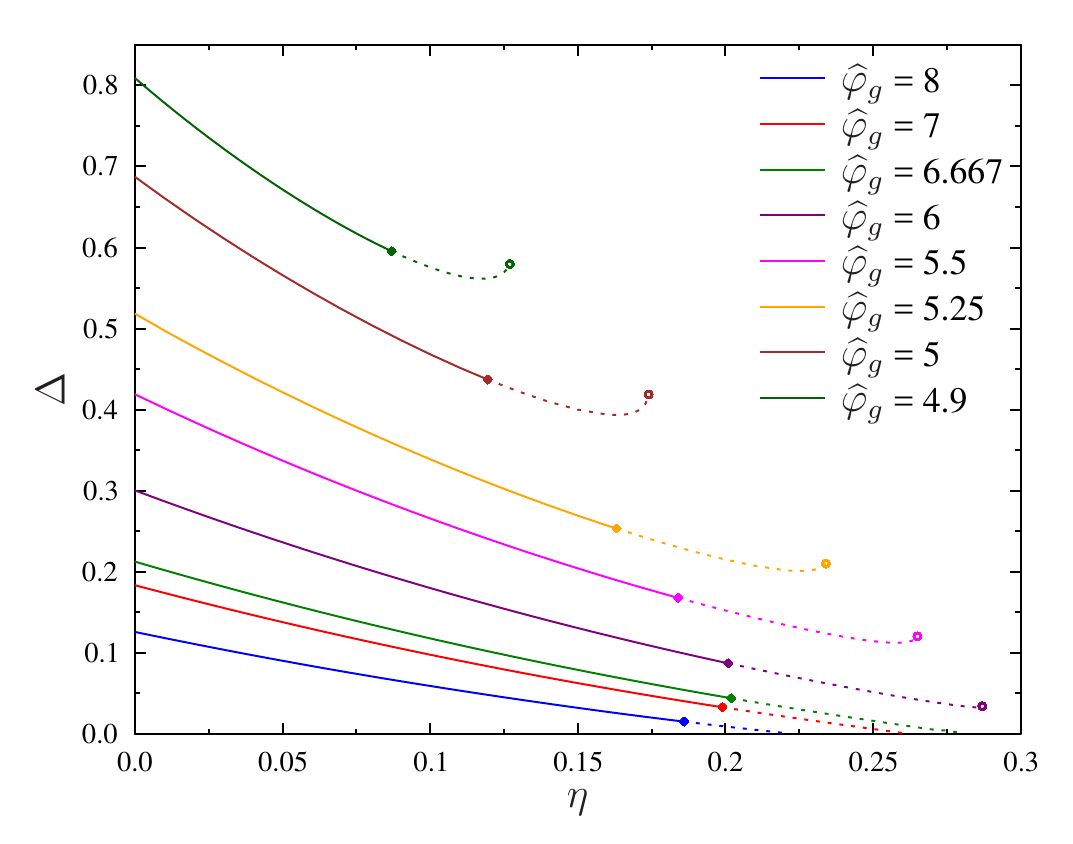}
\includegraphics[width=.5\textwidth]{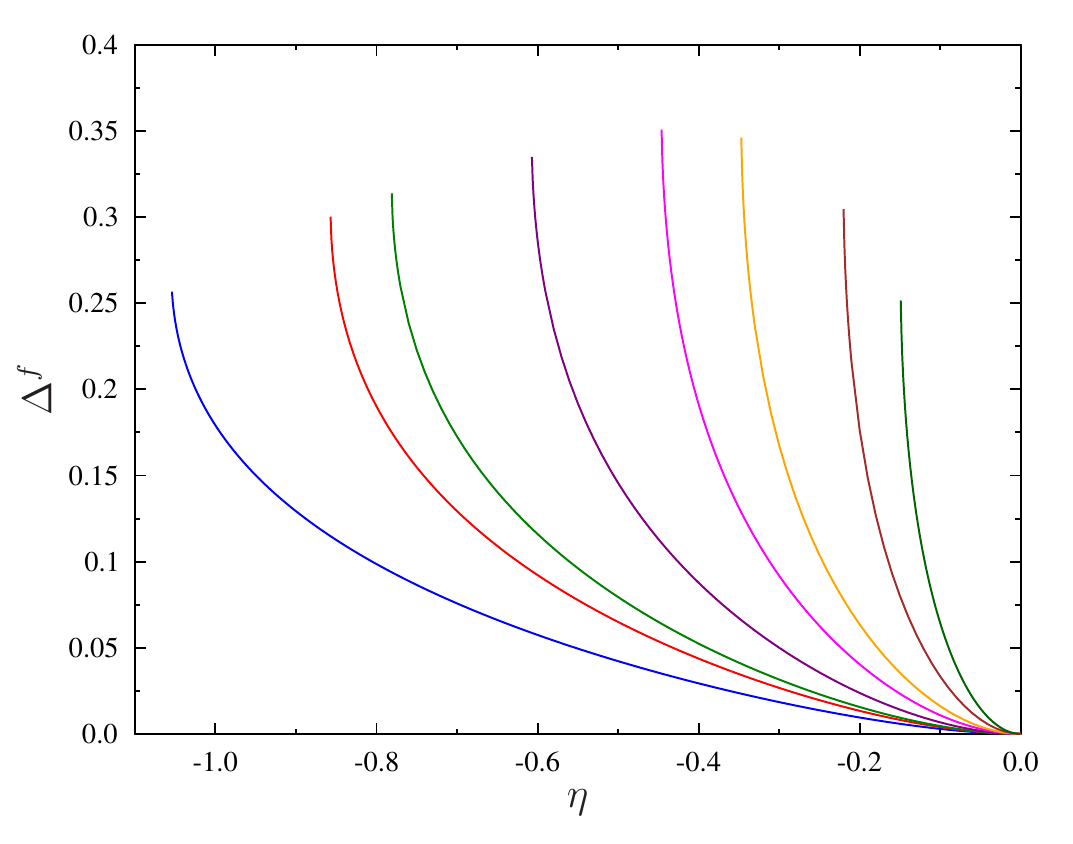}\nolinebreak
\includegraphics[width=.5\textwidth]{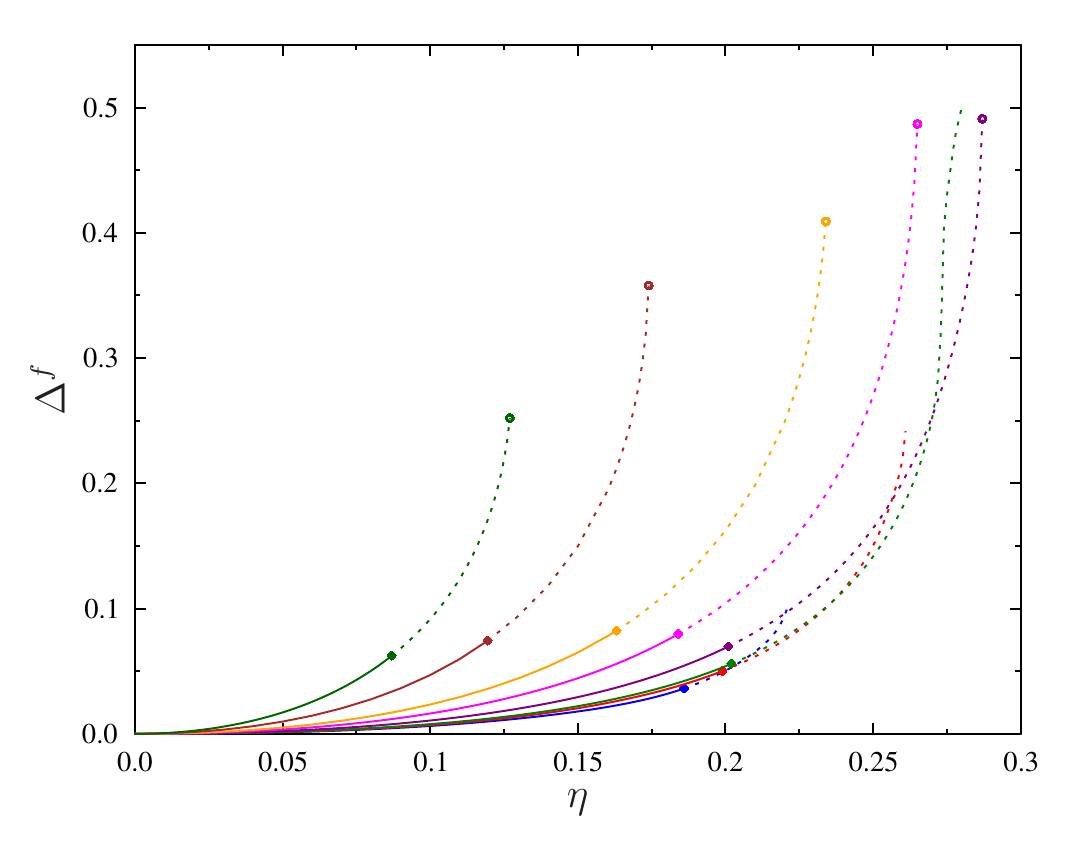}
\caption{
$\De$ ({\it top panels}) and $\D^f$ ({\it bottom panels}), solutions of eqs. 
\eqref{eq:eqD}-\eqref{eq:eqDf},
for different glassy states followed in decompression ($\h<0$, {\it left 
panels}) and compression ($\h>0$, {\it right panels}).
We use separate scales to improve readability of the figures. The dashed lines 
indicate the unstable
region wherein the replicon mode is positive (figure \ref{fig:rep_compr}). 
\label{fig:Delta_compr}}
\end{figure}

In figure \eqref{fig:Delta_compr} we report the evolution of the MSDs (or 
equivalently, Debye-Waller factors) $\De$ and $\D^f$ under compression ($\h>0$) 
or
decompression ($\h<0$) at $\g=0$. In decompression, we find that $\D^f$ 
increases quadratically from zero, while $\De$ also increases; this is 
reasonable, since the spheres in the glass become more free to move
when the system is decompressed and $\D$ increases as a result. We observe that 
at low enough $\h$ a spinodal point is met, whereupon the solution disappears 
with a square-root singularity
in both $\De$ and $\D^f$; this behavior is the usual one for spinodal points. At 
this spinodal point, the glass ceases to exist and melts into the liquid phase: 
within our formalism we recover the onset transition discussed in paragraph 
\ref{subsec:ultrastable} and reported in \cite{Sw07,SEP13}.

Upon compression, again $\D^f$ increases quadratically while $\De$ decreases, as 
expected. If the planting density $\wh\f_g$ is high enough, (see for example 
$\wh\f_g=8$ in figure \ref{fig:Delta_compr}), $\De$ vanishes linearly at $\h_j$ 
while $\D^f$ stays finite, as predicted by the asymptotic analysis of the 
preceding paragraph. The values of $\h_j$ and $\D^f_j$ coincide with the ones
obtained from the equations in the jamming limit reported in section 
\ref{subsec:speclimits}.\\ 
At low density, however (see for example $\wh\f_g=5$ in figure 
\ref{fig:Delta_compr}) another spinodal point is met (signaled again by a square 
root singularity) before jamming occurs (we mark it with a symbol in figure 
\ref{fig:Delta_compr}). We will see in the following that this spinodal point is 
unphysical (it does not correspond to a true loss of stability within the glass 
state), and it is an artifact of the theory, originated by the fact that we are 
using the RS ansatz in a region where it is unstable. This unphysical spinodal 
point has also been found in spin glasses in a similar setting, see 
\cite{KZ10b}.

\subsubsection{The replicon and the Gardner transition}

\begin{figure}[ht!]
\includegraphics[width=.5\textwidth]{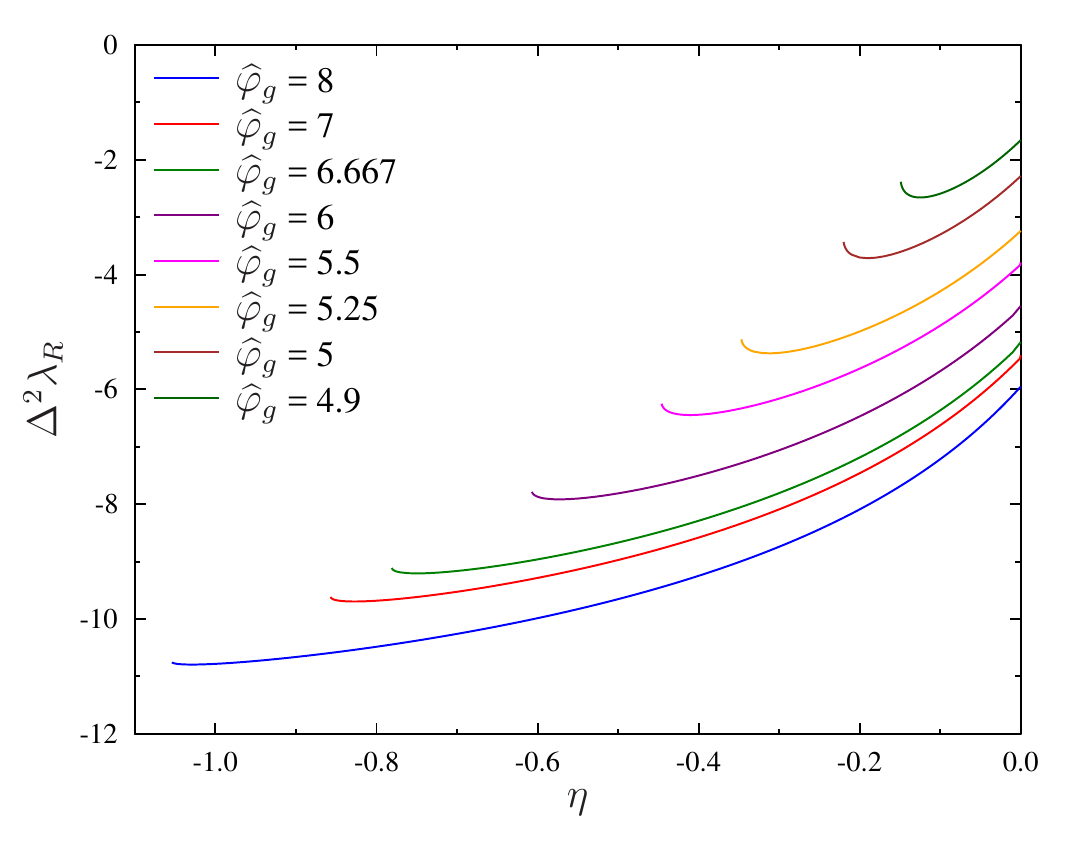}
\includegraphics[width=.5\textwidth]{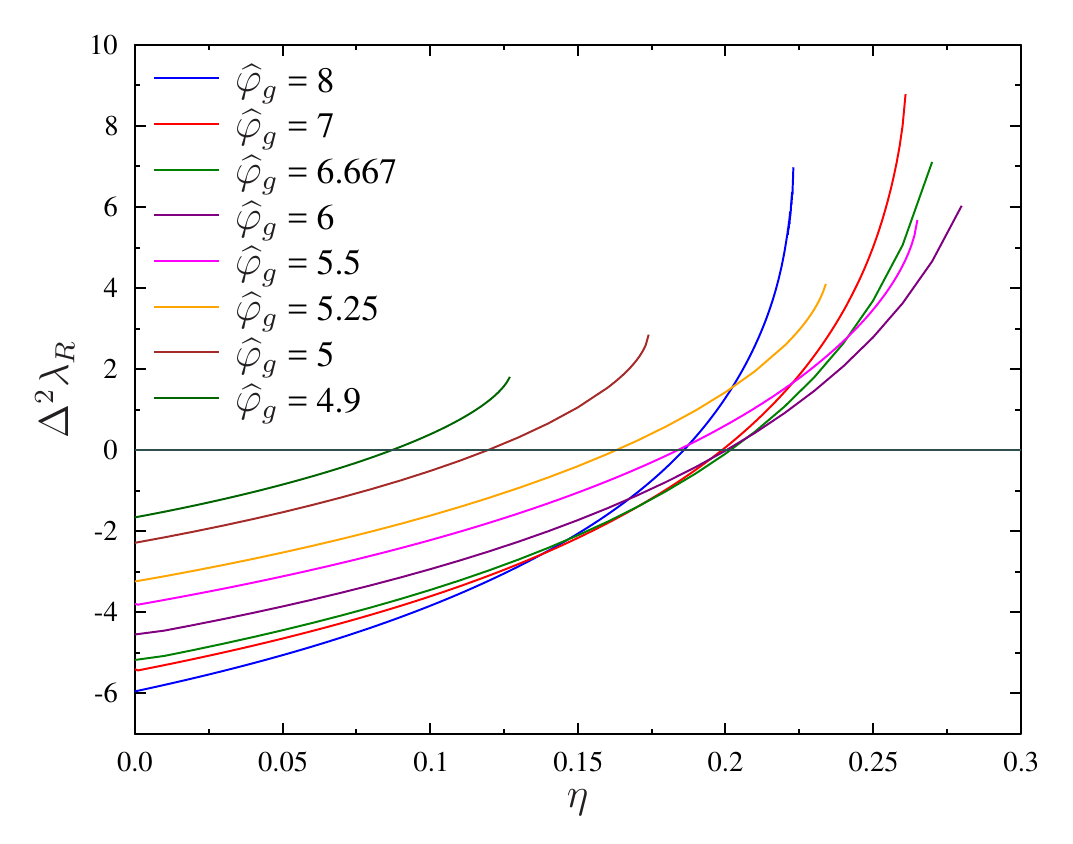}
\caption{
The replicon mode given by equation \eqref{eq:lR}, for the same glasses as in 
figures \ref{fig:Delta_compr} and \ref{fig:dinfstatefo}.
Upon decompression the replicon is always negative and the RS solution is always 
stable; upon compression, the replicon vanishes at the Gardner
transition signaling an instability of the RS solution. \label{fig:rep_compr}}
\end{figure}

We now focus on the replicon mode. We observe from figure \ref{fig:rep_compr} 
that upon compression, before either jamming or the unphysical spinodal point is 
reached, the replicon mode becomes positive,
signaling that the glass state undergoes a phase transition as previously 
discussed, whereupon a more complicated structure of sub-minima manifests inside 
it~\cite{Ga85,CKPUZ13}.\\
In principle, the structure appearing within the state could correspond to an 
arbitrary level of RSB, and one could check only \emph{a posteriori} which is 
the correct number of steps of RSB of the ansatz corresponding to a stable 
solution.
However, based on the analogy with spin glasses, and also the results of 
\cite{nature,CKPUZ13} and the discussion of section \ref{sec:replicongen}, we 
expect that replica symmetry is broken towards a fullRSB ansatz \cite{parisiSG}, 
corresponding to an infinite number of RSBs and a fractal hierarchy of 
sub-states appearing within the original glassy minimum. Such a transition had 
been discovered by Gardner in \cite{Ga85} in the context of spin glass models, 
and it is fittingly referred to as the \emph{Gardner transition}. In paragraph 
\ref{subsec:gardner} we discuss it in more detail. In all figures, we report the 
unstable part of the curves with a dashed line, to remember that the RS ansatz 
is unstable in that region.

\subsubsection{Phase diagram and observables}
\begin{figure}[t]
\begin{center}
\includegraphics[width=0.75\textwidth]{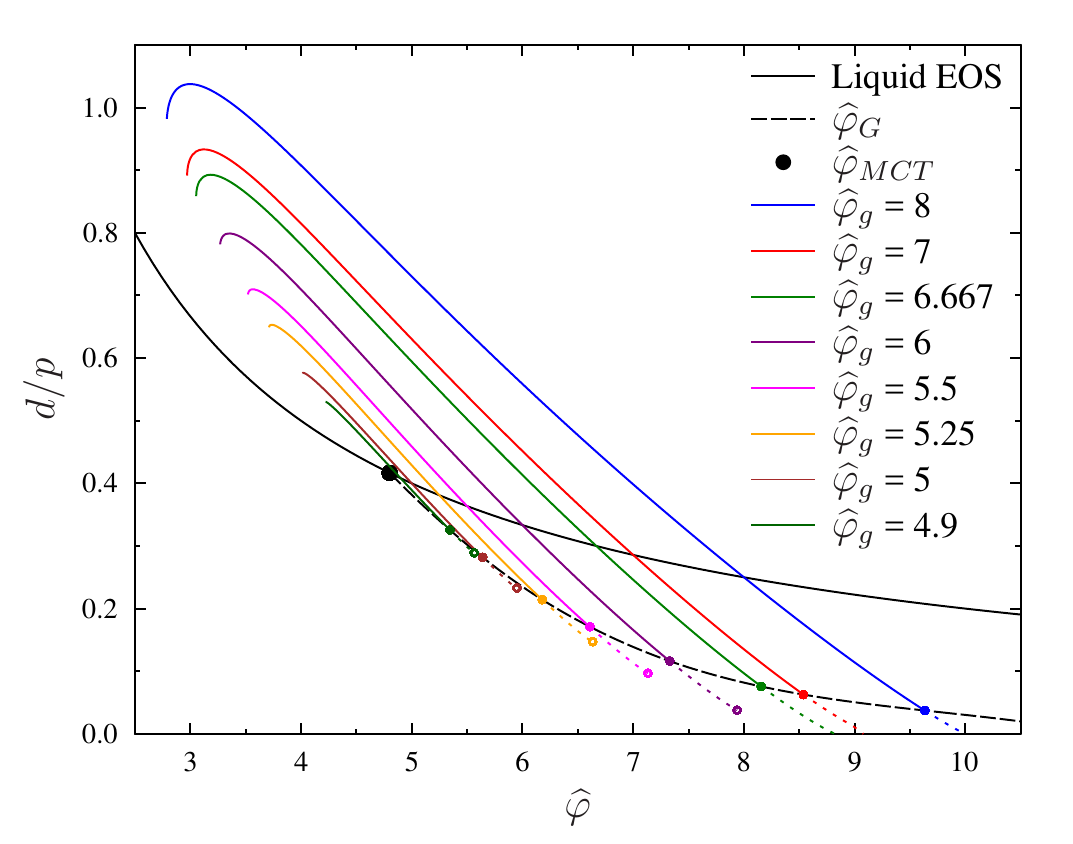}
\caption{Following glasses in (de)compression. Inverse 
reduced pressure $d/p$ is plotted
versus packing fraction $\wh\f = 2^d \f/d$. 
Both quantities are scaled to have a finite limit for $d\to\io$. 
The liquid EOS is $d/p = 2/\wh\f$. The MCT transition
$\wh\f_{MCT}$ is marked by a black dot. The glassy EOS are reported as full 
colored lines, that intersect the liquid EOS at the glass transition density (or 
equivalently, fictive density \cite{Tool46}) $\wh\f_g$. Upon compression, 
a glass prepared at $\wh\f_g$ undergoes a Gardner transition at 
$\wh\f_G(\wh\f_g)$ (full symbols 
and long-dashed black line); beyond $\wh\f_G$ our computation is not correct and 
glass EOS are reported 
as dashed lines. For low $\wh\f_g$, they end at an unphysical spinodal point 
(open symbol) before jamming occurs.\\ 
Upon decompression, the glass pressure falls below the liquid one, until it 
reaches a minimum,
and then grows again until a physical spinodal point whereupon the glass melts 
into the liquid \cite{Sw07,SEP13}. \label{fig:dinfstatefo}}
\end{center}
\end{figure}

We can finally report (figure \ref{fig:dinfstatefo}) the phase diagram of HSs 
computed from the SF construction.
The phase diagram in figure \ref{fig:dinfstatefo} is to be compared with phase 
diagrams like those shown in figures \ref{fig:summary}, \ref{fig:ultraexp}, 
\ref{fig:ultranum} and \ref{fig:diagcomparison}, and the reader can appreciate 
how the phenomenology is qualitatively well reproduced by the SF construction. 
The system is prepared at low density $\r$ and then particle volume $V_s$ is 
slowly increased (or equivalently, container volume is decreased), and the 
reduced pressure $p$ is
monitored. We plot the reduced pressure $p = \b P/\r$ versus the packing 
fraction $\f = \r V_s$, and they can be seen as playing the roles of the 
temperature and enthalpy for the purpose of comparison with figure 
\ref{fig:ultraexp}. As long as the system is equilibrated, it follows the liquid 
equation of state (EOS), which in the MF limit is just the Van Der Waals EOS 
\cite{simpleliquids}. At the MCT transition density $\f_{ MCT}$ glasses appear, 
and the system can fall out of equilibrium, starting to age in a glass state 
selected by an equilibrium configuration at $\f_g > \f_{MCT}$.\\
The slope of the glass EOS at $\f_g$ is different from that of the liquid EOS, 
indicating that when the system
falls out of equilibrium at $\f_g$, the compressibility has a jump, as discussed 
in paragraph \ref{subsec:ultrastable} in the case of the heat capacity. 
Following glasses in compression, the pressure increases faster than in the 
liquid (compressibility is smaller)
and diverges at a finite {\it jamming} density $\f_j(\f_g)$. Before jamming is 
reached, the glass undergoes the Gardner transition \cite{Ga85,nature}, and we 
can compute precisely
the transition point $\f_G(\f_g)$ for all $\f_g$.
Interestingly, the Gardner transition line ends at $\f_{MCT}$, i.e. 
$\f_G(\f_g=\f_{MCT}) = \f_{MCT}$.
This implies that the first glasses appearing at $\f_{\rm MCT}$ (which are the 
easiest to probe experimentally) are marginally stable towards breaking into 
sub-states, while glasses
appearing at $\f_g > \f_{\rm d}$ remain stable for a finite interval of 
pressures. Yet, all glasses undergo the Gardner transition at finite pressure 
before jamming occurs, in agreement with the results of \cite{nature}.\\
A given glass prepared at $\f_g$ can be also followed in {\it decompression}, by 
decompressing it a relatively fast rate $t_{\rm dec}$ such that $\t_{\beta} \ll 
t_{\rm dec} \ll t_{\rm exp}$. In this case we observe hysteresis, again 
consistently with experimental results \cite{Dy06,Sw07,SEP13}. In fact, the 
glass pressure becomes {\it lower} than the liquid
one, until the spinodal point whereupon the glass becomes unstable and melts 
into the liquid is met. Note that pressure ``undershoots'' (it has a local 
minimum, see figure \ref{fig:dinfstatefo}) 
before the spinodal is reached \cite{corrado}, and the compressibility becomes 
infinite: this is a result of the MF nature of our approach. A Maxwell 
construction should be performed at the onset point in order to get the right 
finite-$d$ behavior.

\begin{figure}[htb]
\includegraphics[width=.5\textwidth]{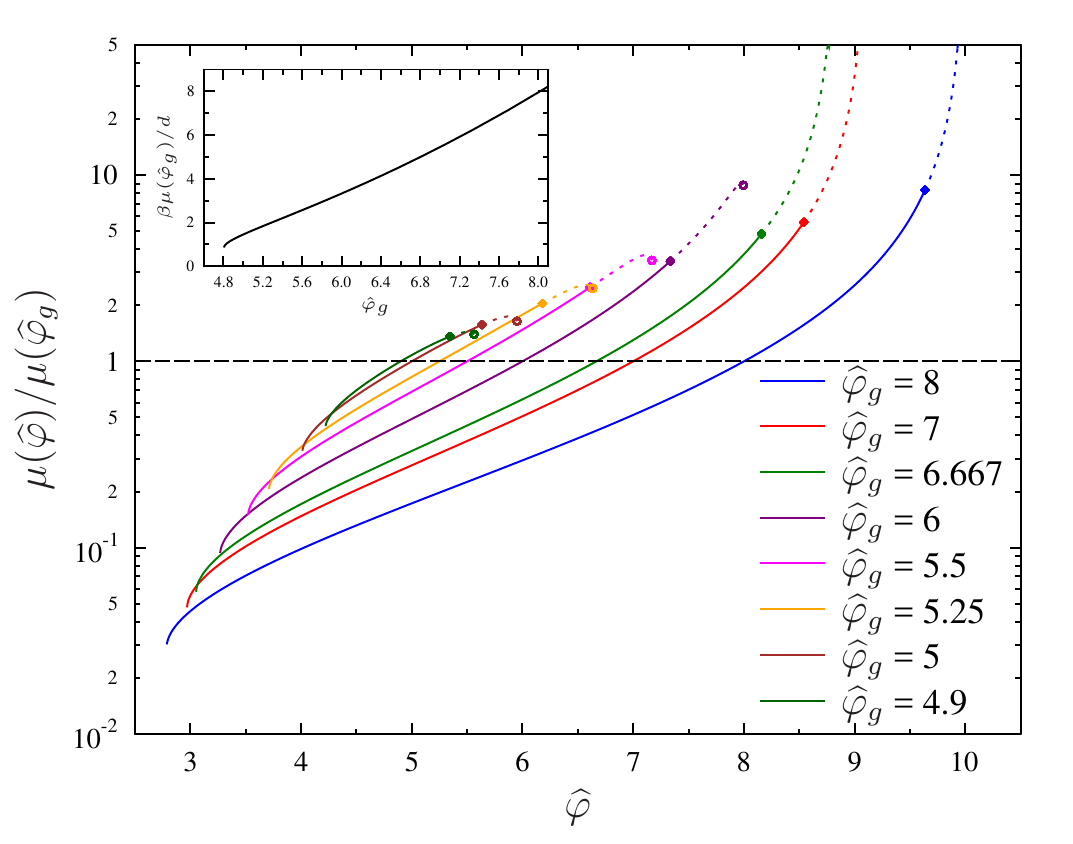}\nolinebreak
\includegraphics[width=.5\textwidth]{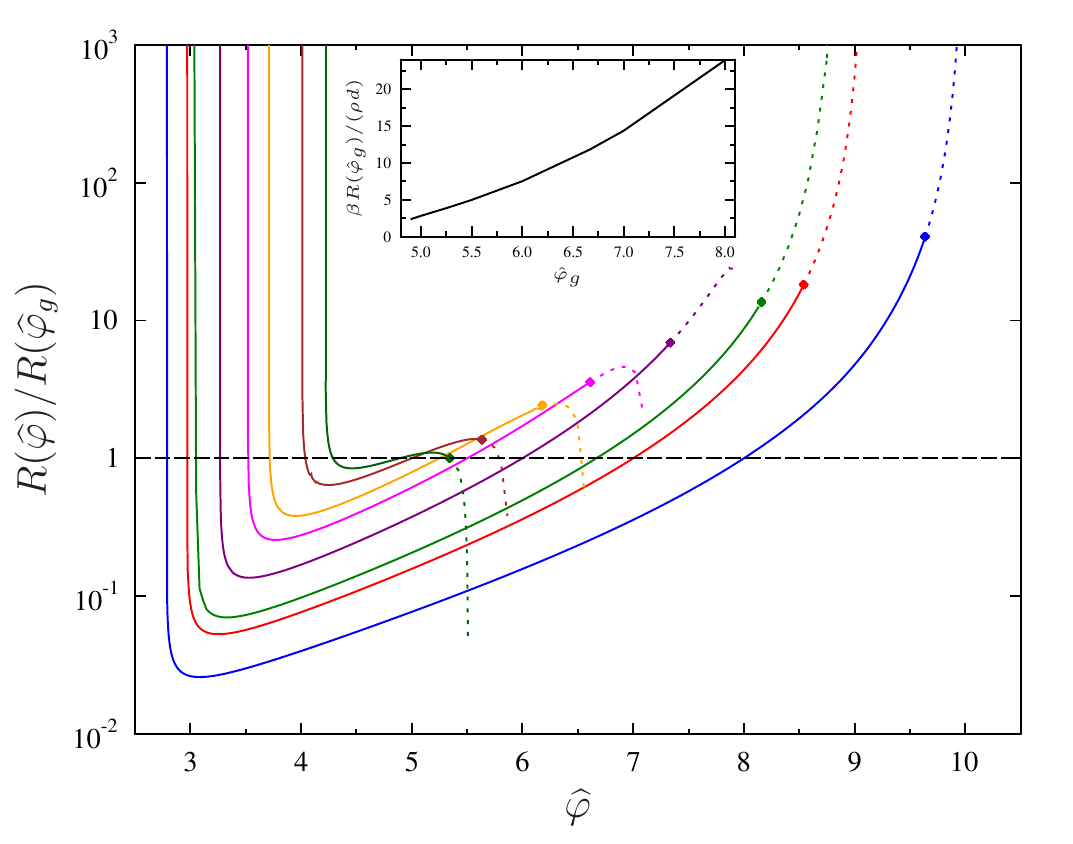} 
\caption{\emph{Left panel}: Shear modulus versus density for different glasses. 
Same styles as figure \ref{fig:dinfstatefo}.
In the inset we report $\mu(\wh\f_g)$ versus $\wh\f_g$. 
Note that the dilatancy $R/\rho=(1/2)\wh\f\partial \mu/\partial \wh\f$ diverges 
both at jamming and at
the low density spinodal point where the glass melts. \emph{Right panel}: 
Dilatancy $R$ as a function of density for different glasses. Recall that 
$R/\rho=(1/2)\wh\f\partial \mu/\partial \wh\f$.
In the inset, the evolution of $R(\wh\f_g)$ with $\wh\f_g$ is reported. 
Note that the dilatancy diverges both at jamming and at
the low density spinodal point whereupon the glass melts. 
\label{fig:sheardilatancy}}
\end{figure}

We also report results for the shear modulus, easily deduced from the results 
for $\De$ reported in figure \ref{fig:Delta_compr} using equation 
~\eqref{eq:mu}, and we can also compute the dilatancy from eq.~\eqref{eq:R}. The 
results are reported in figure \ref{fig:sheardilatancy}. Note that 
$R/\rho=(1/2)\wh\f\partial \mu/\partial \wh\f$ as it can be deduced by combining 
equation \eqref{eq:R} and \eqref{eq:mu}.
From this last relation one can easily notice that the singularities in the 
shear modulus also impact the dilatancy $R$,  as pointed out in  \cite{Tighe14}. 
As a result of this, the dilatancy diverges both at the spinodal point whereupon 
$\De$ has a square-root singularity (hence infinite derivative) and at the 
jamming point where $\De\to 0$.

\subsection{Shear strain \label{subsec:shearRS}}
We now focus on shear-strain. Our typical protocol will be as follows: we again 
anneal the glass former down to a glass density $\wh \f_g$, however, this time, 
we apply a simple shear strain to it instead of quenching it down to a target 
density $\wh \f = \wh\f_g e^{\h}$. This is different from the athermal AQS 
protocols reviewed in paragraph \ref{subsec:qsshear}, and we will discuss the 
differences in the final section.

\subsubsection{Mean square displacements and replicon mode}

\begin{figure}[htb]
\includegraphics[width=.5\textwidth]{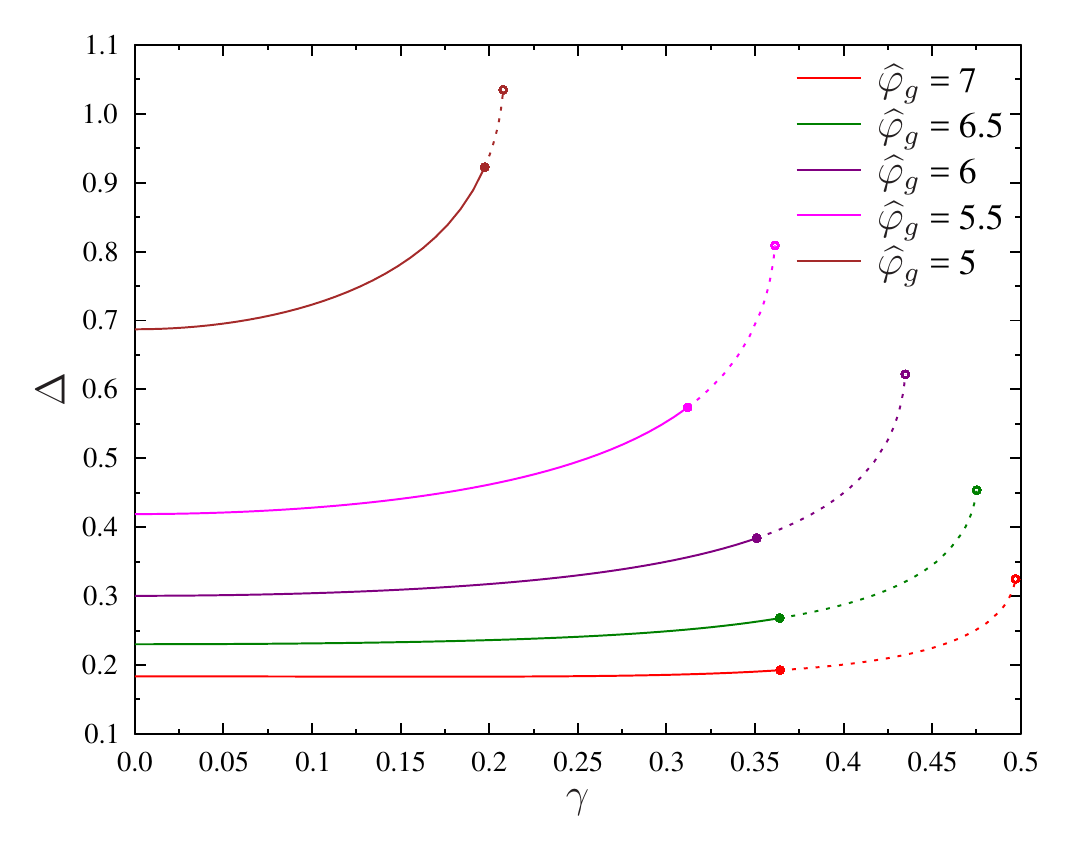}
\includegraphics[width=.5\textwidth]{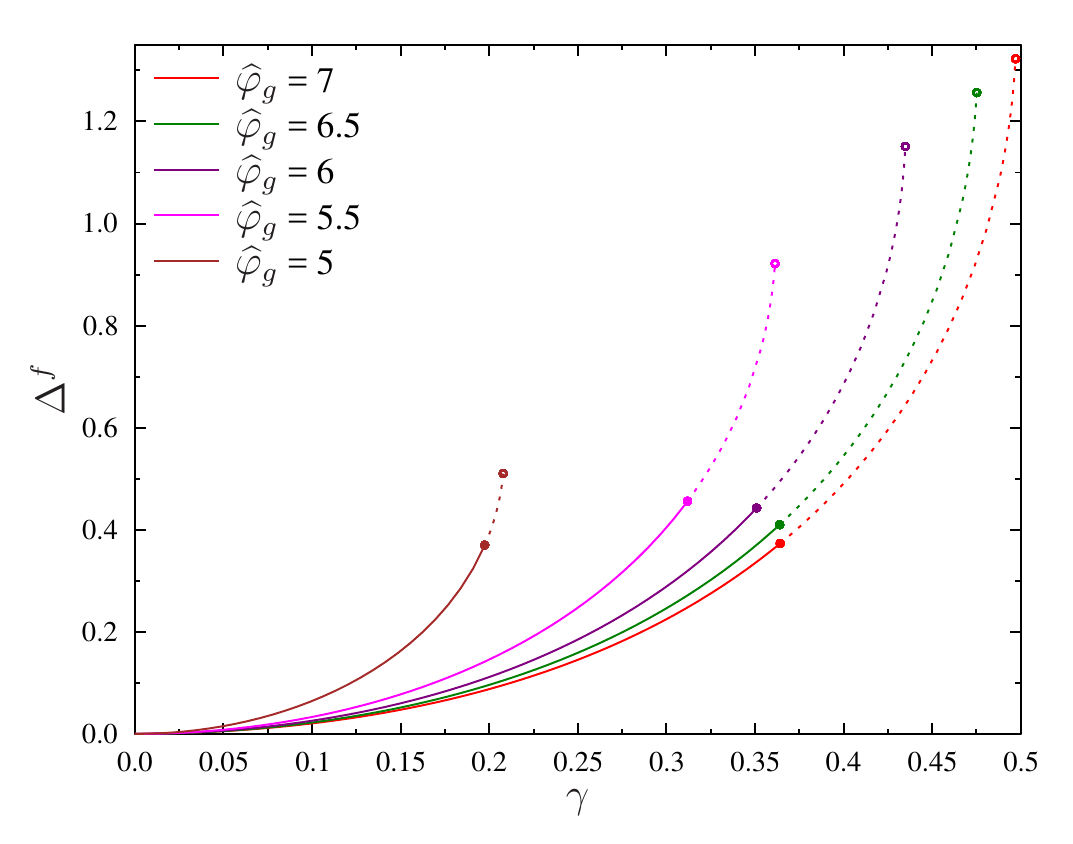}
\caption{
Values of $\D$ ({\it left panel}) and $\D^f$ ({\it right panel}) as functions of 
shear strain $\g$. The dashed lines again indicate the unstable
region wherein the replicon mode is positive (figure \ref{fig:rep_shear}).}
\label{fig:Delta_shear}
\end{figure}
We report the results for $\De$ and $\D^f$ in figure \ref{fig:Delta_shear}. We 
observe that upon increasing
$\g$ both $\De$ and $\D^f$ increase, until a spinodal point is reached, 
whereupon they both display a square root singularity, which is shared by both 
the shear stress $\s$ and the glass pressure $p_g$ (see figure \ref{fig:shear}). 
Interestingly, their behavior is somewhat specular to the one found in 
compression: $\De$ stays almost constant for a fairly long range of $\g$s, while 
$\D_f$ immediately grows rapidly. This is reasonable, since the $\D$ is nothing 
but the Debye-Waller factor of the glass, so we do not expect it to change much 
upon shearing, shear strain being a volume-preserving perturbation.\\
\begin{figure}[htb]
\begin{center}
\includegraphics[width=.6\textwidth]{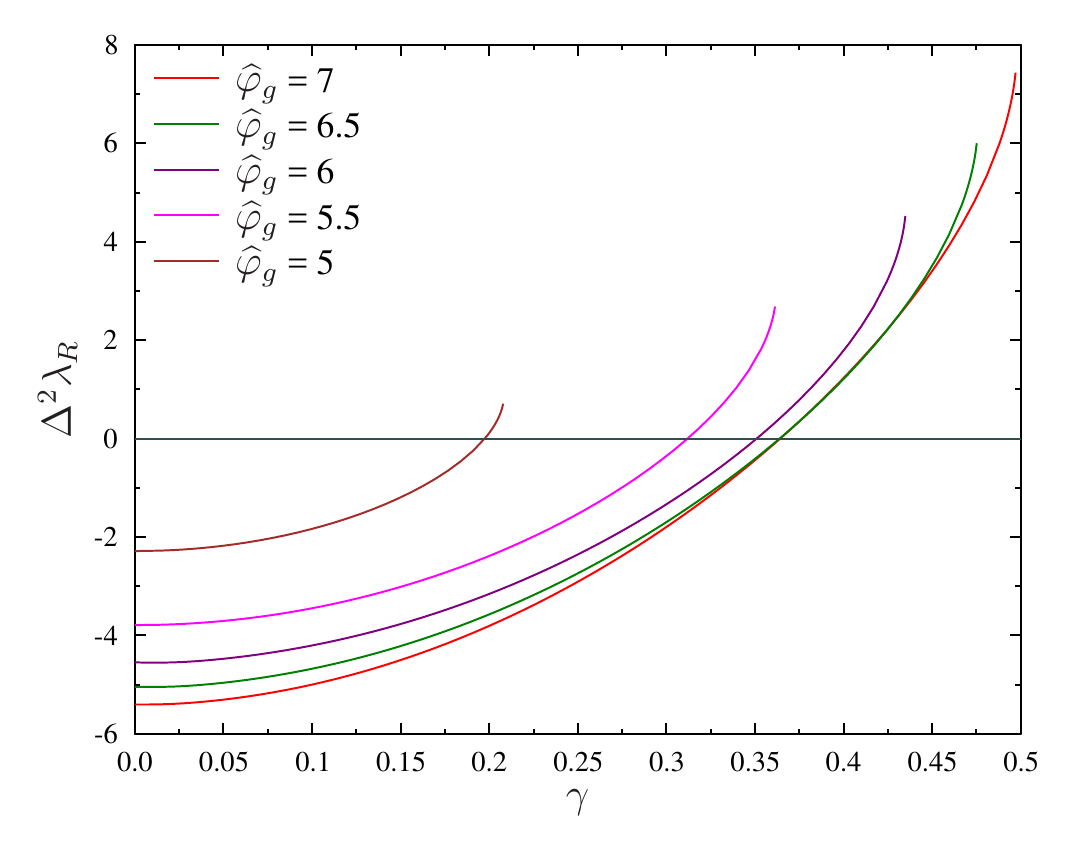}
\caption{
The replicon mode, eq.~\eqref{eq:lR}, for the same glasses as in 
figures~\ref{fig:Delta_shear} and  \ref{fig:shear}.
The replicon vanishes at the Gardner
transition signaling an instability of the RS solution. \label{fig:rep_shear}}
\end{center}
\end{figure}
However, before the spinodal is met, the replicon mode becomes positive again 
(figure \ref{fig:rep_shear}) and the system undergoes, again, a Gardner 
transition.
The fact that that Gardner transition is met when the system is subject to a 
shear strain might be surprising at first sight, because one would intuitively 
think that straining a glass state amount to a simple deformation of the state, 
without inducing its breaking into sub-states. Moreover, the effect of a 
mechanical drive should intuitively amount to injecting energy into the system 
(i.e., a heating or a decompression) as argued for example in 
\cite{BerthierBarrat00}.\\
However, note first that on general grounds, the free energy landscape 
\emph{can} change once perturbations are added \cite{parisiSGandbeyond}. 
Moreover, we also find (see figure \ref{fig:shear}) that the pressure of the 
glassy state increases when the shear strain is applied, because of dilatancy 
\cite{Reynolds85,Tighe14}. This means that the particles in the glass basins 
may, upon shearing, become more constrained, triggering an ergodicity breaking 
inside the state, and with it a Gardner transition. We discuss further this 
issue in paragraph \ref{subsec:gardner}

\subsubsection{Stress-strain curves}
\begin{figure}[t!]
\begin{center}
\includegraphics[width=.75\textwidth]{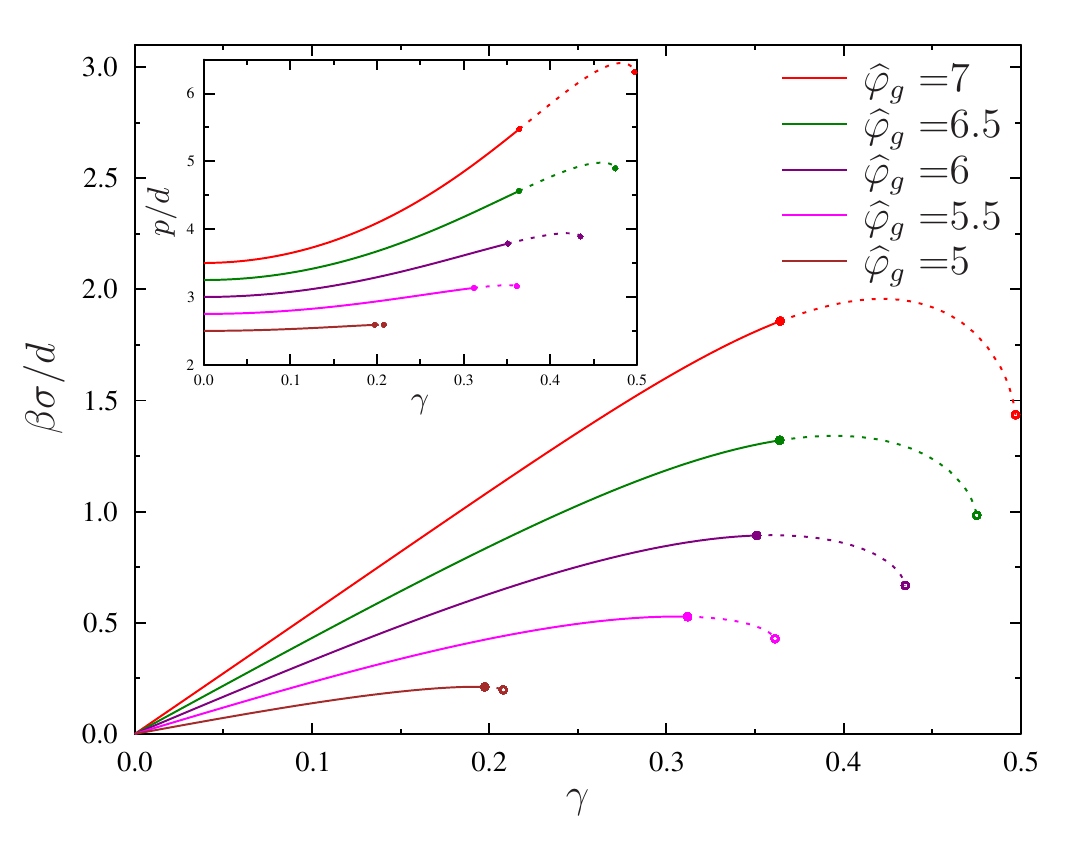}
\caption{Following glassy states prepared at $\wh\f_g$ upon applying a 
shear-strain $\g$.
Shear-stress $\s$ {\it (main panel)} and reduced pressure $p$ 
{\it (inset)} as a function of strain for different $\wh\f_g$. Same styles as 
fig.~\ref{fig:dinfstatefo}. Upon increasing the strain,
the states undergo a Gardner transition at $\g_G(\wh\f_g)$. For $\g > \g_G$ our 
RS computation is unstable but predicts a stress overshoot followed by a 
spinodal point. \label{fig:shear}}
\end{center}
\end{figure}

We can now draw the stress-strain curves of the glass. We report the behavior of 
shear-stress $\s$ and pressure $p$ versus $\g$, see figure \ref{fig:shear}, and 
observe how they well reproduce qualitatively the phenomenology shown in figure 
\ref{fig:shearstrain}.\\
At small $\g$, we observe a {\it linear response} elastic regime wherein $\s 
\sim \mu \g$, as expected; the pressure increases quadratically above the 
equilibrium liquid value, $p(\g) \sim p(\g=0) + (\b R/\r) \g^2$. Both the shear 
modulus $\mu$ and the dilatancy $R>0$ increase with $\wh\f_g$, indicating that 
glasses prepared by slower annealing are more rigid, as discussed in paragraph 
\ref{subsec:qsshear}

Upon further increasing $\g$, glasses enter a non-linear regime, and undergo a 
Gardner transition at $\g_G(\f_g)$. As it happened in the compression protocol, 
we find $\g_G(\f_{\rm d})=0$, and $\g_G$ increasing rapidly with $\f_g$. For $\g 
> \g_G(\f_{\rm d})$, the glass breaks into sub-states and the RS calculation 
becomes unstable, however, we can anyway keep following the state. Then, we 
notice that the RS computation correctly predicts the stress overshoot, followed 
by a spinodal point where the glass basin loses stability and disappears.\\
The spinodal point corresponds to the yielding point whereupon the glass starts 
to flow, as discussed in paragraph \ref{subsec:yielding}, so within  the SF 
formalism we have ad unambiguous definition for the yielding point $\g_Y$; the 
values of yield strain $\g_{\rm Y}$ and of yield stress $\s_{\rm Y}$ are found 
to increase with $\f_g$, as expected. These results are qualitatively consistent 
with the experimental and numerical observations of \cite{RTV11,Kou12}, and we 
will see in the next chapter that the fRSB computation gives similar results in 
terms of stress overshoot and yielding point.

\section{Discussion}

We have performed the computation of the Franz-Parisi potential \eqref{eq:FP}, 
with the simplest RS ansatz. We show that the state following method is able to 
give predictions for many physical observables of experimental interest, and
reproduces a quite large number of observations. These include: {\it (i)} the 
pressure as a function of density for different glasses 
(figure~\ref{fig:dinfstatefo}), which
displays a jump in compressibility at $\f_g$~\cite{parisizamponi,CIPZ11}; {\it 
(ii)} the presence of hysteresis and of a spinodal point in decompression
in the pressure-density curves, whereupon we show that more stable glasses 
(those with higher $\f_g$) display a larger hysteresis, consistently with the 
experimental observations of~\cite{Dy06,Sw07,SEP13};
the behavior of pressure and shear-stress under a startup shear perturbation 
(figure~\ref{fig:shear}), where we show that {\it (iii)}
the shear modulus and the dilatancy increase for more stable glasses (higher 
$\f_g$), and {\it (iv)} that the shear-stress overshoots before a spinodal 
(yielding)
point is reached where the glass yields and starts to flow \cite{RTV11,Kou12}. 
Note however that the spinodal (yield) point falls beyond the Gardner transition 
and therefore
its estimate, reported in figure~\ref{fig:shear}, is only approximate. 
Furthermore, {\it (v)} we predict that glasses undergo a Gardner transition both 
in compression (fig ~\ref{fig:dinfstatefo}) and in shear (fig~\ref{fig:shear}), 
and
we locate the Gardner transition point. Finally, we {\it (vi)} compute the 
dilatancy and the shear modulus everywhere in the glass phase 
(figure~\ref{fig:sheardilatancy}) and study their behavior close to the jamming 
transition.

\subsection{The Gardner transition \label{subsec:gardner}}
Both in compression and shear strain, we detect an instability in the RS 
computation from the study of the replicon mode \eqref{eq:lR}. This instability 
corresponds to a second order critical point which is referred to as 
\emph{Gardner transition}, and corresponds to a breaking of the glassy state in 
a fractal hierarchy of micro-states, described by a fRSB ansatz 
\cite{parisiSG,parisiSGandbeyond}, in a spirit similar to the ferromagnetic 
transition in figure \ref{fig:curieweiss}. It has been discovered in \cite{Ga85} 
in the context of the $p$-spin Ising model (essentially the Ising spin version 
of the PSM), and is essentially analogue to the transition found in the 
Sherrington-Kirkpatrick spin glass model \cite{SK} in presence of a magnetic 
field, by de Almeida and Thouless in \cite{TAK80}. The only difference is that 
in the SK model the state that undergoes the transition is the ergodic 
paramagnet, while for the Gardner transition it is a glass state dynamically 
selected by an annealing protocol as previously discussed. The relevant 
phenomenology is the same in both cases.

The presence of a Gardner transition in a realistic model of glass former is not 
a surprise, as it had been already reported in \cite{nature,CKPUZ13}, always for 
hard spheres. As of today, the nature of the Parisi fRSB solution is still not 
completely understood, but some of its phenomenology is well known. Without any 
doubt, the most physically relevant trait of the fRSB micro states is that they 
are \emph{marginal} \cite{BM79}, which means that their replicon mode is zero 
everywhere in the fRSB phase \cite{KPUZ13}.\\
This marginality of the fRSB solution has all kinds of implications. In 
\cite{CKPUZ13}, in particular, it was shown that the marginality condition of 
the fRSB solution directly implies the isostaticity property of jammed packings 
\cite{Mo98,Ro00}, i.e. packings are \emph{predicted} to be isostatic, with 
$z=2d$. This way, the isostaticity property is recovered, within replica theory, 
as the manifestation of a critical mode (or equivalently, an infinite 
susceptibilty), revealing the jamming transition as a critical phenomenon as 
anticipated in paragraph \ref{subsec:jamming}. Moreover, in \cite{CKPUZ13}, the 
fRSB ansatz is also show to be necessary to compute the critical exponents 
$\kappa, \gamma, \theta$ of the jamming transition (paragraph 
\ref{subsec:jamming}), while the RS ansatz would for example predict $\kappa = 
1$ as mentioned before. These findings show the relevance of the Gardner 
transition for glasses at low temperatures and high pressure, at least within 
RFOT. We must however stress the fact that the ``mechanical'' marginality 
related to isostaticity in jammed packings, and the marginality of fRSB micro 
states related to the vanishing of the replicon are not the same thing, although 
it is clear that they must be connected in some way. Work is still ongoing to 
better understand their relation. The results of \cite{CKPUZ13} were obtained 
within the real replica method \cite{monasson} and the isocomplexity 
approximation, but in the next chapter we show that they can be more 
satisfactorily re-derived in the state following setting.

The presence of a Gardner transition in shear, however, is a novelty which 
deserves further investigation and may open new theoretical scenarios for the 
study of the yielding transition. As we mentioned, fRSB micro-states are 
marginal, which means that just a little perturbation will kick the system 
outside of a microstate to another, producing a very intermittent and rough 
response. This kind of physics looks well suited to the stress-strain curves in 
figures \ref{fig:shearstrain} and \ref{fig:shearstrainlemaitre}. For example, 
the fact that the interval $\thav{\D\g}$ before the first avalanche is met 
scales as $N^\beta_{iso}$, with $N$ the system size and $\beta_{iso} < 1$ 
\cite{KarmakarLerner10b} (i.e. in the thermodynamic limit an infinitesimal 
strain will destabilize the system), could be well explained in terms of 
marginality.\\
There is however a cardinal difference between the AQS protocols discussed in 
paragraph \ref{subsec:qsshear} and the ones reproduced here in the SF setting: 
in AQS protocols the system is quenched to zero temperature \emph{before} strain 
is applied, which means that the glass is already in the fRSB phase, and 
fittingly, it immediately responds very roughly like in figures 
\ref{fig:shearstrain} and \ref{fig:shearstrainlemaitre}, while in figure 
\ref{fig:shear} the strain is immediately applied after preparation at $\wh 
\f_g$; as a result of this, the system is still equipped with thermal energy 
when subjected to strain, and the roughness of the PEL responsible for the 
intermittent response in AQS protocols is smoothed away by thermal fluctuations, 
producing a perfectly elastic response all the way to $\g_G$. This prediction is 
very relevant and should in principle be easily verifiable, although there is 
relative paucity of numerical results for shear of thermal amorphous systems. We 
will discuss this issue in more detail in section \ref{sec:futureresearch}.

At the end of the day, though, the Gardner transition first and foremost implies 
that a fRSB computation must be performed, in order to follow the less dense 
glasses (which we recall are the easiest to prepare) down to the jamming limit, 
compute their jamming density, and the critical exponents of jamming. We also 
need it to check whether the phenomenology of yielding predicted by the RS 
ansatz remains the same when the fRSB one is employed.\\
In the next chapter we perform the fRSB computation in the state following 
setting.

\chapter{The full replica symmetry breaking ansatz \label{chap:SFfRSB}}

In the preceding chapter, we have computed the FP potential assuming the 
simplest possible replica symmetric ansatz for the slave replicas sampling the 
bottom of the followed glassy state. We have then used the so obtained 
thermodynamic potential to compute many quantities of interest, including 
equations of state for the glass, stress-strain curves, dilatancy and shear 
modulus, obtaining in all these cases a good qualitative agreement between the 
results of our computation and the phenomenology of glasses reviewed in the 
first chapters. However, we have also detected, for a sufficiently large value 
of both perturbations considered, a Gardner transition inside glassy states, 
rendering the RS ansatz unstable in a region of the phase diagram both in 
compression and strain. As a result of this, the approach is unable to provide 
predictions in a whole region of the phase diagram, as it happens for example 
for the EOS of the least dense glasses (corresponding, coincidentally, to 
shorter and thus easier to realize preparation protocols) in compression that 
terminate in an unphysical spinodal point before jamming can occur. And even 
where the RS equation of state exists, it can be at best considered an 
approximation of the results which must be derived with the correct ansatz. In 
this chapter we assume the ``correct'' ansatz to be the full replica symmetry 
breaking one, and perform again the computation of the FP potential and physical 
observables within this ansatz.

\section{The potential}
Our starting point is again the replicated entropy \eqref{eq:gauss_r}. The 
entropic and interaction (equations \eqref{FFreplica},\eqref{eq:FF0binomial}) 
terms have the same definition as before. The difference with respect to the 
previous chapter is that this time we choose a matrix $\hat \D_{ab}$ (or 
equivalently, $\hat \a_{ab}$) in the form
\beq
\hat \Delta=\begin{pmatrix}
\hat\Delta^g & \hat \Delta^r\\
(\hat\Delta^r)^T & \hat\Delta^s
\end{pmatrix},
\eeq
where the matrices $\hat \D^g$ and $\hat \D^r$ are defined as in the 
\eqref{eq:DeltaRS}, and the matrix $\hat\D^s$ is now a matrix with an infinite 
number of RSBs \cite{parisiSGandbeyond,parisiSG}.

\subsection{The fRSB parametrization \label{subsec:fRSBpar}}

\begin{figure}
\begin{center}
\includegraphics[width=0.5\textwidth]{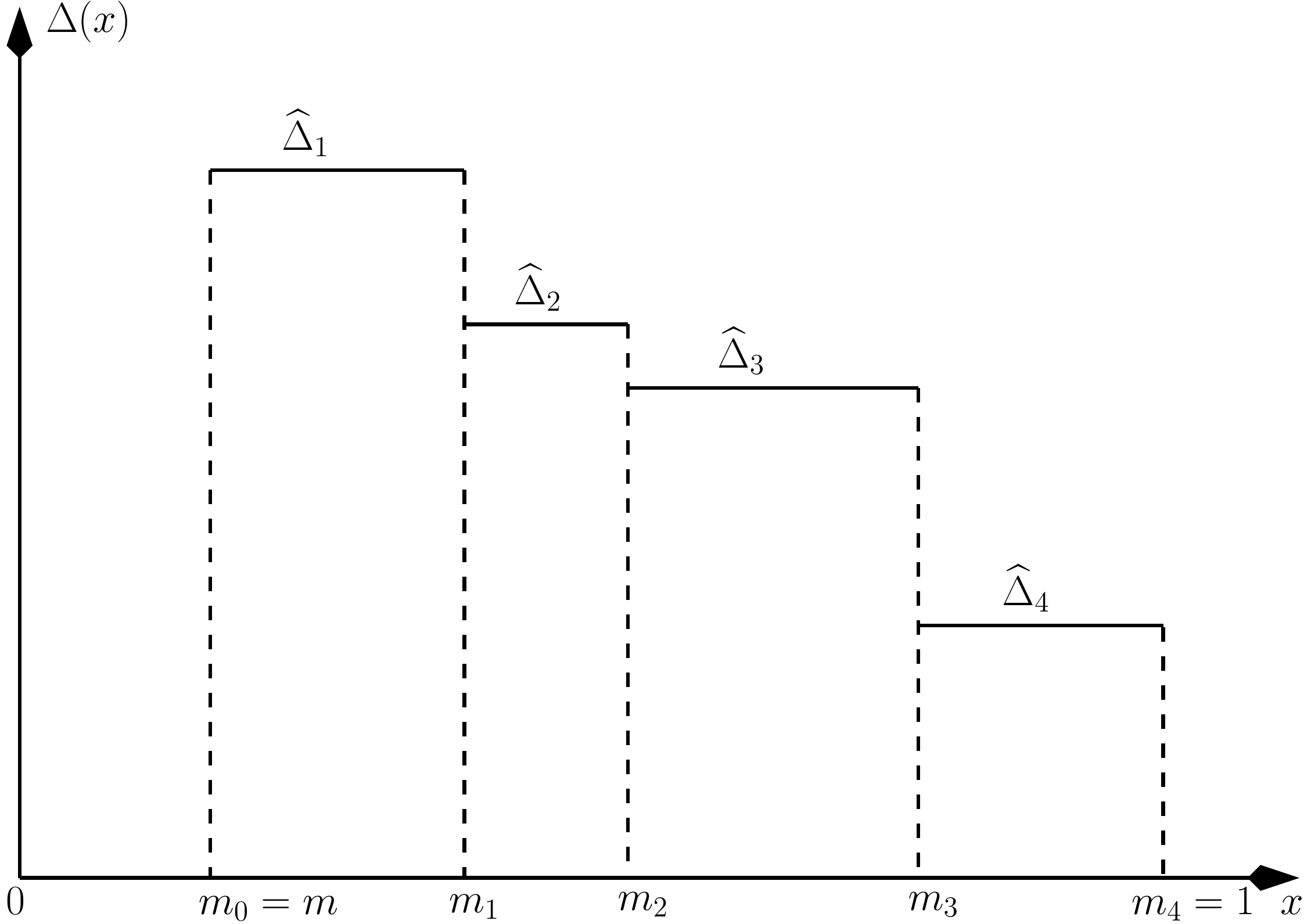} \nolinebreak
\includegraphics[width=0.5\textwidth]{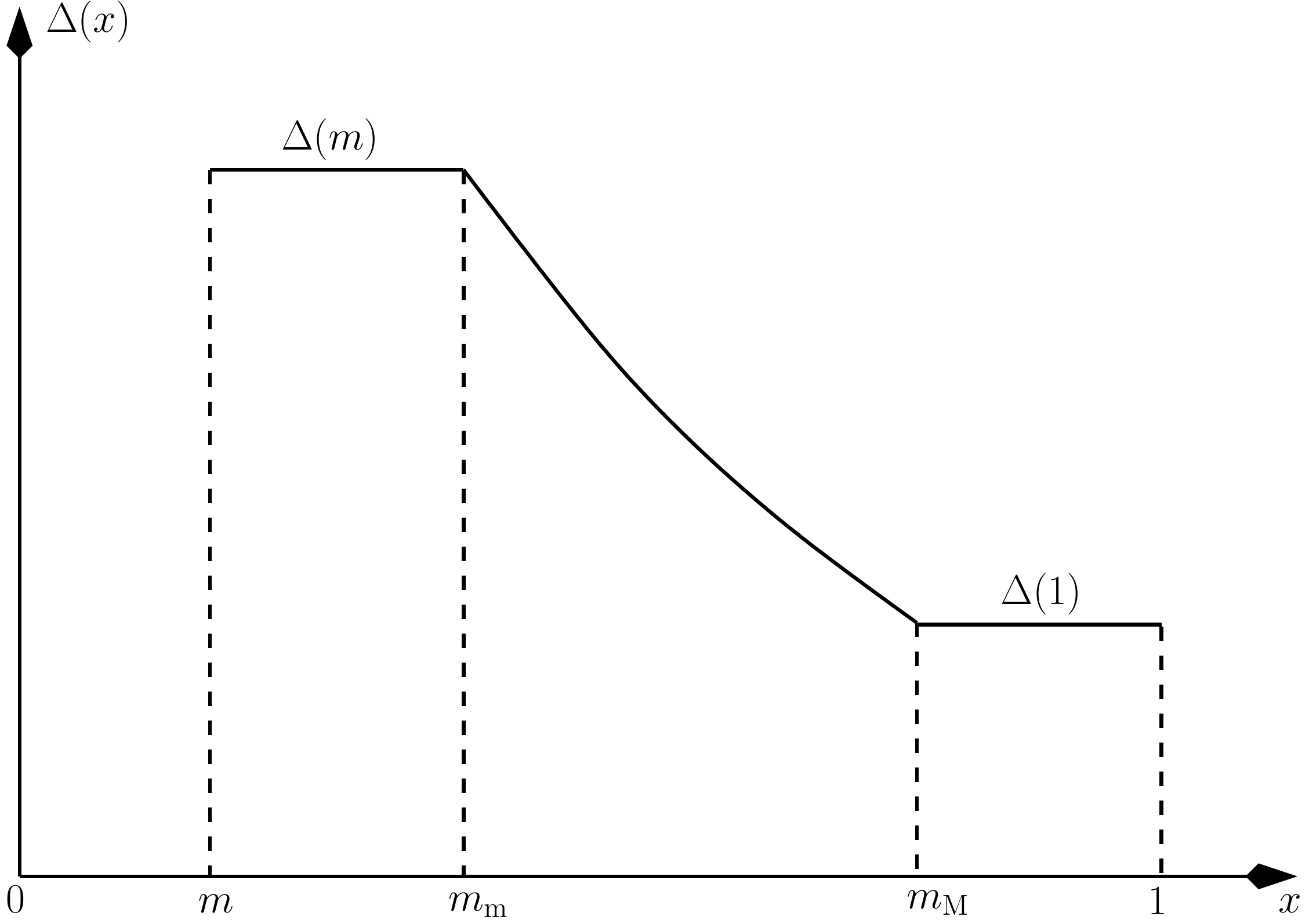}
\caption{The function $\D(x)$ for $k=4$ (\emph{left}) and its expected form for 
$k=\infty$ (\emph{right}). Reprinted from \cite{CKPUZ13}. \label{fig:hier}}
\end{center}
\end{figure}

Let us sketch rapidly how a matrix with an infinite number of RSBs can be 
parametrized in practice. An RS matrix is parametrized by a single element $\D$; 
as we said in paragraph \ref{subsec:replicateds}, a 1RSB matrix corresponds to 
having two relevant parameters $\D_1$ and $\D_2$, one for each level in the 
hierarchy of states. Moreover, we also need to specify how replicas are grouped 
in the states, i.e. we need to say how many replicas $s_1$ of the total $s$ we 
have, end up in the same state at the lowest level of the hierarchy; in 
paragraph \ref{subsec:replicateds} we had $s=9$ and $s_1 = 3$, which would 
correspond to a matrix $\hat\D^s$
\beq
\D_{ab}^{1RSB} = 
\begin{bmatrix}
0 & \D_2 & \D_2 & \D_1 &\D_1 &\D_1 & \D_1 &\D_1 &\D_1 &\\
\D_2 & 0 & \D_2 &\D_1 &\D_1 &\D_1 & \D_1 &\D_1 &\D_1 &\\
\D_2 &\D_2 & 0 & \D_1 &\D_1 &\D_1 & \D_1 &\D_1 &\D_1 &\\
\D_1 &\D_1 &\D_1 & 0 & \D_2 &\D_2 & \D_1 &\D_1 &\D_1 &\\
\D_1 &\D_1 &\D_1 &\D_2 & 0 & \D_2 & \D_1 &\D_1 &\D_1 &\\
\D_1 &\D_1 &\D_1 &\D_2 &\D_2 & 0 & \D_1 &\D_1 &\D_1 &\\
\D_1 &\D_1 &\D_1 &\D_1 &\D_1 &\D_1 & 0 & \D_2 &\D_2 &\\
\D_1 &\D_1 &\D_1 &\D_1 &\D_1 &\D_1 & \D_2 & 0 &\D_2 &\\
\D_1 &\D_1 &\D_1 &\D_1 &\D_1 &\D_1 & \D_2 & \D_2 & 0\\
\end{bmatrix}; 
\eeq
and of course one has also the diagonal elements $\D_d$, corresponding to 
considering the same replica, whose value depends on how the ``similarity'' 
between replicas is defined\footnote{For HSs one has $\D_d = 0$, while for spin 
glasses one would have $q_d = 1$.}.
A generic $(k-1)$RSB matrix will then be parametrized by a set of MSD parameters 
$\D_d;\D_1,\D_2,$ $\dots,\D_k$ and a set of ``block'' parameters 
$s_1,s_2,\dots,s_{k-1}$\footnote{Obviously $s>s_1>s_2>\dots>s_{k-1}>1$ when 
$s>1$, and conversely $s<s_1<s_2<\dots<s_{k-1}<1$ when $s$ is analytically 
continued to real $s<1$ \cite{pedestrians}.}, with $s_k=1$ and $s_0 = s$ by 
definition; every generic kRSB matrix can be reconstructed from knowledge of 
this set of parameters.\\
One can then construct a function $\D(x)$, $1<x<s$, in the following way
\beq
\D(x) \equiv \D_k\ \ \ {\rm if}\ \ \ x\in ]s_{k-1},s_k],
\eeq
which essentially describes the profile of the first row of a generic 
hierarchical matrix. When the number of breakings $k$ is finite, this function 
has a step structure. When $k$ goes up, the function will more and more look 
like a continuous function, and in the limit $k\to \infty$ corresponding to the 
fRSB ansatz, the function will be a continuous function of $x$. In figure 
\ref{fig:hier} we give a demonstrative cartoon of the function $\D(x)$ for $k=4$ 
and $k=\infty$.\\
In summary, every hierarchical matrix, with finite or infinite $k$, will be 
parametrized by a couple
\beq
\{\D_d,\D(x)\}.
\label{eq:fRSBpar}
\eeq
It can be shown \cite{MP91} that this parametrization preserves all the 
properties of the algebra of hierarchical matrices, and formulas for the product 
and the inverse in terms of the parametrization \eqref{eq:fRSBpar} can be 
derived. We refer to \cite{MP91,parisiSGandbeyond,CKPUZ13} for further reading 
on the issue. 

\subsection{Expression of the potential and the observables}

Now that we are equipped with a parametrization for the fRSB ansatz, we are 
ready to compute the FP entropy within the fRSB ansatz. As in the RS case, the 
computation is long and not very instructive, so we report it in appendix 
\ref{app:replicatedfRSB}, and skip, again, to the final result. The expression 
of the entropy of the followed state is 
\beq
\begin{split}
s_g[\a] =\ &\frac{d}{2} + \frac{d}{2}\log\left(\frac{\pi\la\D\ra}{d^2}\right) - 
\frac{d}{2}\int_0^1\frac{\de y}{y^2}\log\left(\frac{\la\D\ra + 
[\D](y)}{\la\D\ra}\right) + \frac{d}{2}\frac{m\D^f+\D^g}{m\la\D\ra}\\ 
&+\frac{d\wh\varphi_g}{2}\int_{-\infty}^\infty\DD \z \int_{-\infty}^\infty \de h 
\, \eee^{h}\Th\left(  \frac{h + \D^g/2}{\sqrt{2\D^g}} \right)^m\\
&\times \int_{-\infty}^\infty \de x' \,  f (0,x' + h - \eta +\D(0)/2) 
\frac{ \eee^{- \frac1{2 \D_\g(\z)} \left( x' - \D_\g(\z)/2  \right)^2  } 
}{\sqrt{2\pi \D_\g(\z)}},
\end{split}
\label{eq:sfull}
\eeq
with the definitions
\ba
[\D](x) &\equiv& x \D(x) - \int_0^x\de y\, \D(y),\\ 
 \langle \D \rangle &\equiv& \int_0^1\de x\, \D(x),\label{eq:Dvdef}\\
 \D^f &\equiv& 2\D^r - \D_g - \D(0),
\ea
and the function $f(x,h)$ obeys the Parisi equation \cite{parisiSGandbeyond}
\beq
\frac{\partial f}{\partial x}=\frac 12\frac{\de \Delta(x)}{\de 
x}\left[\frac{\partial^2 f}{\partial h^2}+x\left(\frac{\partial f}{\partial 
h}\right)^2\right],
\label{Parisi_eq}
\eeq
with the boundary condition
\beq
f(1,h)=\log \Theta \left(\frac{h}{\sqrt{2\Delta(1)}}\right).
\label{Parisi_initial}
\eeq
Again, this form is valid for a generic matrix (that is, a generic profile 
$\D(x)$) and must be optimized over $\D(x)$ and $\D^r$ as detailed in appendix 
\ref{app:variationalfRSB}.\\
The definitions of the main observables, pressure and shear strain, are the same 
as before. For the pressure we have 
\beq
\frac{p_g}{d} = \frac{\wh\varphi_g}{2}\int dh\ e^{h+\eta-\D(0)/2}\int 
\DD\z\Th\left(\frac{h+\eta + 
\D_r+\zeta^2\gamma^2/2-\D(0)}{\sqrt{2(2\D_r+\zeta^2\gamma^2-\D(0))}}\right) 
f'(0,h),
\label{eq:pressureP0}
\eeq
and for the shear strain we get
\beq
\frac{\beta\s}{d} = \gamma\frac{\wh\varphi_g}{2}\int dh\ e^{h-\D(0)/2}\int 
\DD\zeta\ \zeta^2\ \frac{e^{-\frac{(h + 
\D_r(\g)-\D(0))^2}{2(2\D_r(\g)-\D(0))}}}{\sqrt{2\pi(2\D_r(\g)-\D(0))}}
\left(\frac{\D_r(\g)-h}{2\D_r(\g)-\D(0)}\right)f(0,h)\:,
\label{eq:stressfull}
\eeq
with the definition
\beq
\D^r(\g) \equiv \D^r + \frac{\g^2\zeta^2}{2}.
\eeq
In the following section we report the results obtained by solving the 
optimization equations for $\D(x)$ and $\D_r$ and computing the physical 
observables defined above.

\section{Results}

\subsection{Phase diagrams and MSDs}

\subsubsection{Compression-decompression}

\begin{figure}[tb]
\begin{center}
\includegraphics[width=0.75\textwidth]{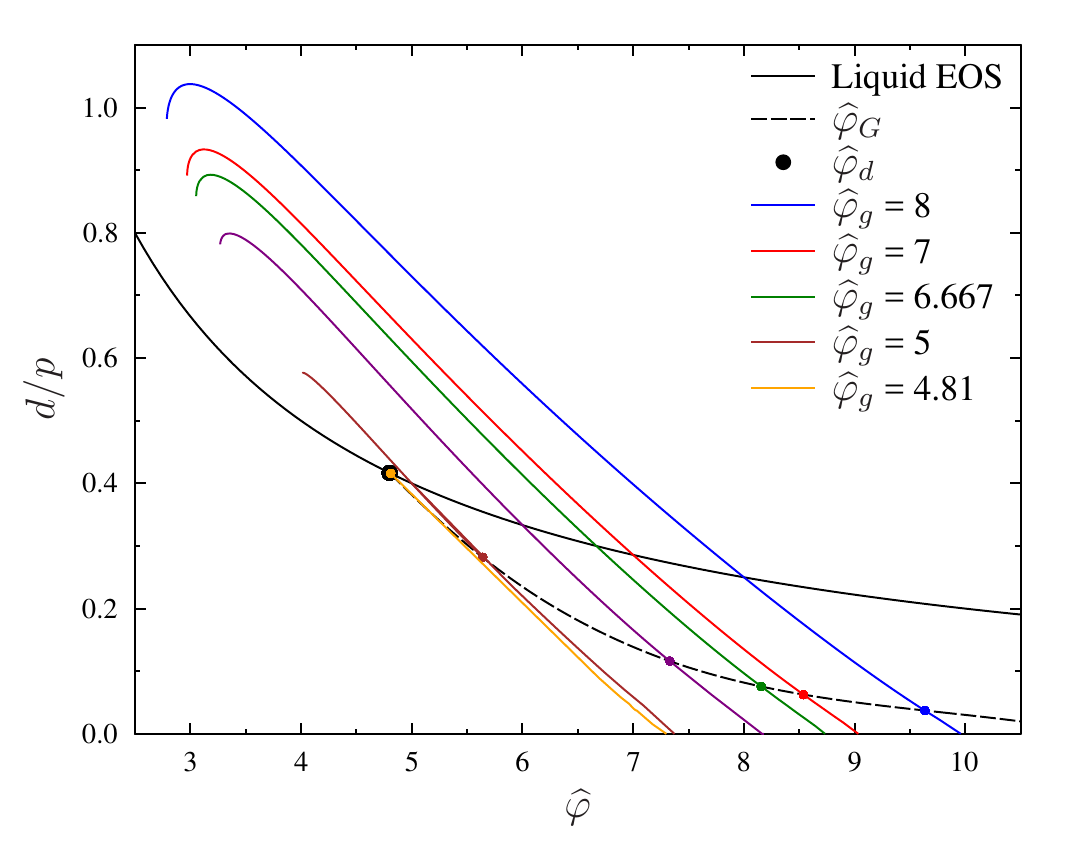}
\caption{Following glasses in (de)compression within the fRSB ansatz. Glassy 
states can now be followed all the way to the jamming point, for all planting 
densities $\wh\f_g$. \label{fig:dinfstatefofull}}
\end{center}
\end{figure}

We focus first on the compression-decompression phase diagram, as done in the 
preceding chapter. We report it in figure \ref{fig:dinfstatefofull}. We can now 
see that within the fRSB ansatz, glasses (including those with $\wh \f_g = \wh 
\f_{MCT}$) can be followed beyond the Gardner point, all the way to the jamming 
transition; the unphysical spinodal points that were predicted by the RS ansatz 
for the less dense glasses have now disappeared. So, at least for hard spheres, 
there is no need to use potentials with three or more replicas to prevent 
unphysical spinodal points from appearing within the theory, differently for 
what was argued in \cite{KZ10,SunCrisanti12}. In particular, we can now compute 
the jamming densities $\wh \f_j$ for every glass; for the glasses prepared 
through an annealing down to $\wh \f_g = \wh \f_d$, corresponding to the 
shortest (and thus easiest) possible annealing protocols, one has 
$$
\wh \varphi_j(\wh\varphi_d)  \simeq 7.30. 
$$
This is the value of the density of the least dense packings that can be 
constructed through an annealing protocol (or equivalently, the most dense that 
con be constructed without needing an exponentially long annealing time, eq. 
\eqref{eq:anntime}). Less dense amorphous packings can in principle be obtained 
with procedures reproducing quenching protocols, but the computation of their 
jamming density cannot be performed within the SF setting, which is conceived 
for the study of annealing protocols. We will return to the issue in section 
\ref{sec:futureresearch}.

It can also be interesting to study how the MSDs behave near the Gardner point. 
Up to the the Gardner transition, the state is a simple minimum and only a 
single MSD $\D$ is present; afterwards, the fRSB structure manifests, the state 
becomes a metabasin of sub-minima, and the MSD becomes a more complicated object 
$\D(x)$, depending on how deep we go in the fRSB hierarchy as $x$ is varied. In 
particular, $x=0$ corresponds to the MSD $\D(0)$ of replicas which are farthest 
apart and sample the big glassy metabasin at the top of the hierarchy, while 
$\D(1)$ is the MSD of replicas which are in the smallest micro states at the 
bottom of the hierarchy; for historic reasons and the analogy with the fRSB 
solution of the SK model, it is commonly called $\D_{EA}$ 
\cite{parisiSGandbeyond}. The $\D_{EA}$ is the Debye-Waller factor of the glass 
in the fRSB phase.\\
\begin{figure}[htb]
\begin{center}
\includegraphics[width=0.7\textwidth]{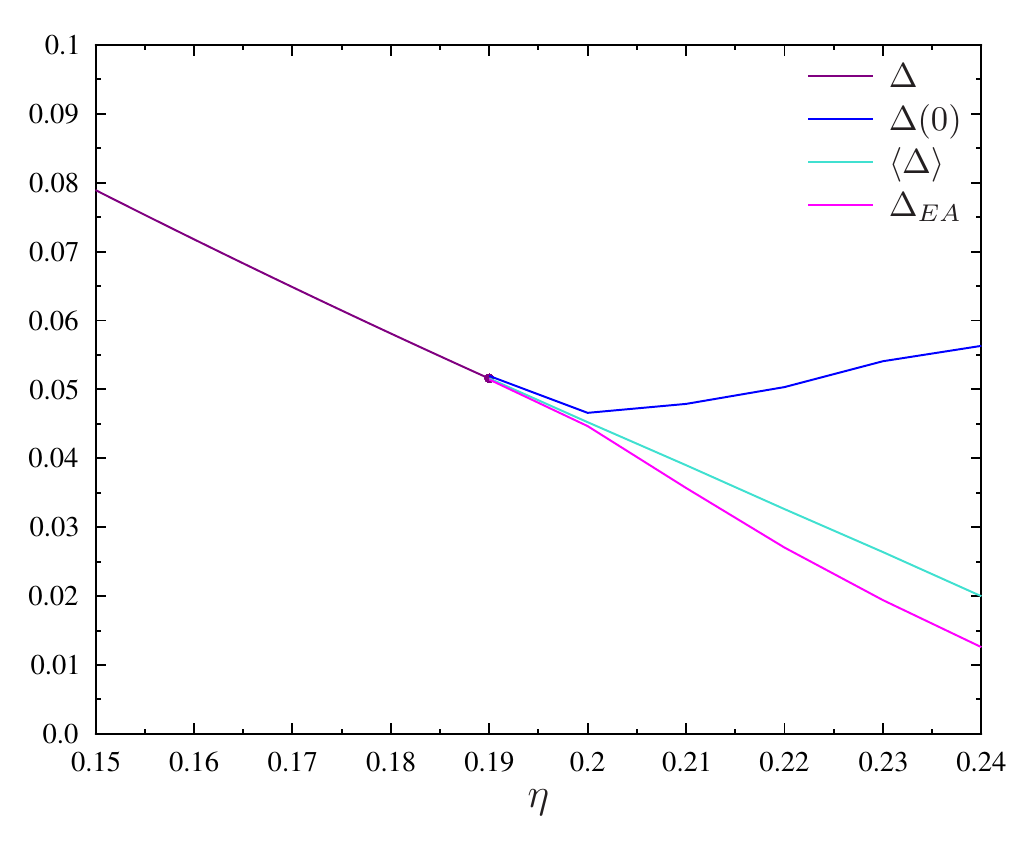}
\caption{MSDs for the glass around the Gardner point for $\wh \f_g = 6.667$. 
\label{fig:MSDfull}}
\end{center}
\end{figure}
In figure \ref{fig:MSDfull} we plot the MSD $\D$ up to $\wh \f_G$, and then the 
values of $\D(0)$, $\D$ and $\thav{\D}$ (equation \eqref{eq:Dvdef}). We can see 
how the three quantities bifurcate at the Gardner point, with $\D_{EA}$ 
decreasing in preparation for the jamming transition, whereupon the spheres 
enter in contact and $\D_{EA} \to 0$. We had mentioned in paragraph 
\ref{subsec:jamming} and in the preceding chapter that the scaling of $\D_{EA}$ 
is that situation is supposed to be $\D_{EA} \simeq p^{-\kappa}$, where $\kappa$ 
is a nontrivial exponent. In order to compute it, and the other two exponents of 
jamming $\gamma$ and $\theta$, one must perform a scaling analysis near jamming 
of the optimization equations in the fRSB ansatz (appendix 
\ref{app:variationalfRSB}). The analysis itself is quite technical in nature and 
very similar to the one already performed in \cite{CKPUZ13}, so we refer the 
interested reader to section \ref{subsec:scalingJ} in the appendix, and here we 
limit ourselves to saying that the results of \cite{CKPUZ13} are recovered. In 
particular, one has
\begin{eqnarray*}
\kappa &\simeq& 1.41574\\ 
\theta &\simeq& 0.42311 \\
\gamma &\simeq& 0.41269 
\end{eqnarray*}
as in \cite{CKPUZ13,nature}.

\subsubsection{Shear strain}
\begin{figure}[tb]
\begin{center}
\includegraphics[width=0.75\textwidth]{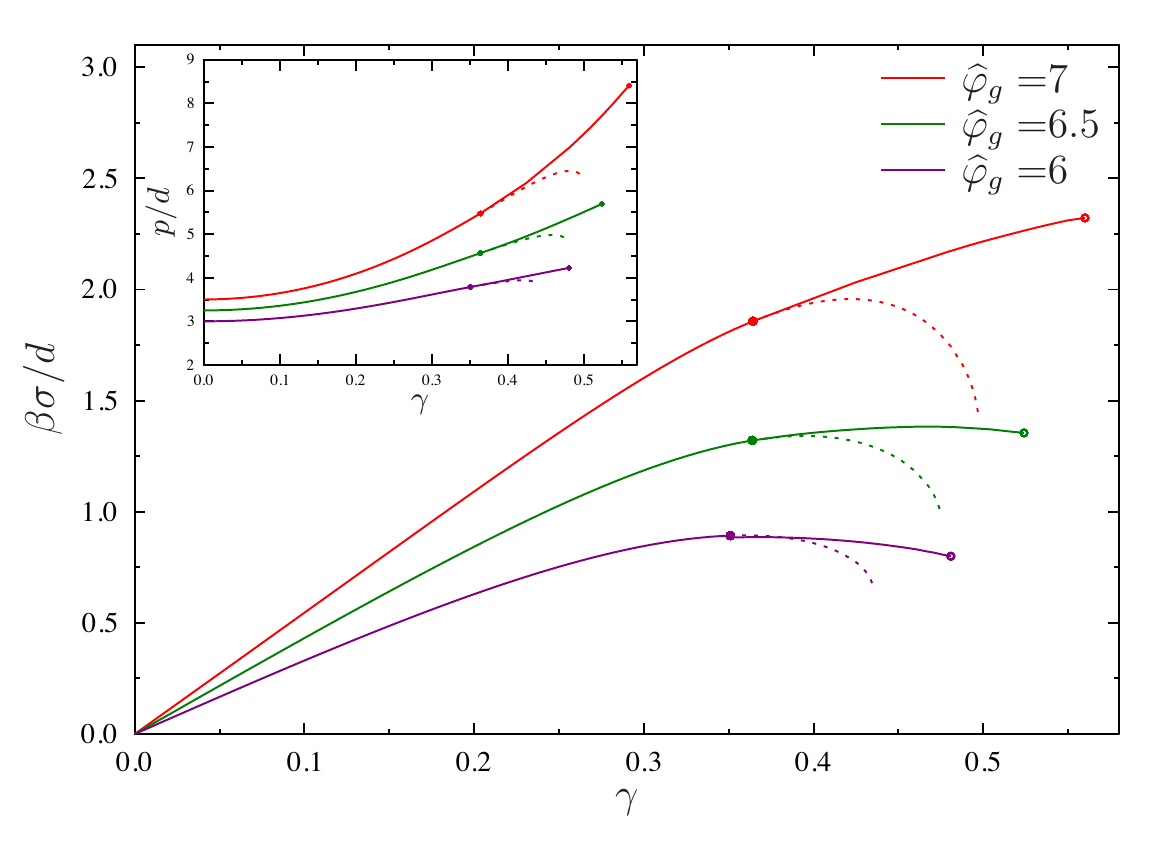}
\caption{Following glassy states prepared at $\wh \varphi_g$ upon applying a 
shear-strain $\g$. With respect to the RS computation (dashed lines) the 
yielding point $\g_Y$ is shifted to larger values of the strain for all glasses. 
However,  we are unable to follows states all the way to the yielding points due 
to instabilities in the numerical code for the solution of the variational 
equations. Despite this, the stress overshoot is still present (see $\wh\f_g = 
6$) and a yielding point, tough not approachable, is clearly present at higher 
values of the strain. An empty circle marks the end of curves beyond which the 
code is unable to follow the solution.\label{fig:shearfull}}
\end{center}
\end{figure}

The phase diagram in strain is reported in figure \ref{fig:shearfull}. One can 
see that, with respect to the RS computation of the preceding chapter, the 
yielding point $\g_Y$ is shifted to higher values of $\g$ (we replot the RS 
solution with a dashed line for comparison), and again a stress overshoot is 
detected.\\
However, we are not able to follow glassy states all the way to the yielding 
point. The reason for this is that the code we use to solve the variational 
equations is unable to approach the yielding point, and it loses track of the 
solution beyond a certain value of $\g$ thet we mark with an empty circle in 
figure \ref{fig:shearfull}. Differently from the spinodal points to the RS 
solution, which were a genuine artifact of the theory, we believe that this 
problem is only technical in nature,  and can be in principle solved just by 
using a more refined code for solving the variational equations in shear. 
Nevertheless, we argue that the presence of a yielding point also within the 
fRSB ansatz is difficult to refute, despite the technical difficulties.

\begin{figure}[htb]
\begin{center}
\includegraphics[width=0.6\textwidth]{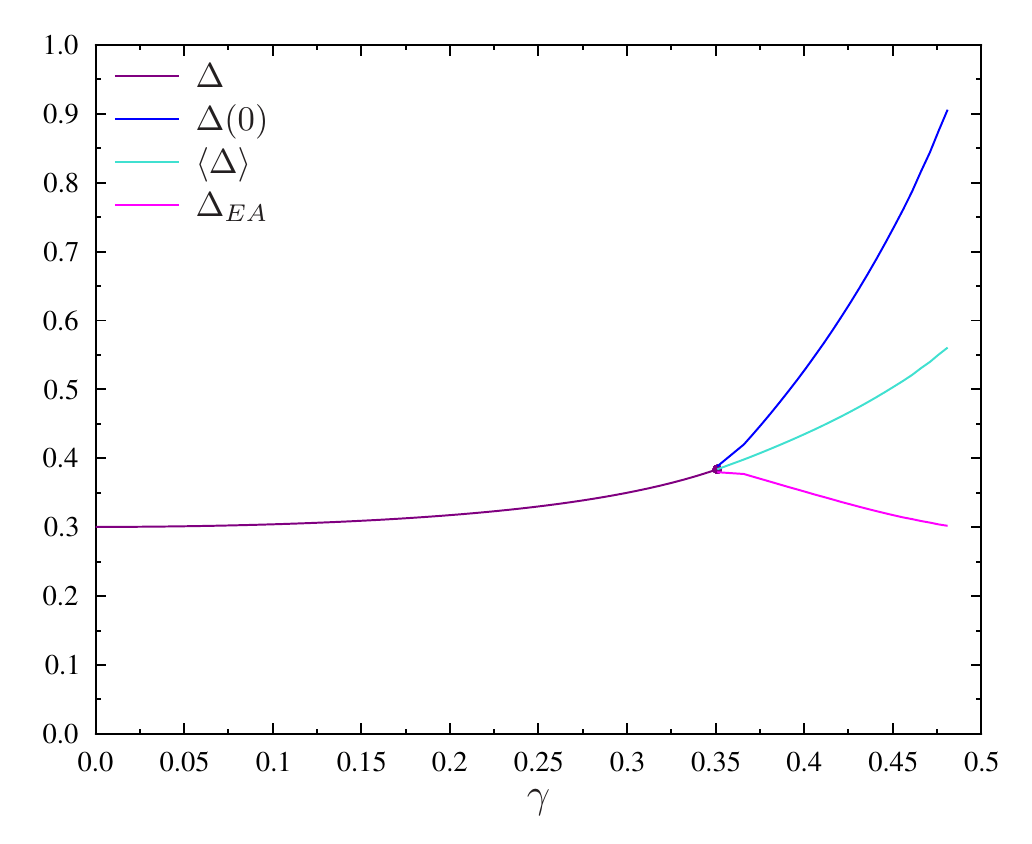}
\caption{MSDs for the glass at $\wh \f_g = 6$ as shear is applied. The MSD of 
the metabasin $\D(0)$ grows rapidly beyond the Gardner point, while the MSD of 
fRSB micro-states $\D_{EA}$ decreases.  \label{fig:MSDfullshear}}
\end{center}
\end{figure}

This view of things is corroborated by the behavior of the MSDs beyond the 
Gardner point, which we report in figure \ref{fig:MSDfullshear}. We can see 
that, beyond the Gardner point, the MSD $\D(0)$ of the glassy metabasin shoots 
up very rapidly, signaling that the glassy state is being widened more and more 
by the shear. Despite this, the MSD $\D_{EA}$ of the micro states within it 
decreases, signaling that they become more tight. This is no surprise, since the 
pressure is actually increasing because of dilatancy as one can see in the inset 
of figure \ref{fig:shearfull}, and we already know that $\D_{EA}$ is supposed to 
be inversely proportional to the pressure $p$. 

In summary, we argue that the yielding transition is always present in the fRSB 
ansatz, and that the picture of yielding as a spinodal point that emerged from 
the RS computation remains true in this case, with the only caveat that is must 
be applied to \emph{metabasin} level. The yielding transition corresponds to a 
loss of stability (or equivalently, a saddle node bifurcation) at the metabasin 
level as we had already surmised in paragraph \ref{subsec:yielding}. This means 
that the MSD $\D(0)$ is supposed to have, near the yielding point, the 
square-root behavior
\beq
\D(0) - \D(0)^{max} = -C\sqrt{\g_Y-\g},
\eeq
just like the MSD $\D$ at the onset transition. This however is true only for 
$\D(0)$. We argue that $\D_{EA}$ will stay finite (and with it, the pressure) at 
the yielding point $\g_Y$, as the plot in figure \ref{fig:MSDfullshear} seems to 
indicate.

\subsection{Critical slowing down \label{subsec:lambdaMCT}}
The Gardner transition, despite its peculiarities, is still a second order 
critical point, whereupon a phenomenology typical of continuous transitions is 
supposed to manifest. First of all, suppose to perform a quenching dynamics with 
an initial condition within the state, as in \cite{BarratBurioniMezard}. We want 
to investigate how the dynamics relaxes towards equilibrium within the 
metastable state.
In the stable glass phase the metastable state is ergodic (just like the liquid 
phase above the MCT transition density $\wh \varphi_{MCT}$) and an exponential 
relaxation is consequently observed 
\cite{BarratBurioniMezard,CharbonneauJin15,corrado}, as anticipated in paragraph 
\ref{subsec:RFOTaging}. However as the Gardner point is approached, we expect a 
dynamical slow down due to the appearance of an internal structure of substates 
within the metastable basin. 
Indeed, at this transition point, the relaxation becomes power law instead of 
exponential, a phenomenon well known within the theory of critical phenomena and 
dubbed \emph{critical slowing down} \cite{zinnjustin}. 
To fix ideas, let us define a dynamical mean square displacement
\beq
\Delta_{D} (t)=\frac{d}{N}\sum_{i=1}^N|x_i(t)-x_i(0)|^2
\eeq
being $x_i(t)$ the position of the sphere $i$ at time $t$.
At the Gardner transition point we have
\beq
\Delta_D(t)\sim \Delta - A t^{-a}
\eeq
being $\Delta$ the solution of the RS saddle point equations \eqref{eq:Dg} to 
\eqref{eq:eqDf}; the constant $A$ is expected to be positive. This in turn 
implies that the relaxation time within the state obey the scaling \cite{PR12}
\beq
\t_\beta \propto (\wh \f_G - \wh \f)^{-1/a}
\eeq
as studied in \cite{CharbonneauJin15}.\\
We want to compute the exponent $a$. This is related to the so called 
\emph{exponent parameter} $\l$ \cite{PR12}
\beq
 \l = \frac{\Gamma(1-a)^2}{\Gamma(1-2a)}.
\eeq
It has been shown in \cite{PR12} that the exponent parameter can be computed 
from the replica approach. Indeed it is given by
\beq
\l=\frac{w_2}{w_1}
\eeq
and $w_1$ and $w_2$ are two cubic terms in the expansion of the free entropy 
around the RS solution at the Gardner point, defined in \cite{KPUZ13}. 
As proven in paragraph \ref{subsec:repliconint}, all the expressions for 
quadratic and cubic terms reported in \cite{KPUZ13} can be reused in the state 
following setting, just by redefining suitably the integral measure for 
computing averages
\beq
\la\mathcal O(\l)\ra_{RS} \equiv \int d\l\ 
\mathcal{O}(\l)\frac{e^{-\frac{(\l+\sqrt{\D})^2}{2}}}{\sqrt{2\pi}} 
\longrightarrow \la\mathcal O(\l)\ra_{SF} \equiv \int d\l\ 
\mathcal{O}(\l)G(\l), 
\eeq
where $G(\l)$ is defined in equation \eqref{eq:Glambdadef}. We can thus 
effortlessly write the expression for $\l$
\beq
\l = \frac{-8\wh\varphi_g w_2^{(I)}}{16/\D^3-8\wh\varphi_g w_1^{(I)}}
\eeq
where $w_1^{(I)}$ and $w_2^{(I)}$ are defined in \cite[Eq.(79)]{KPUZ13}. 
With some algebra (see \cite{RainoneUrbani15}), we get the final result:
\beq
\l = \frac{-4\wh\varphi_g A}{16+8\wh\varphi_g(1+B)},
\eeq
with
\begin{eqnarray}
A &=& \la\Th_0^{-1}(\l)\G_2(\l,0)\ra, \label{eq:defA}\\
B &=& \la\Th_0^{-1}(\l)\G_1(\l,0)\ra, \label{eq:defB}
\end{eqnarray}
and
\beq
\begin{split}
\Gamma_2(\l,s)&=\left[ 2\left(\frac{\Th_1(\l)}{\Th_0(\l)}\right)^3 
-3\frac{\Th_1(\l)\Th_2(\l)}{\Th_0^2(\l)}+\frac{\Th_3(\l)}{\Th_0(\l)}  
\right]\left[  2\l^3+2(s-6)\left(\frac{\Th_1(\l)}{\Th_0(\l)}\right)^3  \right.  
+\\
&\left.+ 3\frac{\Th_1(\l)}{\Th_0(\l)}\left[ 
4\l\frac{\Th_1(\l)}{\Th_0(\l)}-(s-4)\frac{\Th_2(\l)}{\Th_0(\l)}\right] 
-6\l\left(\l\frac{\Th_1(\l)}{\Th_0(\l)}+\frac{\Th_2(\l)}{\Th_0(\l)}
\right)+(s-2)\frac{\Th_3(\l)}{\Th_0(\l)}\right] \ , \\
\G_1(\l,s) &=\left[ 
1+\frac{\Th^2_1(\l)}{\Th_0^2(\l)}-\frac{\Th_2(\l)}{\Th_0(\l)} \right]^2 
\left[(s-3\l^2)+(s-6)\frac{\Th_1^2(\l)}{\Th_0^2(\l)}+6\l\frac{\Th_1(\l)}{
\Th_0(\l)}-(s-3)\frac{\Th_2(\l)}{\Th_0(\l)}\right],
\end{split}
\eeq
and the $\Th_k(\l)$ functions are defined as \cite{KPUZ13}
\beq
\Th_k(x) \equiv \frac{1}{\sqrt{2\pi}}\int_x^\infty dy\ y^k \eee^{-\frac12 y^2}.
\eeq
The quantities $A$ and $B$ can be easily computed numerically;
we report the results of the numerical evaluation of the exponents in table 
\ref{table:lambdaMCT}.
\begin{table}[htb!]
\begin{center}
\begin{tabular}{c@{\hspace{1cm}} c@{\hspace{1cm}}c}
\hline\hline
$\wh\varphi_g$ & $\l$ & $1/a$\\
\hline
4.8 & 0.702666 & \\
4.9 & 0.560661 & 2.65122\\
5 & 0.509074 & 2.54665\\
5.25 & 0.437754 & 2.42661\\
5.5 & 0.393779 & 2.36344\\
5.87 & 0.351157 & 2.30848 \\
6 & 0.339808 & 2.29475\\
6.667 & 0.295692 & 2.24461\\
7 & 0.280148 & 2.22805\\
8 & 0.246892 & 2.19440\\
10.666 & 0.204280 & 2.15441\\
\hline
\end{tabular}
\caption{Our results for $\l$ and $1/a$ for various planting densities 
$\wh\f_g$. \label{table:lambdaMCT}}
\end{center}
\end{table}

\subsection{Fluctuations \label{subsec:fluctuations}}
Besides critical slowing down, another phenomenology typical of second order 
critical points is the manifestation of an infinite susceptibility, which as a 
rule of thumb is usually linked to the inverse of the zero mode that signals the 
transition \cite{zinnjustin} (in our case, the replicon).\\
In paragraph \ref{subsec:coop}, we discussed the dynamical susceptibility 
$\chi_4(t)$ and how it reaches a maximum at a time $t^* \simeq \t_\al$ 
corresponding to cooperative relaxation of the system. In the case of the 
Gardner point, the situation is similar but there is a very relevant difference, 
namely the fact that the transition is second order, and differently from the 
MCT transition, is not avoided: the susceptibility converges to a stable static 
value, $\lim_{t \to \infty} \chi_4(t) = \chi_4$, which goes to infinity as the 
transition is approached, $\lim_{ \f \to \f_G^-} \chi_4 = \infty$. This 
divergence can be also used in practice to locate numerically the Gardner 
transition, see \cite{CharbonneauJin15}.\\
To fix ideas, let us define 
\beq
\chi(t)=\langle \Delta^2(t)\rangle-\langle\Delta(t)\rangle^2\:, 
\eeq
where the brackets are used to denote the average over the thermal history of 
the system. The $\chi(t)$ is a dynamical quantity, but we need to focus on its 
large time behavior when $\langle\Delta(t)\rangle\to \Delta$.
In this case $\chi(t) \to \chi_4$, where $\chi_4$ is a static susceptibility
\beq
\chi_4 = \langle\D_{ab}^2\rangle- \thav{\D_{ab}}^2, 
\label{eq:chi4static}
\eeq
already well known in the context of spin glasses, that can be computed from a 
static approach \cite{FP00,FranzJacquin12}.\\ 
Within a mean field theory, fluctuations like the \eqref{eq:chi4static} are 
encoded in the quadratic term (the so-called mass term) of the field theory that 
is obtained by expanding the free energy around the critical point (the Gardner 
point in this case) \cite{zinnjustin,FranzJacquin12}. Within this field theory, 
the order parameter field is $\D_{ab}$ and the ``mass matrix'' is nothing but 
the tensor $M_{a<b;c<d}$ defined in the \eqref{eq:defhessian}. We need to study 
the inverse of this quadratic operator in order to obtain the value of the 
$\chi_4$ susceptibility.\\
In principle, and as already discussed, the whole tensor $M$, with all indices 
running form 1 to $s+1$ (we always assume $m=1$), has to be taken into account. 
However, here we are mostly interested with the most divergent part, and not on 
the finite corrections. As a result of this, we can again focus only on the 
tensor $M$ restricted to the sector of $s$ replicas, equation \eqref{stab_matr}, 
that we had already studied in appendix \ref{app:stabilityRS} for locating the 
Gardner transition. The inversion of the tensor $M^s$ is just a matter of 
standard linear algebra, so we refer the interested reader to 
\cite{RainoneUrbani15} and just quote the the final result
\beq
\chi_4^\textrm{div}=\frac{24}{\l_R},
\eeq
where $\l_R$ is the replicon mode, equation \eqref{eq:lR}. Unsurprisingly, the 
diverging susceptibility is inversely proportional to the critical mode at the 
transition. 

\subsection{Shear moduli \label{subsec:shearmodulifull}}
The appearance of a fRSB hierarchy within the glassy states has very interesting 
consequences in terms of shear modulus. The reason for this is the following: 
suppose that our system is into one of the bottom fRSB microstates, with MSD 
$\D(1)= \D_{EA}$. We now apply a small strain to the system. If the shear is 
small enough that the system does not leave the microstate, it will respond 
elastically with a shear modulus $\mu = \frac{1}{\D_{EA}}$. However, if the 
shear is large enough that the system is kicked out of the microstate, goes all 
the way to the top of the hierarchy, and then falls in another microstate (but 
anyway within the same glass basin), the response will be described by a modulus 
$\mu = 1/\D(0)$, where $\D(0)$ is the MSD of the whole metabasin as previously 
discussed.
In summary, the presence of a hierarchy of sub-states makes it necessary to keep 
track of the fact that the strain may not only act as a deformation of the 
minimum the system is in (as it was the case above the Gardner transition where 
the glassy state has no structure), but may also cause the system to escape from 
the minimum and end up into another \cite{YoshinoZamponi14}. As a result of 
this, the response of the system will be described by a generalized, 
protocol-dependent shear modulus
\beq
\mu(x) \equiv \frac{1}{\D(x)}
\eeq
where the relevant value of $x$ depends on how high in the hierarchy the system 
has to go to make the jump. This fact is well known in the context of spin 
glasses where an external magnetic field and the magnetic susceptibility play 
the role of the strain and shear modulus, respectively, but the physical picture 
is the same \cite{YoshinoZamponi14}.\\
It is obvious that the most important moduli correspond to the two extreme cases 
$\mu(0) = \frac{1}{\D(0)}$ and $\mu(1) = \frac{1}{\D_{EA}}$. We are interested 
in studying their behavior near the Gardner point. In order to achieve this, we 
just go for the easiest route and assume that near enough to the Gardner point, 
the full profile $\D(x)$ can be reasonably approximated with a two-step (1RSB) 
profile with two MSDs $\D_1$ and $\D_2$ and two shear moduli
\beq
\begin{split}
\mu_1&=\frac{1}{\Delta_1},\\
\mu_2&=\frac{1}{\Delta_2}.
\end{split}
\eeq
We are interested in the behavior of $\D_1-\D_2$ as a function of the density 
$\wh\f$ near the Gardner point $\wh\f_G$, which requires the study of the saddle 
point equations for the replicated free energy truncated at cubic order, as done 
in paragraph \ref{sec:replicongen}. The computations are just a matter of 
standard algebra and are again very similar to the study of \cite[sec. 
VII]{CKPUZ13}, so we skip again to the final result and refer to 
\cite{RainoneUrbani15} for details. One gets:
\beq
\frac{1}{\mu_1}-\frac{1}{\mu_2}=\Delta_1-\Delta_2\propto \wh \f-\wh \f_G.
\eeq
So the difference between the two moduli is linear in the distance from the 
Gardner point. This prediction should be easy to check numerically.

\section{Discussion}

In this chapter we have performed the computation of the Franz-Parisi entropy 
within the fRSB ansatz, describing a glassy minimum wherein a hierarchical, 
fractal structure of microstates manifests at the bottom of the glassy minimum, 
thereby making it a metabasin of fractal sub-minima \cite{nature}. Using this 
ansatz, we have been able to cure the unphysical spinodal points that appeared 
within the RS ansatz, making it possible to follow glassy states in compression, 
all the way to the jamming point. For what concerns yielding, we have been 
unable to follow the state all the way to the yielding point because of 
technical difficulties, but we are anyway able to detect a stress overshoot and 
the presence of a yielding point also within the fRSB ansatz appears to be 
irrefutable. We believe that the technical problems can be fixed simply by using 
a more refined code to solve our state following equations. Notice however that 
the fRSB equations (appendix \ref{app:variationalfRSB}) are Partial Differential 
Equations (PDEs), so their numerical treatment is anyway a hard task and an open 
field of study in numerical analysis as of today. There may also be another 
possibility, namely that the equations near yielding develop a scaling regime as 
in the jamming case, with an associated set of critical exponents, rendering 
possible an analytic study near the yielding point. However, the presence of 
such a criticality near yielding is far from established, and still an open 
problem 
\cite{KarmakarLerner10,KarmakarLerner10b,HentschelKarmakar11,IlyinProcaccia15}, 
so more studies will be needed in the future.

There is one very important point to discuss. Throughout this thesis we have 
stated multiple times that the result we derive are valid on a timescale 
$t_{exp}$ such that
$$
\t_\be \ll t_{exp} \ll \t_\al,
$$
where $\t_\be$ is the time needed to equilibrate within the glassy states. 
Within state following with the two-replica potential, we are \emph{always} 
looking at equilibrium within the glassy state, no matter which ansatz we use.\\
Now, in the stable glass phase above $\wh\f_G$, relaxation within the glassy 
state is exponential \cite{BarratBurioniMezard} and the timescale $\t_\be$ is 
short enough that any reasonable protocol has a $t_{exp}$ such that the above 
requirement is met. This however is not true anymore beyond the Gardner point. 
We recall the reader that the Gardner transition \cite{Ga85} is equivalent to 
the ferromagnetic transition in the SK model \cite{parisiSG,parisiSGandbeyond}; 
this means that the dynamics of the system beyond the Gardner point will be 
characterized by an aging phenomenology \cite{Ogielski85} much like the one 
exhibited by the SK model \cite{parisiSGandbeyond}.\\
This aging phenomenology is extremely rich and as of now still not fully 
understood. What is certain is that it is a lot more complicated than the aging 
phenomenology commonly exhibited by structural glasses. Instead of a 
two-timescale scenario like the one commonly found in RFOT models, a very 
complicated hierarchy of timescales (and with them, effective temperatures) 
manifests within the dynamics of systems with a fRSB transition 
\cite{parisiSGandbeyond,FranzVirasoro00,CugliandoloDyn}; as a result of this, 
the system takes a long time (though not as large as $\t_\al$ of course) to 
attain equilibration, even within the glassy metabasin.\\
In such a situation, a theory aiming to reproduce a realistic laboratory 
protocol would focus on the state following of \emph{a single fRSB microstate}, 
instead to the whole metabasin as we have done in this chapter. This is however 
extremely difficult (and maybe impossible) to realize in practice, because fRSB 
microstates merge, bifurcate, and cross even when the perturbation is 
infinitesimal, an effect well explained in \cite{YoshinoZamponi14}; as a result 
of this, a two-replica (or even a three-replica) potential is not sufficient for 
following a single microstate, and a continuous chain, with an infinite number 
of replicas as detailed in section \ref{sec:chain}, is in principle needed. 
However, a computation with a continuous replica chain within the fRSB ansatz 
looks extremely difficult and perhaps not even doable.\\
This effect can be easily understood by looking at the stress strain curves in 
figure \ref{fig:shearfull}, beyond the Gardner point, and comparing then with 
the ones in figure \ref{fig:shearstrain}. The response in figure 
\ref{fig:shearstrain} is very rough, as one should expect from a system whose 
dynamics takes place in a very disordered landscape. The curves in 
\ref{fig:shearfull}, however, are smooth, despite the fact thet the system is 
now moving in a very rough FEL characterized by a fractal hierarchy of 
microstates. The reason for this is that in figure \ref{fig:shearfull} we are 
are looking at equilibrium inside the metabasin, so we are effectively giving 
the system enough time to explore ergodically all the fRSB sub-minima after each 
strain step; from a practical point of view, we are taking various rough shear 
histories like the ones in figure \ref{fig:shearstrain} and we are averaging 
their stress-strain curves, getting back a smooth response.

In summary, beyond the Gardner point, our SF calculation is only an 
approximation of the real dynamics of the system. However, as the numerical 
results of \cite{CharbonneauJin15} indicate, the deviations between theory and 
simulation, in compression-decompression protocols, are anyway very small. For 
what concerns shear, our approach is not able to reproduce the state following 
of a single inherent structure, as done in AQS protocols (paragraph 
\ref{subsec:qsshear}), since it looks always at equilibrium within the 
metabasin. It is however well suited to the rheology of thermal systems like 
foams, pastes and soft matter in general.

\subsection{Yielding within the fRSB ansatz}
We conclude the chapter with some comments about the picture that emerges from 
our calculation for what concerns the yielding transition. As in the preceding 
chapter, the yielding point looks very much like a spinodal point whereupon the 
state loses stability and opens up along an unstable direction, becoming a 
saddle. This is signaled by a behavior of the intra-metabasin MSD $\D(0)$ that 
looks very much like the one exhibited at the onset transition (see figure 
\ref{fig:Delta_compr}, top left panel), with a square root singularity
$$
\D_{1} - \D_1^{max} = -C\sqrt{\g_Y-\g},
$$
as already discussed. However, from figure \ref{fig:MSDfullshear} we can see 
that while $\D(0)$ shoots up, the behavior of the other MSDs is milder, and 
$\D_{EA}$ is even decreasing. Thus, as the metabasin is being widened more and 
more by the shear, the fractal hierarchy of state within it survives, and the 
bottom states with $\D = \D_{EA}$ are even getting tighter. It is safe to assume 
that even when the basin finally opens and loses stability because of the shear, 
the bottonmost microstates anyway maintain their form.

This fact is very interesting if one views it in term of shear moduli instead of 
MSDs, as detailed in paragraph \ref{subsec:shearmodulifull}. As explained in 
paragraph \ref{subsec:yielding}, the onset of an instability (and with it, an 
avalanche) is signaled by a square root-like singularity in the shear modulus. 
We have also mentioned, figure \ref{fig:shearstrainlemaitre}, how the curves 
typically found in AQS protocols have a sort of scale-invariant structure, and 
are made of small segments which have a general trend similar to the whole 
stress-strain curve.\\
This fact can be nicely interpreted in terms of the $\mu(1)$ and $\mu(0)$ shear 
moduli: the shear modulus $\mu(1)$ is the one relative to the small segments and 
to a single inherent structure, while the shear modulus $\mu(0)$ is the one 
relative to the ``elastic'' part of the whole stress-strain curve (see figure 
\ref{fig:shearstrain} for small $\g$). The small plastic events would then 
correspond to a loss of stability at fRSB microstate level and a singularity in 
the $\mu(1)$ modulus, while the yielding transition whereupon the system starts 
to flow corresponds to a loss of stability within the \emph{whole metabasin} and 
a singularity in the $\mu(0)$ shear modulus. In addition to this, one can also 
see that the curves after yielding are on average flat (no $\mu(0)$ can be 
defined) but do show an elastic response on small scales ($\mu(1)$ can still be 
defined). This seems to be in accordance with the approach to the yielding point 
within our state following calculation, wherein only $\D(0)$ exhibits a 
square-root like behavior at yielding while $\D(1)$ stays finite.\\ 
It is worth of note that the basic phenomenology of yielding seems to be well captured by our HS model in the MF limit. This appears surprising since one of the defining traits of MF models is their lack of space structure, while the phenomenology of yielding (and in particular of avalanches) seems to be mainly ruled by local rearrangement modes (we remind the discussion in paragraph \ref{subsec:yielding}), which should be completely missed by the MF solution. We must however precise that the local modes are certainly relevant for small avalanches like the ones visible in figure \ref{fig:shearstrainlemaitre}, while there is no consensus about the localization properties of the rearrangement modes that separate the elastic part of stress-strain curves from the flowing, steady-state part. While Eshelby-like \cite{EshelbyInclusion} modes still seem to be relevant, it is argued in \cite{DasguptaHentschel12} that a concatenation of many Eshelby modes is necessary to destroy the glass via a shear-banding process; in this case, the localization properties of the rearrangement modes may be much more complicated to figure out then how it would be in the case of a single, simple Eshelby mode, and an effective delocalization effect may appear. We remind the reader that from the discussion of paragraph \ref{subsec:coop}, it transpires that a cooperative, delocalized rearrangement is necessary for glass relaxation. This may also be the case for driven glasses under shear.\\     
In any case, more work is required to better understand the flow regime beyond the yielding 
point.

\chapter{Numerics in the Mari-Kurchan model \label{chap:isocomplexity}}

All the results we presented in the preceding two chapters are valid for hard 
spheres in the $d \to \infty$ limit. Despite the fact that the physics predicted 
by the State Following approach is in qualitative agreement with the general 
phenomenology of glasses reviewed in chapter \ref{chap:metglass}, it would be 
anyway nice to get a quantitative comparison with numerical results. Such a 
program is however difficult to implement as the simulation of a particle system 
with many spatial dimensions is obviously an endeavor of considerable 
computational cost. In this chapter we report a workaround, centered around a 
special HS model which has a MF-like behavior also in finite dimension, in 
particular dimension three. The model has the advantage of being both 
analytically tractable with the replica method and of being very easy to 
implement numerically, allowing for a systematic comparison between theory  and 
simulation.

\section{Model}
The model we employ is referred to as \emph{Mari-Kurchan (MK) model} 
\cite{marikurchan}, and its Hamiltonian is
\begin{equation}
H_{MK} \equiv \sum_{i<j} V(\bx_{i}-\bx_{j}-\bA_{ij}),
\end{equation}
where $V$ is any suitable interaction potential (in our case it will be the HS 
one). The $\bA_{ij}$ are ``random shifts'', i.e. quenched, random 
$d$-dimensional vectors identically, independently and uniformly distributed in 
the $d$-dimensional cube:
$$
P(\bA) = \frac{1}{V};
$$
and of course $\bA_{ij} = \bA_{ji}$.

This model can be seen as MF in multiple ways. First, we can notice that the 
model is devoid of any space structure: despite the fact that every particle 
interacts, given a certain realization of the $\bA$s, with a finite number of 
``neighbors'', those neighbors can be anywhere in the sample, since the shifts 
are uniformly distributed in the whole cube. From this point of view, the model 
is MF because the physical space the model is embedded into plays no role on the 
interactions.\\
A less intuitive, but more profound line of reasoning stems from considering the 
probability of having three particles, say $i$, $j$ and $k$, interact with each 
other \emph{at the same time}, i.e., each of them interacts with \emph{both the 
other two} at the same time. For this to happen, we should have, for the HS 
potential,
\begin{equation}
\begin{split}
|\bx_{i}-\bx_{j} - \bA_{ij}|\simeq &\ D,\\
|\bx_{j}-\bx_{k} - \bA_{jk}|\simeq &\ D,\\
|\bx_{k}-\bx_{i} - \bA_{ki}|\simeq &\ D,
\end{split}
\end{equation}
which would imply
$$
|\bA_{ij} + \bA_{jk}+\bA_{ki}| \simeq D,
$$
which is very unlikely (and, in the thermodynamic limit, outright impossible), 
since the shifts are $O(L)$\footnote{$L$ is the side of the simulation cube.}. 
In this model, three body interactions are effectively forbidden: if $i$ 
interacts with $j$, and $i$ interacts also with $k$, then $k$ and $j$ do not 
interact with each other. Thus, the MK model is mean-field in the sense that the 
network of interactions is tree-like, i.e., there are no \emph{loops} 
\cite{replicanotes}. Actually, it is indeed the disappearance of loops for $d 
\to \infty$ that gives high-$d$ HSs their mean field nature. In that case, 
three-body interactions are made impossible for $d\to\infty$ by the high 
dimensionality itself \cite{frischpercus}.

Besides the fact that the model has a MF behavior also for finite $d$, it has 
two more big advantages. First, the presence of a quenched disorder in the form 
of the random shifts allows one to implement a procedure, dubbed the 
\emph{planting} method \cite{krzakalazdeborovaplanting}, which allows one to 
obtain thermalized configurations also in the glass phase above $\f_{MCT}$, 
wherein it would normally be extremely time-consuming to do so, because of the 
glassy slowdown. So one can effectively implement an annealing protocol with 
arbitrary $\f_g$ with a negligible computational cost.\\
The basic idea of planting is to invert the order according to which initial particle positions $\br_i$ and quenched random shifts $\bA_{ij}$ are chosen:
first one extracts a random configuration of particle positions $\br_i$, and afterwards the random shifts $\bA_{ij}$ are chosen uniformly, with the sole requirement that spheres do not overlap. 
In the liquid phase $\varphi \in [\varphi_{MCT},\varphi_{K}]$, this straightforward process produces an equilibrium configuration that automatically satisfies the liquid EOS (see \cite{krzakalazdeborovaplanting} for more details). 
Since in the MK model the complexity never vanishes and consequently the ideal glass transition density $\varphi_K$ goes to infinity, this procedure allows one to produce equilibrated configurations for any density, with negligible computational
cost. A set of positions $\{\br_i\}$ and shifts $\{\bA_{ij}\}$ identifies a \emph{sample}.

Besides this numerical flexibility, the model can be also easily treated analytically, in the 
sense that its replicated entropy (section \ref{sec:realreplica}) can be 
computed using the techniques discussed in \cite{parisizamponi}; the result is 
exact only for infinite $d$ \cite{KPZ12}, but for finite $d$ it constitutes an 
anyway excellent approximation, as we are going to see and as discussed in 
\cite{CharbonneauJin15}. The computation of the replicated entropy of the MK 
model is reported in full detail in \cite{replicanotes}, so we just report the 
final result, for an RS ansatz:
\beq
s[m,\varphi,A] = -\log\rho + \log N + S_{harm}(m,A) - 2^{d-1}\varphi[1-G_m(A)],
\label{eq:sMK}
\eeq
where $A$ is the \emph{cage radius}, $A  \equiv \frac{D^2}{2d}\D$, $S_{harm}$ is 
defined as
\beq
S_{harm}(m,A) \equiv (m-1)\frac{d}{2}\log (2\pi A) -\frac{d}{2}\log m + 
d\frac{m-1}{2},
\eeq
and the function $G_m(A)$ is defined in \cite{replicanotes,parisizamponi}. The 
reader is invited to appreciate the similarity with the expression 
\eqref{eq:gauss_r} for $s\to 0$. Indeed, for $d\to\infty$ the \eqref{eq:sMK} 
reduces to the \eqref{eq:gauss_r} computed with the RS ansatz \cite{KPZ12}. As 
the \eqref{eq:gauss_r}, this expression must be optimized over $A$ to obtain the 
physical replicated entropy $\Sc(m,\f)$.

\section{Results}
The MK model defined above is a very convenient test-bed for the theory exposed 
in this thesis. We now report some results obtained with numerical simulations 
of the model, contained in references \cite{corrado,CharbonneauJin15}. Those 
studies have a larger scope and contain much more material than reported here, 
so we refer to them for further reading.\\
Simulations are always performed following the same basic guideline: first, an 
equilibrium configuration is produced through planting at a certain planting 
density $\f_g$ ($\f_0$ in the notation of \cite{corrado,CharbonneauJin15}); then 
a compression (decompression) protocol is implemented by inflating (deflating) 
the spheres with the LS algorithm \cite{LS90}. This way, a state following 
procedure in compression/decompression is reproduced. To our knowledge, a 
similar numerical procedure has not been implemented for State Following under 
quasi-static shear, though it is an obvious continuation of the studies 
\cite{corrado,CharbonneauJin15} that we leave for future work.

\subsection{Isocomplexity}
As a warm-up exercise we first derive the equations of state of the glass 
through the isocomplexity approximation \cite{corrado}, paragraph 
\ref{subsec:isocomplexity}, also for the sake of comparison with the more 
refined State Following computation. We then compare the analytical results with numerical simulations of an MK model in $d=3$ with $N=800$ particles of diameter $D=1$ and periodic boundary conditions.
Since we focus only on thermodynamic, self-averaging observables (see the discussion of section \ref{sec:FP}), only a small number of samples, $N_s = 6$, is needed to perform ensemble averages. The dynamics is implemented through a Metropolis
algorithm and compression-decompression is implemented through inflation-deflation of the spheres as in the LS \cite{LS90} algorithm.

\begin{figure}[htb]
\begin{center}
\includegraphics[width=0.5\textwidth,angle=270]{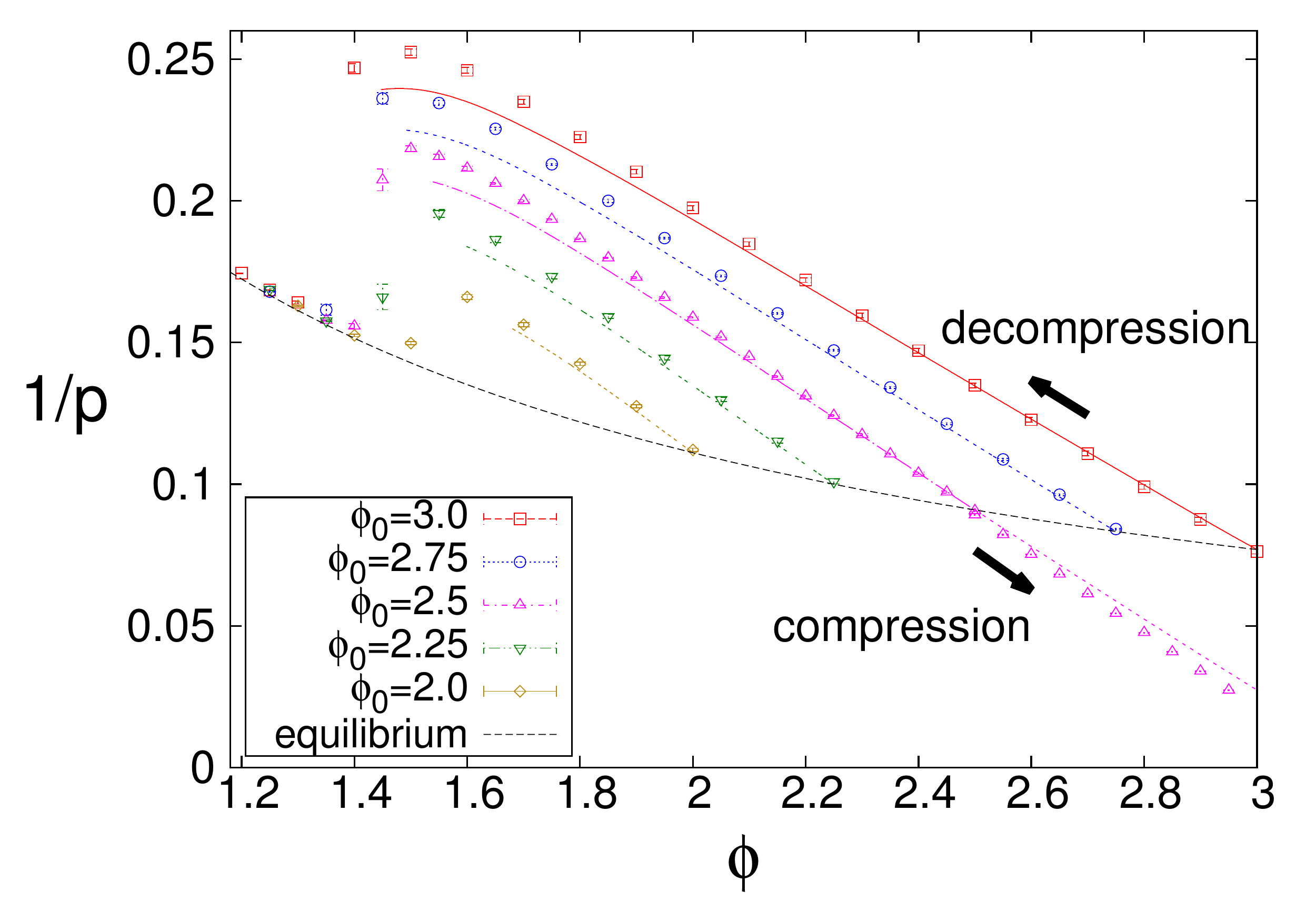}
\caption{Equation of state of various glasses as in figure 
\ref{fig:dinfstatefo}. The lines are theory and the dots are simulation data as 
described in \cite{corrado}. A satisfactory agreement between theory and 
simulation can be observed already with the simplest isocomplexity computation.  
\label{fig:iso}}
\end{center}
\end{figure}

To perform the derivation of glassy EOS, one just needs to compute the complexity and in-state free energy using equations 
\eqref{eq:mdepf},\eqref{eq:mdepcompl} (we recall that the prescription is $\Sc = 
-\beta\Phi$) and then implement the isocomplexity approximation by solving the 
\eqref{eq:isocompl}; more details are given in the SI of \cite{corrado}.\\
In figure \ref{fig:iso} we report the results obtained; as in figure 
\ref{fig:dinfstatefo}, we plot the inverse pressure $1/p$ versus the packing 
fraction $\f$ for various planting densities $\f_0$. Notice that the quantities 
are not rescaled anymore as in chapters \ref{chap:SFRS} and \ref{chap:SFfRSB}, 
as now we are in $d=3$.\\
Overall, one can see that a good agreement with simulation is already visible, 
even with isocomplexity approximation. Despite the fact that it is a lot less 
flexible and rigorous than the SF formalism, the theory is anyway satisfactory 
enough, at least at the level of thermodynamic observables.\\
Since the isocomplexity approximation (and the real replica method in general) 
are a lot less mathematically convoluted than the SF formalism, it is easy to 
study the singular behavior of the cage radius at the onset point whereupon the 
glass state melts back into the liquid. One gets:
\beq
A(\f) = A_{max} - C\sqrt{\f-\f_{on}},
\eeq 
where the coefficient $C$ depends on $m$ and can be computed easily (see the SI 
of \cite{corrado}).

\subsection{State Following and the Gardner transition}
As mentioned before, the MK model for $d \to \infty$ is quantitatively and 
qualitatively identical to mean-field HSs. In finite dimension, some corrections 
have to be taken into account with respect to the $\infty-d$ result.
The most important finite-dimensional effect in the MK model lies in the fact 
that caging is not perfect in the glass phase, and particles are effectively 
free to move in a network of cages as well studied in \cite{CharbonneauJin14}. 
This effect washes away the MCT singularity at $\f_{MCT}$, but since hopping is 
exponentially suppressed in $\f$, it produces relevant effects only for $\f_0\simeq 
\f_{MCT}$. Once the effect of hopping has been removed by focusing on higher 
$\f_0$s, the corrections with respect to $\infty-d$ are effectively only 
quantitative in nature. So, for the sake of comparison, we can just take the SF 
results exposed in the preceding chapters, paying attention to rescaling the 
quantities in the correct way (for example $\f = \frac{d}{2^d}\wh\f$).

\begin{figure}[htb]
\begin{center}
\includegraphics[width=0.75\textwidth]{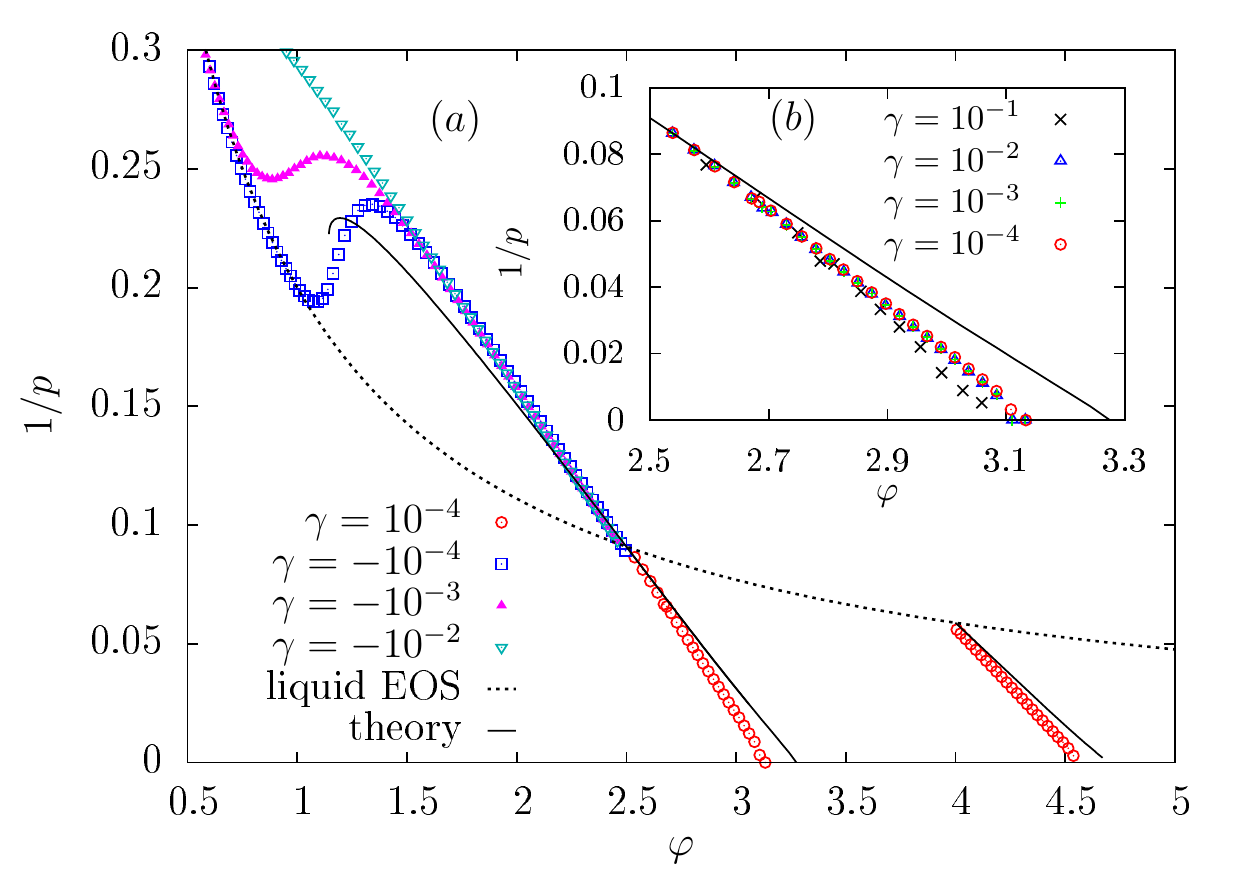}
\caption{Equations of state of a glass planted art $\f_0 = 2.5$, ($\wh\f_g = 
6.667$ in $\infty-d$ units), and $\f_0=4$. Hopping effects are negligible in 
both cases. The lines are theory and the dots are simulation data as described 
in \cite{CharbonneauJin15}. (\emph{Inset}) though a larger compression rate 
smooths away the singular behavior at the onset point, its effects on the 
pressure in a compression protocol are effectively negligible as long as 
$\f<\f_G$. \label{fig:2.5}}
\end{center}
\end{figure}

The study of \cite{CharbonneauJin15} is performed with the same basic numerical setup 
as the one of \cite{corrado}, though with the aim of verifying the prediction of 
the presence of a Gardner transition in compression; comparison with the SF results reported 
in the preceding chapters is also reported (with the proper rescaling as 
discussed above). A first check of the SF procedure is reported in figure 
\ref{fig:2.5}, where the effect of the compression rate on the measured glass 
EOS is also considered. Its effect on the pressure in a compression protocol is 
effectively negligible as one can see in the inset of figure \ref{fig:2.5}. This 
is to be expected as the relaxation time within the state $\t_\be$ is expected 
to be very small as long as $\f<\f_G$.

\subsubsection{Dynamics}
A non-trivial and rich aging phenomenology manifests in the fRSB phase \cite{Ogielski85}. To better explore the complex free-energy landscape structure
that is associated with the fRSB ansatz, it is convenient to define two instantaneous quantities. The first is the MSD between configurations at different times
\beq
\hat\D(t,t_w) \equiv \frac{1}{N}\sum_{i=1}^N |\br_i(t+t_w)-\br_i(t_w)|^2,
\eeq
where the waiting time is the time elapsed since the end of the compression protocol, as in section \ref{sec:aging}. The second is the MSD between two different ``cloned'' configurations $A$ and $B$
\beq
\hat\D_{AB}(t) \equiv \frac{1}{N}\sum_{i=1}^N |\br_i^A(t) - \br_i^B(t)|^2.
\eeq
The ``cloning'' procedure works as follows: before the compression starts, an equilibrated planted configuration is duplicated ($\{\br_A(0)\} = \{\br_B(0)\}$), but each of the two instances is given a different set of initial velocities, randomly drawn from the Maxwell distribution. The two configurations are then compressed independently. This way, A and B will evolve inside the same glass metabasin (which as we recall from section \ref{sec:FP} is selected by a single equilibrated configuration) but will have independent compression histories. They will thus serve as a probe of the metabasin structure, in the same fashion as the replicas discussed in section \ref{subsec:fRSBpar}.\\ 
These instantaneous quantities are to be averaged over the statistical ensemble of samples:
\beq
\D(t,t_w) \equiv \dav{\thav{\hat\D(t,t_w)}} \qquad \D_{AB}(t) \equiv \dav{\thav{\hat\D_{AB}(t)}}.
\eeq
The number of samples varies from $N_s = 500$ to $N_s = 15000$ depending on the statistical convergence properties of the observable under consideration. It is also convenient to define a quantity
\beq
\d\D(t,t_w) \equiv \D_{AB}(t+t_w) - \D(t,t_w).
\eeq

\begin{figure}[htb]
\begin{center}
\includegraphics[width=0.5\textwidth]{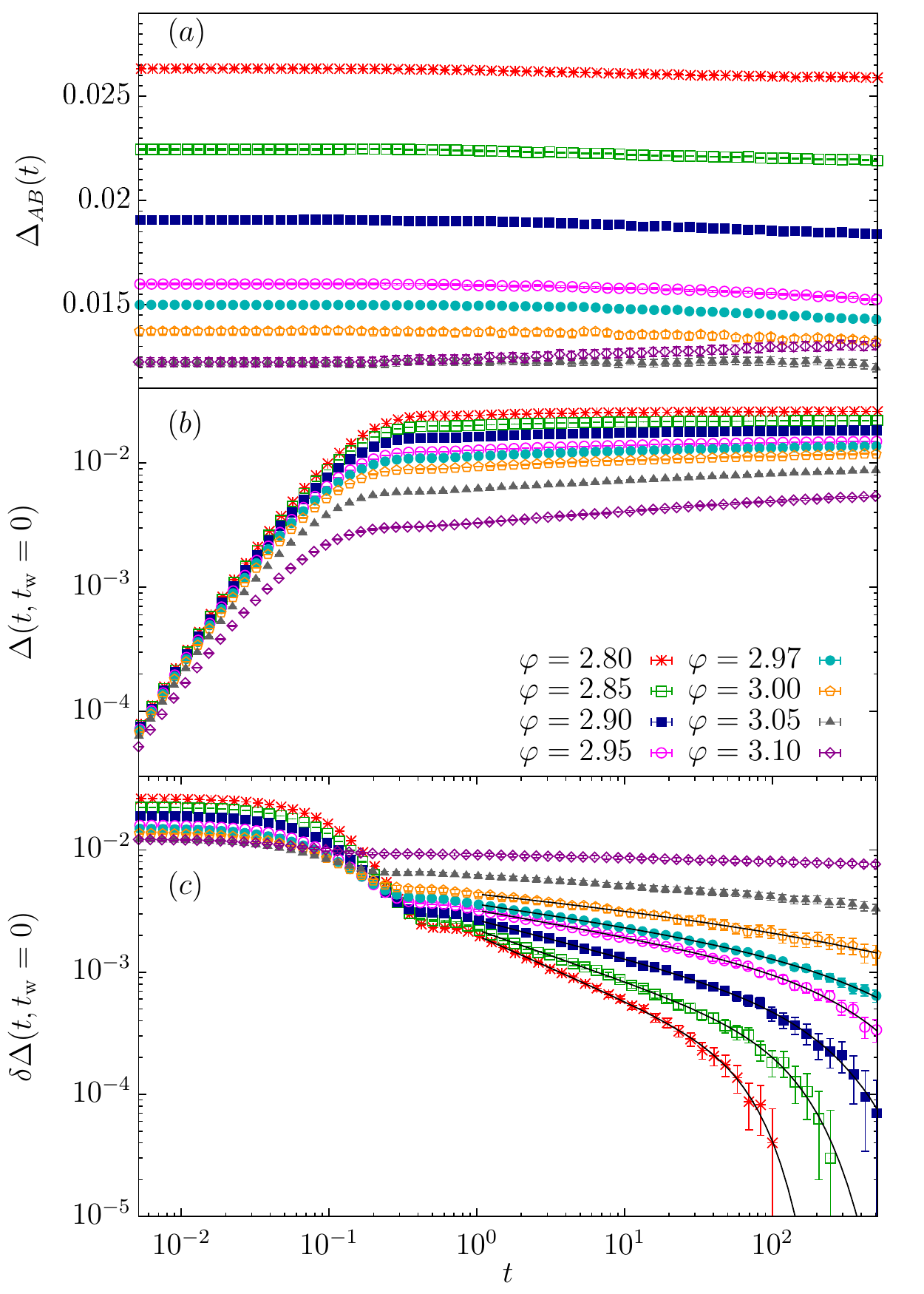}\nolinebreak[4]
\includegraphics[width=0.5\textwidth]{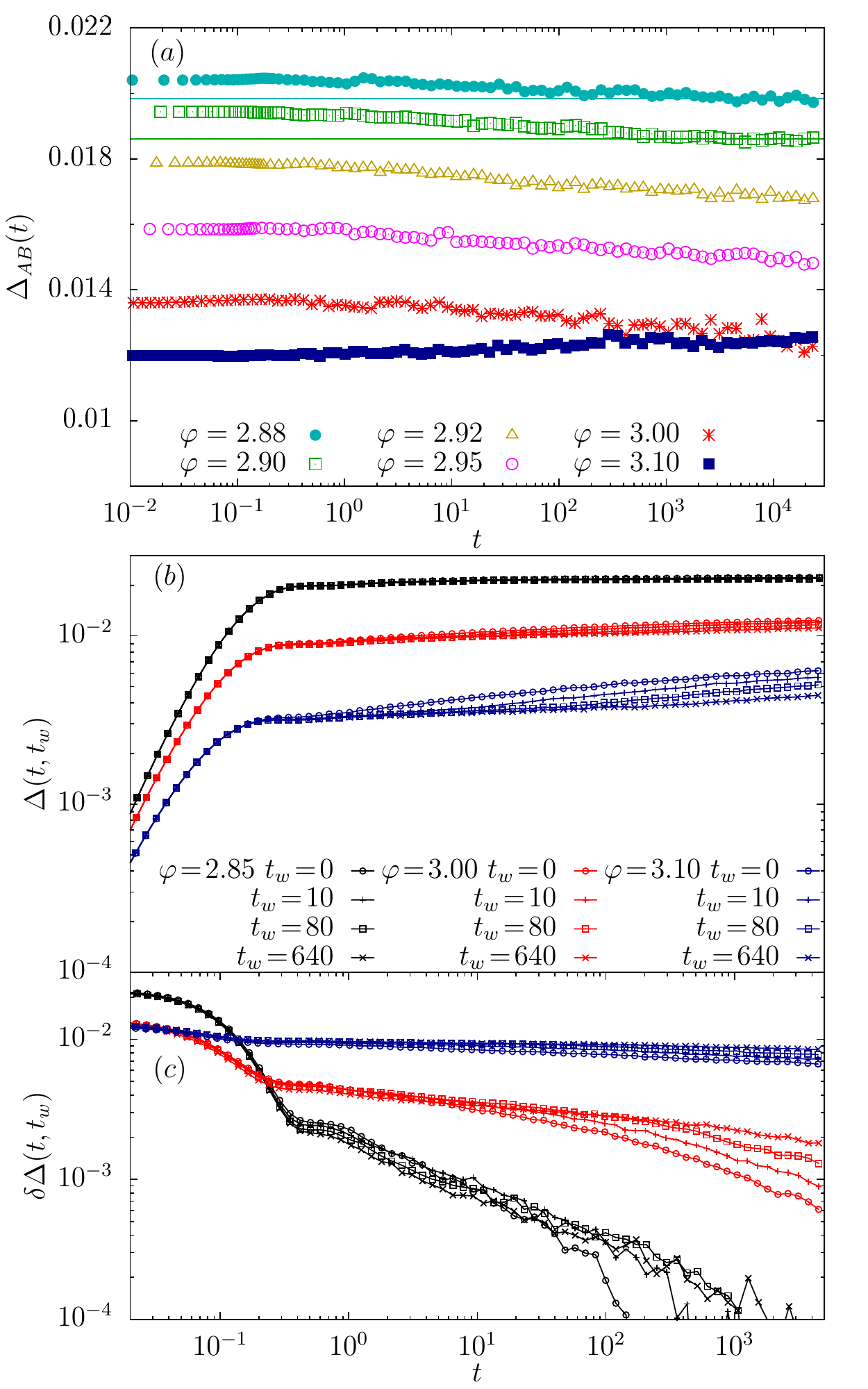}
\caption{Dynamics in a glassy state planted at $\wh \f_g = 6.667$, for densities in the vicinity of the Gardner transition, for $t_w = 0$ (\emph{left panel}) and $t_w\neq 0$ and large times (\emph{right panel}). When $\varphi < \varphi_G$ (here $\varphi_G \simeq 3.0$ as reported in the following), the MSDs $\D(t,t_w)$ and $\D_{AB}(t)$ rapidly attain their equilibrium value ($\D_{AB}(t)$ is effectively independent of time), $\d\D(t,t_w)$ decays to zero at long times and no dependence on $t_w$ is observed: the system is in restricted equilibrium (solid lines are fits to equation \eqref{eq:dynamicsfit}). When $\varphi \gtrsim \varphi_G$, the MSDs drift for all observable times without ever attaining a plateau value, and $\d\D(t,t_w)$ remains positive; a marked dependence on $t_w$ is observed, especially for high density: the system is continuously aging inside the glassy metabasin, signature of a Gardner phase. Notice how the relaxation time $\t_{meta}(t_w)$ grows with $t_w$, another typical signature of aging (fig. 
\ref{fig:aging}). \label{fig:DynamicsMK}}
\end{center}
\end{figure}

If the system is in restricted equilibrium (and so $t_w$ is large enough) in the RS phase above the Gardner transition, we have basically the same caging picture of figure \ref{fig:msd}
\beq
\lim_{t\to \infty} \D(t) = \D,
\eeq
where $\D$ is the Debye-Waller factor computed in chapter \ref{chap:SFRS}. The only difference is that in this case we are looking at the MSD inside the glass (as opposed to the supercooled liquid), and so we are unable to see the diffusive regime as in figure \ref{fig:msd}. Accordingly, since the system is able to explore ergodically the glass basin, one will have
\beq
\lim_{t\to\infty} \D_{AB}(t) = \D,
\eeq
and as a result of this
\beq
\lim_{t\to\infty}\lim_{t_w \to \infty} \d\D(t,t_w) = 0.
\eeq
The picture changes in the fRSB phase $\varphi \gtrsim \varphi_G$, wherein aging manifests and equilibrium within the metabasin is out of computational reach (the time $\tau_{meta}$ needed to achieve restricted equilibration is expected to scale as $\exp(N^{1/3})$ \cite{AspelmeierBlythe06}, well beyond the length of simulations considered in \cite{CharbonneauJin15}). In this case, the system will initially equilibrate in one microstate at the bottom of the fRSB hierarchy, so the MSD $\D(t,t_w)$ will reach a short plateau, $\D(t,t_w=0) = \D_{EA}$ for $t \ll \tau_{meta}(t_w)$ and $t>\t_0$, where $\t_0$ is the typical timescale for the ballistic regime. Thereafter, the system will start to explore the fRSB hierarchy, visiting higher and higher levels of it and hopping over larger and larger free energy barriers. Since the hierarchy of sub-states is fractal ($k=\infty$), the system will continuously surmount barriers until equilibration is reached (\emph{continuous aging}) instead of attaining it in a simple step-like manner as it would with a finite number of RSBs (section \ref{subsec:twostep}). As a consequence of this, $\D(t,t_w)$ will drift upwards for all observable times after leaving the initial plateau corresponding to $\D_{EA}$.\\
For what concerns $\D_{AB}(t)$, the two copies have different compression histories, as we already mentioned; this means that they will usually end up in different microstates, and so the long-time limit of $\D_{AB}(t)$ will correspond to the average distance $\thav{\D}$ between microstates. This means that in the Gardner phase
\beq
\lim_{t\to\infty}\lim_{t_w \to \infty} \d\D(t,t_w) \neq 0,
\eeq
which provides a dynamical order parameter for the Gardner transition. Notice that in general $\D_{AB}(t+t_w) > \D(t,t_w)$ for every $t$. However, since equilibration is never reached $\D_{AB}(t)$ as well slowly drifts (the drift can be observed for every simulation time) without ever reaching a plateau. In figure \ref{fig:DynamicsMK} we report the results obtained, which follow the basic phenomenology described above.   

\begin{figure}[htb]
\begin{center}
\includegraphics[width=0.75\textwidth]{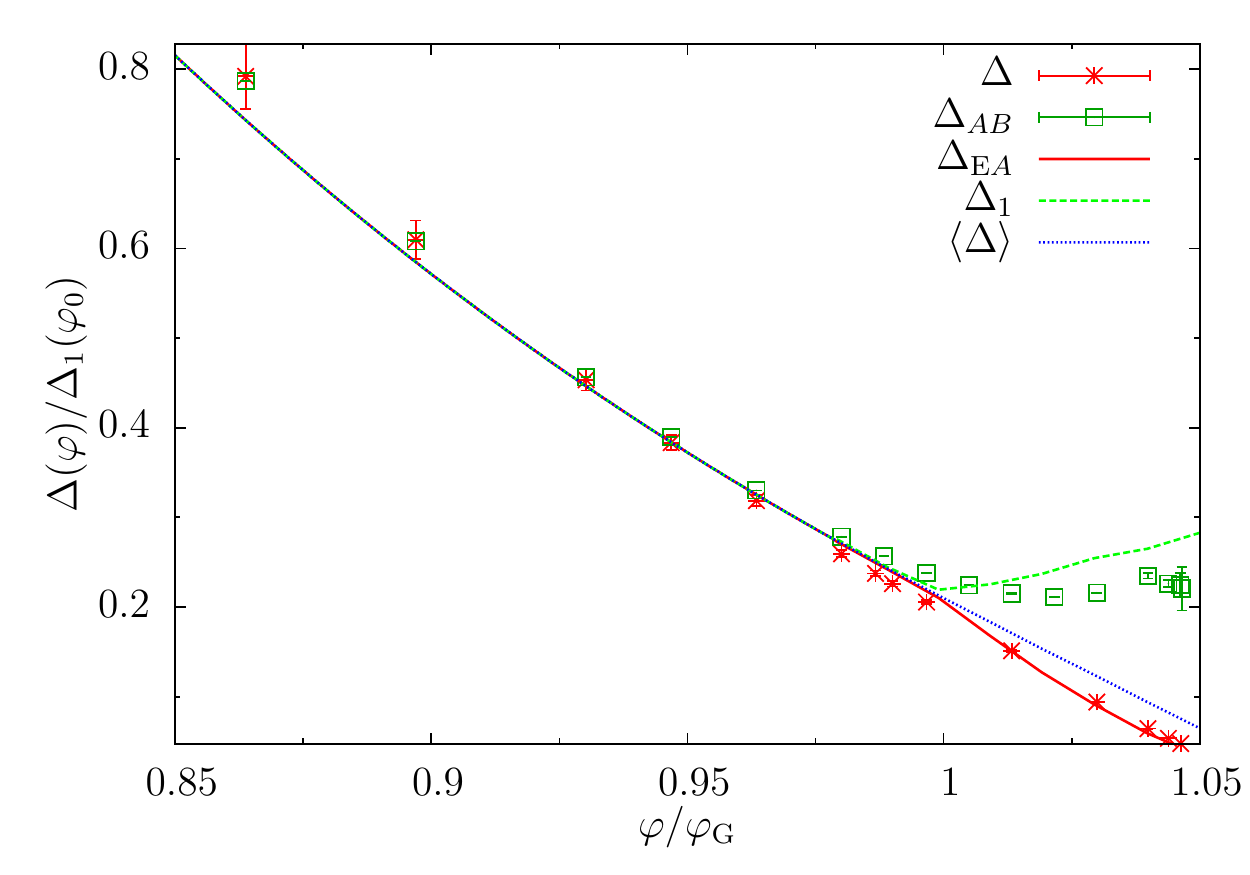}
\caption{MSDs for the glass around the Gardner point for $\wh \f_g = 6.667$. The 
lines are theory and the dots are simulation data as described in 
\cite{CharbonneauJin15}. $\D_1 \equiv \D(0)$ in the notation of 
\cite{CharbonneauJin15}. Both the theoretical and the numerical datasets have 
been rescaled with their value $\D(\f_0)$ on the equilibrium line. 
\label{fig:MSDfullnum}}
\end{center}
\end{figure}

From this picture, it transpires that if one chooses $t_s$ such that $\t_0<t_s\ll \t_{meta}$, then $\D(t_s,t_w=0)$ can be used as an estimator for $\D_{EA}$ and $\D_{AB}(t_s)$ for $\D_1$, and then one can compare the so obtained results with the theoretical prediction for the bifurcation of MSDs presented in figure \ref{fig:MSDfull}. In particular, one 
can verify if $\D(0)$ increases after $\f_G$ as predicted. In figure 
\ref{fig:MSDfullnum} we report the results so obtained, with again a 
satisfactory agreement between theory and simulation.\\
Besides this, it certainly would be appropriate to get also a check of the 
behavior in shear, figure \ref{fig:MSDfullshear}, and in particular the 
decreasing of $\D_{EA}$ when $\g>\g_G$. We leave the issue for future work.

\subsubsection{Locating the Gardner point}
Apart from the check of the equations of state and dynamics of MSDs, the numerics of 
\cite{CharbonneauJin15} are also able to robustly locate the Gardner points for 
various planting densities $\f_0$ and to verify that their position on the glass 
EOSs is compatible with the theoretical prediction in figure 
\ref{fig:dinfstatefo}. Three different methods are used, here we just give the 
basic picture and refer to \cite{CharbonneauJin15} for details.

\begin{figure}[htb]
\begin{center}
\includegraphics[width=0.7\textwidth]{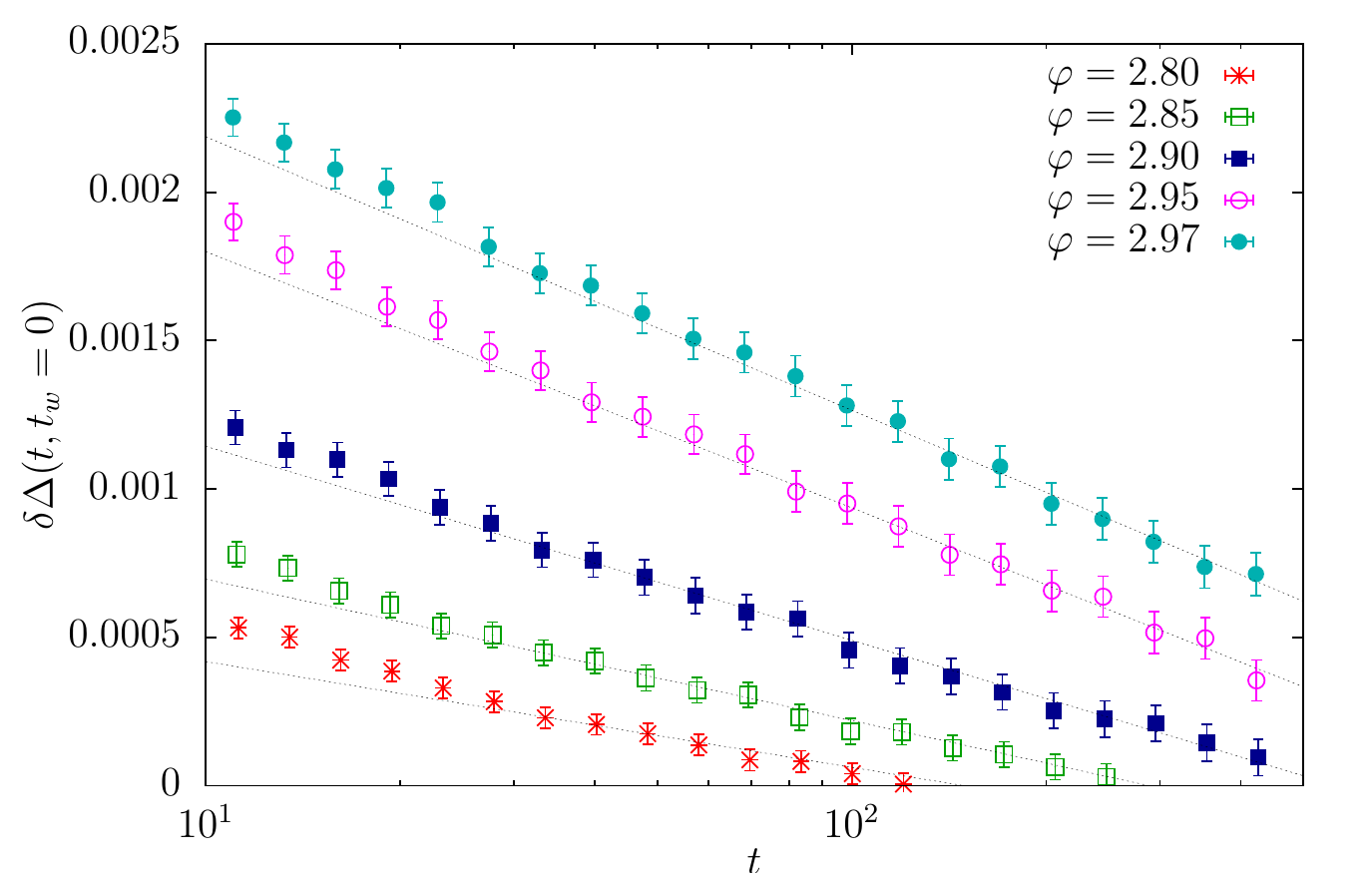}
\caption{Logarithmic decay of $\d\D(t,t_w=0)$ for $\f_0=2.5$. The data are fitted to equation \eqref{eq:logdecay}. \label{fig:logdecay}}
\end{center}
\end{figure}

The first focuses on dynamics above the Gardner point, in particular the 
power-law divergence of the relaxation time due to critical slowing down, 
paragraph \ref{subsec:lambdaMCT}.
Besides the qualitative picture of the dynamics exposed in the preceding paragraph, one can also try a more quantitative analysis following the lines of the classic work \cite{Ogielski85} of Ogielski on spin glasses and the theoretical scheme for glassy dynamics developed in \cite{ParisiRizzo13}. The basic idea is to extract the relaxation timescale $\tau_\be$ from the decay of the $\d\D(t,t_w)$ on approaching the Gardner transition from above. Since this timescale diverges as a power law in the distance $\d\varphi \equiv |\varphi-\varphi_G|$ from the transition as a consequence of critical slowing down, it can be used to locate the Gardner point $\varphi_G$ as shown in the following. Notice that this scheme is then well defined only in the RS phase.
\begin{figure}[t!]
\begin{center}
\includegraphics[width=0.7\textwidth]{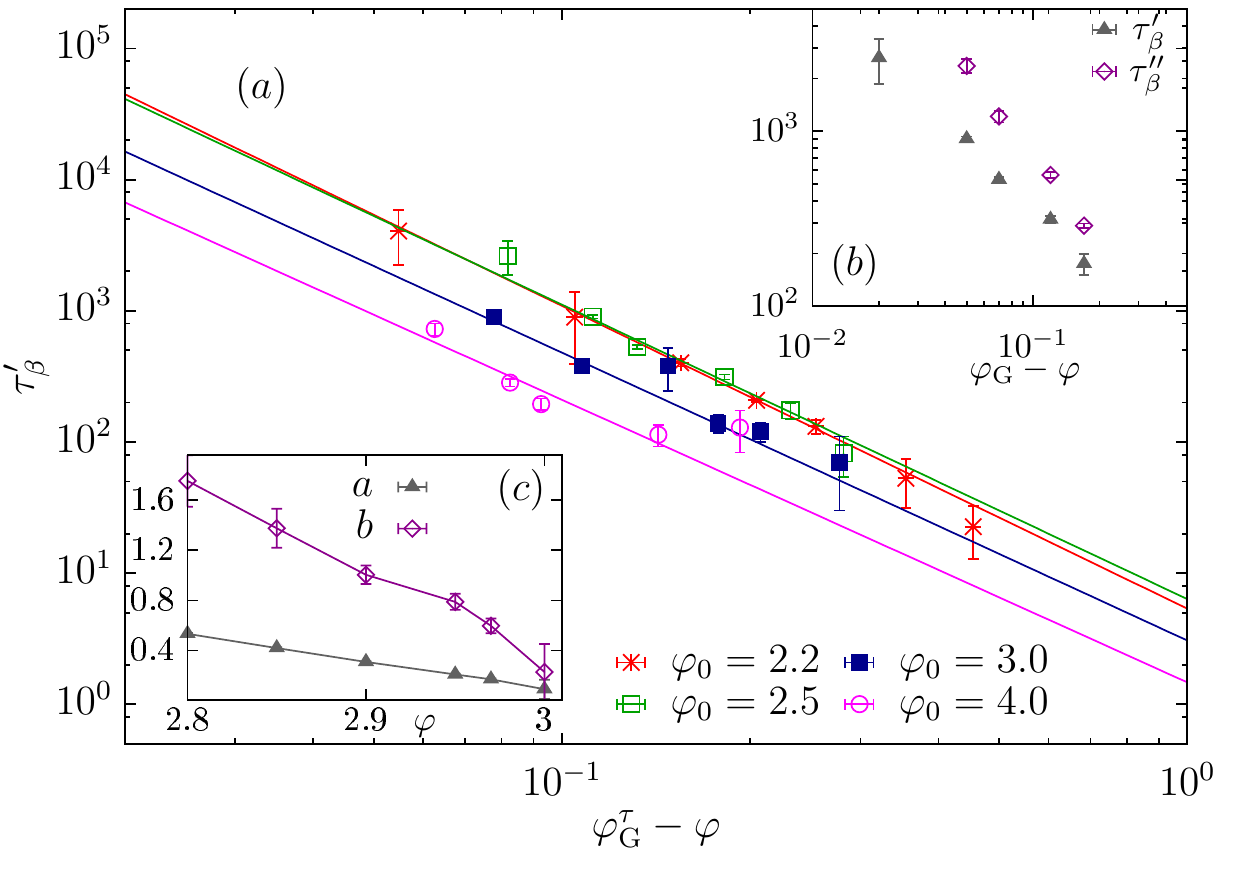}
\caption{(a) Growth of $\tau_\be'$ with $\d\f$ for various planting densities $\f_0$. (b) The two estimates $\tau_\be'$ e $\tau_\be''$ as a function of $\d\f$ for $\f_0=2.5$; both exhibit the same power-law scaling. (c) Evolution of the fitted $a$ and $b$ exponents with $\f$, again for $\f_0 = 2.5$. \label{fig:taubetafit}}
\end{center}
\end{figure}
According to the theory developed in \cite{ParisiRizzo13}, upon approaching the Gardner point from above while in restricted equilibrium in the RS phase, one must have
\beq
\d\D(t) \simeq \d\varphi\mathcal{F}(t/\tau_\beta),\qquad \t_\be \simeq \d\varphi^{-\g},
\eeq
where $\mathcal{F}(x)$ is a function such that $\mathcal{F}(x\ll 1) \simeq x^{-a}$ while $\mathcal{F}(x\gg 1)$ decays exponentially, and the exponents $a$ and $\gamma = 1/a$ are those defined in paragraph \ref{subsec:lambdaMCT}; notice how $\d\D(t)$ does not depend on $t_w$ here, as we are in the RS phase where aging is not present.\\
We can estimate $\t_\be$ by choosing the empirical form \cite{Ogielski85} $\mathcal{F}(x) \propto x^{-a}e^{-x^b}$, which means fitting $\d\D(t,t_w=0)$ to the form
\beq
\d\D(t,t_w=0) = c\frac{\exp[-(t/\t'_\be)^b]}{t^a},
\label{eq:dynamicsfit}
\eeq
where all parameters will depend on $\varphi$ and $\varphi_0 \equiv d/2^d\wh\f_g$, and $\t'_\be$ offers a first estimate of $\t_\be$. We fit the exponent $a$ instead of fixing its value to the analytical prediction reported in table \ref{table:lambdaMCT}, since the value of $a$ away from the transition is quite different from the critical value, as one can appreciate by looking at the linear part of the plots in the lower left panel of figure \ref{fig:DynamicsMK}.\\
The relaxation time $\t_\be$ is then expected to scale as
$$
\tau_\be \propto |\varphi-\varphi_G^\tau|^{-\gamma},
$$
which gives a first estimate of $\varphi_G$. This time, we fix the value of $\g$ to the analytic prediction and we fit $\varphi_G^\t$ to get a first estimate of the transition point (see figure \ref{fig:taubetafit} for the results). To get a better check on our result, we also estimate $\tau_\beta$ through the logarithmic decay of the $\d\D(t)$ at long times (see figure \ref{fig:logdecay}):
\beq
\d\D(t,t_w=0) = k\left[1-\frac{\ln(t)}{\ln(\t_\be'')}\right],
\label{eq:logdecay}
\eeq
where in this case we fit $\tau_\be''$ and $k$. From inset (b) in figure \ref{fig:taubetafit}, one can see how the growth of both estimators $\tau_\beta'$ and $\tau_\beta''$ is compatible with the same power-law scaling, which supports our claim that the slowing down is a manifestation of an underlying second-order transition.

\begin{figure}[ht!]
\begin{center}
\includegraphics[width=0.6\textwidth]{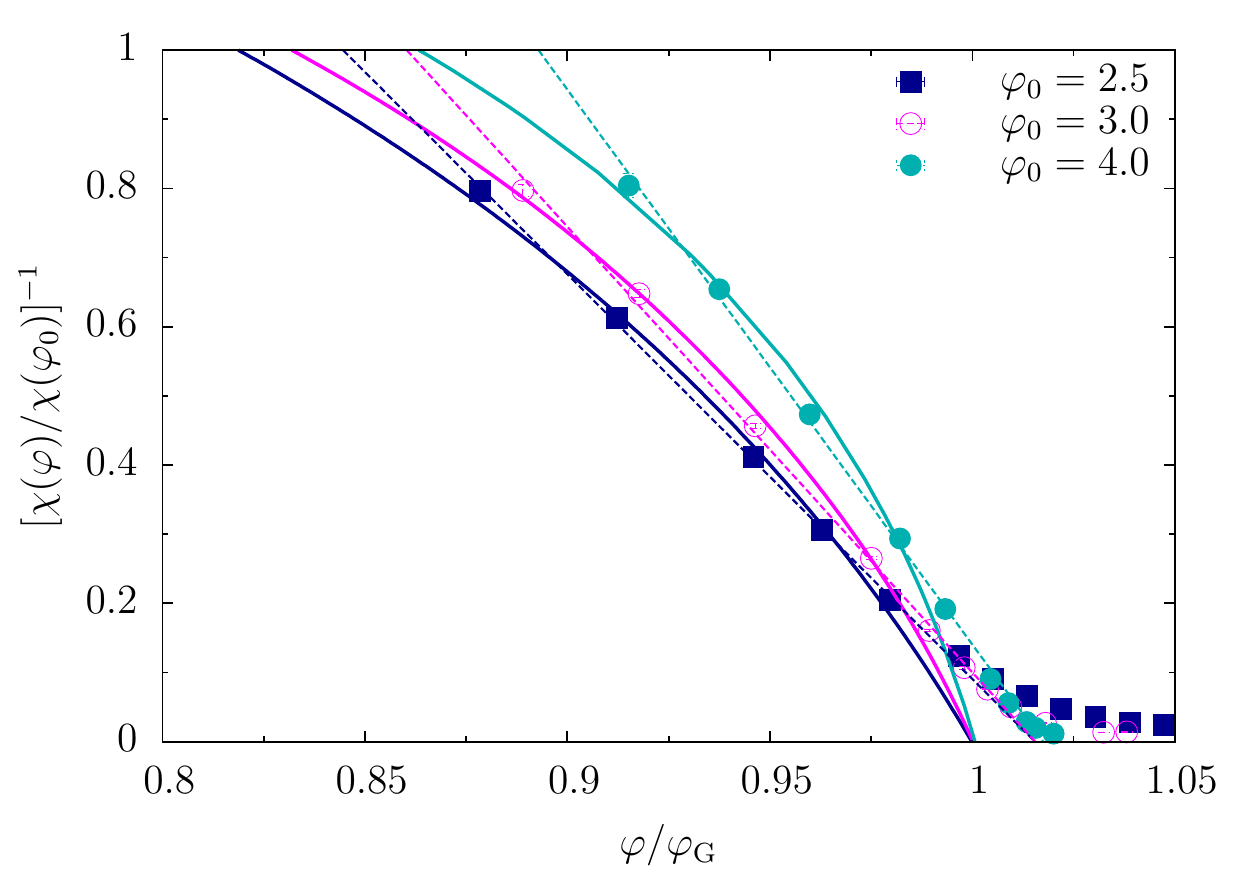}
\caption{The critical susceptibility $\chi_4$ near the Gardner point for three 
different planting densities. Lines are the theoretical prediction while the 
symbols are numerical data. $\f_0 \equiv d/2^d\wh\f_g$ in the notation of 
\cite{CharbonneauJin15}. \label{fig:chi4}}
\end{center}
\end{figure}

The second method focuses on static properties below the Gardner point. Let us suppose 
to prepare two initial configurations for the simulation, each of them 
identified by a $2Nd$-dimensional position-velocity couple 
$(\br,\boldsymbol{v})$. The configuration of the $\br$ is obtained though the 
planting method and is the same for both, while the velocities are extracted at 
random from the Maxwell distribution. This procedure defines two \emph{clones} 
of the same initial configuration. One can then define a MSD between clones
\beq
\hat\D_{AB}(t) = \frac{1}{N}\sum_{i=1}^N\left|\br_i^A(t) - \br_i^B(t) \right|^2.
\eeq
Since in the fRSB phase the glass state breaks up in a fractal hierarchy of 
sub-states, usually the two replicas will not end up in the same state. One can 
measure the long-time limit of $\D_{AB}(t)$ and construct an histogram 
$P(\D_{AB})$ which will describe the structure of the glass metabasin. Proven 
that finite-size effects are accounted for \cite{CharbonneauJin15}, the two 
replicas sample the Parisi probability distribution $P(\D)$ \cite{parisiSG}. One 
can than prove that its variance $\chi_{AB}$ obeys
\beq
\chi_{AB} \equiv N \frac{\dav{\thav{\D_{AB}^2}} - 
\dav{\thav{\D_{AB}}}^2}{\dav{\thav{\D_{AB}^2}}^2} = \frac{ 
\langle\D_{ab}^2\rangle- \thav{\D_{ab}}^2}{\thav{\D_{ab}}^2} = 
\frac{\chi_4}{\D^2}.
\eeq
We recall that the divergent part of the $\chi_4$ susceptibility is the 
$\chi^{div}_4 = \frac{24}{\l^R}$ computed in paragraph 
\ref{subsec:fluctuations}. Thus, the numerical $\chi_{AB}$ is supposed to 
diverge as $\chi_{AB} \simeq (\f_G-\f_G^\chi)^{-1}$ since the replicon vanishes 
linearly at the Gardner point (see figure \ref{fig:rep_compr}). This provides 
another estimator for $\f_G$, and also a way to measure the critical $\chi_4$ 
susceptibility (see figure \ref{fig:chi4}).

\begin{figure}[hb!]
\begin{center}
\includegraphics[width=0.75\textwidth]{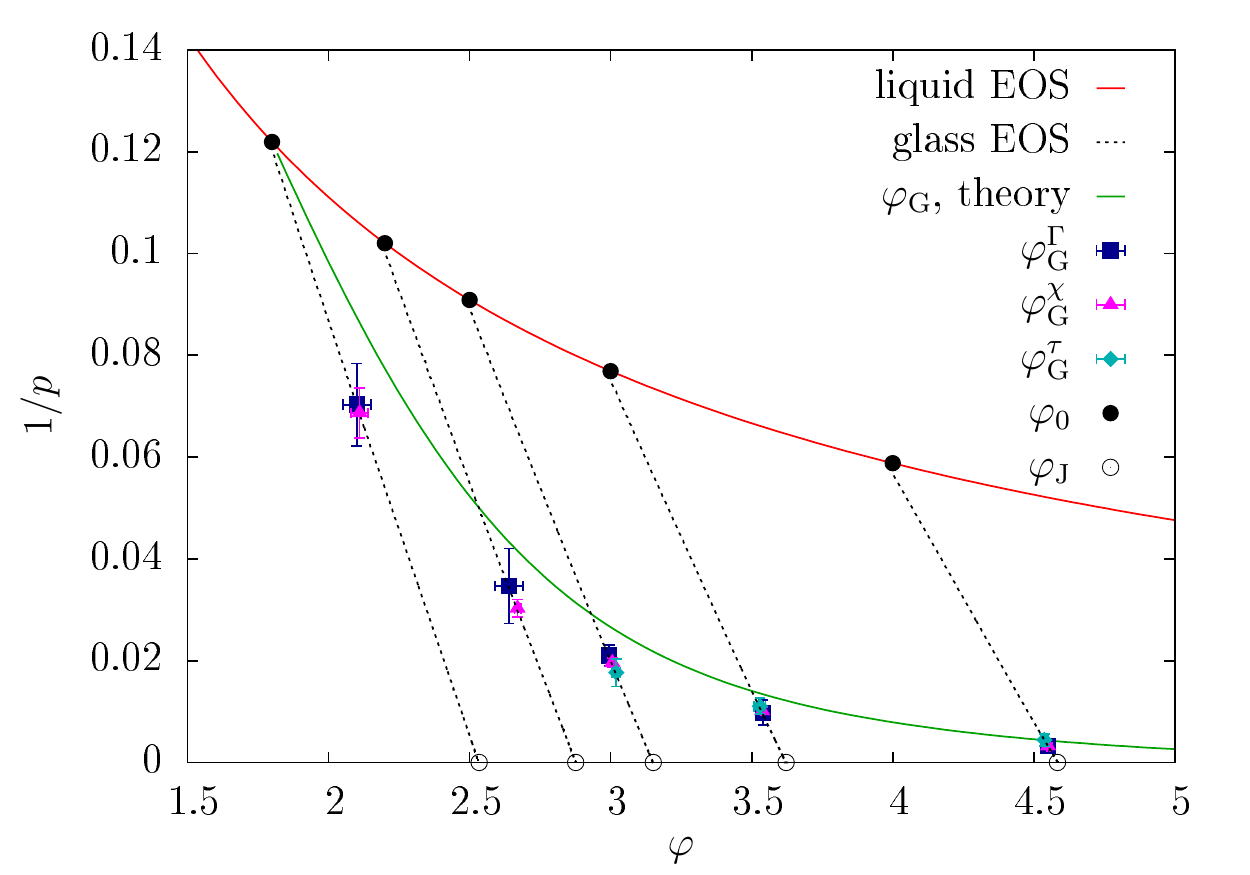}
\caption{The Gardner transition points located with the three different 
estimators $\f_G^\t$, $\f_G^\chi$ and $\f_G^\Gamma$. The numerical estimate is 
robust and consistent with the theoretical prediction reported in chapter 
\ref{chap:SFRS}. \label{fig:gardnernumerics}}
\end{center}
\end{figure}
The third method uses again the $P(\D_{AB})$, but focuses on another parameter, 
its \emph{skewness} $\Gamma$ whose definition is 
\beq
\Gamma \equiv \frac{\dav{\thav{(\D_{AB} - 
\dav{\thav{\D_{AB}}})^3}}}{\dav{\thav{(\D_{AB} - 
\dav{\thav{\D_{AB}}})^2}}^{3/2}},
\eeq
this quantity is supposed to be maximal at the Gardner point, so one can get 
another estimate $\f^\Gamma_G$.

In figure \ref{fig:gardnernumerics} we finally report the results for the Gardner point 
obtained with these three estimates. The reader can appreciate how all three 
estimates give compatible values for different planting densities $\f_0$, and 
how those are compatible with the Gardner line reported in figure 
\ref{fig:dinfstatefo}. Only the estimate for the $\f_0 = \f_{MCT}$ is off 
(although anyway very robust), which can be attributed to the hopping effects 
which show up prominently near $\f_{MCT}$ and may spoil the numerical result.

In summary, the prediction of a Gardner transition in HS under compression 
\cite{nature,CKPUZ13,corrado2} is validated at least for a MF glass former, in 
this case the infinite range MK model. Work is ongoing in this moment to prove 
its existence also for ordinary HSs in finite $d$, and a study in the same 
spirit for HS under shear will be the logical continuation of the research 
effort started in \cite{corrado,CharbonneauJin15}.

\chapter{Conclusions \label{chap:conclusions}}

In this final chapter, we summarize our main results and predictions and give an 
assessment of the strengths and weaknesses of the state following construction, 
and RFOT as a whole. Finally, we conclude the thesis with some perspectives and 
proposals on further research inspired by the present work.

\section{Summary of main predictions}

Our state following approach is able to provide many interesting predictions 
which fit well within the phenomenology of glasses explored in the first two 
chapters. These include:
\begin{enumerate}
 \item The basic phenomenology of the calorimetric glass transition (paragraph 
\ref{subsec:glasstrans}), with a jump in heat capacity (or equivalently, 
compressibility), is captured by out approach.
 \item We are able to compute from first principles the equation of state of a 
generic glass prepared with a generic annealing protocol (paragraph 
\ref{sec:aging}), reproducing the basic phenomenology of glasses as observed in 
DSC experiments (paragraph \ref{subsec:ultrastable}), in particular those on 
ultrastable glasses \cite{Sw07,SEP13,SEP13corr,LEDP13,SepulvedaTylinski14}. We 
are able to observe hysteresis in agreement with the results of 
\cite{Sw07,SEP13,Dy06}, and the onset transition whereupon the glass melts back 
into the liquid at high temperature is well reproduced.
 \item The Gardner transition detected in \cite{nature,CKPUZ13} is recovered 
within the state following approach, along with the results concerning the 
isostaticity and marginality of jammed packings and the critical exponents of 
the jamming transition (paragraph \ref{subsec:jamming}). We are also able to 
compute the jamming density $\varphi_j$ for generic disordered packing 
constructed with an annealing-like (i.e. reasonably slow) procedure 
\cite{SDST06,CBS09}. 
 \item We are able to compute from first principles stress-strain curves of the 
glass and obtain again a phenomenology in agreement with simulations and 
experiments (paragraph \ref{sec:shear}); our approach reproduces the presence of 
a stress overshoot \cite{RTV11,Wang09,Laurati11,FalkLanger04} and a yielding 
point (paragraph \ref{subsec:yielding}) whereupon the glassy state dies and the 
glass former starts to flow \cite{RTV11,Kou12}. We are able to compute the 
Debye-Waller factor, the shear modulus \cite{Cates03} and the dilatancy 
\cite{Reynolds85,Tighe14} everywhere in the glass phase.
\end{enumerate}
These results indicate that the state following approach is at least a good 
starting point for a first principles theory of glasses prepared via slow 
coolings.

In addition to this, we can also provide some novel predictions which make for a 
good test bed of the theory and may allow one to validate/falsify it through 
simulations and experiments:
\begin{enumerate}
\item We detect the presence of a Gardner transition for high enough values 
$\g_G$ of the strain perturbation (section \ref{subsec:shearRS}), for 
\emph{every} preparation density $\wh\f_g$. From the practical point of view, 
this means that the response of a glass to the strain is elastic and solid-like 
only up to $\g_G$. Afterwards, the fRSB structure manifests within the state and 
the system will start to jump from a microstate to the other in a 
non-equilibrium fashion, producing a rough response with avalanches like the one 
that can be seen in figure \ref{fig:shearstrain} for athermal materials. In 
summary, we argue that an avalanche-dominated regime with energy dissipation is 
\emph{bound to appear}, for high enough strains, also in thermal systems, no 
matter which their preparation temperature is. In this regime, as one can see 
from figure \ref{fig:MSDfullshear} the Debye-Waller factor $\D_{EA}$ decreases 
and consequently the $\mu(1)$ shear modulus (which is the slope of the small 
elastic segments between avalanches) increases. All of these are novel 
predictions that should be easily verifiable.
\item We predict the insurgence, in the Gardner phase, of a set $\mu(x)$ of 
shear moduli (paragraph \ref{subsec:shearmodulifull}), with a scaling relation 
$\mu(0) - \mu(1) \propto (\wh\f - \wh\f_G)$. It can be argued that the $\mu(1)$ 
modulus corresponds to the average slope of a stress strain curve, while the 
$\mu(0)$ corresponds to the slope of small elastic segments like in figure 
\ref{fig:shearstrainlemaitre}. This is true both for athermal systems (which are 
already in the Gardner phase when shear is applied) and thermal ones beyond the 
Gardner point induced by the strain itself.
\item We compute critical slowing down exponents and critical fluctuations 
around the Gardner Point for various preparation densities $\wh\f_g$. A first 
encouraging test of these results can be found in \cite{CharbonneauJin15}, for 
the MK model discussed in the preceding chapter. 
\end{enumerate}

\section{Strengths and weaknesses of our approach}

The state following theory presented in this thesis has some very appealing 
traits. First of all, it allows one to perform theoretical physics calculations 
on glasses using only tools of standard statistical mechanics, like partition 
functions, free energies, large deviations and saddle-point methods, etc., and 
only requires, as an input, a microscopic interaction potential. In addition to 
this, it is also completely static in nature, which is a very welcome feature 
considering that the established theory for the first principle dynamics of 
glass formers has been, up to very recently, the Mode Coupling theory, whose 
flaws and weaknesses (especially in the low-temperature/high density regime) are 
well known. There is no need to employ dynamics, or phenomenological arguments, 
or parameter tunings of any sort. The state following construction brings 
everything back to the definition of a suitable Gibbs measure for the 
statistical-mechanical study of metastable, out of equilibrium glassy states. In 
summary, it cleverly exploits the basic picture of RFOT to bring back the 
problem of describing out of equilibrium glass to an equilibrium formalism. And 
despite these aspects of theoretical ``cleanliness'', it still provides 
predictions in remarkable agreement with the phenomenology of glasses.

However, it also has some weaknesses, both technical and conceptual. On a 
conceptual level, the SF construction always assumes restricted equilibration of 
the glass former within a glassy state. As a result of this, is only able to 
provide predictions in regimes where such equilibration is effectively attained, 
i.e. when perturbations are \emph{adiabatically} applied to the glass.\\
However, the experimental literature on glasses (see 
\cite{Dy06,berthierbiroli11} and references therein) is teeming with experiments 
and protocols which do not meet this requirement. This is particularly evident 
in the case of driven dynamics, wherein time dependent shear protocols are 
pretty much the norm. A steady shear rate $\dot\g = const$ must be employed for 
the determination of the flow curve, and oscillatory shear protocols are used 
for the determination of the storage and loss moduli $G'(\omega)$ and 
$G''(\omega)$ \cite{Cates03,Larson99}; in both these cases, the perturbation 
changes with time and is not applied adiabatically.\\
Also in the case of simple aging, one could for example consider applying AC 
currents to measure dielectric relaxation spectra as done for example in 
\cite{LehenyNagel98,BellonCiliberto00}. All such situations cannot be described 
within the state following setting.\\
The main technical difficulty lies in the fact that the state following approach 
is quite heavy from the computational point of view (the size of the appendix is 
probably a giveaway on this point), already at two-replica level; performing 
calculations with the replica chain, for example, looks feasible only for the 
simplest spin glass models like the PSM. This is part of the reason why we had 
to focus on the mean-field limit $d \to \infty$; just by looking at the 
expression \eqref{eq:FP} of the Franz-Parisi potential one can understand how 
going beyond the MF calculation is a hard and perhaps impossible task; some 
proposals on how to do so are formulated in \cite{parisizamponi,corrado2}, but 
there is no systematic perturbation scheme and it is very difficult to 
understand how big of an error one is committing in considering these 
approximations. This weakness is however shared also by other sectors of 
theoretical physics wherein a small perturbation parameter is difficult to 
identify \cite{1N}.

\subsection{The current status of RFOT}
However, the reliance of RFOT on MF concepts is not only a technical, but also a 
conceptual problem. The RFOT is clearly an impressive piece of theoretical 
physics: it provides an elegant and easy to grasp picture of the glassy 
slowdown, and is able to make very different observations and theoretical inputs 
(in some cases produced by completely unrelated scientific communities, such as 
MCT \cite{modecoupling} and spin glass theory \cite{parisiSGandbeyond}) fit 
together in a coherent way. It also comes with a set of tools, in the form of 
the replica method, for performing first principles calculations 
\cite{MP09,replicanotes} (a trait which is not shared by other approaches such 
as Dynamical Facilitation Theory and Frustration Limited Domains), and it has 
proven to be able to provide quantitative predictions for problems which were, 
originally, outside of the domain wherein it was initially conceived, such as 
the jamming problem \cite{liunagel98,Biroli07,LNSW10}.

Despite these many strengths, RFOT is still struggling to break out of the 
mean-field domain wherein it originally appeared; as a matter of fact, much of 
the research work that is produced today by the RFOT community is aimed at 
understanding non-MF effects, and the strongest criticisms which are today moved 
to RFOT are centered around the claim that the theory cannot be valid beyond 
mean-field.\\
Indeed, for much time there were no finite dimensional models wherein the RFOT 
picture could apply (the very first proposal for such a model is very recent 
\cite{AngeliniBiroli14}), a difficulty which is shared by the replica theory of 
spin glasses; in both cases, the existence beyond mean field  of replica 
symmetry breaking, and the Parisi picture that comes with it, is still a matter 
of intense debate and simulations performed on finite-dimensional models are not 
conclusive. This MF-bound character of RFOT can be also found in the scenario 
where RFOT works best, namely the jamming problem. In 
\cite{CharbonneauCorwin15}, it is shown how the RFOT computation of 
\cite{CKPUZ13} fails to account for localized excitations within packings, and 
as a result of this, localized excitations must be numerically removed in order 
to recover from simulations the predicted value of the exponent $\theta$ of the 
force distribution. This incapacity of taking into account localized excitations 
is very much to be expected from a MF theory. However, while in the case of 
jamming it is possible to disentagle localized and extended modes, this does not 
look easy in the case of yielding, whereupon the relevant modes really seem to 
be localized ones \cite{HentschelKarmakar11}.

On a more technical side, importing non-MF effects within RFOT also amounts to 
understanding how the physics of the MCT and Gardner transitions are modified in 
finite dimension. In the case of MCT, it is still not very clear what is the 
actual mechanism that transforms the transition into a crossover; while there is 
consensus on the viewpoint that MCT is a MF-like theory 
\cite{andreanovbiroli,FranzJacquin12} unable to take into account activation 
mechanisms, there are indications that activation may be already at play in the 
regime above $T_{MCT}$ 
\cite{berthierbiroli11,BerthierGarrahan03,heuer2008exploring}. Some attempts 
have been made to include activation effects within MCT, but their results are 
subject of debate, see for example \cite{CatesRamaswamy06}. It is even argued in 
\cite{Rizzo14} that the MCT transition is destroyed by critical fluctuations 
instead of activation, so the problem is still very much open.\\
For what concerns the Gardner transition, its presence in finite dimensional 
systems is still to be proven. Since it is a second order critical point,  a 
renormalization group study of the transition must be performed to understand 
how and if its physics changes in finite dimension. Such an attempt has been 
made in \cite{UrbaniBiroli15} using perturbative tools, but the results are 
quite odd. The upper critical dimension $d_u$, for example, appears to be model 
dependent; the only conclusive statement one can make is that if the Gardner 
transition survives, a perturbative RG approach is unable to predict this fact, 
requiring the use of non-perturbative RG tools. In any case, a solution to all 
these problems does not seem to be forthcoming.

\section{Proposals for further research \label{sec:futureresearch}}

We conclude the chapter with some proposals for further research on the field of 
glass physics. The reader will surely notice how they are all somewhat relative 
to the yielding transition and related problems.

\subsection{The Gardner transition in shear}

The most novel prediction produced by the present work is the existence of a 
Gardner transition in shear strain. As we mentioned, this Gardner transition is 
related to the onset of energy dissipation within the system for large enough 
strain.\\
A possible way to test this prediction is the following. First of all, glassy 
configurations equilibrated at low temperature must be prepared, using for 
example the algorithms of \cite{SEP13} or \cite{GrigeraParisi01}. Then a cycling 
shear strain protocol, like the one of \cite{FioccoFoffi13}, must be applied, 
for example implementing it with an affine transformation on particle 
coordinates like in \cite{GendelmanJaiswal15}. The strain should be applied in 
small steps in such a way that the liquid is able to equilibrate at each step, 
i.e. one should always have $\dot{\g} < 1/\t_\be$. This condition should be easy 
to satisfy since $\t_\be$ is very low.\\
As soon as the amplitude of the cycle exceeds $\g_G$, the Gardner transition is 
met and energy starts to be dissipated, producing a closed hysteresis curve with 
nonzero surface like in \cite{FioccoFoffi13}. The presence or not of a stress 
overshoot (which, if one believes figure \ref{fig:shear}, appears only within 
the Gardner phase) can be used to discriminate the Gardner transition from the 
yielding transition that happens at $\g_Y > \g_G$.\\
The behavior of the $\mu(1) = 1/\D_{EA}$ shear modulus can also be studied 
easily: if the prediction of figure \ref{fig:MSDfullshear} is true, the slope of 
the small elastic segments between avalanches is supposed to increase as $\g_Y$ 
is approached. One could also measure the Debye-Waller factor ($\D$ for 
$\g<\g_G$, $\D_{EA}$ for $\g>\g_G$) in the whole glass phase in order to get a 
double check both on the numerics and the theory, and verify that it has a 
maximum for $\g = \g_G$ as reported in figure \ref{fig:MSDfullshear}.

\subsection{State following in AQS protocols}
The protocols we reproduce in figure \ref{fig:shear} are thermal protocols 
wherein a glass former is ``annealed'' down to a density $\wh\f_g$ and then 
strain is applied. However, most literature on the subject considers Athermal 
Quasi Static protocols as detailed in paragraph \ref{subsec:qsshear}. Since such 
a protocols are anyway quasi-static, they can be described within the state 
following setting. There are two possible ways to do this: one can either choose 
$m=0$ in the expression \eqref{eq:defF} and then follow the state in $\g$ 
(instantaneous quench to $T=0$ followed by strain), or one can keep $m=1$, 
compress a state all the way to $\varphi_j$ and then follow it in strain (slow 
annealing to $T=0$ followed by strain). This way, AQS stress-strain curves 
should be obtained. However, we must again stress that the SF approach looks at 
equilibrium within the glassy metabasin, so the stress strain curves will again 
be smooth curves corresponding to the average of many shear histories like the 
ones in figure \ref{fig:shearstrainlemaitre}, and no stress drops (avalanches) 
will be observed.

\subsection{Avalanches}
It is however possible to study avalanches. Within the fRSB approach, avalanches 
correspond to the rearrangement modes that take place when the system crosses 
over from a microstate to another. In spin glasses, one has a similar phenomenon 
with the hysteresis curve of magnetization $M$ versus magnetic field $H$, that 
exhibits an intermittent response (Barkhausen noise) much like the avalanche 
response seen in AQS protocols (see for example \cite{CombeRoux00} for a study 
in stressed athermal packings).
The statistics of these magnetization drops in the SK model has been extensively 
studied in \cite{LeDoussalMuller12,LMW10}. Their probability distribution has 
been found to have a power-law behavior 
$$
\rho(\D M) \propto \D M^{-\t}, 
$$
where the exponent $\tau$ has been computed in \cite{LeDoussalMuller12} 
($\t=1$). A similar power law behavior is observed for avalanches in jammed 
packings \cite{CombeRoux00} under stress control ($\g$ vs. $\s$) with an 
exponent  $ \t \simeq 1.46$, so one would like to compute again the exponent in 
the case of hard spheres.\\
The calculation of \cite{LeDoussalMuller12} can in principle be ``translated'' 
from spin glasses to hard spheres. In \cite{LeDoussalMuller12} the starting 
point of the calculation is the computation of the cumulants of the 
magnetization in different fields
\beq
\overline{\langle m_{h_1}m_{h_2}\dots m_{h_p}\rangle}^c
\eeq
where $\overline{\bullet}$ denotes an average over the quenched disorder. In the 
case of HSs, one should consider the cumulants of the stress in different 
strains
\beq
\overline{\thav{\s_{\g_1}\s_{\g_2}\dots \s_{\g_p}}}^c,
\eeq
where now the average over the ``disorder'' is the average over the 
configuration $R$ as defined in the \eqref{eq:FP}. The first step for performing 
the calculation is certainly understanding how the relevant quantities scale; in 
the case of the SK model it is known that the magnetic field should scale like 
$H = \frac{h}{\sqrt{N}}$ \cite{LeDoussalMuller12} for a magnetization drop of 
order $\sqrt{N}$ to happen. In the case of strain, it may be that the relevant 
scaling is $\g \simeq N^{\be_{iso}}$, with $\beta_{iso}\simeq -0.62$ as reported 
in \cite{KarmakarLerner10b}.\\
In any case, the study of \cite{LeDoussalMuller12,LMW10} is valid at zero 
temperature in the SK model and a scaling solution of the fRSB equations is 
needed to perform the calculation, so the computation in the case of spheres can 
be only made for the jammed, athermal case. An eventual extension to the case of 
yielding beyond $\g_Y$, wherein a scaling regime is not even guaranteed to 
exists, does not seem like an easy task.

\subsection{Yielding}
The picture that emerges from this work for what concerns the yielding 
transition is that of a spinodal point, akin the the onset transition found in 
decompression/heating (paragraph \ref{subsec:ultrastable}). Having a spinodal 
point within a replica theory means having a zero \emph{longitudinal} mode in 
the Hessian matrix of the replica free energy $s[\hat\al]$ as discusses in 
section \ref{subsec:unstablemode}; intuitively, one should consider the 
longitudinal mode of the outermost block of the fRSB ansatz. A computation of 
the longitudinal mode of the fRSB solution of state following should probably be 
performed in order to better understand yielding.\\
The analogy between yielding and a spinodal point has led some authors to 
proposing a picture of yielding as a phenomenon akin to an inverse glass 
transition by raising the temperature (essentially, an onset transition), a 
point of view strongly criticized in \cite{IlyinProcaccia15}. While it is true, 
at least for what concerns this work, that the yielding transition is a spinodal 
point, there is a big difference between it and the onset point: the onset 
transition \emph{always} takes place at a temperature $T_{on} > T_{MCT}$ (see 
also figure \ref{fig:dinfstatefo}), which means that when the glass former is 
kicked out of the state, the only thing it finds outside is a trivial FEL and an 
equilibrium measure dominated by the ergodic liquid. This need not be true in 
the case of yielding, which can take place at temperatures below $T_{MCT}$; this 
means that the system, when is kicked out the state because of strain, may still 
have a rough FEL to move in and may even end up in a different glassy state.\\
Understanding where the systems ends up after the spinodal point is an obvious 
question whose answer is related to the description of the steady flow regime 
beyond yielding. The bare-to-the-bone problem is essentially answering the question ``Which is the thermodynamic state an RFOT system, with $T<T_d$, end up in when the glassy state it finds itself in is 
killed by an external perturbation? For example, is it another glassy state, or a liquid-like state?''. Such a question could be in principle 
answered by numerical simulations of simple RFOT models such as the PSM 
\cite{pedestrians}, using a magnetic field to induce a spinodal point.

\subsection{Non-linear rheology}
We have mentioned how the state following construction is unable to reproduce 
time dependent shear protocols $\dot\g(t)$; the replica chain discussed in 
section \ref{sec:chain} may provide a solution for this problem, at least if 
certain conditions are met. The idea is to consider a chain of replicas, all of 
them at the same temperature $T_g$, and apply upon each replica a shear strain 
in the following way
\beq
\g_a = \sum_{c<a}\d\g,
\eeq
in summary, we apply the strain progressively with a small increment $\d\g$ on 
each replica, $\g_{a+1} - \g_a = \d\g$. Such a construction can in principle be 
implemented for MF hard spheres using the expression $\eqref{eq:gauss_r}$ of the 
replicated entropy and implementing shear through the $\eqref{FFreplica}$. If 
one is able to successfully perform the infinite-bond limit of the chain, a 
constitutive equation \cite{Cates03} for HS glasses prepared at $\wh\f_g$ should 
in principle be obtained. It is very interesting to notice that since the 
temperature is constant, the chain will automatically yield a TTI dynamics when 
this construction is performed, in agreement with the phenomenon of 
rejuvenation.\\
However, besides the obvious technical difficulties, there is a big conceptual 
limit: pseudodynamics with the replica chain assumes quasi-equilibration, and as 
a result of this all details of the fast relaxation in the $\be$-regime are 
hopelessly lost. Only the slow part of the dynamics is accounted for, as 
detailed in \cite{FranzParisi13,FranzParisi15}.\\
It is therefore necessary to understand how much the fast part of relaxation 
matters for the rheology of glassy materials, for example for the determination 
of the flow curve. Intuitively, in the case of oscillatory protocols, the 
construction should be meaningful as long as the frequency $\omega$ is chosen in 
such a way that 
$$
\omega < 1/\t_\beta,
$$
which is not a particularly restrictive condition, unless the temperature is 
very low.

\appendix

\chapter{The infinite-$d$ solution of hard spheres \label{app:infinited}}

In this appendix we sketch the derivation of the replicated entropy 
\eqref{eq:gauss_r} for hard spheres in the $d \to \infty$ limit. We only give 
the broad strokes, referring to \cite{KPZ12} for further details.

The starting point is a density functional theory \cite{simpleliquids} in the 
form \cite{replicanotes}
\beq
\SS[\rho(\ox)] = \int d\ox \rho(\ox)[1-\log\rho(\ox)] + \frac{1}{2}\int d\ox\ 
d\oy\ \rho(\ox)f(\ox-\oy)\rho(\oy), 
\label{eq:DFTspheres}
\eeq
where $\ox \equiv (\bx_1,\dots,\bx_m)$ is a ``molecular coordinate'' that 
specifies the position of a molecule made of $m$ spheres, and $f(\ox)$ is a 
replicated Mayer function \cite{parisizamponi}. The replicated liquid of HSs is 
nothing but a molecular liquid wherein each molecule is made up of $m$ 
replicas\footnote{Of course we can choose $m$ as we please.} of the same 
original sphere. The equilibrium density profile $\rho(\bx)$ is required to 
minimize the functional
\beq
\dfunc{\SS}{\rho(\ox)} = 0 \Longrightarrow \log\rho(\ox) = \int d\oy 
f(\ox-\oy)\rho(\oy)
\eeq
as usual. Normally, this density functional would be given by a diagrammatic 
expansion \cite{simpleliquids} whose first term would be the second one in the 
\eqref{eq:DFTspheres}, but for $d\to \infty$, only the first diagram survives 
\cite{frischpercus}, yielding the \eqref{eq:DFTspheres}. The $\SS$ is essentially 
the analogue, for HSs and in the continuum, of the TAP free energy \cite{TAP77}.

Passing from the \eqref{eq:DFTspheres} to the \eqref{eq:gauss_r} amounts to 
nothing more than a change of coordinates in the integrals, which is done by 
exploiting the inherent symmetries of the problem. First of all, the liquids is 
homogeneous, so it is invariant by translations. More specifically, if one 
defines
$$
X \equiv \frac{1}{m}\sum_{a=1}^m \bx_a \qquad\qquad u_a \equiv \bx_a -X
$$
which are the center of mass of a molecule and the displacement of atom $a$ with 
respect to it, then the density profile $\rho(\ox)$ can only depend on the 
$u_a$. We can exploit this fact by changing coordinates
\beq
d\ox = dX\ d\ou\ m^d \d\left(\sum_{a=1}^m u_a\right) \equiv dX\DD\ou,
\eeq
so that we can rewrite the \eqref{eq:DFTspheres} as 
\beq
\SS[\rho(\ox)] = V \int \DD \ou \rho(\ou)[1-\log\rho(\ou)] + \frac{V}{2}\int 
\DD\ou\ \DD\ov\ \rho(\ou)\overline{f}(\ou-\ov)\rho(\ov)
\eeq
where
\beq
\overline{f}(\ou-\ov) \equiv \int dX\ f(\ou-\ov + X).
\eeq
Secondly, the liquid is also isotropic, i.e. it is rotationally invariant. So, 
if one defines new coordinates
\beq
q_{ab} \equiv u_a\cdot u_b \qquad p_{ab} \equiv v_a\cdot v_b \qquad r_{ab} 
\equiv u_a\cdot v_b,
\eeq
then the density profile can only depend on the matrix $\hat q = q_{ab}$. We can 
write
\beq
\frac{\SS[\rho(\ox)]}{V} = \int d\hat q J(\hat q)\rho(\hat q) [1-\log \rho(\hat 
q)] + \frac{1}{2}\int d\hat q d\hat p d\hat r K(\hat q, \hat p, \hat r 
)\rho(\hat q)\rho(\hat p) \overline{f}(\hat q + \hat p -\hat r -\hat r^T).
\eeq
We can now see that the integration is not anymore on $md$ variables, but on 
$m(m-1)/2$ variables, all just by exploiting the symmetries of the problem. To 
proceed with the calculation, one must then compute the Jacobians $J$ and $K$. 
For the Jacobian J, for example, one has
\beq
\begin{split}
J(\hat q) =&\ \int \DD\ou \prod_{a\leq b}^{m}\d(q_{ab} -u_a\cdot u_b),\\
 = &\ m^d \int d\ou\ \d\left(\sum_{a=1}^m u_a\right) \prod_{a\leq 
b}^{m}\d(q_{ab} - u_a\cdot u_b),\\
 = &\ m^d \prod_{a=1}^m \d\left(\sum_{b=1}^m q_{ab}\right)\int du_1\dots 
du_{m-1}\prod_{a\leq b}^{m-1}\d(q_{ab} - u_a\cdot u_b).
\end{split}
\eeq
Luckily, the last integral has been already computed. It is nothing but the 
Jacobian that one must compute to infer the probability distribution of a 
Wishart matrix $Q = U^\dagger U$ wherein $U$ is a $M\times N$ matrix (whose 
probability distribution is known) with $M=d$ and $N = m-1$; its value is just 
\cite{MV09}
\beq
C_{m,d}\exp\left(\frac{d-m}{2}\log \det \hat q^{m,m}\right).
\eeq   
So the Jacobian is
\beq
J(\hat q) = m^d \prod_{a=1}^m \d\left(\sum_{b=1}^m q_{ab}\right) C_{m,d} 
\exp\left(\frac{d-m}{2}\log \det \hat q^{m,m}\right),
\eeq
where $C_{m,d}$ is a normalization constant (see \cite[appendix A]{CKPUZ13} for 
its computation); the calculation proceeds on similar lines for the Jacobian 
$K$. From this last equation we can see that the integrand scales like $e^d$, so 
for $d\to\infty$ we can compute its value using the saddle point-method 
\cite{AdvancedMethods} on the three matrices $\hat q$, $\hat p$ and $\hat r$ as 
discussed in section \ref{subsec:replicateds}. At the saddle point one finds 
\cite{KPZ12}
\beq
\hat q_{ab } = \hat p_{ab} \qquad\qquad \hat r_{ab} = 0. 
\eeq
so that the entropy must be optimized only on the matrix $\hat q$. Afterwards, 
one defines
\beq
\hat \a \equiv \frac{d}{D_g} \hat q,
\eeq
and after some more manipulations, one finally gets
\begin{equation*}
\begin{split}
\frac{\SS[\rho(\bx)]}{N}= \ &1 - \log\r + d \log (m) + \frac{(m-1)d}{2} \log(2 
\pi e D_g^2/d^2)\\
&+ \frac{d}2 \log \det(\hat \a^{m,m}) -  \frac{d}2 \wh \f_g \,
 \FF\left( 2 \hat \a \right) \ ,
 \end{split}
\end{equation*}
which is the \eqref{eq:gauss_r}. We stress again the fact that this derivation 
is completely exact, without any approximations.

\chapter{Computation of the replicated entropy in the RS ansatz 
\label{app:replicatedRS}}

\section{Entropic term \label{app:entrRS}}

We want to compute $\det \hat \a^{m+s,m+s}$, where we recall that 
$\hat \a^{a,a}$ is the matrix obtained from $\hat\a$
by deleting the $a$-th row and column, i.e. it is the $(a,a)$-cofactor of 
$\hat\a$.
Being a Laplacian matrix, $\hat \a$ has a vanishing determinant. 
Also, the ``Kirchhoff's matrix tree theorem'' states that for Laplacian 
matrices,
all the cofactors are equal, hence $\det \hat\a^{a,a}$ is independent of $a$.
Therefore, if $\mathbf{1}$ is the identity in $m+s$ dimensions, we have
\beq\label{eq:detmm}
\det(\hat\a + \e \mathbf{1}) = \det\hat\a + \e \sum_{a=1}^{m+s} \det\hat\a^{a,a} 
+ O(\e^2)
\ \ \ 
\Rightarrow 
\ \ \ 
\det\hat\a^{a,a} = \lim_{\e\to 0} \frac{1}{\e(m+s)} \det(\hat\a + \e \mathbf{1}) 
\ .
\eeq
We then define $\hat \b(\e)=\hat \a + \e \mathbf{1}$ and we note that
\beq\label{eq:beta}
\hat \b(\e)=\begin{pmatrix}
A & B\\
B^T & D
\end{pmatrix}
\eeq
where
$A$ is a $m\times m$ matrix with components
$A_{ab}=(\d^g+\a^g + \e) \d_{ab}-\a^g$,
$D$ is a $s\times s$ matrix 
with $D_{ab}=(\d+\a + \e)\delta_{ab}-\a$, and
$B$ is a $m\times s$ matrix with $B_{ab} = \chi$. 

We can use the following general formula
\beq\label{eq:betae}
\det \hat\b(\e)=(\det A)\det(D-B^TA^{-1}B) \ ,
\eeq
recalling that a $m\times m$ matrix $M_{ab} = M_1 \d_{ab} + M_2$ has determinant 
$\det M =  M_1^{m-1} (M_1 + m M_2)$ and
its inverse is $M^{-1}_{ab} = (M^{-1})_1 \d_{ab} + (M^{-1})_2$ with 
\beq\label{eq:GM}
\begin{split}
(M^{-1})_1 &= \frac1{M_1} \ , \\
(M^{-1})_2 &= -\frac{M_2}{M_1 ( M_1 + m M_2 )} \ .
\end{split}\eeq
The matrix $A^{-1}$ has this form, with
\beq\label{eq:A}
\begin{split}
(A^{-1})_1 &= \frac{1 }{ \alpha^g +\delta^g +\e } \ , \\
(A^{-1})_2&= \frac{\alpha^g }{(\alpha^g (1- m) + \delta^g  +\e ) (\alpha^g 
+\delta^g +\e )} \ , \\
\det A &= \left(\d^g +\a^g + \e\right)^{m-1}\left(\d^g+(1-m)\a^g + \e\right) \ .
\end{split}
\eeq
The matrix $\Omega = D - B^T A^{-1} B$ has the same form with
\beq\label{eq:Omega}
\begin{split}
\Omega_1 &= \delta + \e + \a \ , \\
\Omega_2 &= -\a- \c^2 [ m (A^{-1})_1 + m^2 (A^{-1}_2) ] \ , \\
\det \Omega &= (\d + \a + \e)^{s-1} \{ \d + \a (1-s) + \ee -s \c^2[ m (A^{-1})_1 
+ m^2 (A^{-1}_2) ] \} \ .
\end{split}\eeq
Using Eqs.~\eqref{eq:A}, \eqref{eq:Omega}, \eqref{eq:betae} and 
\eqref{eq:detmm}, we obtain the final result
\beq\label{entropic_final}
\begin{split}
\det \hat \a^{(m+s,m+s)}&= \c ( m \a^g + s \c)^{m-1} ( s\a + m \c)^{s-1}  
\ .
\end{split}
\eeq
By taking the logarithm and replacing the $\chi$, $\a$, $\a_g$ with $\D_g$, $\D$ 
and $\D_f$ using the \eqref{eq:Deltaer}, we get the \eqref{eq:entrRS}.

\section{Interaction term \label{app:intRS}}
Here we compute the interaction function $\FF(2\hat\a)$. 
This function has been computed in~\cite{KPZ12}, but only for $\h=0$ and $\g=0$. 
Here we need
to generalize the calculation to non-zero perturbations.

\subsection{General expression of the replicated Mayer function}

We follow closely the derivation of~\cite{KPZ12} which has been generalized 
in~\cite{YoshinoZamponi14}
to the presence of a strain.
The replicated Mayer function is
\beq\begin{split}
\ol f(\bar u ) &= \int dX \left\{  -1 + \prod_{a=1}^m \theta(|X + u_a | - 
D_g)\prod_{b=m+1}^{m+s} \theta(|S(\g)(X + u_b) | - D) \right\} \\
&= - \int dX \,  \theta\left( \max_{a=1,m+s} \{ D_a -  | S(\g_a) ( X + u_a ) | 
\} \right) \ ,
\end{split}\eeq
where we introduced $D_a = D_g ( 1 + \h_a/d)$ with $\h_a = \g_a=0$ for $1\leq a 
\leq m$, and $\h_a = \h$ and $\g_a = \g$ for
$m+1 \leq a \leq m+s$.

The $u_a$ are $m+s$ vectors in $d$ dimensions and define a hyper-plane in the 
$d$-dimensional space.
It is then reasonable to assume that this $(m+s)$-dimensional plane
is orthogonal to the strain
directions $\mu=1,2$ with probability going to 1 for $d\to\io \gg m+s$.
Hence, the vector $X$ can be decomposed in a two dimensional
vector $\{X_1,X_2\}$ parallel to the strain plane, 
a $(d-m-s-2)$-component
vector $X_\perp$, orthogonal to the plane $\mu=1,2$ and to the plane defined by 
$u_a$,
and a $m+s$-component vector $X_\parallel$ parallel to that plane.
Defining $\Omega_d$ as the $d$-dimensional solid angle and recalling that 
$V_d=\Omega_d/d$,
and following the same steps as in~\cite[Sec.~5]{KPZ12},
we have, calling $k = m+s$
\beq\begin{split}
\ol f(\bar u ) 
=& -\int dX_1 dX_2 \ d^{k}X_\parallel \ d^{d-k-2}X_\perp \\ 
&\times\theta\left( \max_a \{ D_a^2 - (X_1 + \g_a X_2 )^2 - X_2^2 - |X_\parallel 
+ u_a |^2 - | X_\perp |^2 \} \right) \\
=& - \Omega_{d-k-2} \int  dX_1 dX_2 \ d^{k}X_\parallel \int_0^\io dx \, 
x^{d-k-3} \\  
&\times\theta\left( \max_a \{ D_a^2 - x^2 -  (X_1 + \g_a X_2 )^2 - X_2^2 - | 
X_\parallel + u_a |^2 \} \right) \\
=& - \Omega_{d-k-2} \int  dX_1 dX_2 \ d^{k}X_\parallel \int_0^{ \sqrt{\max_a \{ 
D_a^2 -  (X_1 + \g_a X_2 )^2 - X_2^2 - | X_\parallel + u_a |^2 \}} }  dx \, 
x^{d-k-3} \\
=& - V_{d-k-2}  \int  dX_1 dX_2 \ d^{k}X_\parallel \\
&\times\Theta_{d-k-2}\left(\max_a \{ D_a^2 -  (X_1 + \g_a X_2 )^2 - X_2^2 - | 
X_\parallel + u_a |^2 \} \right)
\end{split}\eeq
where we defined the function
$\Th_{p}(x) = x^{p/2} \th(x)$.

It has been shown in~\cite{KPZ12} that
the region where $\overline{f}(\bar u)$ has a non-trivial dependence on the 
$u_a$ is 
where $u_a \sim 1/\sqrt{d}$.
Here we use $D_g$ as the unit of length, hence we define
$u_a = x_a D_g/\sqrt{d}$, $X_{1,2} = \z_{1,2} D_g/\sqrt{d}$
and $X_\parallel = \ee D_g/\sqrt{d}$. 
Using that $\lim_{n\to\io} \Th_n(1+y/n) = e^{y/2}$, 
and that for large $d$ and finite $k$ we have $V_{d-k}/V_d \sim d^{k/2} / 
(2\pi)^{k/2}$,
we have
\beq\begin{split}
\ol f(\bar u ) 
 =& - \frac{V_{d-k-2}}{V_d} \frac{V_d D_g^{d}}{d^{(k+2)/2}} \int  d\z_1 d\z_2  
d^{k}\ee \\ 
&\times\Theta_{d-k-2}\left(1  - \frac1d \min_a\{ -2\h_a + (\z_1 + \g_a \z_2 )^2 
+ \z_2^2 + | \ee + x_a |^2 \} \right) \\
\sim& - V_d D_g^d  \int \frac{ d\z_1 d\z_2  d^{k}\ee }{(2\pi)^{(k+2)/2}} 
\, e^{  - \frac12 \min_a \{ -2 \h_a + (\z_1 + \g_a \z_2 )^2 + \z_2^2 + | \ee + 
x_a |^2 \}  }\\ 
\equiv& - V_d D_g^d \FF(\bar x )
\ ,
\end{split}\eeq
where the function $\FF$ has been introduced following~\cite{KPZ12,KPUZ13}.

We can then follow the same steps as in~\cite[Sec.V C]{KPUZ13} and 
in~\cite{YoshinoZamponi14} to obtain
\beq
\begin{split}
\mathcal F(\bar x) &= 
\int \frac{ d\z_1 d\z_2  d^{k}\ee }{(2\pi)^{(k+2)/2}} \, e^{  - \frac12 \min_a 
\{ -2 \h_a + (\z_1 + \g_a \z_2 )^2 + \z_2^2 + | \ee + x_a |^2 \}  } \\
&=
\int \frac{ d\z_1 d\z_2  d^{k}\ee }{(2\pi)^{(k+2)/2}}
\lim_{n\to 0} \left( \sum_{a=1}^k e^{-\frac1{2n} [  -2 \h_a +(\z_1 + \g_a \z_2 
)^2 + \z_2^2 + | \ee + x_a |^2 ] }
\right)^n
\\
&=\lim_{n\to 0}\sum^*
\int \frac{ d\z_1 d\z_2  d^{k}\ee }{(2\pi)^{(k+2)/2}}
e^{ - \sum_a \frac{n_a}{2n} [ -2 \h_a  + (\z_1 + \g_a \z_2 )^2 + \z_2^2 + | \ee 
+ x_a |^2 ]     }
\\
&=\lim_{n\to 0}\sum^*\ 
e^{ \sum_{a=1}^k \frac{n_a}{n} \h_a
-\frac 12 \sum_{a=1}^k \frac{n_a}{n}|x_a|^2+\frac{1}{2} 
\sum_{a,b}^{1,k}\frac{n_an_b}{n^2}x_a\cdot x_b } 
\int \frac{ d\z_1 d\z_2 }{2\pi}
e^{ - \sum_a \frac{n_a}{2n} [ (\z_1 + \g_a \z_2 )^2 + \z_2^2  ]     }
\\
&=\lim_{n\to 0}\sum^*\ e^{ \sum_{a=1}^k \frac{n_a}{n} \h_a- \frac{1}{4} 
\sum_{a,b}^{1,k}\frac{n_an_b}{n^2} (x_a - x_b)^2 } 
\int \frac{ d\z }{\sqrt{2\pi}}
e^{ - \frac{\z^2}2 \left[ 1 + \frac12 \sum_{ab} \frac{n_a n_b}{n^2} (\g_a - 
\g_b)^2 \right]    } \ .
\end{split}\eeq
where 
\beq
\sum^* \equiv \sum_{n_1,\ldots, n_k; \sum_{a=1}^kn_a=n}\frac{n!}{n_1!\ldots 
n_k!}.
\eeq
We now introduce the matrix $\hat\D$ of mean square displacements between 
replicas
\beq
\D_{ab} = (x_a - x_b)^2 = \frac{d}{D_g^2} (u_a - u_b)^2 \ .
\eeq
We should now recall that the Mayer function is evaluated in $\bar u - \bar v$,
hence after rescaling the function $\FF$ is evaluated in $\bar x - \bar y$.
For $d\to\io$, the interaction term is dominated by a saddle point on $\bar u$ 
and $\bar v$, such that
$(x_a - x_b)^2 = (y_a - y_b)^2 = \D_{ab}$ and $x_a \cdot y_b 
=0$~\cite{KPZ12,KPUZ13,CKPUZ13}, hence
$(x_a - y_a -x_b + y_b)^2 = (x_a - x_b)^2 + (y_a - y_b)^2 = 2 \D_{ab}$. This is 
also why the function $\FF$ is evaluated in $2\hat\a$ in 
Eq.~\eqref{eq:gauss_r}. 
The contribution of the interaction term to the free energy~\eqref{eq:gauss_r} 
is~\cite{KPZ12}
\beq
\frac12 \frac{N}V \ol f(\bar u - \bar v) = - \frac{N V_d D_g^d}{2V} \FF(\bar x - 
\bar y) = - 2^{d-1} \f_g \FF(2\hat\a) = - \frac{d \wh\f_g}2 \FF(2\hat\a) \ .
\eeq
With an abuse of notation, we now call $\FF(\hat\D) = \FF(2\hat\a)$.

We therefore
obtain at the saddle point
\beq\label{FFreplica}
\begin{split}
\mathcal F(\hat\D) 
&=\lim_{n\to 0}\sum^*\ 
e^{\sum_{a=1}^k \frac{n_a}{n} \h_a- \frac{1}{2} 
\sum_{a,b}^{1,k}\frac{n_an_b}{n^2} \D_{ab} } 
\int \frac{ d\z }{\sqrt{2\pi}}
e^{ - \frac{\z^2}2 \left[ 1 + \frac12 \sum_{ab} \frac{n_a n_b}{n^2} (\g_a - 
\g_b)^2 \right]    } \\
&=\int \frac{ d\z }{\sqrt{2\pi}}
e^{ - \frac{\z^2}2 }
\left[\lim_{n\to 0}\sum^*\ 
e^{\sum_{a=1}^k \frac{n_a}{n} \h_a- \frac{1}{2} 
\sum_{a,b}^{1,k}\frac{n_an_b}{n^2} \left( \D_{ab} + \frac{\z^2}{2} (\g_a - 
\g_b)^2 \right) }\right]
 \\
&=\int \frac{ d\z }{\sqrt{2\pi}}
e^{ - \frac{\z^2}2 }
\FF_0\left( \D_{ab} + \frac{\z^2}{2} (\g_a - \g_b)^2 \right) \ , 
\end{split}\eeq
where $\FF_0(\hat\D)$ is the interaction function in absence of strain and is 
given by
\beq\label{eq:FF0binomial}
\FF_0( \hat\D )  = \lim_{n\to 0}\sum_{n_1,\ldots, n_k; \sum_{a=1}^k 
n_a=n}\frac{n!}{n_1!\ldots n_k!}
e^{\sum_{a=1}^k \frac{n_a}{n} \h_a- \frac{1}{2} 
\sum_{a,b}^{1,k}\frac{n_an_b}{n^2}  \D_{ab}  } \ .
\eeq

\subsection{Computation of the interaction term for a RS displacement matrix}

We now compute the function $\FF_0(\hat\D)$ for the replica structure encoded by 
the matrix~\eqref{eq:DeltaRS}.
Defining $\Si_m = \sum_{a=1}^m \frac{n_a}n$ and $\Si_s = \sum_{a=m+1}^{m+s} 
\frac{n_a}n$, keeping in mind that
$\Si_m + \Si_s = 1$, and recalling that $\h_a = \h$ for $m+1 \leq a \leq m+s$ 
and $\h_a=0$ otherwise,
we can then write with some manipulations
\beq
\begin{split}
\FF_0(\hat \D) & =\lim_{n\to 0}\sum^* e^{- \left(\frac{\Dg}2 + \frac{\D^f}2 
\right) \Si_m - \left( \frac{\De}2- \h \right) \Si_s  + 
 \frac{ \D^f }2 \Si_m^2 
 + \frac\Dg{2} \sum_{a=1}^m \frac{n_a^2}{n^2} + \frac{\De}2 \sum_{a=m+1}^{m+s} 
\frac{n_a^2}{n^2} } \ . 
\end{split}\eeq

We now introduce Gaussian multipliers (Hubbard-Stratonovich transformation 
\cite{pedestrians}) to decouple the quadratic terms
and introduce the notation
$\DD \l = \frac{d\l}{\sqrt{2\pi}} e^{-\l^2/2}$.
Note that $\Dg \geq 0$ and $\De \geq 0$. {\it Under the assumption that $2 \D^f 
\geq 0$} (to be discussed
later on),
we get
\beq\begin{split}
\FF_0(\hat \D)  & = 
\int \DD \l_a \DD \m \lim_{n\to 0}\left[ 
\sum_{a=1}^m
e^{-\frac1{n} \left(\frac{\Dg}2 + \frac{\D^f}2 + \l_a \sqrt{\Dg} + \m \sqrt{ 
\D^f} \right)} 
+\sum_{a=m+1}^{m+s} e^{-\frac1{n} \left( \frac{\De}2- \h + \l_a \sqrt{\De}  
\right)}
\right]^n \\
& = 
\int \DD \l_a \DD \m \,
e^{
-\min\left\{ 
\frac{\Dg}2 + \frac{\D^f}2  + (\min_{a\leq m} \l_a) \sqrt{\Dg} + \m \sqrt{\D^f}, 
\,
\frac{\De}2- \h + (\min_{a>m} \l_a) \sqrt{\De}  
\right\}
}
\end{split}\eeq
Now we use that for any function $f(x)$,
\beq\begin{split}
\int \DD \l_a f\left(\min_{a\leq m} \l_a \right) =& 
m \int \DD \l f(\l) \left( \int_\l^\io \DD\l' \right)^{m-1}\\
=&-\int d\l f(\l) \frac{d}{d\l}  \Th\left(  -\frac{\l}{\sqrt{2}} \right)^m\\
\equiv& \int  \ol\DD_m[ \l] f(\l)
\end{split}
\eeq
with
\beq
\ol\DD_n[ \l] = - d\l \frac{d}{d\l} \Th\left(  -\frac{\l}{\sqrt{2}} \right)^n.
\label{eq:measuredef}
\eeq
We therefore obtain
\beq\begin{split}
\FF_0(\hat \D)  & = 
\int \ol\DD_m \l \, \ol \DD_s \l' \, \DD \m \,
e^{
-\min\left\{ 
\frac{\Dg}2 + \frac{\D^f}2  + \l \sqrt{\Dg} + \m \sqrt{\D^f}, \,
\frac{\De}2- \h + \l' \sqrt{\De}.  
\right\}
}
\end{split}\eeq
The integral over $\mu$ can be done explicitly, and we obtain
\beq\label{eq:Kdef}
\begin{split}
K(\l,\l') =& \int \DD \m \,
e^{
-\min\left\{ 
\frac{\Dg}2 + \frac{\D^f}2  + \l \sqrt{\Dg} + \m \sqrt{\D^f}, \,
\frac{\De}2- \h + \l' \sqrt{\De}  
\right\}
} \\
 =&\ 
e^{-\frac{\De}2+ \h - \sqrt{\De} \l'} \Th\left(\frac{ \h + \frac{\D^f + \Dg - 
\De}2 + \sqrt{\Dg} \l - \sqrt{\De} \l' }
{\sqrt{2\D^f}}
\right)\\
&\ + e^{-\Dg/2 - \sqrt{\Dg} \l} \Th\left(\frac{-\h + \frac{\D^f - \Dg + \De}2 - 
\sqrt{\Dg} \l + \sqrt{\De} \l' }
{\sqrt{2 \D^f }}
\right)
\end{split}\eeq

Now, integrating by parts, we can write
\begin{align*}
\FF_0(\hat \D)   =& 
\int  d\l \frac{d}{d\l} \left[1 - \Th\left(  -\frac{\l}{\sqrt{2}} \right)^m 
\right] \, 
\int d\l' \frac{d}{d\l'} \left[ 1 - \Th\left(  -\frac{\l'}{\sqrt{2}} \right)^s 
\right] 
K(\l,\l') \\
=& \int d\l \frac{d}{d\l} \left[1 - \Th\left(  -\frac{\l}{\sqrt{2}} \right)^m 
\right]
\Bigg\{
K(\l,\l'=\io)\\
& - \int d\l' \left[ 1 - \Th\left(  -\frac{\l'}{\sqrt{2}} \right)^s \right] 
\frac{\partial K}{\partial \l'}(\l,\l')
\Bigg\} \\
=& \int d\l \frac{d}{d\l} \left[1 - \Th\left(  -\frac{\l}{\sqrt{2}} \right)^m 
\right]
e^{-\Dg/2 - \sqrt{\Dg} \l} \displaybreak\\
&- \int d\l' \left[ 1 - \Th\left(  -\frac{\l'}{\sqrt{2}} \right)^s \right] 
\int d\l \frac{d}{d\l} \left[1 - \Th\left(  -\frac{\l}{\sqrt{2}} \right)^m 
\right]
\frac{\partial K}{\partial \l'}(\l,\l') \\
=&
\sqrt{\Dg} \int d\l \left[1 - \Th\left(  -\frac{\l}{\sqrt{2}} \right)^m \right]
e^{-\Dg/2 - \sqrt{\Dg} \l}\\
& - \int d\l' \left[ 1 - \Th\left(  -\frac{\l'}{\sqrt{2}} \right)^s \right] 
\Bigg\{
\frac{\partial K}{\partial \l'}(\l=\io,\l')\\
& - \int d\l \left[1 - \Th\left(  -\frac{\l}{\sqrt{2}} \right)^m \right]
\frac{\partial^2 K}{\partial \l \partial \l'}(\l,\l')
\Bigg\} \\
=&\sqrt{\Dg} \int d\l \left[1 - \Th\left(  -\frac{\l}{\sqrt{2}} \right)^m 
\right]
e^{-\Dg/2 - \sqrt{\Dg} \l}\\
& +\sqrt{\De} \int d\l' \left[ 1 - \Th\left(  -\frac{\l'}{\sqrt{2}} \right)^s 
\right] 
e^{-\De/2 + \h - \sqrt{\De} \l'}
\\
& + \int d\l \, d\l' 
 \left[1 - \Th\left(  -\frac{\l}{\sqrt{2}} \right)^m \right]
\left[ 1 - \Th\left(  -\frac{\l'}{\sqrt{2}} \right)^s \right] 
\frac{\partial^2 K}{\partial \l \partial \l'}(\l,\l') \ .
\end{align*}
We also have
\beq\label{K}
\frac{\partial^2 K}{\partial \l \partial \l'}(\l,\l') = - \sqrt{\Dg \De} \, 
e^{\h - \De/2 -\sqrt{\De} \l'} \,
\frac{ e^{- \frac1{2 \D^f} \left( -\h + \frac{ \De - \Dg - \D^f}2 - \sqrt{\Dg} 
\l + \sqrt{\De} \l' \right)^2 }  }{\sqrt{2\pi \D^f}} \ .
\eeq
We remark that the function $K$ does not depend explicitly on $m$ and $s$,
therefore the derivative with respect to $s$ can be computed straightforwardly.
Also, using Eq.~\eqref{K} one can write
\beq\begin{split}
\FF_0(\hat \D)  
=&\sqrt{\Dg} \int d\l \left[1 - \Th\left(  -\frac{\l}{\sqrt{2}} \right)^m 
\right]
e^{-\Dg/2 - \sqrt{\Dg} \l}\\
& + \sqrt{\De} \int d\l' 
\left[ 1 - \Th\left(  -\frac{\l'}{\sqrt{2}} \right)^s \right] 
e^{-\De/2 + \h - \sqrt{\De} \l'}\\ 
  &\times\int d\l \, 
 \Th\left(  -\frac{\l}{\sqrt{2}} \right)^m \,
\sqrt{\Dg} \frac{ e^{- \frac1{2 \D^f} \left(- \h + \frac{ \De - \Dg - \D^f}2 - 
\sqrt{\Dg} \l + \sqrt{\De} \l' \right)^2  } }{\sqrt{2\pi \D^f}} \ .
\end{split}\eeq
We can also change to variables $y = -\Dg/2 -\sqrt{\Dg} \l$ and $y' = \h -\De/2 
-\sqrt{\De} \l'$, and $x= y' - y$. 
Then we have
\beq
\begin{split}
\FF_0(\Dg,\De,\D^f)  
=& \int dy \, e^{y} \, \Bigg\{  1 
- \Th\left(  \frac{y + \Dg/2}{\sqrt{2\Dg}} \right)^m\\
&\times \int dx \,  \Th\left(  \frac{ x+ y - \h +\De/2}{\sqrt{2 \De}} \right)^s 
\frac{ e^{- \frac1{2 \D^f} \left( x - \D^f/2  \right)^2  } }{\sqrt{2\pi \D^f}} 
\Bigg\} \ .
\end{split}\eeq
From Eq.~\eqref{FFreplica}, recalling that $\g_a = \g$ for the $s$ replicas and 
zero otherwise, we have
\beq
\FF(\Dg,\De,\D^f)  =
\int \frac{ d\z }{\sqrt{2\pi}}
e^{ - \frac{\z^2}2 }
\FF_0\left( \Dg,\De,\D^f + \z^2 \g^2 \right) \ .
\eeq
\nopagebreak[4]
Which is the \eqref{eq:FF_final}.

\chapter{Computation of the replicon mode \label{app:stabilityRS}}

In this appendix we discuss in detail the stability of the replica symmetric 
ansatz \eqref{eq:DeltaRS} for the calculation of the Franz-Parisi entropy.
We want to compute the stability matrix of the small fluctuations around the RS 
solution and from that extract the replicon eigenvalue~\cite{KPUZ13}.
The calculation is very close to the one given in \cite{KPUZ13} and we will use 
many of the results reported in that work.

\section{The structure of the unstable mode \label{sec:replicongen}}

The general stability analysis of the RS solution can be done on the following 
lines. 
In principle, we have to take the general expression (\ref{eq:gauss_r}) and 
compute the Hessian matrix 
obtained by varying at the second order the replicated entropy with respect to 
the full matrix $\hat \a$. We can then compute the Hessian on the RS saddle 
point.
The task here is complicated by the fact that the entropy (\ref{eq:gauss_r}) is 
not symmetric under permutation of all replicas. The symmetries are restricted
to arbitrary perturbations of the $m$ replicas and the $s$ replicas separately. 
Hence the structure of the Hessian matrix is more complicated than the one
studied in~\cite{KPUZ13}. 

However, here we are mostly interested in studying the problem when the $m$ 
replicas are at equilibrium in the liquid phase, hence $m=1$, and in that
case we already know that the RS solution is stable in the sector of the $m$ 
replicas~\cite{KPUZ13}. 
Moreover, the $m$ reference replicas only select the glass state and do not 
evolve when the state is followed, differently from the $s$ replicas.
Hence, on physical grounds, we expect that replica symmetry will be broken in 
the sector of the $s$ replicas and that the unstable mode in that sector
will have the form of a ``replicon'' mode similar to the one studied 
in~\cite{KPUZ13}. Based on this reasoning, we conjecture the following form for 
the unstable mode:
\beq\label{eq:replicon}
\d \hat\D = \left[
\begin{matrix}
\d \Dg (I^{m}_{ab} - \d_{ab}) &    \d \D^r I_{ab}^{m,s} \\
\d \D^r I_{ab}^{s,m}    & \d \D_R \, r_{ab} \\
\end{matrix} \right] \ ,
\eeq
where $\hat I^m$ is a $m\times m$ matrix and $\hat I^{m,s}$ is a $m\times s$ 
matrix with all elements equal to $1$, 
and $\hat r$ is a $s\times s$ ``replicon'' matrix such that $\sum_{ab} 
r_{ab}=0$~\cite{KPUZ13,CKPUZ13}.
In other words, we look for fluctuations around the RS matrix \eqref{eq:DeltaRS} 
where the matrix elements of the $m$
replicas $\D^g$ and the matrix elements connecting the $m$ and $s$ replicas 
$\D^r$ are varied uniformly, while in the
$s$ block we break replica symmetry following the replicon mode.

Let us write the variation of the entropy \eqref{eq:gauss_r} around the RS 
solution, along the unstable mode \eqref{eq:replicon}. We have
\beq
\d s = \frac12 \sum_{a\neq b, c\neq d} M_{ab;cd} \d\D_{ab} \d\D_{cd} + \frac16 
\sum_{a\neq b, c\neq d, e\neq f} W_{ab;cd;ef} \d\D_{ab} \d\D_{cd} \d\D_{ef} 
+ \cdots
\ .
\eeq
The mass matrix $M_{ab;cd}$ and the cubic term $W_{ab;cd;ef}$ are derivatives of 
the entropy 
$s$ (which is replica symmetric) computed in a RS point, and
therefore they must satisfy certain symmetries which are simple extensions of 
the ones discussed in~\cite{KPUZ13}.
Let us call $(ab)^m$ a pair of indices $a\neq b$ that both belong to the $m$ 
block. Similarly $(ab)^s$ belong to the $s$ block, and 
$(ab)^r$ are such that one index belong to the $m$ block and the other to the 
$s$ block.
At the quadratic order, we obtain
\beq\begin{split}
\d s =\ & \frac12 (\d\Dg)^2 \sum_{(ab)^m, (cd)^m} M_{ab;cd} 
+\frac12 (\d\D^r)^2 \sum_{(ab)^r, (cd)^r} M_{ab;cd} \\
& + \frac12 \d\D_R^2 \sum_{(ab)^s, (cd)^s} M_{ab;cd} r_{ab} r_{cd} + \d\Dg 
\d\D^r \sum_{(ab)^m,(cd)^r} M_{ab;cd}\\
&+ \d\Dg \d\D_R \sum_{(ab)^m,(cd)^s} M_{ab;cd} r_{cd}
+ \d\D^r \d\D_R \sum_{(ab)^r,(cd)^s} M_{ab;cd} r_{cd} \ .
\end{split}\eeq
It is easy to show that the cross-terms involving the replicon mode vanish. In 
fact,
the sum $\sum_{(ab)^m} M_{ab;cd}$ must be a constant independent of the choice 
of indices $(cd)^s$, which are all
equivalent due to replica symmetry in the $s$-block. Hence $\sum_{(ab)^m,(cd)^s} 
M_{ab;cd} r_{cd} = const. \sum_{(cd)^s} r_{cd} =0$
because of the zero-sum property of the matrix $\hat r$. The same property 
applies to the other cross-term.
The quadratic term has therefore the form
\beq
\d s^{(2)} = \frac12 A (\d\Dg)^2 
+\frac12 B (\d\D^r)^2 + C \d\Dg \d\D^r 
+ \frac12 \d\D_R^2 \sum_{(ab)^s, (cd)^s} M_{ab;cd} r_{ab} r_{cd} \ ,
\eeq
and the stability analysis of the replicon mode in the $s$-block can be done 
independently of the presence of the $m$ replicas.

A similar reasoning can be applied to the cubic terms. Let us write only the 
terms that involve the replicon mode:
\begin{align*}
\d s^{(3)} =\ & \frac16 \d\D_R^3 \sum_{(ab)^s, (cd)^s, (ef)^s} W_{ab;cd;ef} 
r_{ab} r_{cd} r_{ef}\\
&+\frac12 \d\D_R^2 \d\Dg \sum_{(ab)^s, (cd)^s, (ef)^m} W_{ab;cd;ef} r_{ab} 
r_{cd} \\
&+\frac12 \d\D_R^2 \d\D^r \sum_{(ab)^s, (cd)^s, (ef)^r} W_{ab;cd;ef} r_{ab} 
r_{cd}\\
&+ \d\D_R \d\D^r \d\Dg \sum_{(ab)^s, (cd)^r, (ef)^m} W_{ab;cd;ef} 
r_{ab}\displaybreak \\
&+\frac12 \d\D_R (\d\D^r)^2 \sum_{(ab)^s, (cd)^r, (ef)^r} W_{ab;cd;ef} r_{ab}\\
&+\frac12 \d\D_R (\d\Dg)^2 \sum_{(ab)^s, (cd)^m, (ef)^m} W_{ab;cd;ef} r_{ab}\\
&+ \text{terms without  } \d\D_R
\end{align*}
Clearly, all terms that are linear in $\d\D_R$ vanish. In fact, for example 
\beq
\sum_{(ab)^s, (cd)^r, (ef)^m} W_{ab;cd;ef} r_{ab} = \sum_{(ab)^s} r_{ab} 
\sum_{(cd)^r, (ef)^m} W_{ab;cd;ef} = \text{const.} \times  \sum_{(ab)^s} r_{ab} 
=0 \ ,
\eeq
because once again $\sum_{(cd)^r, (ef)^m} W_{ab;cd;ef}$ must be a constant 
independent of the choice of $(ab)^s$ which are all equivalent thanks to
replica symmetry in the $s$-block.
Collecting all non-vanishing terms that involve the replicon mode, we obtain
\beq\begin{split}
\d s  &= \frac12 A (\d\Dg)^2 
+\frac12 B (\d\D^r)^2 + C \d\Dg \d\D^r 
+ \frac12 \d\D_R^2 \sum_{(ab)^s, (cd)^s} M_{ab;cd} r_{ab} r_{cd}  \\
 &+ \frac16 \d\D_R^3 \sum_{(ab)^s, (cd)^s, (ef)^s} W_{ab;cd;ef} r_{ab} r_{cd} 
r_{ef} +
\frac12 \d\D_R^2 \d\Dg \sum_{(ab)^s, (cd)^s, (ef)^m} W_{ab;cd;ef} r_{ab} r_{cd} 
\\
&+\frac12 \d\D_R^2 \d\D^r \sum_{(ab)^s, (cd)^s, (ef)^r} W_{ab;cd;ef} r_{ab} 
r_{cd}  \ .
\end{split}\eeq
This perturbative entropy must be optimized over $\d\Dg, \d\D^r, \d\D_R$, in 
order to check if one can construct a perturbative saddle point solution with 
more RSBs \cite{KPUZ13}.
The above equation clearly shows that for a fixed $\d\D_R$, the optimization 
over $\d\Dg, \d\D^r$ given $\d\Dg \sim \d\D^r \sim \d\D_R^2$. Hence we conclude 
that all the terms that involve 
$\d\Dg$ and $\d\D^r$ are at least of order $\d\D_R^4$ and can be neglected in 
the linear stability analysis. We finally obtain at the leading
order
\beq
\d s  =  \frac12 \d\D_R^2 \sum_{(ab)^s, (cd)^s} M_{ab;cd} r_{ab} r_{cd} 
+ \frac16 \d\D_R^3 \sum_{(ab)^s, (cd)^s, (ef)^s} W_{ab;cd;ef} r_{ab} r_{cd} 
r_{ef} 
\eeq
and all the couplings between the $s$-block and the $m$-block disappear. 
This shows that the stability analysis of the replicon mode
can be performed by restricting all the derivatives to the $s$-block, both at 
the quadratic and cubic orders. The Gardner transition corresponds
to the appearance of a negative mode in the quadratic term for a particular 
choice of the matrix $r_{ab}$ that corresponds to a 1RSB
structure in the $s$-block, characterized by a Parisi parameter $s_1$, as 
discussed
in~\cite[Sec.~VII]{CKPUZ13}. The unstable quadratic mode is stabilized by the 
cubic term leading to a fullRSB phase~\cite{Ri13,CKPUZ13}.
Note that, according to the analysis of~\cite{Ri13,CKPUZ13}, in the ``typical 
state'' calculation done with $m$ replicas with $m\in[0,1]$ taken as
a free parameter, the fullRSB phase
can only be stabilized if the parameter $m_1 > m$, and this only happens at low 
enough temperature or large
enough densities, hence the fullRSB phase can only exist at sufficiently low 
temperatures and high densities~\cite{Ri13,CKPUZ13}: unless these conditions are 
met, the replicon instability only
means that the replica calculation is generally unstable and only the liquid 
phase exists.\\
However, the situation is crucially different here because the state following 
construction requires $s\to 0$. The perturbative analysis
gives $s_1 = \l(s)$, where $\l(s) > 0$ is the MCT parameter discussed in 
paragraph \ref{subsec:lambdaMCT}, hence one always has 
$s_1 = \l(s) > s =0$ and the fullRSB phase exist at all temperatures and 
densities when the RS phase becomes unstable. This a big difference between the 
State Following calculation and the real replica one.

Summarizing, we have shown that one can define the following stability matrix
\beq
M^s_{a \neq b;c\neq d} = 
M_1\left(\frac{\d_{ac}\d_{bd}+\d_{ad}\d_{bc}}{2}\right)+M_2\left(\frac{\d_{ac}
+\d_{ad}+\d_{bc}+\d_{bd}}{4}\right)+M_3
\eeq
where the indices $a,\ b,\ c,\ d$ run between $m+1$ and $m+s$.
The fact that the replica structure of this stability matrix is the one defined 
in eq. (\ref{stab_matr}) is due to
replica symmetry under permutation of the $s$ replicas as already discussed. 
When a zero mode appears in this matrix, the RS solution becomes unstable and 
transform continuously in a fullRSB one, signaling that the glass state
sampled by the $s$ replicas undergoes a Gardner transition.

We now compute the replicon mode. We divide the problem of computing that 
stability matrix in the part coming from the derivatives of the entropic term 
and the part relative to the interaction term. 
We will first derive the stability matrix in the case of absence of shear, and 
the generalize it to a shear-strain situation.

\section{Entropic term}
We want to compute the contribution of the entropic term to the stability 
matrix.
Note that under a variation of $\d\a_{ab}$, we have an identical variation of 
$\d\a_{ba} = \d\a_{ab}$, 
and the diagonal terms vary by minus the same amount, $\d\a_{aa} = \d\a_{bb} = 
-\d\a_{ab}$ to
maintain the Laplacian condition of $\hat\a$.
Hence we have
\beq
\frac{\d}{\d\a_{a<b}} = \frac{\d}{\d\a_{ab}} +\frac{\d}{\d\a_{ba}} - 
\frac{\d}{\d\a_{aa}} - \frac{\d}{\d\a_{bb}} \ .
\eeq
From Eq.~\eqref{eq:detmm}, recalling that $\hat \b(\e)=\hat \a + \e \mathbf{1}$,
we have $\log\det\hat\a^{m+s,m+s} = \log\det \hat\b - \log(\e) - \log(m+s) + 
O(\e)$, therefore, using (for symmetric matrices)
\beq
\frac{\d}{\d\b_{ab}} \log\det\hat \b = \b^{-1}_{ab}, \longrightarrow
\frac{\d^2}{\d\b_{ab} \d \b_{cd}} \log\det\hat \b = 
\frac{\d\b^{-1}_{ab}}{\d\b_{cd}} = -\b^{-1}_{ac} \b^{-1}_{bd} \ ,
\eeq
we obtain
\beq\label{eq:MEapp}
\begin{split}
M_{ab;cd}^{(E)} &=\frac{\d^2}{\d \a_{a<b}\d \a_{c<d}} \log \det \hat 
\a^{m+s,m+s}= \lim_{\e\to 0} \frac{\d^2}{\d \b_{a<b}\d \b_{c<d}}\log \det \hat 
\b \\
 &= \lim_{\e\to0}  \left[  -2 \b^{-1}_{ac}\b^{-1}_{bd} - 2 
\b^{-1}_{ad}\b^{-1}_{bc}  + 2 \b^{-1}_{ac}\b^{-1}_{bc}  +2 
\b^{-1}_{ad}\b^{-1}_{bd} + 2\b^{-1}_{ac}\b^{-1}_{ad}
+ 2 \b^{-1}_{bc}\b^{-1}_{bd} \right. \\
&  \hskip32pt  \left. - (\b^{-1}_{ac})^2 - (\b^{-1}_{bc})^2 - (\b^{-1}_{ad})^2 - 
 (\b^{-1}_{bd})^2 \right]
\ .
\end{split}\eeq
Based on the discussion above,
we are only interested in the matrix elements corresponding to $a,b,c,d$ 
belonging to the block of $s$ replicas.
The matrix $\hat\b$ has the form \eqref{eq:beta}, and using the block-inversion 
formula, its inverse in the $s$ block
is $\Omega^{-1} = ( D - B A^{-1} B^T )^{-1}$. Hence, for $a,b \in [m+1,m+s]$ we 
have
$\b^{-1}_{ab} = \Omega^{-1}_{ab} = (\Omega^{-1})_1 \d_{ab} + (\Omega^{-1})_2$ 
where the coefficients are obtained
from equations \eqref{eq:Omega} and \eqref{eq:GM}. In particular we have $ 
(\Omega^{-1})_1 = 1/(\d+\a+\e) = 1/(\De/2+\e)$.

Plugging this form of $\b^{-1}_{ab}$ in equation \eqref{eq:MEapp}, one can check 
that all terms involving $(\Omega^{-1})_2$ disappear
(as it should, because this term is divergent when $\e\to 0$), and one gets 
(recalling that $a\neq b$ and $c\neq d$):
\beq\begin{split}
M_{ab;cd}^{(E)} &= 
M_1^{(E)}\left(\frac{\d_{ac}\d_{bd}+\d_{ad}\d_{bc}}{2}\right)+M_2^{(E)}
\left(\frac{\d_{ac}+\d_{ad}+\d_{bc}+\d_{bd}}{4}\right)+M_3^{(E)} \\
&= - \frac{16}{\D^2} \left(\frac{\d_{ac}\d_{bd}+\d_{ad}\d_{bc}}{2}\right)- 
\frac{16}{\D^2} \left(\frac{\d_{ac}+\d_{ad}+\d_{bc}+\d_{bd}}{4}\right)
\:.
\end{split}\eeq

\section{Interaction term \label{subsec:repliconint}}

We define
the interaction part of the stability matrix in absence of shear as
\beq
\begin{split}
M_{ab;cd}^{(I)}=&\ \left.\frac{\d^2 \mathcal F_0[\hat \y]}{\delta \y_{a<b}\d 
\y_{c<d}}\right|_{\hat \y=2\hat \a_{RS}}\\
=&\ 
M_1^{(I)}\left(\frac{\d_{ac}\d_{bd}+\d_{ad}\d_{bc}}{2}\right)+M_2^{(I)}
\left(\frac{\d_{ac}+\d_{ad}+\d_{bc}+\d_{bd}}{4}\right)+M_3^{(I)}
\end{split}
\eeq
so that the expression for the matrix coefficients $M_i$ of the full stability 
matrix is given by
\beq
M_{i}=M_i^{(E)}-4\wh \varphi M_i^{(I)} \ .
\label{stab_matrix_full}
\eeq
The calculation of the derivatives of the interaction term can be done on the 
same lines and following the same tricks of \cite{KPUZ13}.
Let us start by writing the general expression for the derivatives using the 
representation~\eqref{eq:FF0binomial} of the function $\FF_0$. 
We have
\beq
\begin{split}
M_{ab;cd}^{(I)}=\ &\lim_{n\to 0}\sum^*f(n_a,n_b)f(n_c,n_d) \exp\Bigg[- 
\left(\frac{\Dg}2 + \frac{\D^f}2 \right) \Si_m - \left( \frac{\De}2- \h \right) 
\Si_s\\
&+  \frac{ \D^f }2 \Si_m^2  + \frac\Dg{2} \sum_{a=1}^m \frac{n_a^2}{n^2} + 
\frac{\De}2 \sum_{a=m+1}^{m+s} \frac{n_a^2}{n^2}\Bigg] \ ,
\end{split}
\eeq
where the function $f$ is defined in \cite[eq. (45)]{KPUZ13}. As a variant 
of~\cite[eq.(46)]{KPUZ13},
we can introduce the following notation
\beq
\begin{split}
\langle O\rangle=\lim_{n\to 0}\sum^* O 
& \exp\left[- \left(\frac{\Dg}2 + \frac{\D^f}2 \right) \Si_m - \left( 
\frac{\De}2- \h \right) \Si_s\right.\\
&\left. + \frac{ \D^f }2 \Si_m^2  + \frac\Dg{2} \sum_{a=1}^m \frac{n_a^2}{n^2} + 
\frac{\De}2 \sum_{a=m+1}^{m+s} \frac{n_a^2}{n^2}\right]\:.
\end{split}
\eeq
The stability matrix can thus be rewritten as~\cite[Eq.(47)]{KPUZ13} where the 
replica indices run from $m+1$ to $m+s$.
Then we have to compute monomials of the form $\langle n_{a_1}\ldots n_{a_k}/ 
n^k\rangle$, which can be done in the following way
\beq
\begin{split}
\langle \frac{n_{a_1}\ldots n_{a_k}}{n^k}\rangle =&\lim_{n\to 0}\sum^* 
\frac{n_{a_1}\ldots n_{a_k}}{n^k}\int_{-\infty}^\infty\frac{\de 
\m}{\sqrt{2\pi}}e^{-\mu^2/2}\int_{-\infty}^\infty\left(\prod_{a=1}^{m+s}\frac{
\de \l_a}{\sqrt{2\pi}}e^{-\l_a^2/2}\right)\\
&\times \exp\left[-\left(\frac{\Dg}2 + \frac{\D^f}2 \right) \Si_m - \left( 
\frac{\De}2- \h \right) \Si_s- \m\sqrt{ \D^f } \Si_m \right. \\
&\left. - \sqrt \Dg \sum_{a=1}^m \frac{n_a \l_a}{n} - \sqrt{\De} 
\sum_{a=m+1}^{m+s} \frac{n_a\l_a}{n}\right]\\
=&\frac{1}{\De^{k/2}}\int_{-\infty}^\infty \DD \m 
\int_{-\infty}^\infty\left(\prod_{a=1}^{m} \DD \l_a \right)
\int_{-\infty}^\infty\left(\prod_{a=m+1}^{m+s}\frac{\de 
\l_a}{\sqrt{2\pi}}\right)\\
&\times \frac{\partial^k}{\partial \l_{a_1}\ldots \partial \l_{a_k}}e^{-\frac 
12\sum_{a=m+1}^{m+s}\l_a^2}\\
&\times e^{-\min\left\{ 
\frac{\Dg}2 + \frac{\D^f}2  + (\min_{a\leq m} \l_a) \sqrt{\Dg} + \m \sqrt{\D^f}, 
\,
\frac{\De}2- \h + (\min_{a>m} \l_a) \sqrt{\De}  
\right\}
} \ .
\end{split}
\eeq
If $O$ is a function that depends only on the $\l_a$ with $a\in[m+1,m+s]$, then 
we can define
\beq
\begin{split}
\langle O\rangle =& \int_{-\infty}^\infty \mathcal 
D\mu\int_{-\infty}^\infty\left(\prod_{a=1}^{m+s}\mathcal D\l_a\right)\\
&\times O \,  e^{-\min\left\{ 
\frac{\Dg}2 + \frac{\D^f}2  + (\min_{a\leq m} \l_a) \sqrt{\Dg} + \m \sqrt{\D^f}, 
\,
\frac{\De}2- \h + (\min_{a>m} \l_a) \sqrt{\De}  
\right\}
}\\
=&\int_{-\infty}^\infty\left(\prod_{a=1}^{m+s}\mathcal D\l_a\right) O \, 
K(\min_{a\leq m} \l_a,\min_{a>m} \l_a)=
\int_{-\infty}^\infty\left(\prod_{a=m+1}^{m+s}\mathcal D\l_a\right) O \ 
G\left(\min_{a>m} \l_a\right) \ ,
\end{split}
\eeq
where
\beq\label{eq:Gdef}
G(\l')=\int_{-\infty}^\infty \overline{\mathcal D}_m \l \,K(\l,\l') \ .
\eeq
In this way we obtain a generalization of~\cite[Eq.(48)]{KPUZ13}, in the form
\beq
\langle n_{a_1}\ldots n_{a_k}/ n^k\rangle = \frac1{\De^{k/2}} \left\langle
e^{\frac 12\sum_{a=m+1}^{m+s}\l_a^2}\frac{\partial^k}{\partial \l_{a_1}\ldots 
\partial \l_{a_k}}e^{-\frac 12\sum_{a=m+1}^{m+s}\l_a^2}
\right\rangle \ .
\eeq
The interaction part of the stability matrix is then given by the same reasoning 
as in~\cite[Eq.(50, 51, 53, 54, 56)]{KPUZ13} where the replica indices must be 
all shifted by $m$.
The only difference with respect to \cite{KPUZ13} is the definition of the 
measure used 
to take the average over the variables $\l$s. In fact instead of 
having~\cite[Eq.(52)]{KPUZ13} we have
\beq
\langle O(\l) \rangle=\int_{-\infty}^\infty \mathcal D \l \, G(\l) O(\l) = 
\int_{-\infty}^\infty \mathcal D \l \bar{\mathcal D}_m \l' \,K(\l',\l) \, O(\l) 
\ .
\eeq
This completes the calculation of the stability matrix without shear.

\subsubsection{The stability matrix in presence of shear}

The result of the previous section is valid when $\g=0$. Here we generalize the 
calculation to the case in which also the shear is present. 
The presence of a non vanishing $\g$ is detectable only in the interaction part 
of the replicated entropy. This means that
the expression of the entropic term does not change, and we need to compute only 
the new interaction part of the stability matrix.
This can be done using the following line of reasoning.
The interaction part of the stability matrix can be written in presence of shear 
as
\beq
M_{ab;cd}^{(I,\g)}=\left.\frac{\d^2}{\delta \y_{a<b}\d\y_{c<d}} \int \mathcal D 
\zeta \mathcal F_0[\D_{ab}+\frac{\z^2}{2}\g^2 \G_{ab}] \right|_{\hat \y=2\hat 
\a_{RS}}
\eeq
where the matrix $\G_{ab} = 1$ if $a$ belongs to the $m$-block and $b$ to the 
$s$-block or viceversa, and zero otherwise.
Recalling that $\D_{ab} =\a_{aa}+\a_{bb}-2\a_{ab}$, we have that the relation 
between $\hat\D$ and $\hat\a$ is linear, therefore
a constant shift of $\hat\D$ induces a constant shift in $\hat\a$, which does 
not affect the derivatives.
We deduce that
\beq
M_{ab;cd}^{(I,\g)}= \int \mathcal D \zeta \, 
M_{ab;cd}^{(I,\g=0)}[\D_{ab}+\frac{\z^2}{2}\g^2 \G_{ab}] 
= \int \mathcal D \zeta \, M_{ab;cd}^{(I,\g=0)}[  \Dg,\De,\D^f + \z^2 \g^2 ]
\ .
\eeq
Because $\D^f$ appears only in the kernel $K$, shifting $\D^f$ amounts to 
nothing more than a change in definition for the measure for the average of 
monomials of $\l$, which will contain a modified kernel
\beq\begin{split}
K^\g(\l,\l') =  \int \mathcal D \zeta & K(\l,\l' ; \Dg,\De,\D^f + \z^2 \g^2) \\ 
 =  \int \mathcal D \zeta & \left[
e^{-\frac{\De}2+ \h - \sqrt{\De} \l'} \Th\left(\frac{ \h + \frac{\D^f + \z^2 
\g^2 + \Dg - \De}2 + \sqrt{\Dg} \l - \sqrt{\De} \l' }
{\sqrt{2( \D^f + \z^2 \g^2)}}
\right) \right. \\
& \left. + e^{-\Dg/2 - \sqrt{\Dg} \l} \Th\left(\frac{-\h + \frac{\D^f + \z^2 
\g^2 - \Dg + \De}2 - \sqrt{\Dg} \l + \sqrt{\De} \l' }
{\sqrt{2 ( \D^f + \z^2 \g^2) }}
\right)  \right] \ ,
\end{split}\eeq
and the functional expression of the interaction part of the stability matrix 
has the same form of the $\g=0$ case.

\chapter{Computation of the replicated entropy in the fRSB ansatz 
\label{app:replicatedfRSB}}
We perform here the computation of the replicated entropy and the FP entropy for 
the fRSB ansatz. As for the RS case, we take care separately of the entropic and 
interaction terms.

\section{Entropic term}
The entropic term for the replicated entropy has the following expression in 
terms of the MSD matrix $\hat \Delta$ \cite{KMZ15}:
\beq
\frac 2d s_{entr} \equiv \log\det\hat \alpha^{m+s,m+s}= 
\log\left[-\frac{2}{(m+s)^2}\left(\sum_{ab}(\hat\Delta)^{-1}_{ab}
\right)\det(-\hat\Delta/2)\right].
\eeq
Let us now compute separately the two terms
\beq
\log \det \hat \Delta \ \ \ \ \ \ \ \ \ \ \ \ \ \ \ 
\sum_{a,b=1}^{m+s}\left[\hat\Delta^{-1}\right]_{ab}\:.
\label{twoterms}
\eeq
$m\times s$ matrix whose elements are all equal to $\Delta^r$, and 
$\Delta^{\infty RSB}$ is a generic hierarchical matrix.
Let us start from the first one that can be rewritten as
\beq
\det \hat \Delta=\left(\det \hat\D^g\right)\det \left(\hat \D^s-\hat\Delta^r 
(\hat\Delta^{g})^{-1} (\hat \Delta^r)^T\right)\:.
\eeq
Using the fact that for a $m\times m$ replica symmetric matrix of the form 
$q_{ab}=\delta_{ab}+(1-\delta_{ab})q$, the expression of the determinant is 
\cite{pedestrians}
\begin{equation}
\det  \hat q = (1-q)^{m-1} [ 1+(m-1)q ]
\end{equation}
then for $\hat\D^g$ we have that
\beq
\begin{split}
\det \hat \D^g =\ &\lim_{\epsilon\to 0}\left(\hat \Delta^g+\epsilon 
\mathbb{1}_{m}\right)=\lim_{\epsilon\to 0}\epsilon^m 
\left(1-\frac{\Delta^g}{\epsilon}\right)^{m-1} \left[ 
1+(m-1)\frac{\Delta^g}{\epsilon} \right]\\
=\ &(1-m)(-\Delta^g)^m
\label{detdeltag}
\end{split}
\eeq
where $\mathbb 1_m$ is the $m$ dimensional identity matrix.
Moreover we have \cite{pedestrians}
\beq
(\Delta^g)^{-1}_{ab} = -\frac{1}{\Delta^g}\left(\delta_{ab}+\frac{1}{1-m}\right)
\eeq
so that
\beq
\left[\hat\Delta^r (\hat\Delta^{g})^{-1} (\hat 
\Delta^r)^T\right]_{ab}=-\frac{(\Delta^r)^2}{\Delta^g}\frac{m}{1-m}
\eeq
This means that the matrix $\hat \Omega=\hat\Delta^s-\hat\Delta^r 
(\hat\Delta^{g})^{-1} (\hat \Delta^r)^T$ will be parametrized within the fullRSB 
ansatz by
\beq
\hat \Omega\to \left\{\Omega_d, \Omega(x)\right\}\ \ \ \ \ \ \ \ \ x\in[s,1]
\eeq
where
\beq
\begin{split}
\Omega_d&\equiv\frac{(\Delta^r)^2}{\Delta^g}\frac{m}{1-m},\\
\Omega(x)&\equiv\Delta(x)+\frac{(\Delta^r)^2}{\Delta^g}\frac{m}{1-m}\:.
\end{split}
\eeq
In this way we can use the result of \cite{MP91,CKPUZ13} to obtain
\beq
\log\det \hat \Omega=  s\log(\Omega_d - \la \Omega \ra)  - s \int_s^1 \frac{\de 
y}{y^2} \log\left(\frac{
\Omega_d - \la \Omega \ra - [\Omega](y)}{\Omega_d-\la \Omega\ra}
\right)
\label{detomega}
\eeq
where
\beq
[\Omega](x)=x \Omega(x) - \int_0^x\de y\, \Omega(y) \ , 
\hskip2cm
 \langle \Omega \rangle = \int_0^1\de x\, \Omega(x) \ ,
\eeq
and we are assuming $\Omega(x)=0\ \ \forall x<s$.
By inserting the fullRSB parametrization for $\hat \Omega$ we get the 
computation of the first term of (\ref{twoterms}).\\
We now turn to the computation of the second term. We need to compute the 
inverse of the matrix $\hat\Delta$. We thus consider a general matrix of the 
following form
\beq
\hat G=\left(
\begin{array}{cc}
\hat q_g & \hat q_r^{(1)} \\
\hat q_r^{(2)} & \hat q\\
\end{array}
\right)
\eeq
where the fullRSB structure is only inside the sub-block $\hat q\to 
\left\{\tilde q, q(x)\right\}$, the entries of the matrices $\hat q_r^{(1)}$ and 
$\hat q_r^{(2)}$ are all equal respectively to $q_r^{(1)}$ and $q_r^{(2)}$, and 
the matrix $\hat q_g$ has an RS form. Again we assume that $q(x)=0$ for 
$x\in[0,s]$. 
We want to solve the inverse matrix problem, namely we want to find the matrix 
\beq
\hat G^{-1}=\left(
\begin{array}{cc}
\hat p_g & \hat p_r^{(1)} \\
\hat p_r^{(2)} & \hat p\\
\end{array}
\label{inverse_general}
\right)
\eeq
such that $\hat G\hat G^{-1}=\mathbb 1_{m+s}$. 
We assume the matrix $G^{-1}$ to have the same general form of the $G$ and be 
likewise parametrized.
Using this form of $G^{-1}$, the equations for the inverse are
\begin{eqnarray}
q_dp_d+(m-1)q_gp_g+sq_r^{(1)}p_r^{(2)}&=&1 \label{inverse_matrix_1}\\
q_dp_g+q_g p_d+(m-2)q_g p_g+sq_r^{(1)} p_r^{(2)}&=&0\label{inverse_matrix_2}\\
q_r^{(2)}p_d+(m-1)p_gq_r^{(2)}+p_r^{(2)}\left(\tilde q-\la 
q\ra\right)&=&0\label{inverse_matrix_3}\\
q_d p_r^{(1)}+(m-1)q_g p_r^{(1)}+q_r^{(1)}\left(\tilde p-\la p\ra \right)&=&0 
\label{inverse_matrix_4}\\
mq_r^{(2)}p_r^{(1)}+\tilde q\tilde p-\int_s^1\de x 
q(x)p(x)&=&1\label{inverse_matrix_5}\\
mq_r^{(2)}p_r^{(1)}-sp(x)q(x)+(\tilde q-\la q\ra)p(x)+(\tilde p-\la p\ra)q(x)&& 
\nonumber\\
-\int_s^x\de y(q(y)-q(x))(p(y)-p(x))&=&0\label{inverse_matrix_6}
\end{eqnarray}
These equations can be solved exactly. Let us focus first on the last two 
equations and let us call $A(x)$ the right hand side of eq. 
(\ref{inverse_matrix_6}). This equation holds for all $x$ in the interval 
$[0,1]$. If we consider its derivative with respect to $x$ we get
\beq
0=\dot A(x)= (\tilde p-\la p\ra)\dot q(x)+(\tilde q-\la q\ra)\dot p(x)-\dot p(x) 
[q](x)-\dot q(x)[p](x)
\eeq
Let us now consider the following quantity:
\beq
B(x)=\left(\tilde p-\la p \ra-[p](x)\right)\left(\tilde q-\la q 
\ra-[q](x)\right)
\eeq
It is simple to show that $\dot A(x)=-x\dot B(x)$ so that we obtain
\beq
\left(\tilde p-\la p \ra-[p](x)\right)\left(\tilde q-\la q 
\ra-[q](x)\right)=\aleph
\label{pqu1}
\eeq
where $\aleph$ is a independent of $x$.
Computing (\ref{pqu1}) in $x=1$ and using eq. (\ref{inverse_matrix_5}) we get
\beq
\left(\tilde p-p(1)\right)\left(\tilde q-q(1)\right)=\aleph
\eeq
Moreover let us consider again eq. (\ref{inverse_matrix_6}) evaluated in $x=1$.
We get
\beq
1=\left(\tilde p-p(1)\right)\left(\tilde q-q(1)\right)
\eeq
so that we have $\aleph=1$.
Let us consider again the equation (\ref{pqu1}) evaluated in $x=s$. We get
\beq
\tilde p-\la p\ra-sp(s)=\frac{1}{\tilde q-\la q \ra-sq(s)},
\eeq
and using again eq. (\ref{inverse_matrix_6}) evaluated at $x=s$ we get
\beq
0=mp_r^{(1)}q_r^{(2)}-sp(s)q(s)+(\tilde p-\la p \ra)q(s)+(\tilde q-\la q 
\ra)p(s)
\eeq
By solving the last two equations with respect to $\tilde p-\la p \ra$ and 
$p(s)$ we get
\beq
\begin{split}
\tilde p -\la p \ra&=-\frac{s}{\tilde q-\la q 
\ra}\left[mp_r^{(1)}q_r^{(2)}\right]+\frac{1}{\tilde q-\la q 
\ra}\label{ptildeminavp}\\
p(s)&=-\frac{1}{\tilde q-\la q \ra}\left[mp_r^{(1)}q_r^{(2)}+\frac{q(s)}{\tilde 
q -\la q \ra-sq(s)}\right]
\end{split}
\eeq
from which we obtain
\beq
[p](x)=-\frac{[q](x)}{(\tilde q-\la q \ra)(\tilde q-\la q 
\ra-[q](x))}-\frac{smp_r^{(1)}q_r^{(2)}}{\tilde q-\la q \ra}\:.
\eeq
Taking the derivative with respect to $x$ we get
\beq
\dot p(x)=-\frac 1x \frac{\de}{\de x}\frac{[q](x)}{(\tilde q-\la q \ra)(\tilde 
q-\la q \ra-[q](x))}
\eeq
so that we have
\beq
\begin{split}
p(x)&=p(s)-\int_s^x\de y \frac 1y \frac{\de}{\de y}\frac{[q](y)}{(\tilde q-\la q 
\ra)(\tilde q-\la q \ra-[q](y))}\\
&=-\frac{1}{\tilde q -\la q \ra}\left[mp_r^{(1)}q_r^{(2)}+\frac 
1x\frac{[q](x)}{\tilde q-\la q \ra-[q](x)}+\int_s^x \frac{\de 
y}{y^2}\frac{[q](y)}{\tilde q-\la q \ra  - [q](y)}   \right]
\end{split}
\eeq
and finally, we get the solution for $\hat p$
\beq
\begin{split}
\la p\ra&=-mp_r^{(1)}q_r^{(2)}\frac{1-s}{\tilde q-\la q \ra}-\frac{1}{\tilde q - 
\la q \ra}\int_s^1\frac{\de y}{y^2}\frac{[q](y)}{\tilde q-\la q \ra-[q](y)}\\
\tilde p&=\frac{1}{\tilde q-\la q 
\ra}\left[1-mp_r^{(1)}q_r^{(2)}-\int_s^1\frac{\de y}{y^2}\frac{[q](y)}{\tilde 
q-\la q \ra-[q](y)}\right]\:.
\end{split}
\eeq
Let us now go back to the first four equations 
(\ref{inverse_matrix_1}-\ref{inverse_matrix_4}). We can use eq. 
(\ref{inverse_matrix_4}) together with \eqref{ptildeminavp} to solve for 
$p_r^{(1)}$, and the remaining three equations 
(\ref{inverse_matrix_1})-(\ref{inverse_matrix_3}) can be used to get $p_d$, 
$p_g$ and $p_r^{(2)}$:
\beq
\begin{split}
p_r^{(1)}=&\ -\frac{q_r^{(1)}}{q_d+(m-1)q_g}\left(\tilde q -\la q 
\ra-\frac{smq_r^{(1)}q_r^{(2)}}{q_d+(m-1)q_g}\right)^{-1}\\
p_r^{(2)}=&\ -\frac{q_r^{(2)}}{q_d+(m-1)q_g}\left(\tilde q -\la q 
\ra-\frac{smq_r^{(1)}q_r^{(2)}}{q_d+(m-1)q_g}\right)^{-1}\\
p_g=&\ 
-\frac{1}{q_d+(m-1)q_g}\left[\frac{q_g}{q_d-q_g}-\frac{sq_r^{(1)}q_r^{(2)}}{
q_d+(m-1)q_g}\left(\tilde q -\la q 
\ra-\frac{smq_r^{(1)}q_r^{(2)}}{q_d+(m-1)q_g}\right)^{-1}\right]\\
p_d=&\ \frac{1}{q_d-q_g}-\frac{1}{q_d+(m-1)q_g}\\
&\ 
\times\left[\frac{q_g}{q_d-q_g}-\frac{sq_r^{(1)}q_r^{(2)}}{q_d+(m-1)q_g}
\left(\tilde q -\la q 
\ra-\frac{smq_r^{(1)}q_r^{(2)}}{q_d+(m-1)q_g}\right)^{-1}\right]
\end{split}
\label{inverse_final_1}
\eeq
Note that the inverse of a symmetric matrix ($q_r^{(1)} = q_r^{(2)}$) is 
symmetric as well.
By inserting the expression of $p_r^{(1)}$ inside $\tilde p$ and $p(x)$ we end 
up with
\beq
\begin{split}
\tilde p=&\ \frac{1}{\tilde q-\la q 
\ra}\left[1+mq_r^{(2)}\frac{q_r^{(1)}}{q_d+(m-1)q_g}\left(\tilde q -\la q 
\ra-\frac{smq_r^{(1)}q_r^{(2)}}{q_d+(m-1)q_g}\right)^{-1}\right.\\
&\ \left.-\int_s^1\frac{\de y}{y^2}\frac{[q](y)}{\tilde q-\la q 
\ra-[q](y)}\right]\\
p(x)=&\ -\frac{1}{\tilde q -\la q 
\ra}\left[-mq_r^{(2)}\frac{q_r^{(1)}}{q_d+(m-1)q_g}\left(\tilde q -\la q 
\ra-\frac{smq_r^{(1)}q_r^{(2)}}{q_d+(m-1)q_g}\right)^{-1}\right.\\
&\ \left.+\frac 1x\frac{[q](x)}{\tilde q-\la q \ra-[q](x)}+\int_s^x \frac{\de 
y}{y^2}\frac{[q](y)}{\tilde q-\la q \ra  - [q](y)}   \right]
\end{split}
\label{inverse_final_2}
\eeq
and this completes the calculation of the inverse.
We can now collect all the results. Firstly we have that 
\beq
\log\det(\hat\Delta)= \log\det(\hat\Delta^g)+\log\det(\hat\Omega)\:.
\label{entropic1}
\eeq
Using \eqref{detdeltag}, the first term of the previous equation is easy and we 
get
\beq
\log\det\hat\Delta^g = \log(1-m) + m \log (-\Delta^g).
\eeq
To evaluate the second term of the right hand side of eq (\ref{entropic1}), we 
use the expression \eqref{detomega}. We have
\beq
\begin{split}
\Omega_d-\la\Omega\ra =\ 
&s\frac{(\Delta^r)^2}{\Delta^g}\frac{m}{1-m}-\la\Delta\ra,\\
\left[\Omega\right](y) =\ & s\frac{(\Delta^r)^2}{\Delta^g}\frac{m}{1-m} + 
\left[\Delta\right](y),
\end{split}
\eeq
so that 
\beq
\log\det(\hat \Omega) = s\log\left(s\frac{(\Delta^r)^2}{\Delta^g}\frac{m}{1-m} - 
\la\Delta\ra \right) - s\int_s^1 \frac{\de 
y}{y^2}\log\left(\frac{\la\Delta\ra+[\Delta](y)}{\la\Delta\ra - 
s\frac{(\Delta^r)^2}{\Delta^g}\frac{m}{1-m}}\right).
\eeq
We now need to perform the sum of the elements of $\hat\Delta^{-1}$. By 
parametrizing the entries of $\hat \Delta^{-1}$ with the same form of 
(\ref{inverse_general}) we get
\beq
\sum_{a,b=1}^{m+s}\left[\hat\Delta^{-1}\right]_{ab}  = msp_r^{(1)} + msp_r^{(2)} 
+ mp_d + m(m-1)p_g + s(\tilde p -\la p \ra).
\label{secondtermexp}
\eeq
Now, using eqs. (\ref{inverse_final_1}) and (\ref{inverse_final_2}) with
\beq
\begin{split}
q_d =\ &0,\\
q_g =\ &\Delta^g,\\
q_r^{(1)} =\ &q_r^{(2)} = \Delta_r,\\
\tilde q = \ & 0,\\
q(x) =\ &\Delta(x),\\
\end{split}
\eeq
we get
\beq
msp_r^{(1)} + msp_r^{(2)}=2ms\left[\frac{\Delta^r}{(m-1)\Delta^g}\left(\la 
\Delta \ra + s\frac{(\Delta^r)^2}{\Delta^g}\frac{m}{m-1}\right)^{-1}\right],
\eeq
while
\beq
mp_d = -\frac{m}{\Delta^g} 
-\frac{m}{(m-1)\Delta^g}\left[-1+\frac{s(\Delta^r)^2}{(m-1)\Delta^g}\left(\la 
\Delta \ra + s\frac{(\Delta^r)^2}{\Delta^g}\frac{m}{m-1}\right)^{-1}\right],
\eeq
and
\beq
m(m-1)p_g=-\frac{m(m-1)}{(m-1)\Delta^g}\left[-1+\frac{s(\Delta^r)^2}{
(m-1)\Delta^g}\left(\la \Delta \ra + 
s\frac{(\Delta^r)^2}{\Delta^g}\frac{m}{m-1}\right)^{-1}\right].
\eeq
The last term is then given by 
\beq
s(\tilde p -\la p \ra)=-\frac{s}{\la \Delta \ra + 
s\frac{(\Delta^r)^2}{\Delta^g}\frac{m}{m-1}}.
\eeq
And summing everything we finally obtain, with some trivial algebra
\beq
\sum_{a,b=1}^{m+s}\left[\hat\Delta^{-1}\right]_{ab} = \frac{m}{(m-1)\Delta^g} 
-s\left(\frac{m\Delta^r}{(m-1)\Delta^g} -1\right)^{2}\left(\la \Delta \ra + 
s\frac{(\Delta^r)^2}{\Delta^g}\frac{m}{m-1}\right)^{-1}.
\eeq
The final expression of the entropic term for the replicated entropy is thus 
given by
\beq
\begin{split}
\frac 2d s_{entr}=\ &(1-m-s)\log2-2\log (m+s) + 
\log\left[\frac{m}{(1-m)\Delta^g}\right.\\
&\left. +s\left(\frac{m\Delta^r}{(m-1)\Delta^g} -1\right)^{2}\left(\la \Delta 
\ra - s\frac{(\Delta^r)^2}{\Delta^g}\frac{m}{1-m}\right)^{-1}\right]\\
&+\log(1-m) + m\log(\Delta^g) + s\log\left(\la\Delta\ra - 
s\frac{(\Delta^r)^2}{\Delta^g}\frac{m}{1-m} \right)\\
&- s\int_s^1 \frac{\de 
y}{y^2}\log\left(\frac{\la\Delta\ra+[\Delta](y)}{\la\Delta\ra - 
s\frac{(\Delta^r)^2}{\Delta^g}\frac{m}{1-m}}\right)\\
=\ &(1-m-s)\log2-2\log (m+s) + (m-1)\log\Delta^g\\
&+ \log[m\la\Delta\ra + 2ms\Delta^r + (1-m)s\Delta^g] - s\int_s^1 \frac{\de 
y}{y^2}\log\left[\la\Delta\ra+[\Delta](y)\right]
\end{split}
\label{entropic_finalfull}
\eeq
This completes the calculation of the entropic term.
One can easily check that this expression reverts back to the result of appendix 
\ref{app:replicatedRS} when a RS profile $\D(x) = \D$ is chosen for the $s$ 
block \cite{RainoneUrbani15}. Taking the linear order in $s$ of this expression 
(with some caution, remember for example that $\thav{\D} \equiv \int_s^1 dx\ 
\D(x)$, so it depends on $s$ as well), one gets the first row of the 
\eqref{eq:sfull}.

\section{Interaction term}

We turn to the interaction term. The general expression we need to compute is 
the \eqref{eq:FF0binomial}, which we recall here
$$
\FF_0( \hat\D )  = \lim_{n\to 0}\sum_{n_1,\ldots, n_k; \sum_{a=1}^k 
n_a=n}\frac{n!}{n_1!\ldots n_k!}
e^{\sum_{a=1}^k \frac{n_a}{n} \h_a- \frac{1}{2} 
\sum_{a,b}^{1,k}\frac{n_an_b}{n^2}  \D_{ab}  } \ .
$$
By introducing Gaussian integrals, we can rewrite this term as \cite{CKPUZ13}
\beq\label{int_gen}
\begin{split}
\mathcal F_0(\hat\Delta) 
= & \int_{-\infty}^\infty \de h\, e^{h} 
 \frac{\de}{\de h} \left\{
\exp\left[-\frac{1}{2}\sum_{a,b=1}^k \Delta_{ab}\frac{\partial^2}{\partial 
h_a\partial h_b}\right] \prod_{a=1}^{m+s}  \th(h_a)  
 \right\}_{\{h_a=h-\eta_a\}} \ . 
\end{split} 
\eeq
where $k=m+s$ and $\theta(x)$ is again the step Heaviside function 
\cite{Abramowitz}. Now we assume that the $s$-sector of the displacement matrix 
has a generic $k$RSB structure.
Thus we have, following the derivation of \cite{Du81} for the $s$ block
\beq
\begin{split}
\mathcal F_0(\hat\Delta) 
=\ &\int_{-\infty}^\infty \de h\, \eee^{h} 
 \frac{\de}{\de h} \left\{
\exp\left[-\frac{1}{2}\sum_{a,b=1}^m \Delta_{ab}\frac{\partial^2}{\partial 
h_a\partial h_b}-\frac 
12\sum_{a=1}^m\sum_{b=m+1}^{m+s}\Delta_{ab}\frac{\partial^2}{\partial 
h_a\partial h_b}\right.\right.\\
&\left.\left.-\frac 
12\sum_{a=m+1}^{m+s}\sum_{b=1}^{m}\Delta_{ab}\frac{\partial^2}{\partial 
h_a\partial h_b} -\frac 
12\sum_{a,b=m+1}^{m+s}\Delta_{ab}\frac{\partial^2}{\partial h_a\partial 
h_b}\right] \prod_{a=1}^k  \th(h_a)  
 \right\}_{\{h_a=h-\eta_a\}}\\
=\ &\int_{-\infty}^\infty \de h\, \eee^{h} 
 \frac{\de}{\de h} \left\{
\exp\left[-\frac{1}{2}\Delta_g\left(\sum_{a=1}^m \frac{\partial}{\partial 
h_a}\right)^2+\frac{\Delta_g}{2}\sum_{a=1}^m\frac{\partial^2}{\partial 
h_a^2}\right.\right.\\
&\left.\left.-\Delta_r\left(\sum_{a=1}^m\frac{\partial }{\partial 
h_a}\right)\left(\sum_{b=m+1}^{m+s}\frac{\partial}{\partial h_b}\right) -\frac 
12\sum_{a,b=m+1}^{m+s}\Delta_{ab}\frac{\partial^2}{\partial h_a\partial 
h_b}\right] \prod_{a=1}^k  \th(h_a)  
 \right\}_{\{h_a=h-\eta_a\}}\\
=\ &\ \int_{-\infty}^\infty \de h\, \eee^{h} 
 \frac{\de}{\de h} \left\{
\exp\left[-\frac{1}{2}\Delta_g\left(\sum_{a=1}^m \frac{\partial}{\partial 
h_a}\right)^2-\Delta_r\left(\sum_{a=1}^m\frac{\partial }{\partial 
h_a}\right)\left(\sum_{b=m+1}^{m+s}\frac{\partial}{\partial 
h_b}\right)\right.\right.\\
&\left.\left.-\frac 12\sum_{a,b=m+1}^{m+s}\Delta_{ab}\frac{\partial^2}{\partial 
h_a\partial 
h_b}\right]\left(\prod_{a=1}^{m}\Theta\left(\frac{h_a}{\sqrt{2\Delta_g}}
\right)\right) \left(\prod_{b=m+1}^{m+s}  \th(h_b)\right)  
 \right\}_{\{h_a=h-\eta_a\}}\\
=\ &\int_{-\infty}^\infty \de h\, \eee^{h} 
 \frac{\de}{\de h} \left\{
\exp\left[-\frac{1}{2}\Delta_g\frac{\partial^2}{\partial 
h'^2}-\Delta_r\frac{\partial }{\partial h'}\frac{\partial}{\partial h''}-\frac 
12\D_1\frac{\partial^2}{\partial h''^2}\right]\right.\\
&\times\left.\left(\Theta\left(\frac{h'}{\sqrt{2\Delta_g}}\right)\right)^m 
g^{s/s_1}(s_1,h''-\eta)  
 \right\}_{h'=h''=h}
\end{split} 
\label{interaction_zero}
\eeq
where the function $g(x,h)$ is defined in terms of $f(x,h)$ as
\beq
f(x,h) \equiv \frac 1x \log g(x,h)
\eeq
Note that we have defined $\Delta(s)=\Delta_1$. At this point we can manipulate 
the last expression to do the final integrals by parts, giving an integral 
representation for the exponential of differential operators.
We consider the differential operator
\beq
\hat O\equiv-\frac{1}{2}\Delta_g\frac{\partial^2}{\partial 
h'^2}-\Delta_r\frac{\partial }{\partial h'}\frac{\partial}{\partial h''}-\frac 
12\D_1\frac{\partial^2}{\partial h''^2}
\eeq
and we introduce also
\beq
\hat H\equiv\frac{\partial}{\partial h'}+\frac{\partial}{\partial h''};
\eeq
we have then
\beq
\hat O=\frac 12 \Delta_f \left(\frac{\partial}{\partial h''}\right)^2-\frac 
12(\D_f+\D_1)\hat H\frac{\partial }{\partial h''}-\frac{\D_g}{2}\hat 
H\frac{\partial}{\partial h'}
\eeq
where we have defined $\D_f=2\D_r-\D_1-\D_g$. By plugging this expression into 
the interaction term we get
\beq
\begin{split}
\mathcal F_0(\hat\Delta) =&\ \int_{-\infty}^\infty \de h\, \eee^{h} 
 \frac{\de}{\de h} \left\{
\exp\left[\frac 12 \Delta_f \left(\frac{\partial}{\partial h''}\right)^2-\frac 
12(\D_f+\D_1)\hat H\frac{\partial }{\partial h''}-\frac{\D_g}{2}\hat 
H\frac{\partial}{\partial h'}\right]\right.\\
&\left.\times \left(\Theta\left(\frac{h'}{\sqrt{2\Delta_g}}\right)\right)^m 
g^{s/s_1}(s_1,h''-\eta)  
 \right\}_{h'=h''=h}\:.
 \end{split}
\eeq
Let us consider now a simple term of the form
\beq
\begin{split}
&\int_{-\infty}^\infty \de h\, \eee^{h} \frac{\de}{\de h} \left\{\exp\left[A 
\hat H \frac{\partial }{\partial h''}\right]f(h',h'')\right\}_{h'=h''=h}\\
=&\left.\int_{-\infty}^\infty \de h\, \eee^{h}\frac{\de}{\de h} 
\sum_{k=0}^\infty\frac{1}{k!}A^k\hat H^k\frac{\partial^k}{\partial 
h''^k}f(h',h'')\right|_{h'=h''=h} \:.
 \end{split}
\eeq
By integrating by parts all the terms of the series expansion we get
\beq
\begin{split}
&\int_{-\infty}^\infty \de h\, \eee^{h} 
 \frac{\de}{\de h} \left\{\exp\left[A \hat H \frac{\partial }{\partial 
h''}\right]f(h',h'')\right\}_{h'=h''=h}\\
 =&\left.\int_{-\infty}^\infty \de h\, \eee^{h}\frac{\de}{\de h} 
\sum_{k=0}^\infty\frac{1}{k!}(-A)^k\frac{\partial^k}{\partial 
h''^k}f(h',h'')\right|_{h'=h''=h} \\
 =&\int_{-\infty}^\infty \de h\, \eee^{h}  \frac{\de}{\de h} \left\{\exp\left[-A 
\frac{\partial }{\partial h''}\right]f(h',h'')\right\}_{h'=h''=h}\\
 =&\int_{-\infty}^\infty \de h\, \eee^{h} 
 \frac{\de}{\de h} \left\{f(h',h-A)\right\}_{h'=h''=h}.
\end{split}
\eeq
Using this result we finally get for the entropic term
\beq
\begin{split}
\mathcal F_0(\hat\Delta) =\ &\int_{-\infty}^\infty \de h\, \eee^{h} 
 \frac{\de}{\de h} \left\{
\exp\left[-\frac 12(\D_f+\D_1)\hat H\frac{\partial }{\partial 
h''}-\frac{\D_g}{2}\hat H\frac{\partial}{\partial h'}\right]\right.\\
&\left.\times\left(\Theta\left(\frac{h'}{\sqrt{2\Delta_g}}\right)\right)^m 
\gamma_{\D_f}\star g^{s/s_1}(s_1,h''-\eta) 
 \right\}_{h'=h''=h}=\\
=\ &\int_{-\infty}^\infty \de h\, \eee^{h} 
 \frac{\de}{\de h} \left\{
\left(\Theta\left(\frac{h+\D_g/2}{\sqrt{2\Delta_g}}\right)\right)^m 
\gamma_{\D_f}\star g^{s/s_1}(s_1,h-\eta+(\D_f+\D_1)/2) \right\}=\\
=\ &\int_{-\infty}^\infty \de h\, \eee^{h}  \left\{1-
\left(\Theta\left(\frac{h+\D_g/2}{\sqrt{2\Delta_g}}\right)\right)^m 
\gamma_{\D_f}\star g^{s/s_1}(s_1,h-\eta+(\D_f+\D_1)/2) \right\}\:.
 \end{split}
\eeq
where we have used the definitions \cite{CKPUZ13}
\beq
\eee^{\frac{a}{2}\frac{\partial^2}{\partial h^2}}f(h)\equiv\g_{a}\star f(h)\ \ \ 
\ \ \ \ \ \ \ \ \ \g_{a}\star f(h)\equiv\int_{-\infty}^\infty\frac{\de 
z}{\sqrt{2\pi a}}\eee^{-\frac{z^2}{2a}}f(h-z).
\label{differential_gaussian}.
\eeq
We now have to take again the the linear order in $s$. We just have to expand 
the function $g$ in the following way
$$
g(s_1,h)^{s/s_1}\simeq 1+ \frac{s}{s_1}\log g(s_1,h) + O(s^2).
$$
and  then we have to send $s \to 0$. In doing so $k \to \infty$ as well, and 
both $\D(x)$ and $f(x,h)$ become continuous functions of $x$. The function $f$ 
can then be shown \cite{Du81,parisiSGandbeyond} to obey the equation 
\eqref{Parisi_eq}
$$
\frac{\partial f}{\partial x}=\frac 12\frac{\de \Delta(x)}{\de 
x}\left[\frac{\partial^2 f}{\partial h^2}+x\left(\frac{\partial f}{\partial 
h}\right)^2\right],
$$
with the boundary condition \eqref{Parisi_initial}
$$
f(1,h)=\log \Theta \left(\frac{h}{\sqrt{2\Delta(1)}}\right),
$$
and we get the interaction part of the \eqref{eq:sfull}.

\section{Simplifications for $m=1$\label{sec:m=1}}
Before proceeding with the variational equations for $\hat \Delta$ we want to 
show that in the case in which the master replicas are taken at equilibrium, 
namely when $m=1$, the form of the state-followed entropy can be much 
simplified. Indeed in this case $\Delta^g$ disappears from the equations.\\
It is quite easy to see this in the case of the entropic term of eq. 
(\ref{eq:sfull}) by remembering that $\Delta^f \equiv 
2\Delta^r-\Delta(s)-\Delta^g$.
It remains to verify that $\Delta^g$ disappears also from the interaction term. 
For $m=1$ its general form is given by 
$$
\mathcal{F}(\wh\D) =\int_{-\infty}^\infty\DD \z \int_{-\infty}^\infty \de y \, 
\eee^{y} \, \left\{  1 
- \Th\left(  \frac{y + \D^g/2}{\sqrt{2\D^g}} \right) \int_{-\infty}^\infty \de x 
\,  \s(x+y) 
\frac{ \eee^{- \frac1{2 \D_\g(\z)} \left( x - \D_\g(\z)/2  \right)^2  } 
}{\sqrt{2\pi \D_\g(\z)}} \right\},
$$
where in our case the function $\s(x)$ is given by
\beq
\s(x)=g^{s/s_1}\left(s_1,x-\eta+\frac{\Delta_1}{2}\right).
\label{sigmafullRSB}
\eeq
This general form is valid for every replica-symmetry-breaking ansatz (the only 
difference is in the specific form of $\s(x)$). We then express the $\Th$ 
function 
with its integral representation
\beq
\Theta\left(\frac{h+\Delta^g/2}{\sqrt{2\Delta^g}}\right)=\int_{-\infty}^\infty 
\frac{\de \l}{\sqrt{2\pi \Delta^g}}\eee^{-\l^2/(2\Delta^g)} 
\theta\left(h+\frac{\Delta^g}{2}-\l\right)
\label{Theta_rep}
\eeq
to get
\beq
\mathcal{F}(\wh\D) =\int_{-\infty}^\infty\DD\z \int_{-\infty}^\infty \de x\de 
y\de \l\ 
\eee^{y}\frac{\eee^{-\frac{(\l+\D^g/2)^2}{2\D^g}}}{\sqrt{2\pi\D^g}}\frac{\eee^{
-\frac{(x-\D_\g(\z)/2)^2}{2\D_\g(\z)}}}{\sqrt{2\pi\D_\g(\z)}}[
1-\theta(y-\l)\s(x+y)]\:.
\eeq 
We now change integration variables in the following way:
\beq
\begin{cases}
u = y+x\\
v = \l+x\\
w = x\:.
\end{cases}
\eeq
Note that the Jacobian of this change of coordinates is one so that we get
\beq
\mathcal{F}(\wh\D) = \int_{-\infty}^\infty\DD \z\int_{-\infty}^\infty \de u \de 
v  \de w\ 
\eee^{u-w}\frac{\eee^{-\frac{(v-w+\D^g/2)^2}{2\D^g}}}{\sqrt{2\pi\D^g}}\frac{
\eee^{-\frac{(w-\D_\g(\z)/2)^2}{2\D_\g(\z)}}}{\sqrt{2\pi\D_\g(\z)}}[
1-\theta(u-v)\s(u)].
\eeq
The integral on $w$ can be easily done analytically, since it is just a 
convolution of two Gaussians. We obtain
\beq
\mathcal{F}(\wh\D) = \int_{-\infty}^\infty\DD \z\int_{-\infty}^\infty \de u \de 
v\ 
\eee^{u}\frac{\eee^{-\frac{(v+\D^g/2+\D_\g(\z)/2)^2}{2(\D^g+\D_\g(\z))}}}{\sqrt{
2\pi(\D^g+\D_\g(\z))}}[1-\theta(u-v)\s(u)]\:.
\eeq
Remembering that $\D_g+\D_f = 2\D_r-\D_1$, we get that $\D_g$ disappears from 
the expression. 
Using again (\ref{Theta_rep}) we get finally
\beq
\mathcal{F}(\wh\D) = \int_{-\infty}^\infty\DD \z\int_{-\infty}^\infty \de u\ 
\eee^{u}\left\{1-\Th\left(\frac{u+(2\D_r+\g^2\z^2-\D_1)/2}{\sqrt{
2(2\D_r+\g^2\z^2-\D_1)}}\right)\s(u)\right\}.
\eeq
This expression is much simpler than the corresponding one with $m \neq 1$. We 
conclude by recalling that with a 1RSB ansatz the function $\s$ would be given 
by
\beq
\s_{\textrm{1RSB}}(u) = \Th\left(\frac{u-\eta+\D_1/2}{\sqrt{2\D_1}}\right)^s,
\eeq
and for the more general fullRSB case it is given by (\ref{sigmafullRSB}).

\chapter{Variational equations in the fRSB ansatz \label{app:variationalfRSB}}
In this appendix we derive the variational equations for the optimization of the 
\eqref{eq:sfull} with respect to $\D(x)$. Two routes are possible: one can start 
from the entropy of the $m+s$ replicas for a finite number of RSBs $k$, take the 
derivatives with respect to the $\D_i$, and then take the $k \to \infty$ limit 
at the end to get the fRSB equations; alternatively, one can start directly from 
the \eqref{eq:sfull}, and obtain the equations by taking functional derivatives 
with respect to $\D(x)$. Here we use only this last procedure, introduced by 
Sommers and Dupont in \cite{SD84}, and refer to \cite{RainoneUrbani15} for the 
computation for finite $k$.

\section{Lagrange multipliers}
In the formalism of \cite{SD84}, one introduces Lagrange multipliers in the 
expression of the entropy in order to enforce the Parisi equation 
\eqref{Parisi_eq} and its boundary condition \eqref{Parisi_initial}.
These Lagrange multipliers are called $P(x,h)$ and $P(1,h)$; we rewrite the 
relevant part of the free energy of the followed system \eqref{eq:sfull} (we 
omit constant terms), adding the Lagrange multipliers
\beq
\begin{split}
\mathcal{S}_\infty =\ &\frac{1}{2}\log\left(\frac{\pi\la\D\ra}{d^2}\right)- 
\frac{1}{2}\int_0^1\frac{dx}{x^2}\log\left(\frac{G(x)}{\la\D\ra}\right) + 
\frac{1}{2}\frac{m\D^f+\D^g}{m\la\D\ra}\\
&+\frac{\wh\varphi_g}{2}\int_{-\infty}^{\infty} dh \ e^{h}\Th\left(  \frac{h + 
\D^g/2}{\sqrt{2\D^g}} \right)^m\\
&\times \int dx' \,  f (0,x' + h - \eta +\D(0)/2) 
\frac{ e^{- \frac1{2 \D^f} \left( x' - \D^f/2  \right)^2  } }{\sqrt{2\pi 
\D^f}}\\
&+\frac{\wh\varphi_g}{2}\int_0^1 dx\ \int_{-\infty}^{\infty}dh\ 
P(x,h)\left\{\dot{f}(x,h)-\frac{\dot{G}(x)}{2x}\left[f''(x,h)+xf'(x,h)^2\right]
\right\}\\
&-\frac{\wh\varphi_g}{2}\int_{-\infty}^{\infty}dh\ 
P(1,h)\left\{f(1,h)-\log\Th\left(\frac{h}{\sqrt{2G(1)}}\right)\right\},
\end{split}
\eeq
where we have defined the function
\beq
G(x)\equiv x\D(x)+\int_x^1 dy\ \D(y).
\eeq
We start from the equation for $\D_f$. We easily get
\beq
\begin{split}
0 =\ &\frac{1}{\la\D\ra} + \frac{\wh\varphi_g}{2}\int dh \, e^{h}\Th\left(  
\frac{h + \D^g/2}{\sqrt{2\D^g}} \right)^m\\
&\times \int dx \,  f' (0,x + h - \eta +\D(0)/2) 
\left(\frac{x+\D_f/2}{\D_f}\right)\frac{ e^{- \frac1{2 \D^f} \left( x - \D^f/2  
\right)^2  } }{\sqrt{2\pi \D^f}}.
\label{eq:dsinddeltaf}
\end{split}
\eeq
We must now take the functional derivatives of the $\mathcal{S}_\infty$. Taking 
the ones with respect to $P(x,h)$ and $f(x,h)$ we get
\begin{eqnarray}
\dot{f}(x,h)&=&\frac{\dot{G}(x)}{2x}\left[f''(x,h)+xf'(x,h)^2\right],\label{eq:eqf}\\
\dot{P}(x,h)&=& 
-\frac{\dot{G}(x)}{2x}\left[P''(x,h)-2x(P(x,h)f'(x,h))'\label{eq:eqP}\right],
\end{eqnarray}
where we have used the apex to denote the derivative with respect to $h$, and 
the dot for the one with respect to $x$. We must now differentiate with respect 
to $f(0,h)$, which is contained in the third line and in the boundary term of 
the fourth line (we can make it explicit with an integration by parts over $x$). 
We get the initial condition for the function $P(x,h)$:
\beq
P(0,h) = e^{h+\eta-\D(0)/2}\int_{-\infty}^{\infty} dx 
\frac{e^{-\frac{(x+\D_f/2)^2}{2\D_f}}}{\sqrt{2\pi\D_f}}\Th\left(\frac{
h-x+\eta-\D(0)/2+\D_g/2}{\sqrt{2\D_g}}\right)^m.
\label{eq:initialP}
\eeq
for generic $m$. For $m=1$ we can use the results of section \ref{sec:m=1} to 
get
\beq
P(0,h) = e^{h+\eta-\D(0)/2}\Th\left(\frac{h+\eta + 
\D_r-\D(0)}{\sqrt{2(2\D_r-\D(0))}}\right),
\eeq
while in presence of shear we get, using the \eqref{eq:FF_final}. 
\beq
P(0,h) = e^{h+\eta-\D(0)/2}\int_{-\infty}^\infty d\z\ 
\frac{e^{-\frac{\zeta^2}{2}}}{\sqrt{2\pi}}\Th\left(\frac{h+\eta + 
\D_r+\zeta^2\gamma^2/2-\D(0)}{\sqrt{2(2\D_r+\zeta^2\gamma^2-\D(0))}}\right)
\eeq
We now differentiate with respect to $G(x)$, $x\neq0,\ x\neq1$. We must 
integrate by parts the term proportional to $\dot G(x)$ in the fourth line, and 
use equations \eqref{eq:eqf} and \eqref{eq:eqP}. We must not forget that the 
second line depends as well on $G(x)$ through $\D(0)$, as \cite{CKPUZ13}
\beq
\D(x) = \frac{G(x)}{x} -\int_x^1 dz\ \frac{G(z)}{z^2}.
\label{eq:DofG}
\eeq
Surprisingly, the terms which contain $\la\D\ra$ give no contribution since 
$\frac{\delta\la\D\ra}{\delta G(x)}=0$, as can be checked using the 
\eqref{eq:DofG}:
\beq
\begin{split}
\frac{\delta \la\D\ra}{\delta G(x)} =\ &\frac{\delta}{\delta G(x)}\left(\int_0^1 
dy\frac{G(y)}{y} - \int_0^1 dy\int_y^1 \frac{dz}{z^2} G(z)\right)\\
=\ &\frac{\delta}{\delta G(x)}\left(\int_0^1 dy\frac{G(y)}{y} - \int_0^1 
dy\int_0^1 \frac{dz}{z^2} G(z)\th(z-y)\right)\\
=\ &\frac{1}{x} - \frac{1}{x^2}\int_0^1 dy\ \th(x-y) = \frac{1}{x} - 
\frac{1}{x^2}x = 0.
\end{split}
\eeq
So we get
\beq
\begin{split}
\frac{1}{G(x)} =\ &-\frac{\wh\varphi_g}{2}\int_{-\infty}^{\infty}dh\ 
P(x,h)f''(x,h) -\frac{\wh\varphi_g}{2}\int_{-\infty}^{\infty} dh\ e^{h}\Th\left( 
 \frac{h + \D^g/2}{\sqrt{2\D^g}} \right)^m\\
& \times\int dy \,  f' (0,y + h - \eta +\D(0)/2) 
\frac{ e^{- \frac1{2 \D^f} \left( y - \D^f/2  \right)^2  } }{\sqrt{2\pi \D^f}}\\
=\ &-\frac{\wh\varphi_g}{2}\int_{-\infty}^{\infty}dh\ P(x,h)f''(x,h) 
-\frac{\wh\varphi_g}{2}\int_{-\infty}^{\infty} dh\ P(0,h)f'(0,h),
\label{eq:gx}
\end{split}
\eeq
where we have made a translational change of coordinates over $h$ and used the 
definition of $P(0,h)$ in the second term. 
Finally, we focus on the boundary term at $x=0$ and we differentiate with 
respect to $G(0) = \la\D\ra$. We remember the reader that
\beq
\frac{\d G(x)}{\d G(y)} = \d(x-y).
\eeq
Let us start with the entropic term. We get
\beq
\begin{split}
\frac{\d\mathcal{S}_{\infty}^{entr}}{\d G(0)} =\ 
&\left[\frac{1}{G(0)}+\int_0^1\frac{dx}{x^2}\frac{1}{G(0)}-\frac{m\D_f+\D_g}{
mG(0)^2}\right]\d(0)\\
=\ &\left[\lim_{x\to 0^+} 
\frac{1}{xG(0)}-\frac{m\D_f+\D_g}{mG(0)^2}\right]\d(0).
\end{split}
\eeq
We stress the fact that the boundary term for $x=0$ in the integral part of the 
entropic term does not depend on $G(0)$, so it gives no contribution.\\
For what concerns the interaction term, we must first differentiate the third 
line with respect to $G(0)$ through its dependence on $\D(0)$. We have, using 
the \eqref{eq:DofG}:
\beq
\frac{\d \D(0)}{\d G(0)} = \lim_{x \to 0^+} \frac{1}{x}\d(0). 
\eeq
then we have to take the boundary term for $x=0$ in the fourth line, again 
integrating it by parts. The final result is, for the interaction term
\beq
\begin{split}
\frac{\d\mathcal{S}_{\infty}^{int}}{\d G(0)} =\ &\left[\lim_{x \to 
0^+}\frac{\wh\varphi_g}{2x}\int_{-\infty}^{\infty}dh\ P(0,h)f'(0,h) + \lim_{x 
\to 0^+}\frac{\wh\varphi_g}{2x}\int_{-\infty}^{\infty}dh\ 
P(0,h)f''(0,h)\right.\\
&\left. +\frac{\wh\varphi_g}{2}\int_{-\infty}^{\infty}dh\ 
P(0,h)f'(0,h)^2\right]\d(0).
\end{split}
\eeq
We can now put the two terms together. Using the \eqref{eq:dsinddeltaf} (or 
equivalently, the \eqref{eq:gx} for $x=0$) to eliminate the term in 
$\frac{1}{G(0)}$, we finally get
\beq
\frac{m\D_f+\D_g}{m\la\D\ra^2} = \frac{\wh\varphi_g}{2}\int_{-\infty}^{\infty} 
dh P(0,h)f'(0,h)^2.
\label{eq:deltaf}
\eeq
This completes the derivation of the variational equations. We now have an 
equation to fix every variable.

In summary, the equations we have to use are
\begin{eqnarray}
\dot{f}(x,h)&=&\frac{\dot{G}(x)}{2x}\left[f''(x,h)+xf'(x,h)^2\right], 
\label{eq:eqfullf}\\
\dot{P}(x,h)&=& -\frac{\dot{G}(x)}{2x}\left[P''(x,h)-2x(P(x,h)f'(x,h))'\right] 
\label{eq:eqfullP}\\
\frac{1}{G(x)} &=& -\frac{\wh\varphi_g}{2}\int_{-\infty}^{\infty}dh\ 
[P(x,h)f''(x,h) + P(0,h)f'(0,h)], \label{eq:eqfullG}\\
\frac{m\D_f+\D_g}{m\la\D\ra^2} &=& \frac{\wh\varphi_g}{2}\int_{-\infty}^{\infty} 
dh\ P(0,h)f'(0,h)^2. \label{eq:eqfullDr}
\end{eqnarray}
The procedure to solve them is as follows:
\begin{itemize}
 \item One starts from a guess for $\D_f$ and $G(x)$.
 \item Then we solve the \eqref{eq:eqf} and \eqref{eq:eqP}, with the boundary 
conditions \eqref{Parisi_initial} and \eqref{eq:initialP}.
 \item Then we compute the new $G(x)$ and $\Delta_f$ using the \eqref{eq:deltaf} 
and \eqref{eq:gx}.
 \item Repeat until the procedure converges.
\end{itemize}

\section{A different equation for $G(x)$\label{subsec:eqk}}
We show here that the equation for $G(x)$ can be recast in the form:
\beq
\begin{split}
\frac{1}{G(x)} =\ &\frac{1}{\la\D\ra} + 
\frac{\wh\varphi_g}{2}\int_{-\infty}^{\infty}dh\ \left[xP(x,h)f'(x,h)^2 - 
\int_0^x dz\ P(z,h)f'(z,h)^2\right],\\
\frac{1}{G(x)} =\ &\frac{1}{\la\D\ra} + x\kappa(x) - \int_0^x dy\ \kappa(y),
\end{split}
\label{eq:Gx2}
\eeq
where we have defined
\beq
\k(x)\equiv \frac{\wh \f_g}{2}\int_{-\infty}^\infty\de h\, 
P(x,h)\left(f'(x,h)\right)^2\:,
\eeq
which is more convenient for the purpose of the scaling analysis near jamming, 
paragraph \ref{subsec:scalingJ}. As in \cite{CKPUZ13}, we show that equations 
\eqref{eq:gx} and \eqref{eq:Gx2} are the same for $x=0$ and then we show that 
their derivatives of every order with respect to $x$ coincide. For $x=0$ it is 
trivial as the integral term in the \eqref{eq:Gx2} is zero, so that
\beq
-\frac{\wh\varphi_g}{2}\int_{-\infty}^{\infty}dh\ [P(0,h)f''(0,h) + 
P(0,h)f'(0,h)] = \frac{1}{\la\D\ra},
\eeq
which is just the \eqref{eq:eqfullG} for $x=0$. For what concerns the 
derivative, we proceed and use the same notation as in \cite{CKPUZ13}. From the 
\eqref{eq:gx} we get
\beq
\dot{P}f''+P\dot{f}'' \sim \dot{P}f''+P''\dot{f} \sim 
\frac{\dot{G}}{2x}[(2x(Pf')'-P'')f''+P''(f''+xf'^2)] \sim \dot{G}P(f'')^2,
\eeq
and from the \eqref{eq:Gx2} we get
\beq
x\dot{P}f'^2 +2xPf'\dot{f}' \sim 
\frac{\dot{G}}{2}[(2x(Pf')'-P'')f'^2+2Pf'(f'''+2xf'f'')] \sim -\dot{G}P(f'')^2.
\eeq
Where $a\sim b \Longrightarrow \int dh\ a(x,h)= \int dh\ b(x,h)$ and we have 
used the equations for $f$ and $P$. We have thus proven that equations 
\eqref{eq:Gx2} and \eqref{eq:gx} are equivalent.

\section{Scaling analysis near jamming \label{subsec:scalingJ}}

We show now that once the glass state is followed in compression up to the 
jamming point, the solution on the fullRSB equations develops a scaling regime 
characterized by a set of critical exponents that coincide with the ones 
computed in \cite{CKPUZ13}. The proof will be given by showing that the scaling 
equations close to jamming, and the asymptotic behavior of their initial 
conditions are the same as those that had been obtained in \cite{CKPUZ13}; the 
values of the critical exponents follow directly from these requirements.

\subsection{Scaling form of the equations}
On approaching the jamming point, the mean square displacement $\D_{EA} \equiv 
\D(1)$ of the fRSB microstates at the bottom of the hierarchy goes to zero. We 
thus define the jamming limit as $\Delta(1)\equiv \Delta_{EA}\to 0$. Moreover we 
expect that $\Delta^f$ stays finite.\\
We want to show that the fullRSB equations develop a scaling regime. At jamming 
the pressure diverges as $1/p\propto\D_{EA}^{1/\k}$ \cite{IBB12,CKPUZ13} and we 
want to determine $\k$. We thereby define the following scaling variables and 
functions (we omit the subscript $EA$ to lighten the notation):
\begin{eqnarray}
 y &\equiv&\D_{EA}^{-\frac{1}{\kappa}} x,\\
 \wh{f}(y,h) &\equiv& \scl f(\scl y,h),\\
 \g(y) &\equiv& \frac{G(\scl y)}{\scl},\\
 \wh{P}(y,h) &\equiv& e^{-h-\eta}P(\scl y,h)\:.
\end{eqnarray}
The initial conditions for the new functions $\wh{f}$ and $\wh{P}$ are therefore
\begin{eqnarray}
\wh{P}(0,h) &=& e^{-\D(0)/2}\int_{-\infty}^{\infty} dx 
\frac{e^{-\frac{(x+\D_f)^2}{2\D_f}}}{\sqrt{2\pi\D_f}}\Th\left(\frac{
h-x+\eta-\D(0)/2+\D_g/2}{\sqrt{2\D_g}}\right)^m,\nonumber\\
\wh{f}(1/\scl,h) &=& 
\scl\log\Theta\left(\frac{h}{\sqrt{2\scl\gamma(1/\scl)}}\right),\nonumber
\end{eqnarray}
and the relation between $\D(y)$ and $\gamma(y)$ becomes
\beq
\D(y)=\frac{\gamma(y)}{y}-\int_{y}^{1/\scl}\frac{dz}{z^2}\gamma(z).
\label{eq:deltaofgamma}
\eeq
Using the same reasoning, the variational equations for the scaling functions 
are:
\begin{eqnarray*}
\frac{\partial\wh{f}(y,h)}{\partial 
y}&=&\frac{\dot{\g}(y)}{2y}\left[\frac{\partial^2\wh{f}(y,h)}{\partial 
h^2}+y\left(\frac{\partial\wh{f}(y,h)}{\partial h}\right)^2\right],\\
\frac{\partial\wh{P}(y,h)}{\partial y}&=& 
-e^{-h}\frac{\dot{\g}(y)}{2y}\left[\frac{\partial^2[e^h\wh{P}(y,h)]}{\partial 
h^2}-2y\frac{\partial}{\partial h}\left(e^h\wh{P}(y,h)\frac{\partial 
\wh{f}(y,h)}{\partial h}\right)\right]\\
\frac{1}{\gamma(y)} &=&\frac{1}{\la\D\ra} + y\kappa(y)-\int_0^y dz\ \kappa(z) 
\\
\frac{m\D_f+\D_g}{m\la\D\ra^2} &=& \kappa(0)\\
\kappa(y) &=& \frac{\wh \varphi_g e^\eta}{2} \int_{-\infty}^{\infty} dh\ e^h 
\wh{P}(y,h)\wh{f}'(y,h)^2,
\label{scaling_eqs}
\end{eqnarray*}
which are very close to those obtained in \cite{CKPUZ13}. The entropy, for its 
part, is rephrased in
\beq
\begin{split}
\mathcal{S} =\ 
&\frac{1}{2}\log\left(\frac{\pi\la\D\ra}{d^2}\right)+\frac{1}{2}\log(\scl) + 
\frac{1}{\scl}\left[-\frac{1}{2}\int_0^{1/\scl}\frac{dy}{y^2}\log\left(\frac{
\gamma(y)}{\la\D\ra}\right)\right.\\
&\left. + \frac{1}{2}\frac{m\D_f+\D_g}{m\la\D \ra} + \frac{\wh\varphi_g 
e^\eta}{2}\int_{-\infty}^{\infty} dh\ e^h\wh{P}(0,h)\wh{f}(0,h)\right],
\label{eq:sclentropy}
\end{split}
\eeq
where now $\la\D\ra$ is defined as
\beq
\la\D\ra = \int_0^{1/\scl}dy\ \D(y). 
\label{entropia_jamming}
\eeq
We expect that the entropy diverges as $\log{1/p} \simeq \log{\scl}$. This means 
that the term between square parentheses on the right hand side of 
(\ref{entropia_jamming}) must vanish. This gives a condition for the jamming 
point $\eta_J$ as in the RS case (paragraph \ref{subsec:speclimits}).

\subsection{Asymptotes and scaling of $\wh{P}$ and $\wh{f}$}
In order to show that the scaling equations (\ref{scaling_eqs}) have the same 
critical exponents as the ones derived in \cite{CKPUZ13} we need to show that 
the asymptotic behavior for $h\to \pm \infty$ of the initial conditions for $\wh 
f$ and $\wh P$ coincides with the one of \cite{CKPUZ13}.
We start from the $\wh{f}$. 
Since the boundary condition for $\wh f$ is the same as the one in 
\cite{CKPUZ13}, it trivially follows that also the asymptotic behavior is the 
same.
Indeed we have
\begin{eqnarray}
 \wh{f}(1/\scl,h\to-\infty) &=& -h^2/(2\gamma(1/\scl)),\\
 \wh{f}(1/\scl,h\to\infty) &=& 0,
\end{eqnarray}
and by inserting this asymptotes in the equation for $\wh{f}$ we get,
\begin{eqnarray}
\wh{f}(y,h\to-\infty) &=& -h^2/(2\gamma(y)),\\
\wh{f}(y,h\to\infty) &=& 0,
\end{eqnarray}
as in \cite{CKPUZ13}. 
Conversely, the boundary condition for $\wh P$ is not the same as in 
\cite{CKPUZ13}. However one can easily see that the asymptotic behavior is still 
the same.
In fact, we have for $y=0$
\begin{eqnarray}
\wh{P}(0,h\to-\infty) &=& A(0)e^{B(0)h-D(0)h^2}\\
\wh{P}(0,h\to\infty) &=& e^{-\D(0)/2},
\end{eqnarray}
thanks to the fact that our $\wh{P}(0,h)$ is the convolution of a $\Theta$ 
function with a normalized Gaussian. 
We can again plug these asymptotes (and those of $\wh{f}$) in the equation for 
$\wh{P}$ in (\ref{scaling_eqs}), to get
\begin{eqnarray}
\wh{P}(0,h\to-\infty) &=& A(y)e^{B(y)h-D(y)h^2}\\
\wh{P}(0,h\to\infty) &=& e^{-\D(y)/2},
\end{eqnarray}
where the equations for $A,B$ and $D$ are the same as in \cite{CKPUZ13}. 

We now look for a solution for $\wh{P}$ and $\wh{f}$ at large $y$. We conjecture 
that $\D(y)\simeq \Delta_\infty y^{-\kappa}$ for large $y$, which through the 
\eqref{eq:deltaofgamma} implies that $\gamma(y)\simeq \gamma_\infty y^{-c}$ with 
$c = \kappa-1$ and $\gamma_\infty = \frac{\kappa}{\kappa-1}\D_\infty$. We can 
then solve the equations for $A,B$ and $D$ for large $y$, and we get for 
$h\to-\infty$
\beq
\wh{P}(y,h) = A_\infty y^c e^{B_\infty h^cy^c-D_\infty h^2y^{2c}} = 
y^cp_0(hy^c).
\eeq
We can thus conjecture for $\wh{P}$ the same exact scaling that was used in 
\cite{CKPUZ13}:
\beq
\wh{P}(y,h)\simeq\begin{cases}
             y^c p_0(hy^c)  & h\simeq-y^{-c}\\
             y^ap_1(hy^b) & |h|\simeq y^{-b}\\
             p_2(h) & h \gg y^{-b}\:.
            \end{cases}
\eeq
This scaling in turn requires that the function $p_1(z)$ must obey the boundary 
conditions
\beq
p_1(z)=\begin{cases}
             z^\theta & z\to\infty \\
             z^{-\alpha} & z\to - \infty
            \end{cases}
\eeq
where $\theta\equiv\frac{c-a}{b-c}$ and $\alpha = \frac{a}{b}$, as in 
\cite{CKPUZ13} and \cite{perceptron}.

For what concerns the $\wh{f}$, we define as in \cite{CKPUZ13} a function
\beq
\wh{j}(y,h) \equiv \wh{f}(y,h) + \frac{h^2\theta(-h)}{2\gamma(y)}. 
\eeq
Using the equation for $\wh{f}$ it is easy to see that, for all $y$
\begin{eqnarray}
\wh{j}(y,h\to-\infty) &=& \int_y^\infty \frac{du}{2u} 
\frac{\dot{\gamma}(u)}{\gamma(u)},\\
\wh{j}(y,h\to \infty) &=& 0.
\end{eqnarray}
For large $y$, again $\gamma(y) \simeq \gamma_\infty y^{-c}$, which means 
$\wh{j}(y,h\to-\infty) \simeq -c/(2y)$. So we can again conjecture the scaling 
form
\beq
\wh{j}(y,h) = -\frac{c}{2y}J(hy^b/\sqrt{\gamma_\infty}).
\eeq
with the boundary conditions $J(-\infty) = 1$ and $J(\infty)=0$. 

Now that we have the boundary conditions for the functions $J$ and $p_1$, all 
that we have to do is to plug them into the equations for $\wh{P}$ and $\wh{f}$ 
in order to get the equations for $p_1$ and $J$: since the scaling equations are 
the same as in \cite{CKPUZ13} we get the same equations for $p_1$ and $J$.
The final step to show that the critical exponents here are the same as in 
\cite{CKPUZ13} is to show that the \emph{marginal stability} equation for the 
replicon mode is the same \cite{CKPUZ13}.
In order to achieve this we can start from equation (\ref{eq:Gx2}) and consider 
the quantity
\beq
\k(x)=\frac{\wh \f_g}{2}\int_{-\infty}^\infty\de h\, 
P(x,h)\left(f'(x,h)\right)^2\:.
\eeq
By taking the derivative with respect to $x$ and assuming that we are in a 
fullRSB region such that $x>0$ and $\dot G(x)\neq 0$ we get
\beq
1=\frac{\wh \f_g}{2}\int_{-\infty}^\infty \de h\, 
P(x,h)\left(G(x)f''(x,h)\right)^2
\eeq
This equation is the same as the starting point that has been used in 
\cite{CKPUZ13} in order to close the system of equations for the critical 
exponents at jamming, and it can be shown \cite{CKPUZ13} that implies both the 
marginality of the replicon mode everywhere in the fRSB phase \cite{BM79}, and 
the isostaticity of jammed packings \cite{CKPUZ13}.\\
We conclude that the scaling behavior of the solution of \eqref{scaling_eqs} is 
the same as the one found in \cite{CKPUZ13}, thus proving that the critical 
exponents
$a, b$ and $c$ and $\k$, $\th$ and $\g$ are the same as in \cite{CKPUZ13}.

\backmatter

\cleardoublepage
\phantomsection

\addcontentsline{toc}{chapter}{\bibname}

\printbibliography

\end{document}